\documentclass[aip,sd,reprint]{revtex4-1}

\usepackage{graphicx}
\usepackage{dcolumn}
\usepackage{bm}
\usepackage{amsmath}
\usepackage{tabulary}
\usepackage{amssymb}
\usepackage[utf8]{inputenc}
\usepackage[T1]{fontenc}
\usepackage{mathptmx}
\usepackage{color}
\usepackage{gensymb}

\begin{document}
\preprint{AIP/Special issue}
\title{An Assessment of Different Electronic Structure Approaches for Modeling Time-Resolved X-ray Absorption Spectroscopy}

\newcommand{\revS}[1]{{\color{red}{#1}}}
\newcommand{\rST}[1]{{\color{blue}{#1}}}

\author{Shota Tsuru}
\altaffiliation{Current address: Arbeitsgruppe Quantenchemie, Ruhr-Universit\"{a}t Bochum, D-44780, Bochum, Germany}
\email{Shota.Tsuru@ruhr-uni-bochum.de}
\author{Marta L. Vidal}
\author{M\'aty\'as P{\'a}pai}
\altaffiliation{Current address: Wigner Research Centre for Physics, Hungarian Academy of Sciences, P.O. Box 49, H-1525 Budapest, Hungary}
\affiliation{DTU Chemistry, Technical University of Denmark, Kemitorvet Bldg 207, DK-2800, Kgs. Lyngby, Denmark}
\author{Anna I. Krylov}
\affiliation{Department of Chemistry, University of Southern California, Los Angeles, California 90089, United States}
\author{Klaus B. M{\o}ller}
\author{Sonia Coriani}
 \email{soco@kemi.dtu.dk}
\affiliation{DTU Chemistry, Technical University of Denmark, Kemitorvet Bldg 207, DK-2800, Kgs. Lyngby, Denmark}

\date{\today}

\begin{abstract}
We assess the performance of different protocols
for simulating excited-state X-ray absorption spectra.
We consider three different protocols based on equation-of-motion coupled-cluster singles and doubles, two of them combined with the maximum overlap method. The three protocols differ in the choice of a reference configuration used to compute target states. 
Maximum-overlap-method time-dependent density functional theory is also considered.
The performance of the different approaches is illustrated using  uracil, thymine, and acetylacetone as benchmark systems.
The results provide a guidance for selecting an electronic  structure method for modeling time-resolved X-ray absorption spectroscopy.
\end{abstract}

\maketitle

\section{Introduction}
Since the pioneering study by Zewail's group in the mid-eighties,~\cite{Zewail85}
ultrafast dynamics has been an active area of experimental research. Advances in light sources
provide new means for probing dynamics by utilizing core-level transitions.
X-ray free electron lasers (XFELs) and instruments based on high-harmonic generation (HHG) enable spectroscopic measurements on the femtosecond~\cite{Roadmap_ultrafast_Xray_Young,Chergui2017,Ueda} 
and attosecond\cite{Calegari_2016,Ramasesha,Ischenko,Villeneuve} timescales.
Methods for investigating femtosecond dynamics can be classified into two categories:
$(i)$ methods that track the electronic structure as parametrically dependent on the nuclear dynamics,
such as time-resolved photoelectron spectroscopy (TR-PES),~\cite{Schuurman_Stolow,Adachi_Suzuki,TSuzuki19,TRPES_UED} and $(ii)$ methods that directly visualize the nuclear dynamics, such as ultrafast X-ray scattering~\cite{Bucksbaum2016,Weber2019,Ruddockeaax6625,Stankus_2020} and ultrafast electron diffraction.~\cite{Centurion16,TRPES_UED}
Time-resolved X-ray absorption spectroscopy (TR-XAS) belongs to the former category. Similarly to X-ray photoelectron spectroscopy
(XPS), XAS is also element and chemical-state specific~\cite{Stohr}
but is able to resolve  the underlying electronic states better than TR-XPS.
On the other hand, TR-XPS allows photoelectron detection from all the involved electronic states with higher yield. XAS has been used to probe the local structure of bulk-solvated systems, such as in most chemical reaction systems in the lab and in cytoplasm. 
TR-XAS has been employed to track photo-induced dynamics in organic molecules~\cite{Pertot,Attar,Wolf,acac_ultrafast_ISC}
and transition metal complexes.~\cite{Chen,Chergui2016,Chergui2017,Wernet2019}
With the aid of simulations,~\cite{Katayama}
nuclear dynamics can be extracted from experimental TR-XAS spectra.

Similar to other time-resolved experimental methods from category $(i)$, interpretation of TR-XAS relies on 
computational methods for simulating electronic structure and nuclear wave-packet dynamics.
In this context, electronic structure calculations should be able to provide:
(1) XAS of the ground states;
(2) a description of the valence-excited states involved in the dynamics;
(3) XAS of the valence-excited states.

Quantum chemistry has made a major progress in simulations of XAS spectra of ground states.~\cite{X-ray_calc_review,Bokarev_Kuhn}
Among the currently available methods, the transition-potential density functional theory (TP-DFT) with the half core-hole approximation~\cite{Triguero,Leetmaa} is widely used to interpret the XAS spectra of ground states.~\cite{Vall-llosera,Perera} 
Ehlert {\em{et al.}} extended the TP-DFT method to  core excitations from valence-excited states,~\cite{TPDFT_TRNEXAFS} and implemented it in 
PSIXAS,~\cite{PSIXAS}
a plugin to the Psi4 code.
TP-DFT is capable of simulating (TR-)XAS spectra of large molecules with reasonable accuracy, as long as the core-excited states can be described by a single electronic configuration. 
Other extensions of Kohn-Sham DFT, suitable for calculating the XAS spectra of molecules in their ground states, also exist.~\cite{Michelitsch} Linear response (LR) time-dependent (TD) DFT, a  widely used method for excited states,~\cite{review_TDDFT_Dreuw_Head-Gordon,Luzanov2012,Laurent_Jacquemine,Ferre16} has 
been extended to the calculation of core-excited states~\cite{Stener2003,core_TDDFT} by means of 
the core-valence separation (CVS) scheme,~\cite{Cederbaum1980} a specific type of truncated single excitation space (TRNSS)
approach.~\cite{BESLEY2004124} 
In the CVS approach, configurations that do not involve core orbitals are excluded from the excitation space; this is justified  because the respective matrix elements are small, owing to  
the localized nature of the core orbitals and the large energetic gap  between the core and the valence orbitals.

Core-excitation energies calculated using TDDFT show errors up to $\approx$20 eV when standard exchange-correlation (xc)
functionals such as B3LYP~\cite{Becke_B3LYP} are used. The errors can be reduced by using specially designed
xc-functionals, such as those reviewed in Sec. 3.4.4. of Ref.~\citenum{X-ray_calc_review}. 
Hait and Head-Gordon recently
developed a square gradient minimum (SGM) algorithm for excited-state orbital optimization to obtain spin-pure restricted open-shell Kohn-Sham (ROKS) energies of core-excited states;  they reported sub-eV errors in XAS transition energies.~\cite{Hait_Head-Gordon} 

The maximum overlap method (MOM)~\cite{Gilbert2008} provides access to excited-state self-consistent field (SCF) solutions and, therefore, could be used to compute core-level states. 
More importantly, MOM can be also combined with TDDFT to compute core excitations from a valence-excited state.~\cite{Attar,acac_ultrafast_ISC,Northey2} 
MOM-TDDFT is an attractive method for simulating TR-XAS spectra, because it is computationally cheap and may provide
excitation energies consistent with the TDDFT potential energy surfaces, which are
often used in the nuclear dynamics simulations.  
However, in  MOM calculations  the initial valence-excited states are independently optimized and thus not orthogonal to each other. 
This non-orthogonality may lead to flipping of the energetic order of the states.
Moreover, open-shell Slater determinants provide a spin-incomplete description of excited states (the initial state in an excited-state XAS calculation), which results in severe spin contamination of all states and may affect the quality of the computed spectra. Hait and Head-Gordon have presented SGM as 
an alternative  general excited-state orbital-optimization method~\cite{SGM} and applied it to compute XAS spectra of radicals.~\cite{Hait_radical}

Applications of methods  containing some empirical component, such as TDDFT, require benchmarking against the spectra computed with a reliable wave-function method, whose accuracy can be systematically assessed. Among various post-HF methods, coupled-cluster (CC) theory yields a hierarchy of size-consistent ans{\"a}tze for the ground state, with the CC singles and doubles (CCSD) method being the most practical.~\cite{Bartlett2007} 
CC theory has been extended to excited states via the linear response~\cite{koch1990,Christiansen_IJQC,Sneskov_Christiansen}
and the equation-of-motion for excited states (EOM-EE)~\cite{Stanton1993,krylov_eom_2008,CC_EOMCC_Bartlett,corianiEOMRSP} formalisms. 
Both approaches have been adapted to treat core-excited states by using the CVS scheme,~\cite{coriani2015jcp} including
calculations of transition dipole moments and other properties.~\cite{Vidal1,tsuru,Faber19,Nanda:RIXS:2020,Coriani:RIXS-CVS:2020,vidal2,Vidal:L-edge} The benchmarks illustrate that CVS-enabled EOM-CC methods 
describe well the relaxation effects caused by the core hole as well as differential correlation effects.
Given their robustness and reliability, CC-based methods provide high-quality XAS spectra, 
which can be used to benchmark other methods.
Beside several CCSD investigations,~\cite{coriani2012pra,coriani2015jcp,Frati:2019,Carbone:2019,Wolf,Peng2015,tsuru,fransson2013jcp,Vidal1,xpsbasis2020,Bruno-cvs,Moitra:Vib:core,Vidal:L-edge,Vidal:Benzene+} core excitation and ionization energies have also been reported at the
CC2,~\cite{coriani2012pra,Carbone:2019,Frati:2019,martina2019,Moitra:Vib:core}
CC3~\cite{Wolf,N2_Myhre_2018,et2020,Alex:newcc3},
CCSDT,\cite{Carbone:2019,N2_Myhre_2018,daMatthews}
CCSDR(3),~\cite{coriani2012pra,daMatthews,Moitra:Vib:core} and EOM-CCSD*\cite{daMatthews} levels of theory. XAS spectra have also been simulated with a linear-response (LR-) density cumulant theory (DCT),~\cite{Peng_LR-DCT} which is closely related to the LR-CC methods.


The algebraic diagrammatic construction (ADC) approach~\cite{ADC82,ADC_review}
has also been  used 
to model inner-shell spectroscopy. The second-order variant ADC(2)~\cite{Barth1985} yields valence-excitation energies with
an accuracy and a computational cost [$O(N^5)$] similar to CC2 
(coupled-cluster singles and approximate
doubles),~\cite{CHRISTIANSEN_CC2} but within the Hermitian formalism. 
ADC(2) was extended to core excitations by the CVS scheme.~\cite{NEXAFS_H2O_NH3_CH4_Exp,wenzel2014} 
Because ADC(2) is inexpensive and is capable of accounting for dynamic correlation when calculating potential energy
surfaces,~\cite{Plasser_Barbatti} it holds promise of delivering reasonably accurate time-resolved XAS spectra
at a low  cost at each step of the nuclear dynamic simulation.
Neville {\em{et al.}} simulated TR-XAS spectra with ADC(2),~\cite{Neville_ADC_TRNEXAFS,Neville_TRXAS,Neville_TRXAS_at_CI} while
using multi-reference first-order configuration interaction (MR-FOCI) in their nuclear dynamics simulations.
Neville and Schuurman also reported an approach to simulate XAS spectra using electronic wave packet autocorrelation functions
based on TD-ADC(2).~\cite{Neville_Schuurman_autocorrelation} 
An {\em{ad hoc}} extension of ADC(2), ADC(2)-x,~\cite{Trofimov_1995} is known to give ground-state XAS spectra with relatively high accuracy (better than ADC(2)) 
employing small basis sets such as 6-31+G,~\cite{nexafs_thymine_adenine} but the improvement comes with a higher
computational cost $[O(N^6)]$.
List {\em{et al.}} have recently used ADC(2)-x, along with RASPT2, to study competing relaxation pathways in malonaldehyde by TR-XAS simulations.~\cite{NHList:TRXAS}

An important limitation of the single-reference methods 
(at least those only including singles and double excitations) is that they can reliably treat only singly excited states. 
While transitions to the singly occupied molecular orbitals (SOMO) result in  target states that are
formally singly excited from the ground-state reference state,  
other final states accessible by core excitation from a valence-excited state 
can be dominated by configurations of double or higher excitation character relative to the ground-state reference.
Consequently, these states are not well described by conventional response methods such as TDDFT, LR/EOM-CCSD, or ADC(2) (see Fig.~\ref{fig:schematic_ccsd} in \ref{subsec:protocols}).~\cite{tsuru,NHList:TRXAS}
This is the main rational of using MOM within TDDFT. 

To overcome this problem while retaining a low computational cost, Seidu {\em{et al.}}~\cite{Seidu19} suggested to combine DFT and multireference configuration interaction (MRCI) with the CVS scheme, which led to the CVS-DFT/MRCI method. 
The authors demonstrated that the semi-empirical Hamiltonian adjusted to describe the Coulomb and exchange interactions of the valence-excited states~\cite{Lyskov_DFT/MRCI} works well for the core-excited states too. 

In the context of excited-state nuclear
dynamics simulations based on complete active-space SCF (CASSCF) or CAS second-order perturbation theory (CASPT2), popular choices for computing core excitations from a given valence-excited state
are restricted active-space SCF (RASSCF)~\cite{RASSCF88,RASSCF90} or RAS
second-order perturbation theory (RASPT2).~\cite{RASPT2_Malmqvist} Delcey {\em{et al.}} have clearly summarized how to apply
RASSCF for core excitations.~\cite{Delcey2019} 
XAS spectra of valence-excited states computed by RASSCF/RASPT2 have been presented by various
authors.~\cite{Mukamel_uracil,Segatta,Northey2} RASSCF/RASPT2 schemes are sufficiently flexible 
and even work in the vicinity of conical intersections; they also can tackle different types of excitations, including, for example, those with multiply excited character.\cite{Schweigert} However, the  accuracy of these methods depends strongly on an appropriate selection of the active space, which makes their application system-specific.
In addition, RASSCF simulations might suffer from insufficient description of dynamic correlation whereas 
the applicability of RASPT2 may be limited by its computational cost.

Many of the methods mentioned above are available in standard quantum chemistry packages. Hence, the assessment of their performance
would be valuable help for computational chemists who want to use these methods to analyze the experimental TR-XAS spectra.
Since experimental TR-XAS spectra are still relatively scarce, we set out assessing the performance of four selected single-reference methods
from the perspective of the three requirements stated above.
That is, they should be able to accurately describe the core and valence excitations from the ground state, to give the transition strengths between the core-excited and valence-excited states, and yield the XAS spectra of the valence-excited states over the entire pre-edge region, i.e., describe the spectral features due to the transitions of higher excitation character.  
More specifically, we extend the use of the MOM approach to the CCSD framework and evaluate its accuracy relative to standard fc-CVS-EOM-EE-CCSD and to MOM-TDDFT. 
We note that MOM has been used in combination with CCSD to calculate double core excitations.~\cite{Lee_Head-Gordon} 
For selected ground-state XAS simulations, we also consider ADC(2) results.

\begin{figure}[h]
    \centering
    \includegraphics[width=8cm,height=8cm,keepaspectratio]{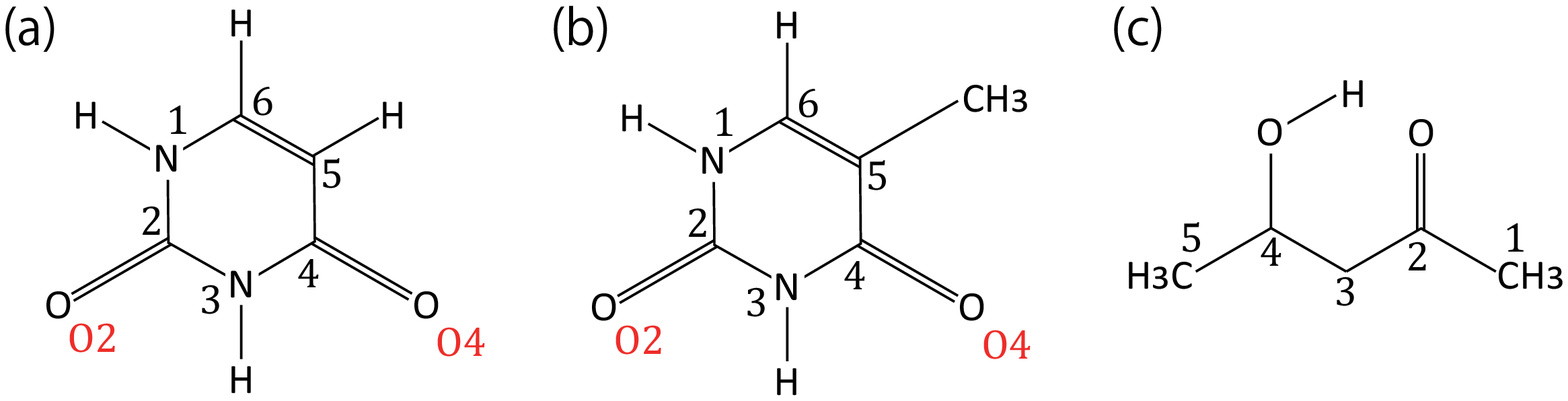}
    \caption{Structures of (a) uracil, (b) thymine, and (c) acetylacetone. Atom numbering follows the IUPAC standard.}
    \label{fig:ms}
\end{figure}

We use the following  systems to benchmark the methodology: uracil, thymine, and acetylacetone (Fig.~\ref{fig:ms}). 
Experimental TR-XAS spectra have not been recorded for uracil yet, but its planar symmetry at the Franck-Condon (FC) geometry and its similarities with thymine 
make it a computationally attractive model system. Experimental TR-XAS data are available at the O K-edge of thymine and at the C K-edge of acetylacetone.

The paper is organized as follows: First, we describe the methodology and computational details.
We then compare the results obtained with the CVS-ADC(2), CVS-EOM-CCSD, and TDDFT methods
against the experimental ground-state XAS spectra.~\cite{nexafs_cytosine_uracil,Wolf,Attar,acac_ultrafast_ISC}
We also compare the computed valence-excitation energies with UV absorption and electron energy loss spectroscopy (EELS, often called electron impact spectroscopy when it is applied to gas-phase molecules).~\cite{Trajmar} 
We then present the XAS spectra of the valence-excited states obtained with different CCSD-based protocols
and compare them with experimental TR-XAS spectra when available.~\cite{Wolf,Attar,acac_ultrafast_ISC}
Finally, we evaluate the performance of MOM-TDDFT.

\section{Methodology}

\subsection{Protocols for Computing XAS}
\label{subsec:protocols}

\begin{figure}[h]
    \centering
    \includegraphics[width=8cm,height=15cm,keepaspectratio]{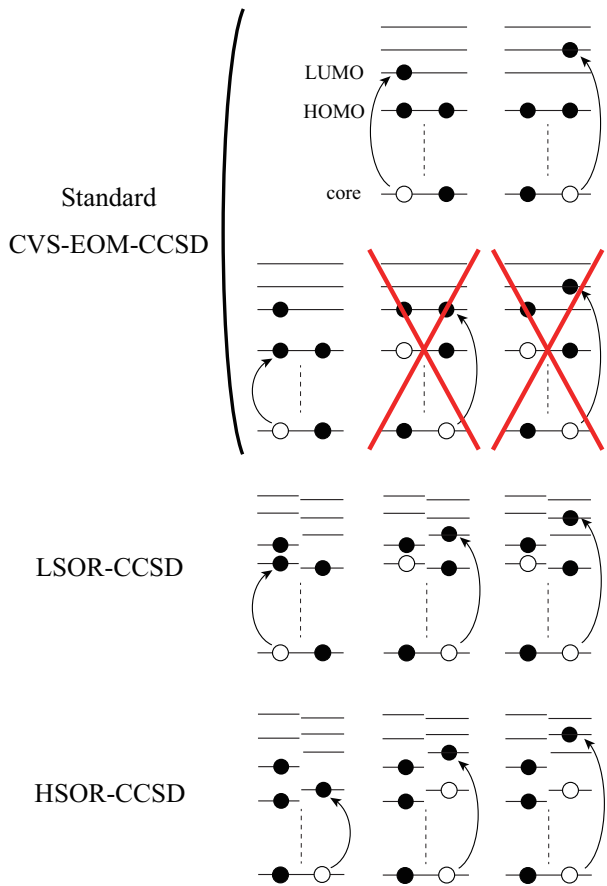}
    \caption{Schematics of the standard CVS-EOM-CCSD, LSOR-CCSD, and HSOR-CCSD protocols.
    The crossed configurations are formally doubly excited with respect to the ground-state reference.
    \label{fig:schematic_ccsd}}
\end{figure}

We calculated the energies and oscillator strengths for core and valence excitations from the ground states
by standard linear-response/equation-of-motion methods: ADC(2),~\cite{ADC82,Trofimov_1995,ADC_review}
EOM-EE-CCSD,~\cite{Christiansen_benzene,Hald,Bartlett2007,Stanton1993,krylov_eom_2008,CC_EOMCC_Bartlett,corianiEOMRSP} and TDDFT.  
In the ADC(2) and CCSD calculations of the valence-excited states, we employ the frozen core (fc) approximation.
CVS~\cite{wenzel2014,coriani2015jcp,Vidal1} was applied to obtain the core-excited states within all methods.
Within the fc-CVS-EOM-EE-CCSD framework,~\cite{Vidal1} we explored three different strategies to obtain the excitation energies
and oscillator strengths for selected core-valence transitions, as summarized in Fig.~\ref{fig:schematic_ccsd}.
In the first one, referred to as standard CVS-EOM-CCSD, we assume that the final core-excited states belong to the set of excited states that can be reached by core excitation from the ground states (see Fig.~\ref{fig:schematic_ccsd}, top panel).
Accordingly, we use the HF Slater determinant, representing the ground state ($|\Phi_0\rangle$) as the reference ($|\Phi_{\mathrm{ref}}\rangle$) for the CCSD calculation; the (initial) valence-excited and 
(final) core-excited states  are then computed with EOM-EE-CCSD and fc-CVS-EOM-EE-CCSD, respectively.
The transition energies for core-valence excitations are subsequently computed as the energy differences between the final core states and the initial valence state. 
The oscillator strengths for the transitions between the two excited states are obtained from the transition moments between the EOM states, according to EOM-EE theory.~\cite{Stanton1993,Bartlett2007,Vidal1}
In this approach, both the initial and the final states are spin-pure states. 
However, the final core-hole states that have multiple excitation character with respect to the ground state are either not accessed or described poorly by this approach (the respective configurations are crossed in Fig. \ref{fig:schematic_ccsd}).

In the second approach, named high-spin open-shell reference (HSOR) CCSD, we use as a reference ($|\Phi_{\mathrm{ref}}\rangle$) for the CCSD calculations
a high-spin open-shell HF Slater determinant that has the same electronic configuration as the initial singlet valence-excited state to be probed in the XAS step.~\cite{tsuru,vidal2,vidal2:correction} 
This approach is based on the assumption that the exchange interactions, which are responsible for the energy
gap between singlets and triplets, cancel out in calculations of the transition energies and oscillator strengths.  
An attractive feature of this approach is that the reference is spin complete (as opposed to a low-spin open-shell determinant of the same occupation) and that the convergence of the SCF procedure is usually robust. 
A drawback of this approach is the inability to distinguish between the singlet and triplet states with the same electronic configurations.

In the third approach, we use low-spin (M$_s$=0) MOM-reference for singlet excited states and high-spin (M$_s$=1) MOM-reference for triplet excited states.
We refer to this approach as low-spin open-shell reference (LSOR) CCSD. 

In both HSOR-CCSD and LSOR-CCSD, the calculation begins with an SCF optimization targeting the dominant configuration of the  initial valence-excited state by means of the MOM algorithm,
and the resulting Slater determinant is then used as the reference $|\Phi_{\mathrm{ref}}\rangle$ in the subsequent CCSD calculation. 
Core-excitation energies and oscillator strengths from the high-spin and the low-spin 
references are computed with standard CVS-EOM-EE-CCSD.
Such MOM-based CCSD calculations can describe all target core-hole states, provided that they have  singly excited character with respect to the chosen reference. Furthermore, in principle, initial valence-excited states of different spin symmetry can be selected.
However, in calculations using low-spin open-shell references  (LSOR-CCSD states), variational collapse might occur. Moreover, the LSOR-CCSD treatment of singlet excited states suffers from spin contamination, as the underlying open-shell reference is not spin complete (the well known issue of spin-completeness in calculations using open-shell references is discussed in detail in recent review articles.~\cite{krylov_open-shell_2017,casanova_krylov_2020})
 
We note that the HSOR-CCSD ansatz for a spin-singlet excited state is identical to the LSOR-CCSD ansatz of a 
(M$_s$ = 1) spin-triplet state having the same electronic configuration as the spin-singlet excited state (see Fig.~\ref{fig:schematic_ccsd}). 

In addition to the three CCSD-based protocols described above, we also considered MOM-TDDFT, which is often used for simulation of the TR-NEXAFS spectra.~\cite{Attar,acac_ultrafast_ISC,Northey2} We employed the B3LYP xc-functional,~\cite{Becke_B3LYP} as in Refs.~\citenum{Attar,acac_ultrafast_ISC,Northey2}. 

\subsection{Computational Details}

The equilibrium  geometry of uracil was optimized at the MP2/cc-pVTZ level. 
The equilibrium geometries of thymine and acetylacetone were taken from the literature;~\cite{Wolf,Faber19} they were optimized at the  CCSD(T)/aug-cc-pVDZ and CCSD/aug-cc-pVDZ level, respectively. These structures represent the molecules
at the Franck-Condon (FC) points. 
The structures of the T$_1$($\pi\pi^{\ast}$) and S$_1$(n$\pi^{\ast}$) states of acetylacetone, and of the S$_1$(n$\pi^{\ast}$) state of thymine were optimized at the EOM-EE-CCSD/aug-cc-pVDZ level.~\cite{Faber19} 

We calculated near-edge X-ray absorption fine structure (NEXAFS) of the ground state of all three molecules using CVS-ADC(2), CVS-EOM-CCSD, and TDDFT/B3LYP. The excitation energies of the valence-excited states 
were calculated with ADC(2), EOM-EE-CCSD, and TDDFT/B3LYP. 
The XAS spectra of the T$_1$($\pi\pi^{\ast}$), T$_2$(n$\pi^{\ast}$), S$_1$(n$\pi^{\ast}$) and S$_2$($\pi\pi^{\ast}$) states of
uracil were calculated at the FC geometry. 
We used the FC geometry for all states in order to make a coherent comparison of the MOM-based CCSD methods with the standard CCSD method, and to ensure that the final core-excited states are the same in the ground state XAS and transient state XAS calculations using standard CCSD.
The spectra of thymine in the S$_1$(n$\pi^{\ast}$) state were calculated at the potential energy minimum of the S$_1$(n$\pi^{\ast}$) state. 
The spectra of acetylacetone in the T$_1$($\pi\pi^{\ast}$) and S$_2$($\pi\pi^{\ast}$) states were calculated at the
potential energy minima of the T$_1$($\pi\pi^{\ast}$) and S$_1$(n$\pi^{\ast}$) states, respectively. 
Our choice of geometries for acetylacetone is based on the fact that the S$_2$($\pi\pi^{\ast}$)-state spectra were measured during wave packet propagation from the S$_2$($\pi\pi^{\ast}$) minimum (planar) towards the S$_1$(n$\pi^{\ast}$) minimum (distorted) and the ensemble was in equilibrium when the T$_1$($\pi\pi^{\ast}$)-state spectra were measured.~\cite{acac_ultrafast_ISC}

The XAS spectra of the valence-excited states were computed with CVS-EOM-CCSD, HSOR-CCSD, and LSOR-CCSD.
Pople's 6-311++G** basis set was used throughout. In each spectrum, the oscillator strengths were convoluted with a Lorentzian function (FWHM = 0.4 eV, unless 
otherwise specified).
We used the natural transition orbitals (NTOs)~\cite{Luzanov1976,Martin2003,Luzanov2012,Dreuw:ESSAImpl:14,Dreuw:ESSAImpl-2:14,Dreuw:ESSA:14,Krylov:Libwfa:18,Plasser:NTOfeature:2020,Krylov:Orbitals} to determine the character of the excited states. 

All calculations were carried out with the Q-Chem 5.3 electronic structure package.~\cite{Qchem_MP_paper}
The initial guesses [HOMO($\beta$)]$^{1}$[LUMO($\alpha$)]$^{1}$ and [HOMO($\alpha$)]$^{1}$[LUMO($\alpha$)]$^{1}$ were used in MOM-SCF for the spin-singlet and triplet states dominated by (HOMO)$^{1}$(LUMO)$^{1}$ configuration, respectively. The SOMOs of the initial guess in a MOM-SCF process are the canonical orbitals (or the Kohn-Sham orbitals) which resemble the hole and particle NTO of the transition from the ground state to the valence-excited state. One should pay attention 
to the order of the orbitals obtained in the ground-state SCF, especially when the basis set has diffuse functions. In LSOR-CCSD calculation, the SCF convergence threshold had to be set to $10^{-9}$ Hartree. To ensure convergence to the dominant electronic configuration of the desired electronic state, we used the initial MOM (IMOM) algorithm~\cite{IMOM} instead of regular MOM; this is  especially important for cases when the desired state belongs to the same irreducible representation as the ground state.

\section{Results and Discussion}

\subsection{Ground-State NEXAFS}
\label{subsec:gs_to_core}
\begin{figure}[h]
    \centering
    \includegraphics[width=8cm,height=8cm,keepaspectratio]{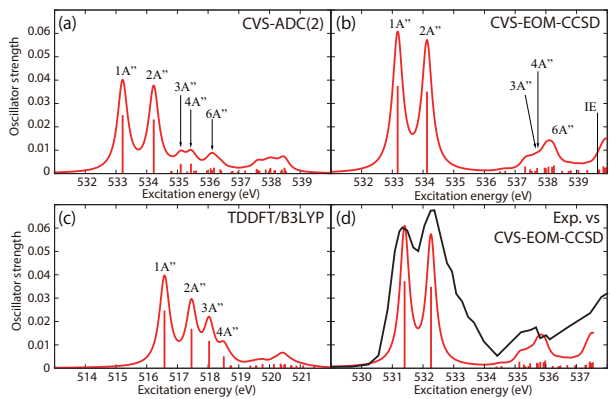}
    \caption{Uracil. Ground-state NEXAFS at the oxygen K-edge calculated with: (a)~ADC(2); 
    (b)~CVS-EOM-CCSD; (c)~TDDFT/B3LYP. The calculated IEs are 539.68 and 539.86 eV (fc-CVS-EOM-IP-CCSD/6-311++G**). In panel (d) the computed spectrum of (b)
      is shifted by $-$1.8 eV and superposed with the experimental spectrum\cite{nexafs_cytosine_uracil} (black curve).
    Basis set: 6-311++G**.}
    \label{fig:static_uracil}
\end{figure}

Fig.~\ref{fig:static_uracil} shows
the O K-edge NEXAFS spectra of uracil in the ground state computed by  CVS-EOM-CCSD, CVS-ADC(2), and TDDFT/B3LYP.
Table~\ref{tab:uracil_NTO_S0} shows NTOs of the core-excited states calculated at the CVS-EOM-CCSD/6-311++G** level, where $\sigma_K$ are the singular values for a given NTO pair (their  renormalized squares give the weights of  the  respective configurations in  the
transition).~\cite{Luzanov1976,Martin2003,Luzanov2012,Dreuw:ESSAImpl:14,Dreuw:ESSAImpl-2:14,Dreuw:ESSA:14,Krylov:Libwfa:18,Plasser:NTOfeature:2020,Krylov:Orbitals}
The NTOs for the other two methods are collected in the SI.
Panel (d) of Fig.~\ref{fig:static_uracil} shows the experimental spectrum (digitized from Ref.~\citenum{nexafs_cytosine_uracil}). The experimental spectrum has two main peaks at 531.3 and 532.2 eV, assigned to core excitations to the $\pi^{\ast}$ orbitals from O4 and O2,  respectively. Beyond these peaks, the intensity remains low up to 534.4 eV. The next notable spectral feature, attributed to Rydberg excitations, emerges at around 535.7 eV, just before  the first core-ionization onset (indicated as IE). 
The separation of $\sim$0.9 eV
between the two main peaks is reproduced at all three levels of theory. 
The NTO analysis at the CCSD level (cf. Table~\ref{tab:uracil_NTO_S0})
confirms that the excitation to the 6A" state has Rydberg character and,
after the uniform  shift,
the peak assigned to this excitation falls in the Rydberg region of the experimental spectrum. 
ADC(2) also yields a 6A" transition of Rydberg character, but it is significantly 
red-shifted relative to the experiment. No Rydberg transitions are found at the TDDFT level.
Only CVS-EOM-CCSD
reproduces the separation between the 1A" and the 6A" peaks 
with reasonable accuracy, 4.91 eV versus 4.4 eV in the experimental spectrum. 
The shoulder structure of the experimental spectrum in the region between 532.2 and 534.4 eV is attributed to vibrational excitations or shake-up transitions.~\cite{Stohr,Rehr1978} 

\begin{table}[h]
 \centering
 \caption{Uracil. CVS-EOM-CCSD/6-311++G** energies, strengths and NTOs of the O$_{1s}$ core excitations from the ground state at the FC geometry (NTO isosurface is 0.04 for the Rydberg transition and 0.05 for the rest).}
 \begin{tabular}{c|c|c|ccc}
     \hline
     Final state & $E^{\mathrm{ex}}$ (eV) & Osc. strength & Hole & $\sigma_K^2$ & Particle 
     \\
     \hline
     1A" & 533.17 & 0.0367 &
     \begin{minipage}{0.06\textwidth}
         \centering
         \includegraphics[scale=0.05]{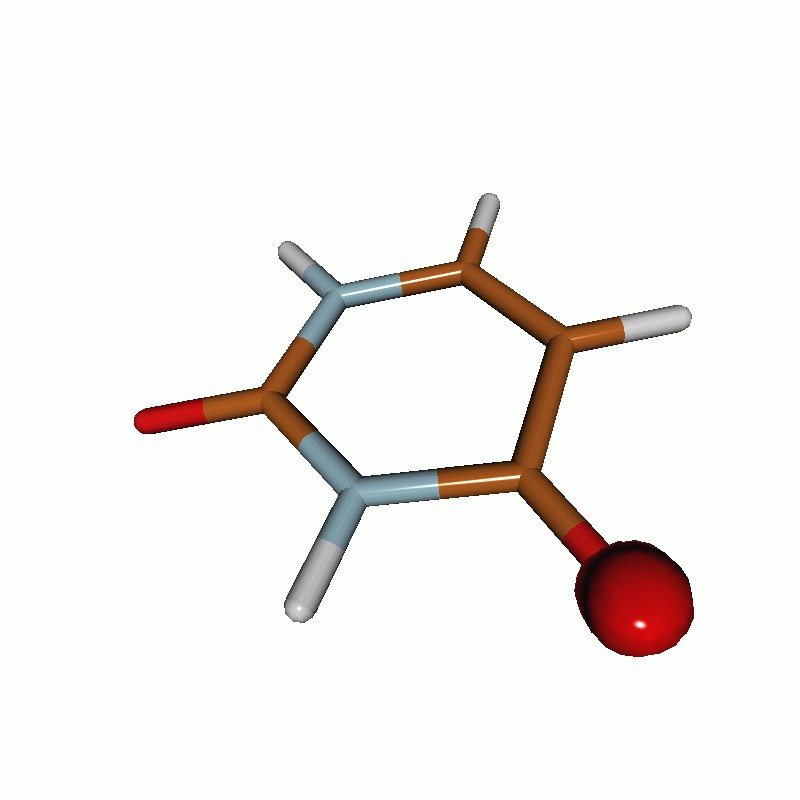}
     \end{minipage}
     & 0.78
     &  \begin{minipage}{0.06\textwidth}
         \centering
         \includegraphics[scale=0.05]{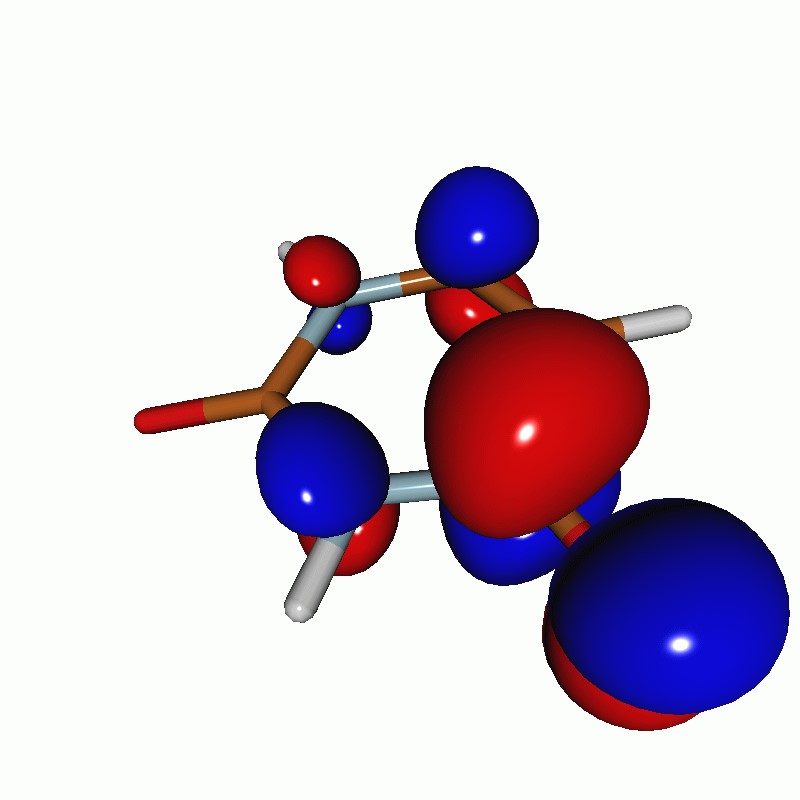}
     \end{minipage}
     \\
     \hline
     2A" & 534.13 & 0.0343 &
     \begin{minipage}{0.06\textwidth}
         \centering
         \includegraphics[scale=0.05]{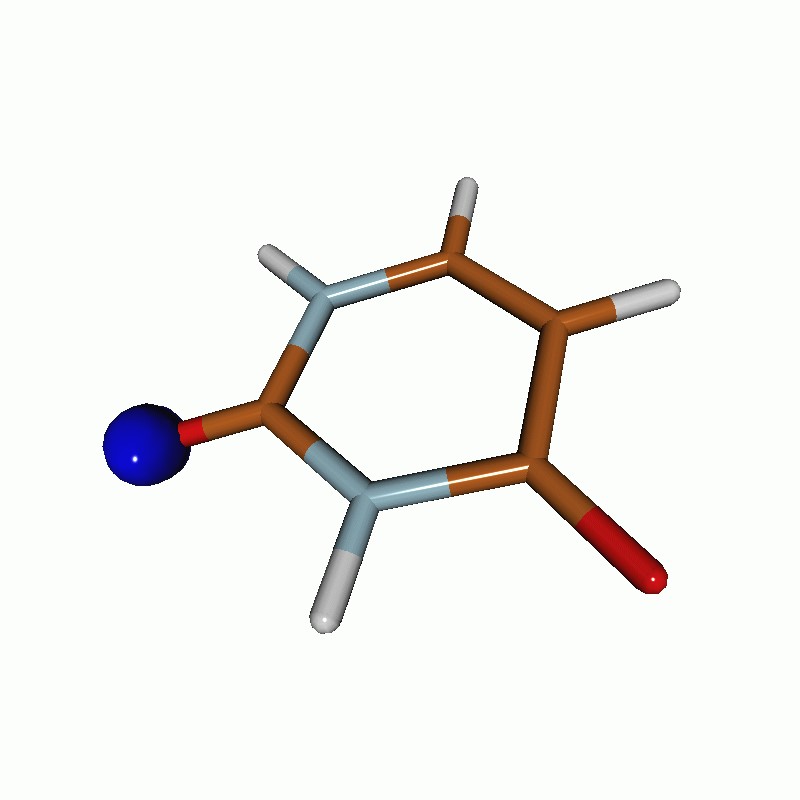}
     \end{minipage}
     & 0.79
     &  \begin{minipage}{0.06\textwidth}
         \centering
         \includegraphics[scale=0.05]{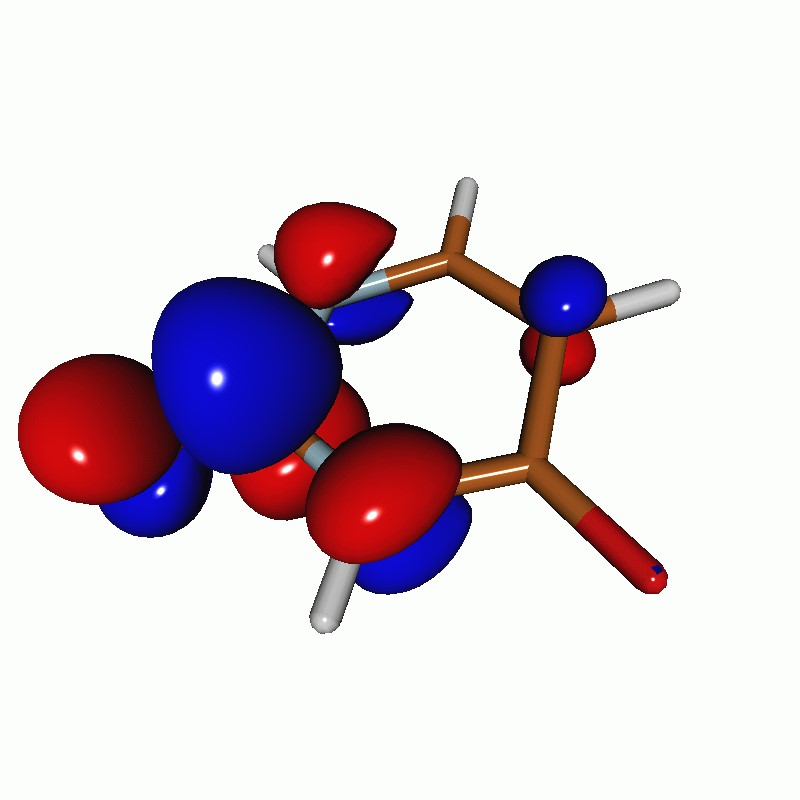}
     \end{minipage}
     \\
     \hline
     3A" & 537.55 & 0.0003 &
     \begin{minipage}{0.06\textwidth}
         \centering
         \includegraphics[scale=0.05]{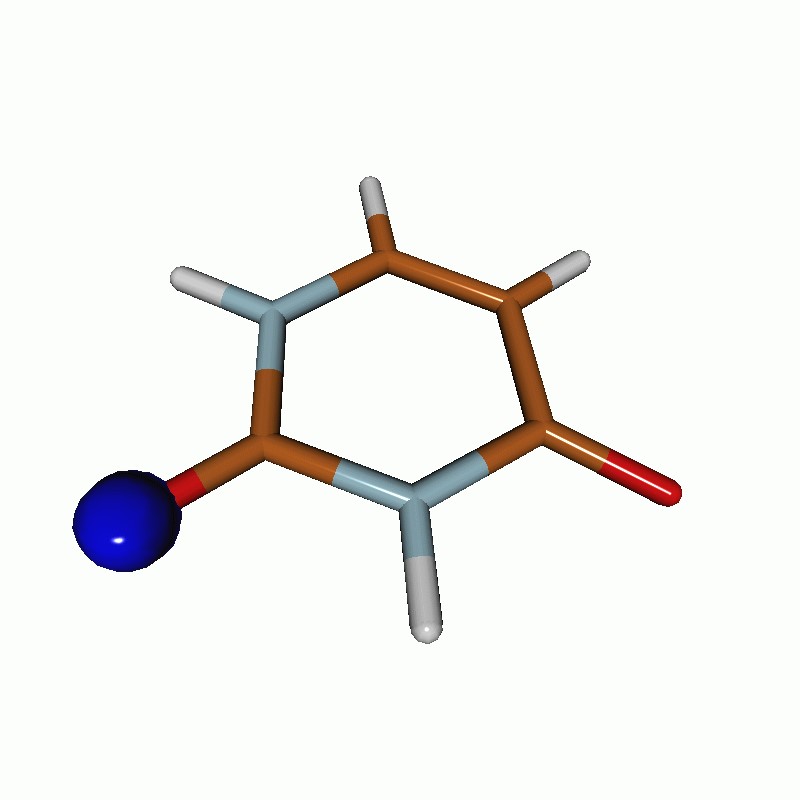}
     \end{minipage}
     & 0.76
     &  \begin{minipage}{0.06\textwidth}
         \centering
         \includegraphics[scale=0.05]{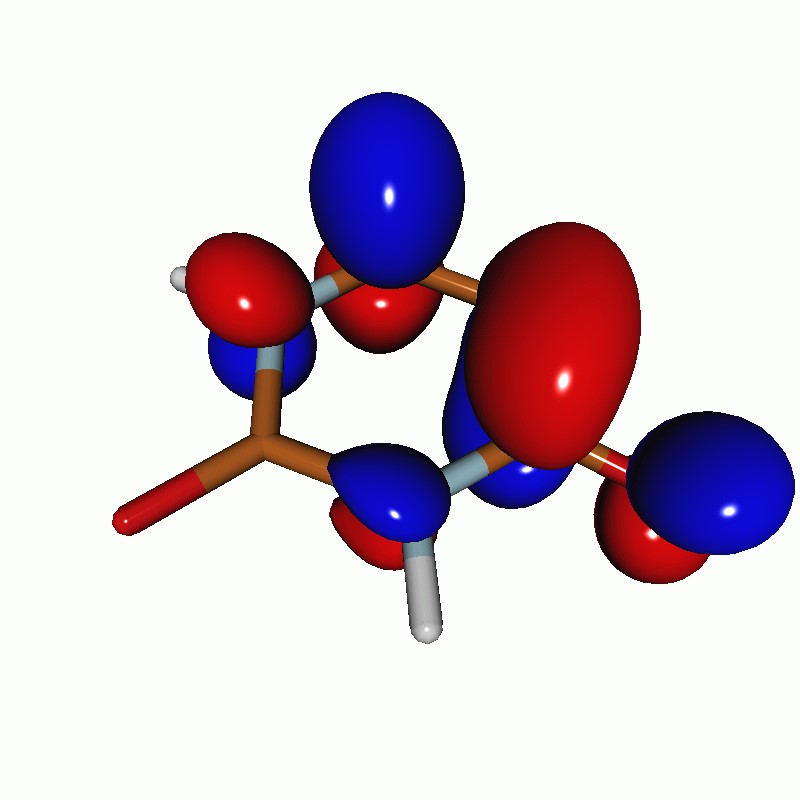}
     \end{minipage}
     \\
     \hline
     4A" & 537.66 & 0.0004 &
     \begin{minipage}{0.06\textwidth}
         \centering
         \includegraphics[scale=0.05]{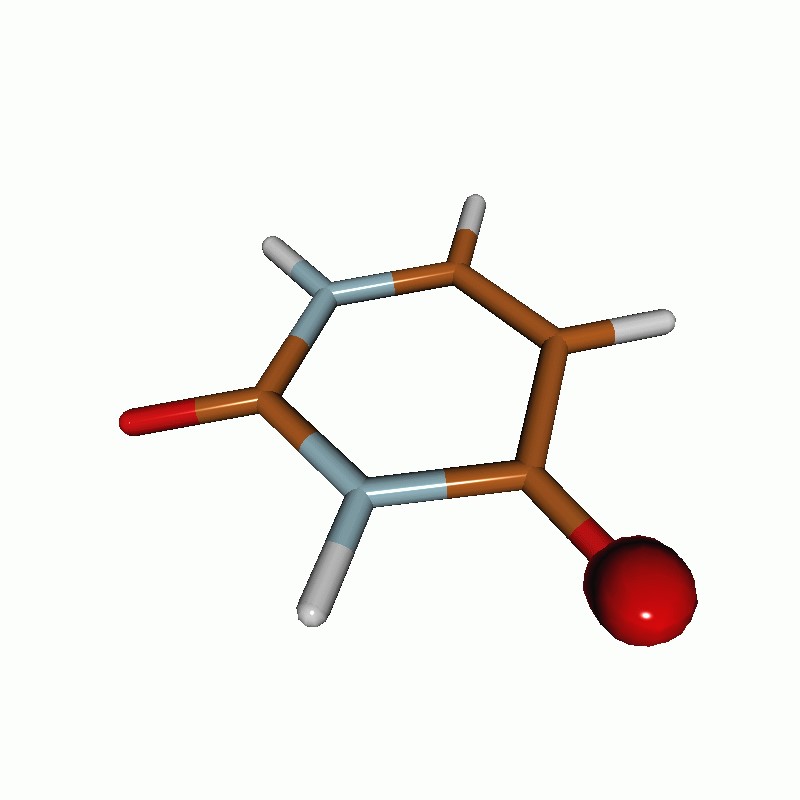}
     \end{minipage}
     & 0.78
     &  \begin{minipage}{0.06\textwidth}
         \centering
         \includegraphics[scale=0.05]{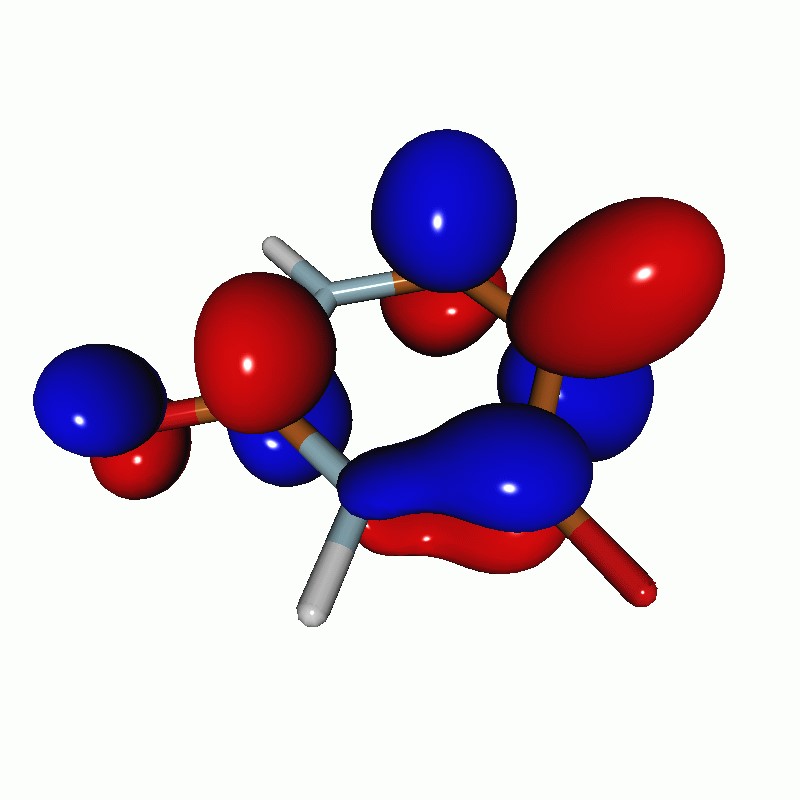}
     \end{minipage}
     \\
     \hline
     6A" & 538.08 & 0.0022 &
     \begin{minipage}{0.06\textwidth}
         \centering
         \includegraphics[scale=0.035]{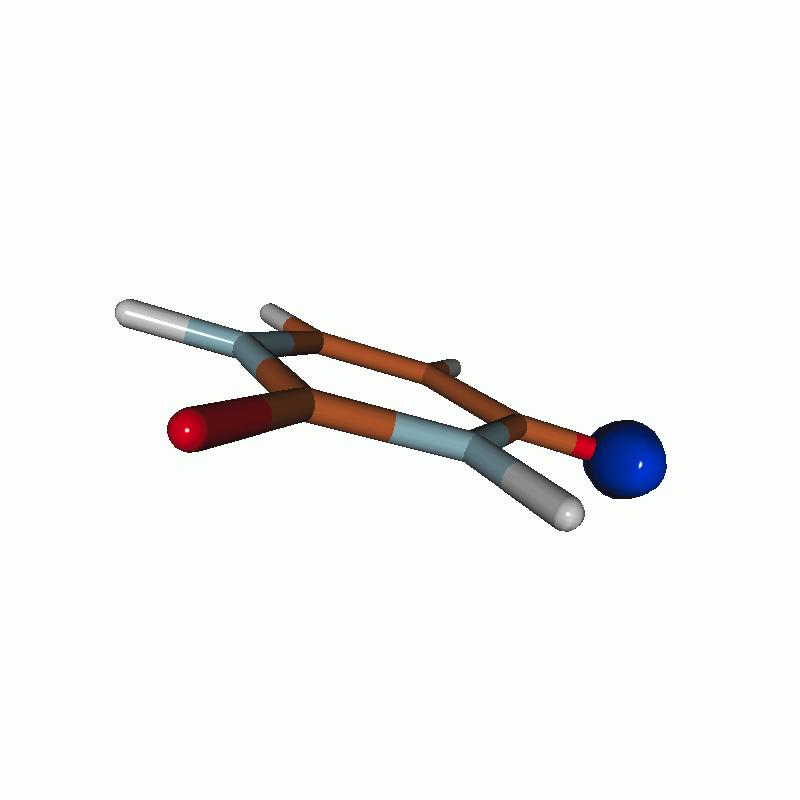}
     \end{minipage}
     & 0.82
     &  \begin{minipage}{0.06\textwidth}
         \centering
         \includegraphics[scale=0.035]{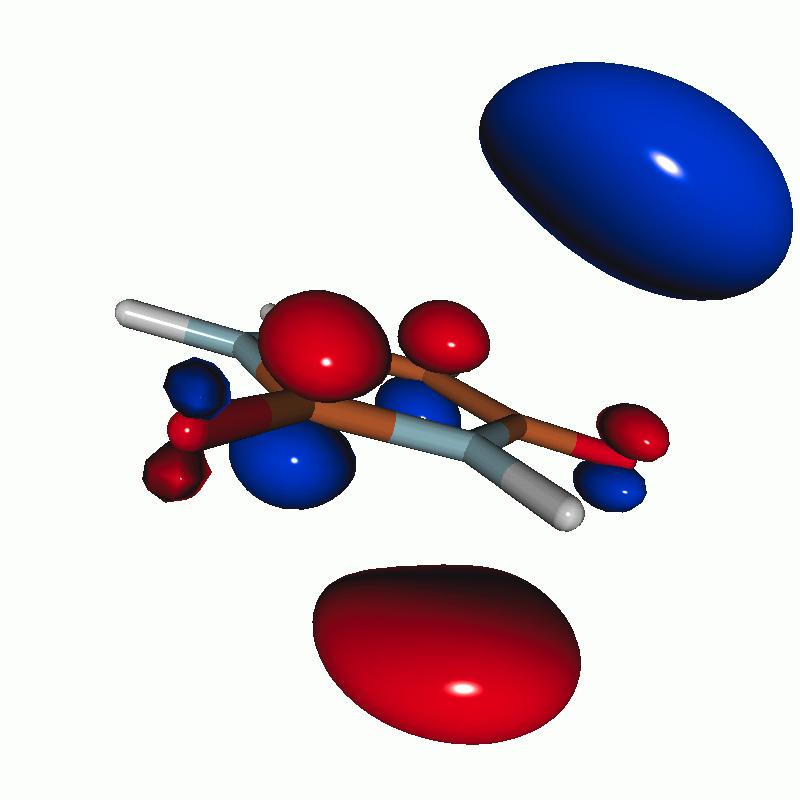}
     \end{minipage}
     \\
     \hline
     \end{tabular}
 \label{tab:uracil_NTO_S0}
 \end{table}

\begin{figure}[h]
    \centering
    \includegraphics[width=8cm,height=8cm,keepaspectratio]{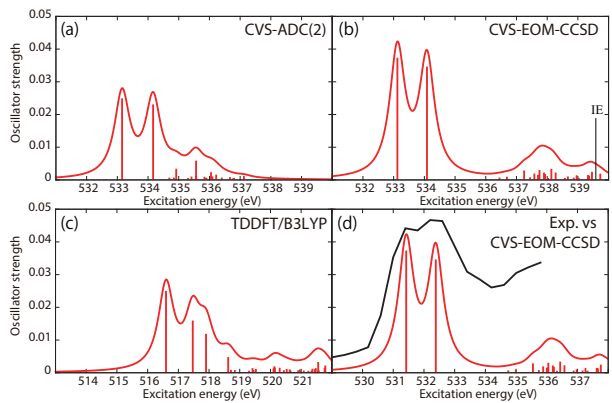}
    \caption{Thymine. Ground-state oxygen K-edge NEXAFS calculated with: (a) ADC(2), (b) CVS-EOM-CCSD, (c) TDDFT/B3LYP. The computed ionization energies (IEs) are 539.67 and 539.73 eV (fc-CVS-EOM-IP-CCSD). In panel (d), the CVS-EOM-CCSD spectrum of (b) is shifted by $-$1.7 eV and superposed with the experimental one\cite{Wolf} (black curve). Basis set: 6-311++G**. FWHM of the Lorentzian convolution function is 0.6 eV.}
    \label{fig:static_thymine}
\end{figure}

Fig.~\ref{fig:static_thymine} shows the ground-state NEXAFS spectra of thymine at the O K-edge. For construction of the theoretical absorption spectra, we used FWHM of 0.6 eV for 
the Lorentzian convolution function.
Panel (d) shows the experimental spectrum (digitized from Ref.~\citenum{Wolf}). 
Both the experimental and calculated spectra exhibit fine structures, similar to those of uracil. Indeed, the first and second peaks at 531.4 and 532.2 eV of the experimental spectrum were assigned to O$_{1s}$-hole states having the same electronic configuration characters as the two lowest-lying O$_{1s}$-hole states of uracil.
The NTOs of thymine can be found in the SI.
Again, only CVS-EOM-CCSD reproduces with reasonable accuracy the Rydberg region after 534 eV. The separation of the two main peaks is well reproduced at all three levels of theory.

\begin{figure}[h]
    \centering
    \includegraphics[width=8cm,height=8cm,keepaspectratio]{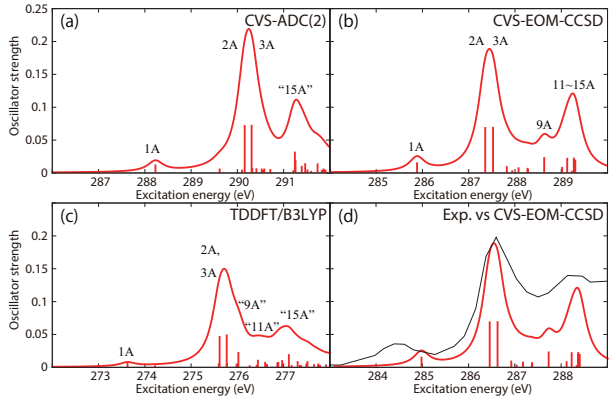}
    \caption{Acetylacetone. Ground-state NEXAFS at carbon K-edge calculated with: (a) ADC(2); (b) CVS-EOM-CCSD; (c) TDDFT/B3LYP. The ionization energies (IEs) are 291.12, 291.88, 292.11, 294.10, and 294.56 eV (fc-CVS-EOM-IP-CCSD). In panel (d), the computational result of (b) is shifted by $-0.9$ eV and superposed with the experimental spectrum\cite{acac_ultrafast_ISC} (black curve).
    Basis set: 6-311++G**. \label{fig:static_AcAc}}
\end{figure}

Fig.~\ref{fig:static_AcAc} shows the C K-edge ground-state NEXAFS spectra of acetylacetone; the NTOs of the core excitations obtained at the CVS-EOM-CCSD/6-311++G** level 
are collected in Table~\ref{tab:AcAc_NTO_S0}. 
The experimental spectrum, plotted in panel (d) of Fig.~\ref{fig:static_AcAc}, was digitized from Ref.~\citenum{acac_ultrafast_ISC}. Table~\ref{tab:AcAc_NTO_S0} shows that the first three core excitations are dominated by the transitions to the LUMO from the 1$s$ orbitals of the carbon atoms C2, C3, and C4.
Transition from the central carbon atom, C3, appears as the first relatively weak peak at 284.4 eV. We note that acetylacetone may exhibit keto--enol tautomerism. In the keto form, atoms C2 and C4 are equivalent. Therefore, transitions from these carbon atoms appear as quasi-degenerate main peaks at $\approx$286.6 eV. The region around 288.2 eV is attributed to Rydberg transitions.
The $\sim$2 eV separation between the first peak and the main peak due to the two quasi-degenerate transitions is well reproduced by ADC(2) and TDDFT/B3LYP, and slightly underestimated by CVS-EOM-CCSD (1.6 eV). On the other hand, the separation of $\sim$1.6 eV between the main peak and the Rydberg resonance region is accurately reproduced only by CVS-EOM-CCSD.

\begin{table}[h]
 \centering
 \caption{Acetylacetone. CVS-EOM-CCSD/6-311++G** NTOs of the C$_{1s}$ core excitations from the ground state at the FC geometry (NTO isosurface is 0.03 for the Rydberg transition and 0.05 for the rest).}
 \begin{tabular}{c|c|c|ccc}
     \hline
     Final state & $E^{\mathrm{ex}}$ (eV) & Osc. strength & Hole & $\sigma_K^2$ & Particle 
     \\
     \hline
     1A & 285.88 & 0.0133 &
     \begin{minipage}{0.06\textwidth}
         \centering
         \includegraphics[scale=0.035]{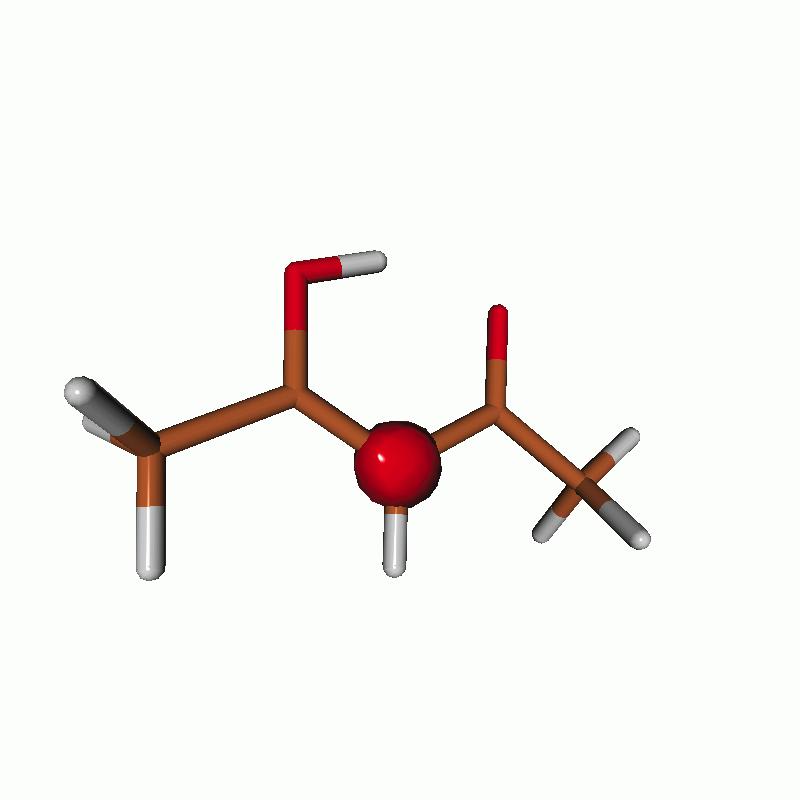}
     \end{minipage}
     & 0.76
     &  \begin{minipage}{0.06\textwidth}
         \centering
         \includegraphics[scale=0.035]{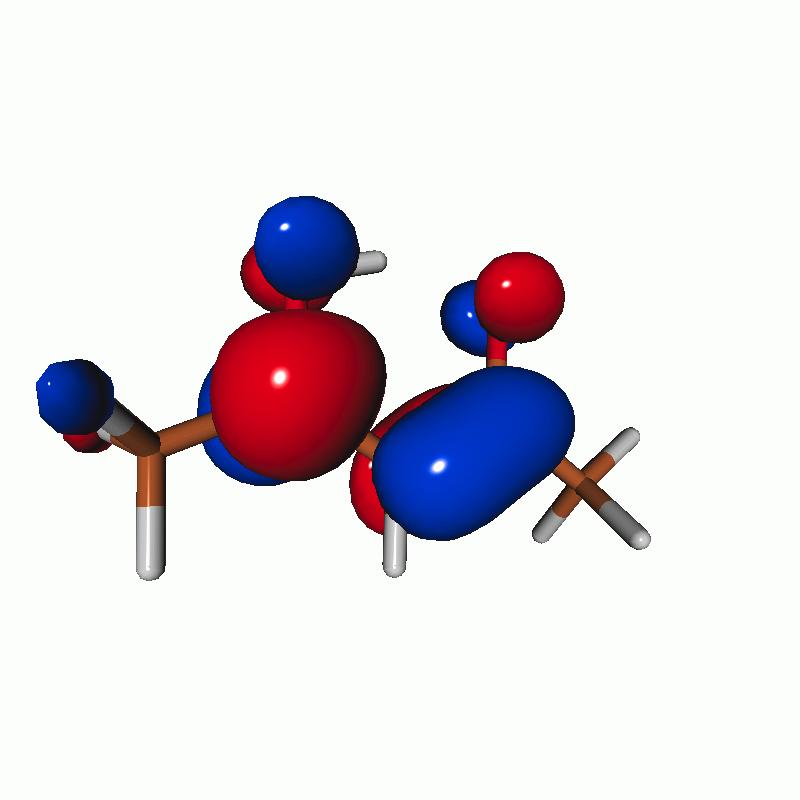}
     \end{minipage}
     \\
     \hline
     2A & 287.36 & 0.0671 &
     \begin{minipage}{0.06\textwidth}
         \centering
         \includegraphics[scale=0.035]{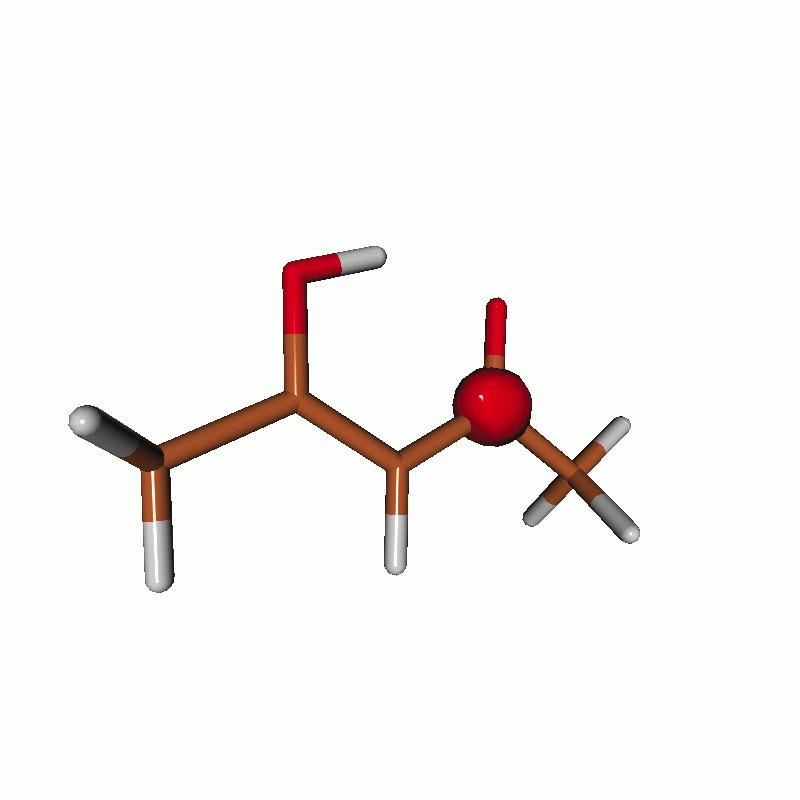}
     \end{minipage}
     & 0.82
     &  \begin{minipage}{0.06\textwidth}
         \centering
         \includegraphics[scale=0.035]{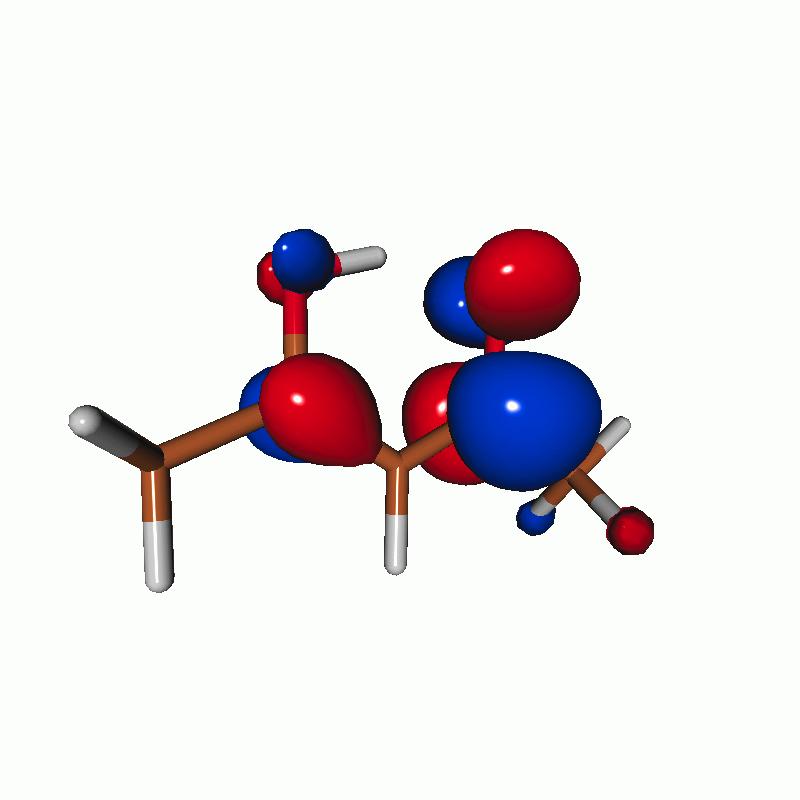}
     \end{minipage}
     \\
     \hline
     3A & 287.53 & 0.0673 &
     \begin{minipage}{0.06\textwidth}
         \centering
         \includegraphics[scale=0.035]{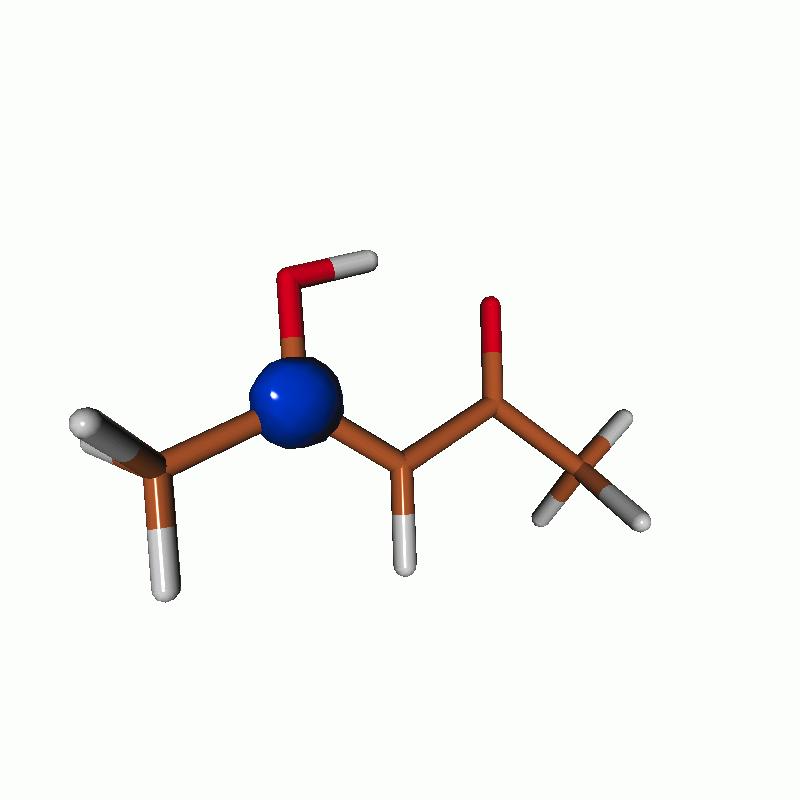}
     \end{minipage}
     & 0.81
     &  \begin{minipage}{0.06\textwidth}
         \centering
         \includegraphics[scale=0.035]{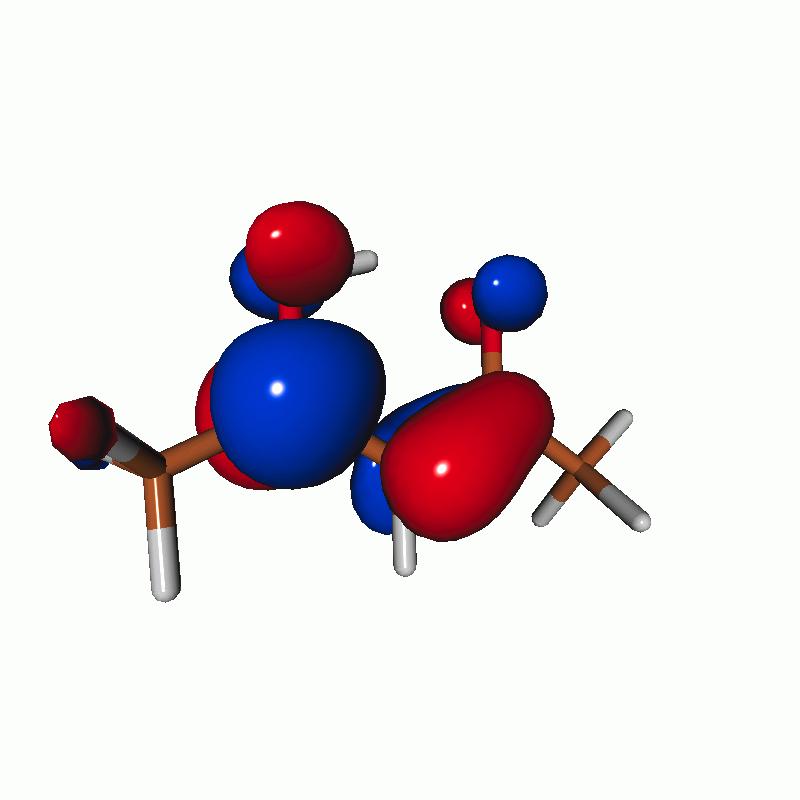}
     \end{minipage}
     \\
     \hline
     9A & 288.63 & 0.0213 &
     \begin{minipage}{0.06\textwidth}
         \centering
         \includegraphics[scale=0.035]{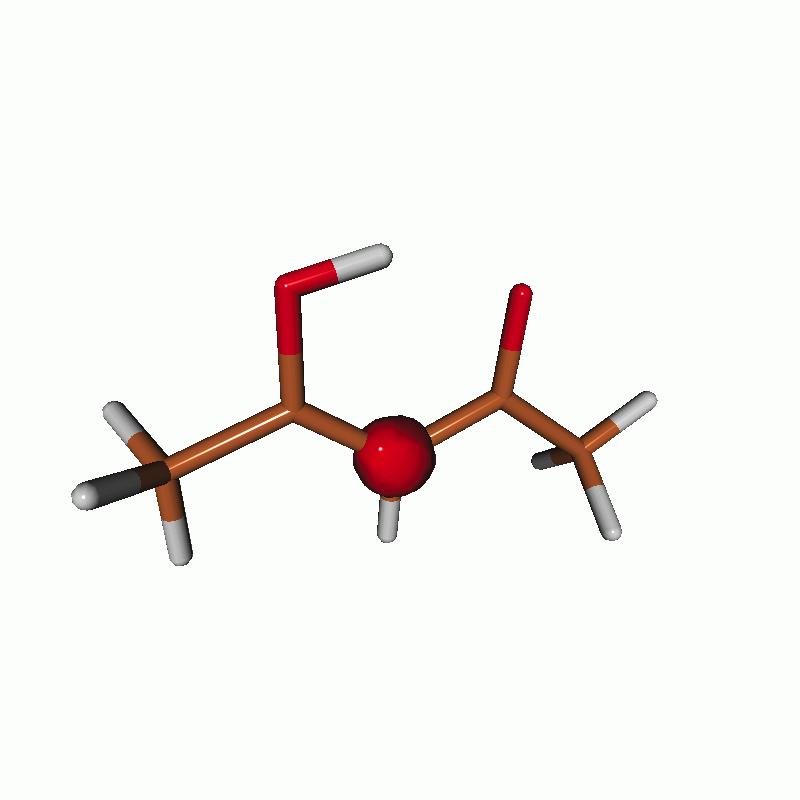}
     \end{minipage}
     & 0.79
     &  \begin{minipage}{0.06\textwidth}
         \centering
         \includegraphics[scale=0.035]{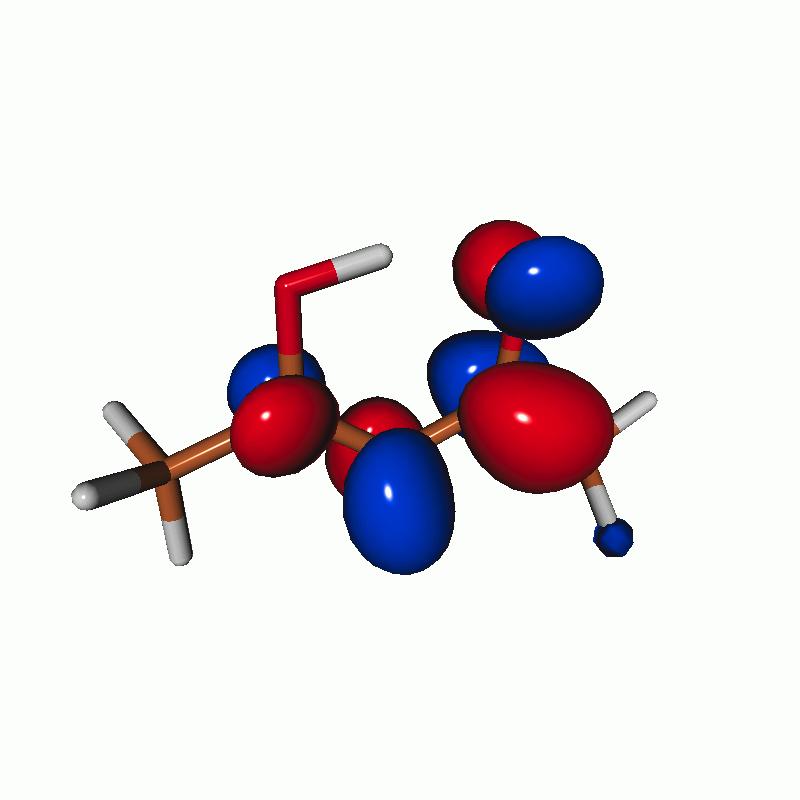}
     \end{minipage}
     \\
     \hline
     11A & 289.13 & 0.0202 &
     \begin{minipage}{0.06\textwidth}
         \centering
         \includegraphics[scale=0.035]{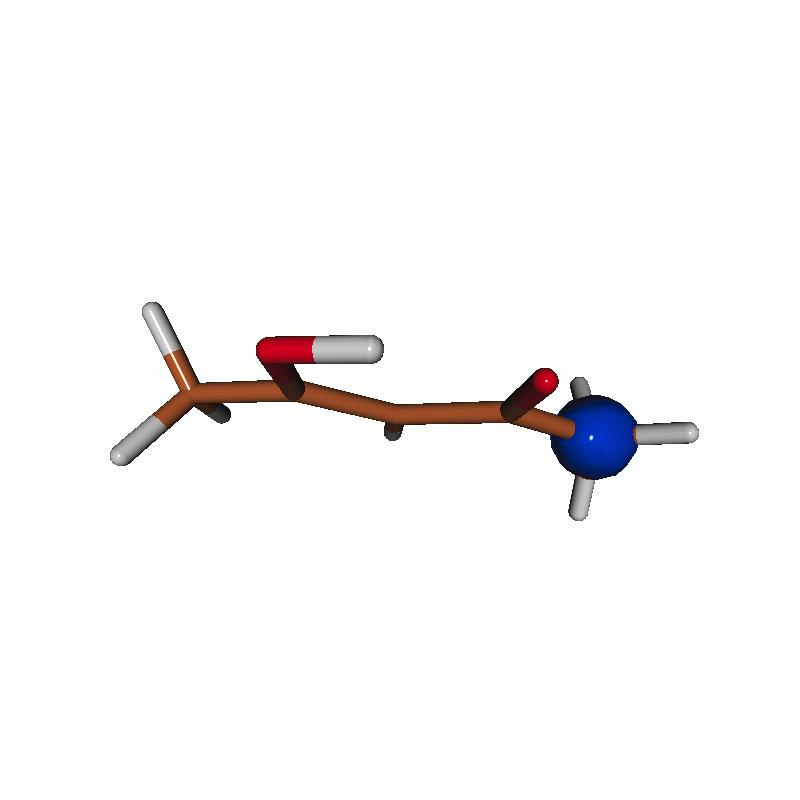}
     \end{minipage}
     & 0.82
     &  \begin{minipage}{0.06\textwidth}
         \centering
         \includegraphics[scale=0.035]{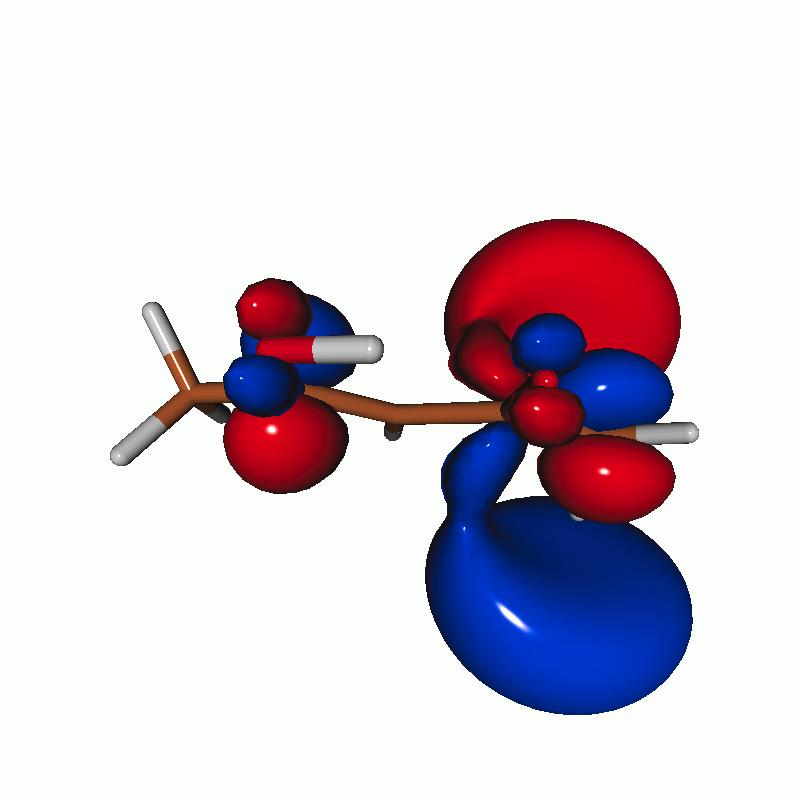}
     \end{minipage}
     \\
     \hline
     13A & 289.27 & 0.0205 &
     \begin{minipage}{0.06\textwidth}
         \centering
         \includegraphics[scale=0.035]{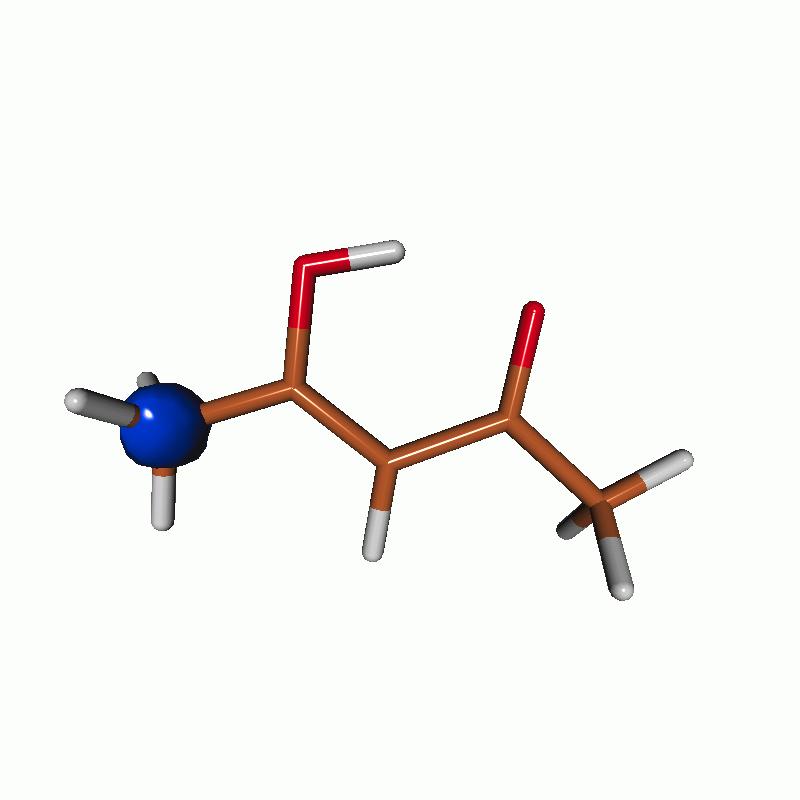}
     \end{minipage}
     & 0.83
     &  \begin{minipage}{0.06\textwidth}
         \centering
         \includegraphics[scale=0.035]{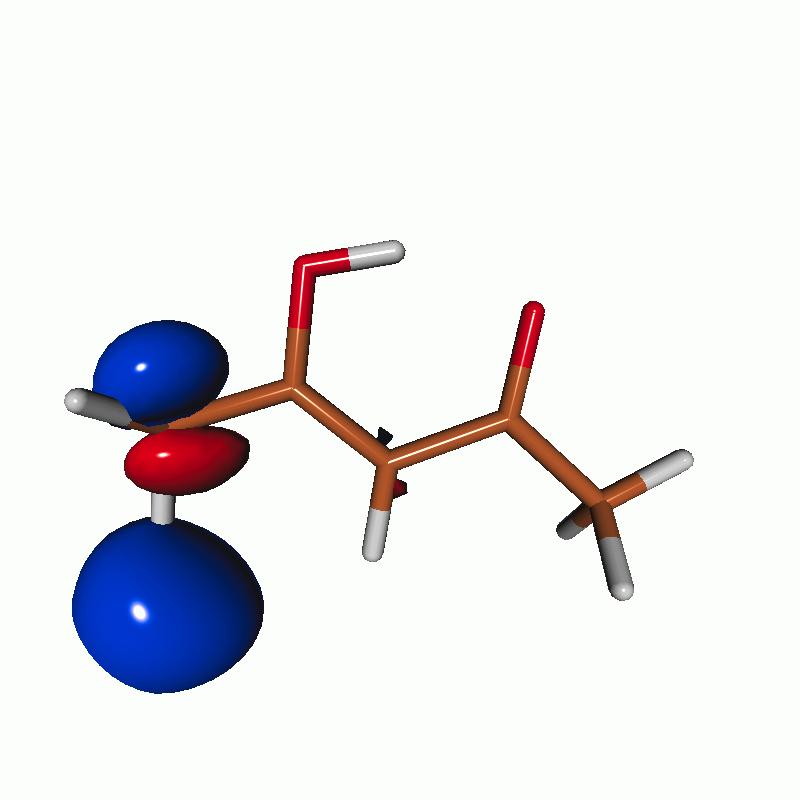}
     \end{minipage}
     \\
     \hline
     14A & 289.28 & 0.0175 &
     \begin{minipage}{0.06\textwidth}
         \centering
         \includegraphics[scale=0.035]{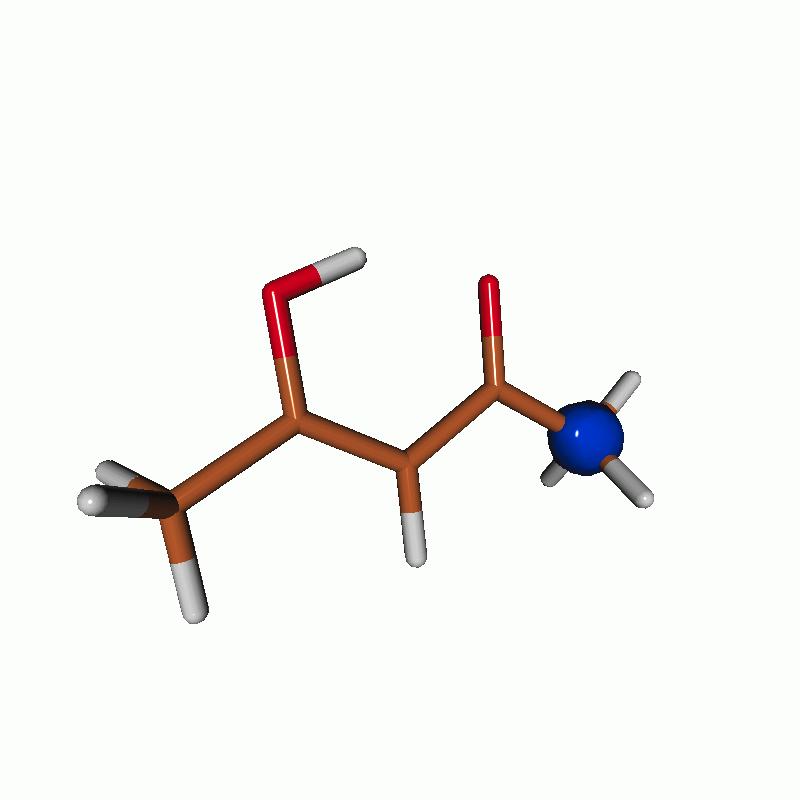}
     \end{minipage}
     & 0.82
     &  \begin{minipage}{0.06\textwidth}
         \centering
         \includegraphics[scale=0.035]{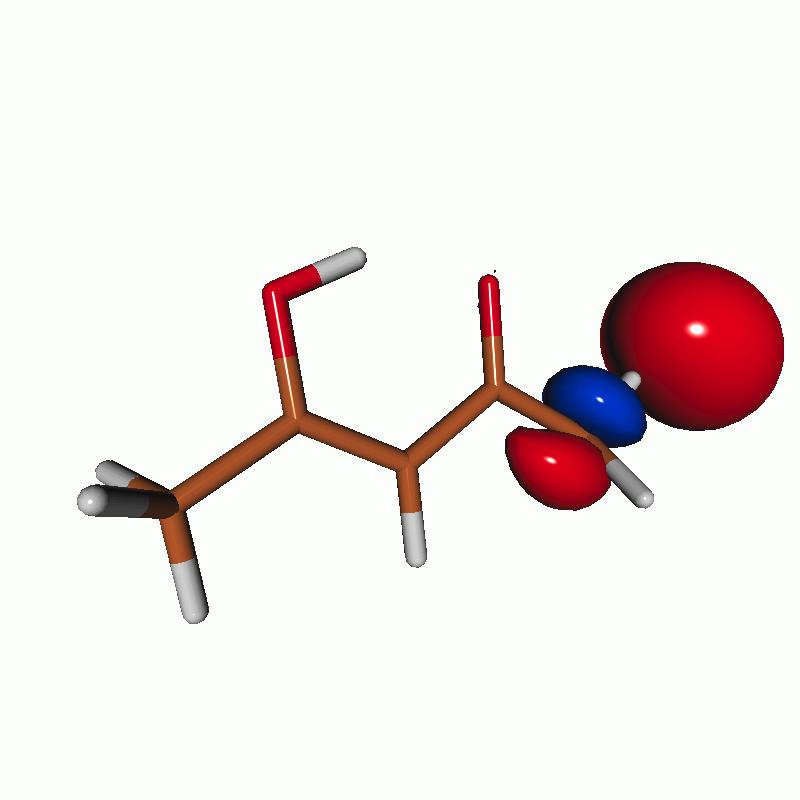}
     \end{minipage}
     \\
     \hline
     15A & 289.30 & 0.0174 &
     \begin{minipage}{0.06\textwidth}
         \centering
         \includegraphics[scale=0.035]{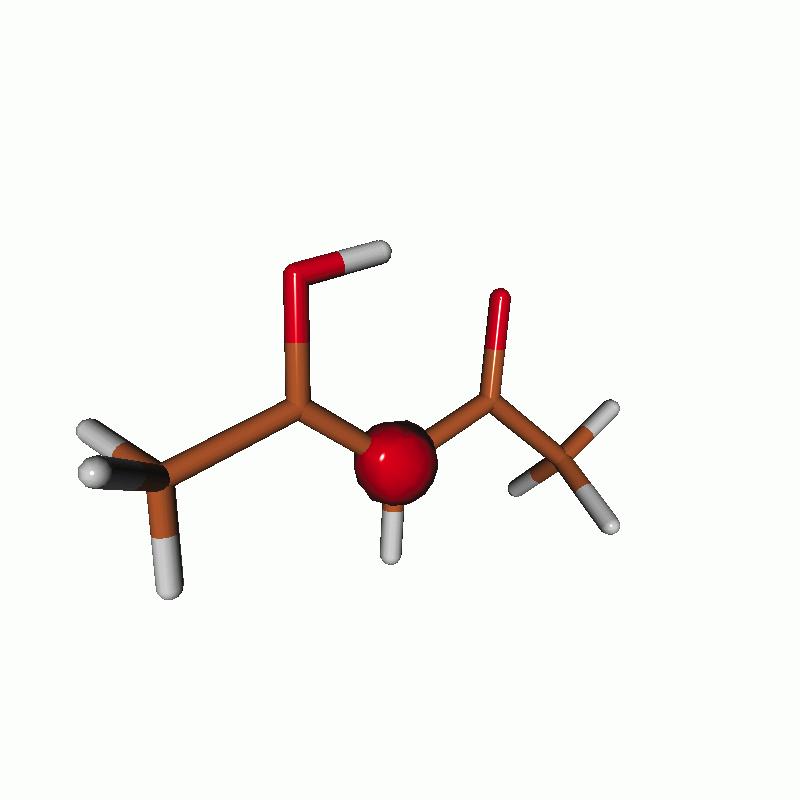}
     \end{minipage}
     & 0.81
     &  \begin{minipage}{0.06\textwidth}
         \centering
         \includegraphics[scale=0.035]{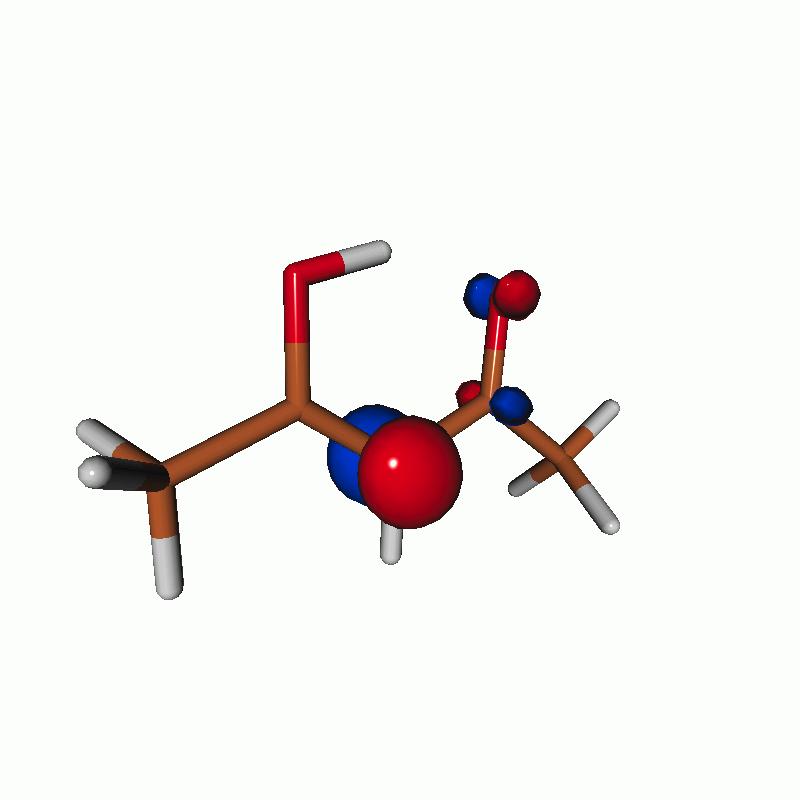}
     \end{minipage}
     \\
     \hline
     \end{tabular}
 \label{tab:AcAc_NTO_S0}
 \end{table}

The results for the three considered molecules illustrate  that CVS-EOM-CCSD describes well the entire pre-edge region of the NEXAFS spectrum. CVS-ADC(2) and TDDFT/B3LYP
describe well the core excitations to the LUMO and LUMO+1 (apart from an overall systematic shift), but generally  fail to describe the transitions at higher excitation energies.

\subsection{Valence-Excited States}
\label{subsec:gs_to_valence}

Table~\ref{tab:uracil_valence_energy} shows the excitation energies of the two lowest triplet states, the three lowest singlet states, plus the S$_5$($\pi\pi^{\ast}$) state of uracil, calculated at the FC geometry, along 
with the values derived from the EELS~\cite{Chernyshova2012} and UV absorption experiments.~\cite{UV_uracil}
The EOM-EE-CCSD/6-311++G** NTOs are collected in Table~\ref{tab:uracil_NTO_gs2ex} and 
the NTOs for other methods are given in the SI.
We refer to Ref.~\citenum{Fedotov:uracil}
for an extensive benchmark study of the one-photon absorption and excited-state absorption of uracil.

In EELS, the excited states are probed by measuring the kinetic energy change of a beam of electrons after inelastic collision with the probed molecular sample.~\cite{Trajmar} In the limit of high incident energy or small scattering angle, the transition amplitude takes a dipole form and the selection rules are same as those of UV-Vis absorption. Otherwise, the selection rules are different and optically dark states can be detected. Furthermore, spin-orbit coupling enables excitation into triplet states. Assignment of the EELS spectral signatures is based on theoretical calculations. Note that excitation energies obtained with EELS may be blue-shifted compared to those from UV-Vis absorption due to momentum transfer between the probing electrons and the probed molecule.

EOM-EE-CCSD excitation energies for all valence states of uracil agree well with the experimental values from EELS.
Both the EOM-EE-CCSD and EELS values slightly overestimate the UV-Vis results. 
For the two triplet states and the S$_1$(A$^{\prime\prime}$,n$\pi^{\ast}$) and S$_2$(A$^{\prime}$,$\pi\pi^{\ast}$) states, 
ADC(2) also gives fairly accurate excitation energies. ADC(2)-x, on the other hand, seems unbalanced for the valence excitations (regardless of the basis set). 
The TDDFT/B3LYP excitation energies are 
red-shifted with respect to the EELS values, but the energy differences between the T$_1$(A$^\prime$, $\pi\pi^{\ast}$), T$_2$(A$^{\prime\prime}$, n$\pi^{\ast}$), S$_1$(A$^{\prime\prime}$, n$\pi^{\ast}$), and S$_2$(A$^{\prime}$, $\pi\pi^{\ast}$) states are in reasonable agreement with the corresponding experimentally derived values.

\begin{table}[h]
    \centering
    \caption{Uracil. Excitation energies (eV) at the FC geometry and comparison with experimental values from EELS~\cite{Chernyshova2012} and UV absorption spectroscopy.~\cite{UV_uracil}}
    \begin{tabular}{l|cccc|c|c}
        \hline
         & ADC(2) & ADC(2)-x & EOM-CCSD & TDDFT & EELS & UV
         \\
         \hline
         T$_1$($\pi\pi^{\ast}$) & 3.91 & 3.36 & 3.84 & 3.43 & 3.75 & 
         \\
         \hline
         T$_2$(n$\pi^{\ast}$) & 4.47 & 3.79 & 4.88 & 4.27 & 4.76 & 
         \\
         \hline
         S$_1$(n$\pi^{\ast}$) & 4.68 & 3.93 & 5.15 & 4.65 & 5.2 & 
         \\
         \hline
         S$_2$($\pi\pi^{\ast}$) & 5.40 & 4.70 & 5.68 & 5.19 & 5.5 & 5.08
         \\
         \hline
         S$_3$($\pi\mathrm{Ryd}$) & 5.97 & 5.39 & 6.07 & 5.70 & - & 
         \\
         \hline
         {S$_5$($\pi\pi^{\ast}$)} & 6.26 & 5.32 & 6.74 & 5.90 & 6.54 & 6.02
         \\
         \hline
    \end{tabular}
    \label{tab:uracil_valence_energy}
\end{table}

\begin{table}[h]
 \centering
 \caption{Uracil. EOM-EE-CCSD/6-311++G** NTOs for the transitions from the ground state to the lowest valence-excited states at the FC geometry (NTO isosurface is 0.05).}
 \begin{tabular}{l|c|c|ccc}
     \hline
     Final state & $E^{\mathrm{ex}}$ (eV) & Osc. strength & Hole & $\sigma_K^2$ & Particle 
     \\
     \hline
     T$_1$(A$^{\prime}$,$\pi\pi^{\ast}$) & 3.84 & - &
     \begin{minipage}{0.06\textwidth}
         \centering
         \includegraphics[scale=0.05]{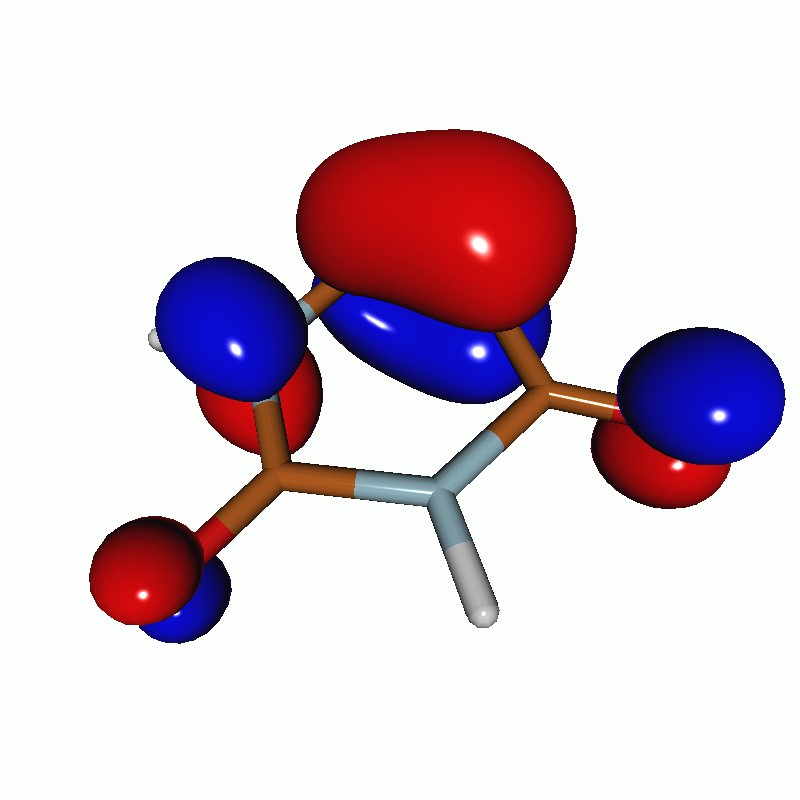}
     \end{minipage}
     & 0.82
     &  \begin{minipage}{0.06\textwidth}
         \centering
         \includegraphics[scale=0.05]{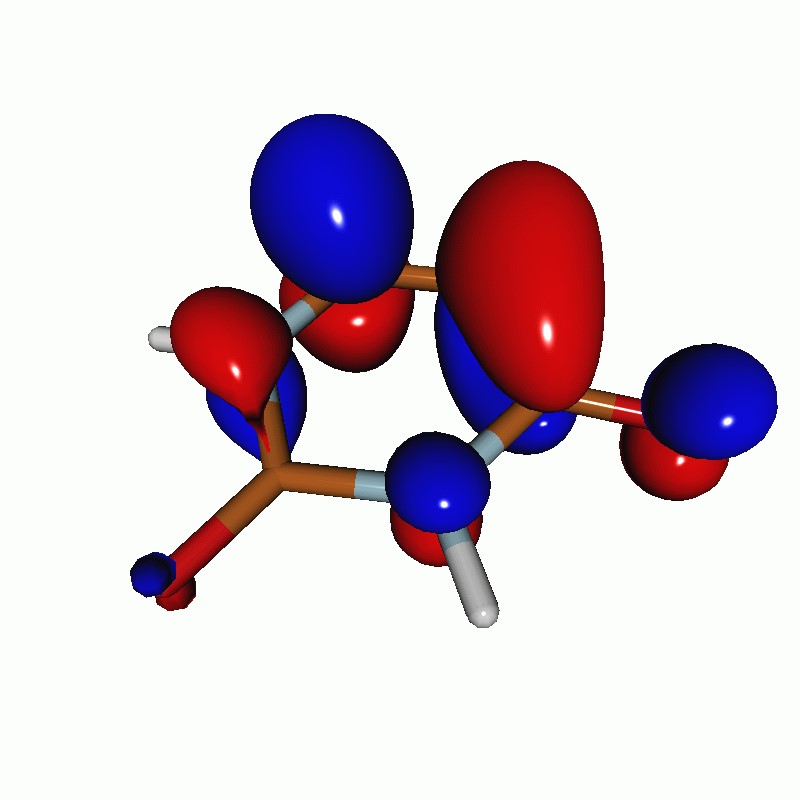}
     \end{minipage}
     \\
     \hline
     T$_2$(A$^{\prime\prime}$,n$\pi^{\ast}$) & 4.88 & - &
     \begin{minipage}{0.06\textwidth}
         \centering
         \includegraphics[scale=0.05]{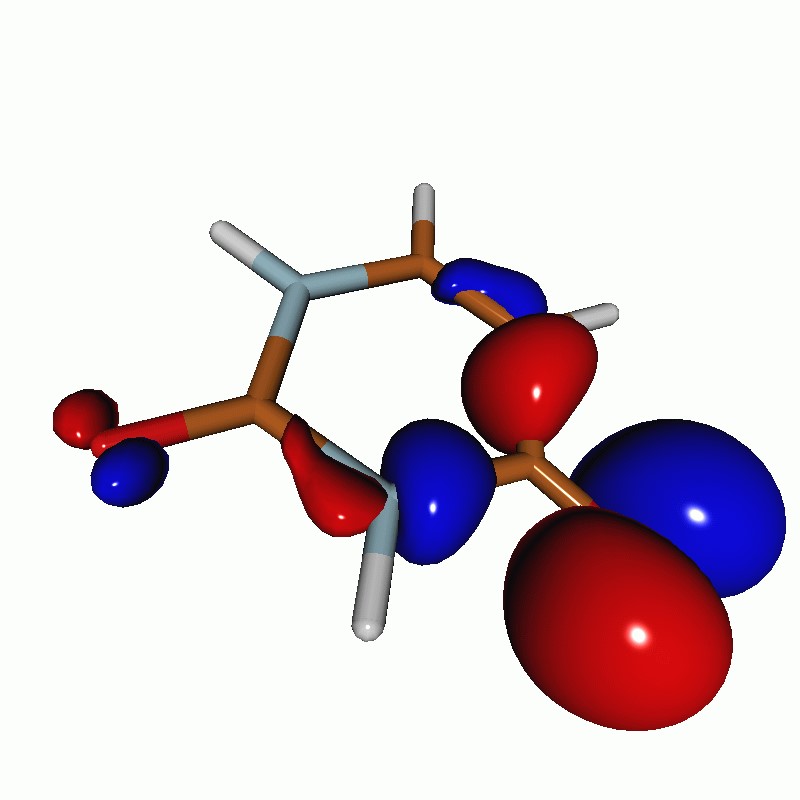}
     \end{minipage}
     & 0.82
     &  \begin{minipage}{0.06\textwidth}
         \centering
         \includegraphics[scale=0.05]{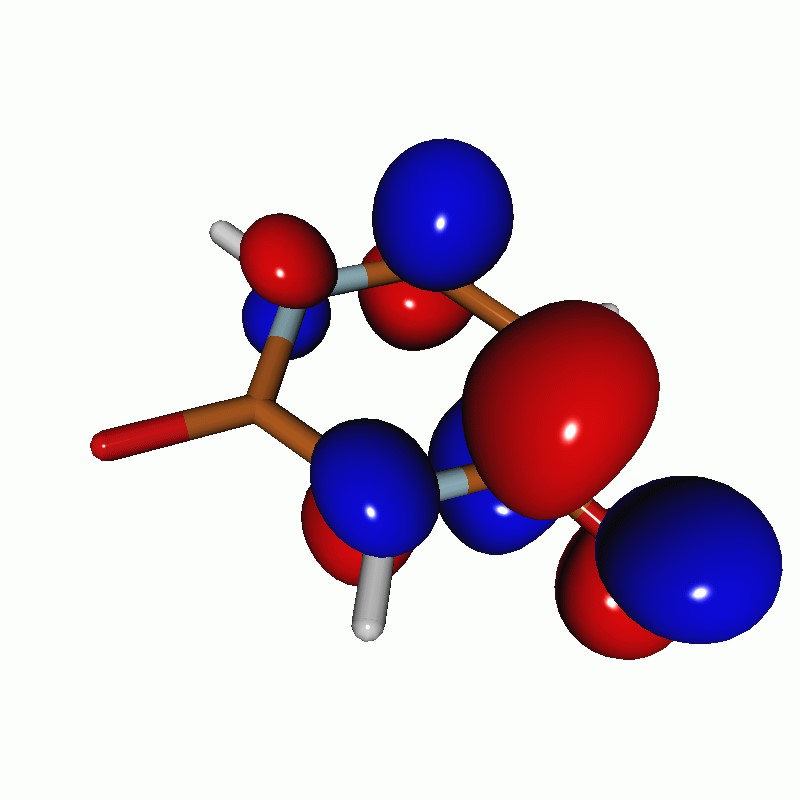}
     \end{minipage}
     \\
     \hline
     S$_1$(A$^{\prime\prime}$,n$\pi^{\ast}$) & 5.15 & 0.0000 &
     \begin{minipage}{0.06\textwidth}
         \centering
         \includegraphics[scale=0.05]{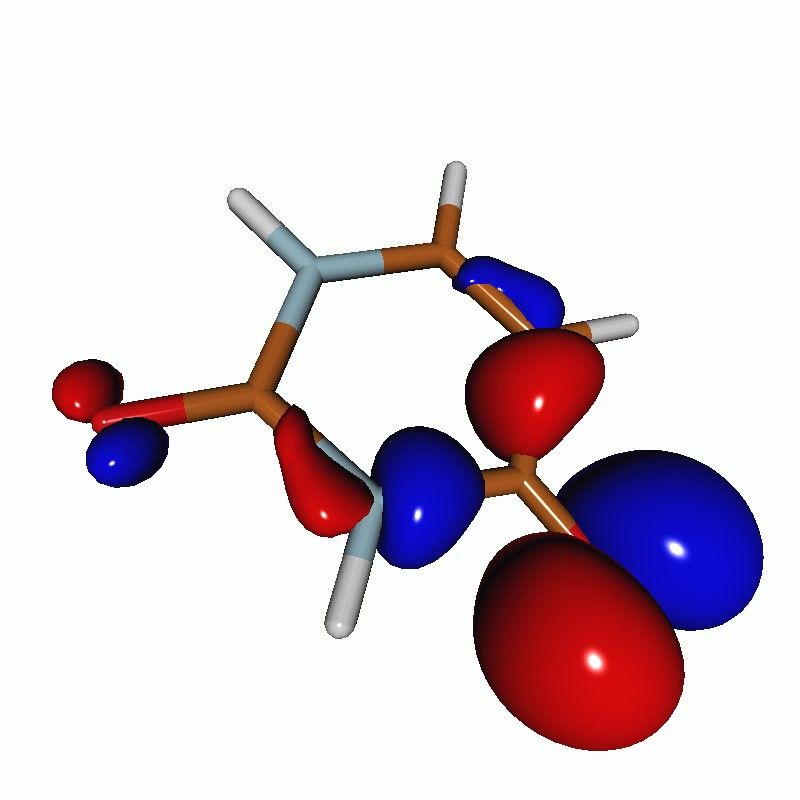}
     \end{minipage}
     & 0.81
     &  \begin{minipage}{0.06\textwidth}
         \centering
         \includegraphics[scale=0.05]{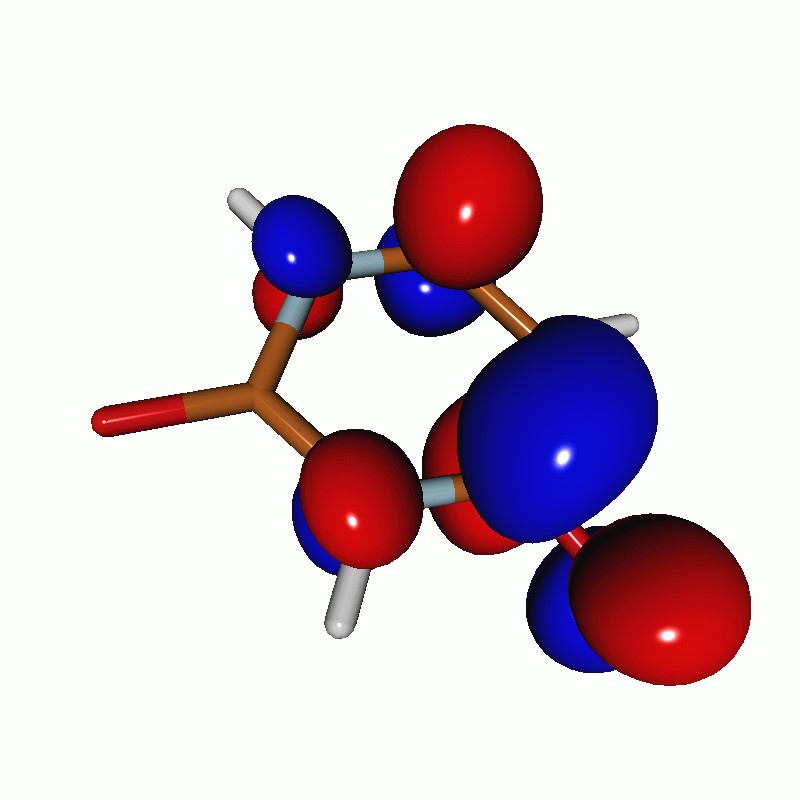}
     \end{minipage}
     \\
     \hline
     S$_2$(A',$\pi\pi^{\ast}$) & 5.68 & 0.2386 &
     \begin{minipage}{0.06\textwidth}
         \centering
         \includegraphics[scale=0.05]{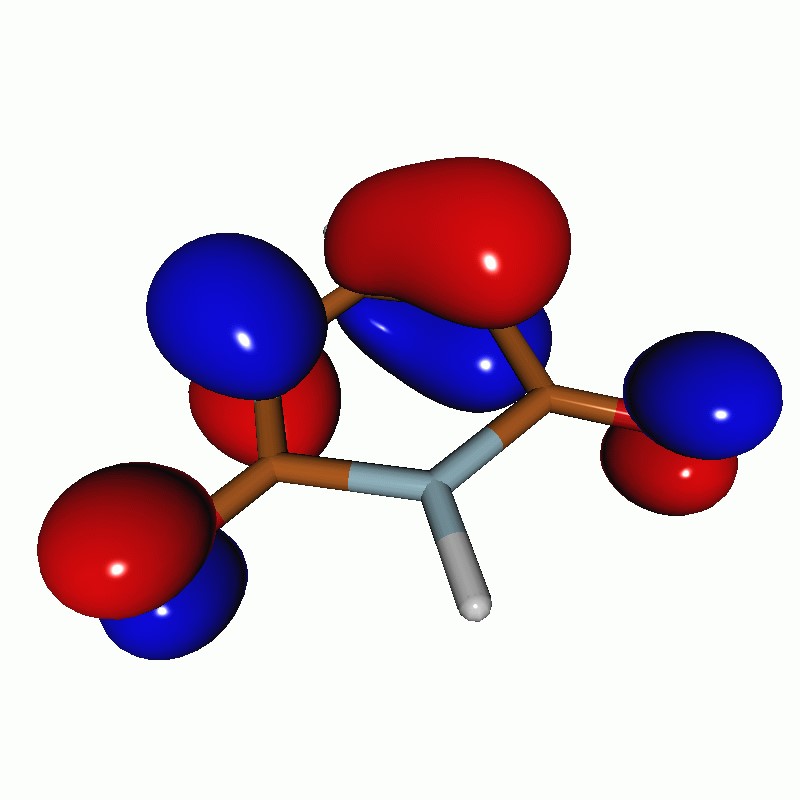}
     \end{minipage}
     & 0.75
     &  \begin{minipage}{0.06\textwidth}
         \centering
         \includegraphics[scale=0.05]{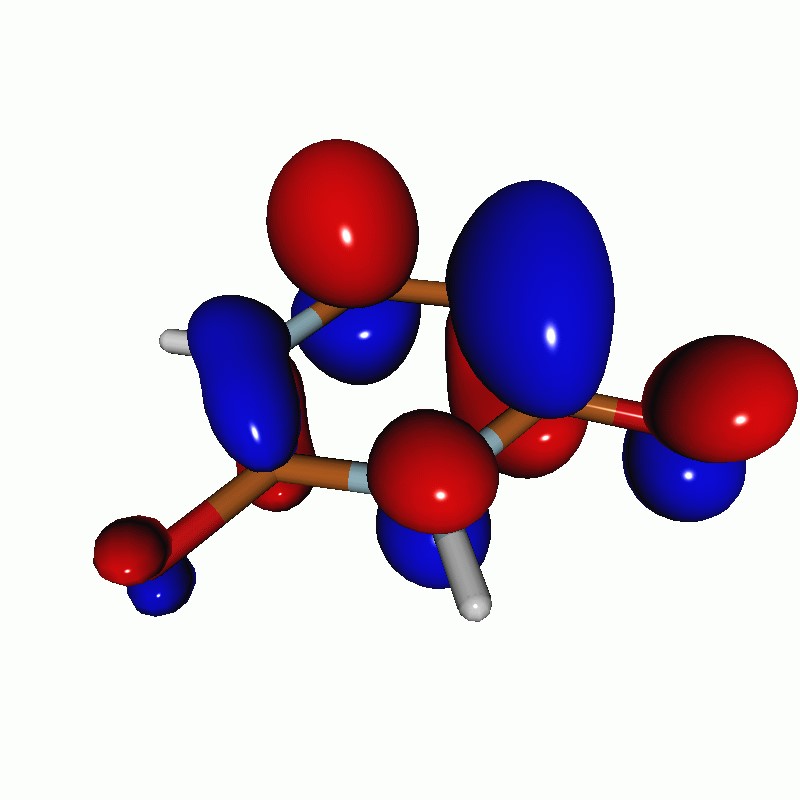}
     \end{minipage}
     \\
     \hline
     S$_3$(A$^{\prime\prime}$,$\pi\mathrm{Ryd}$) & 6.07 & 0.0027 &
     \begin{minipage}{0.06\textwidth}
         \centering
         \includegraphics[scale=0.05]{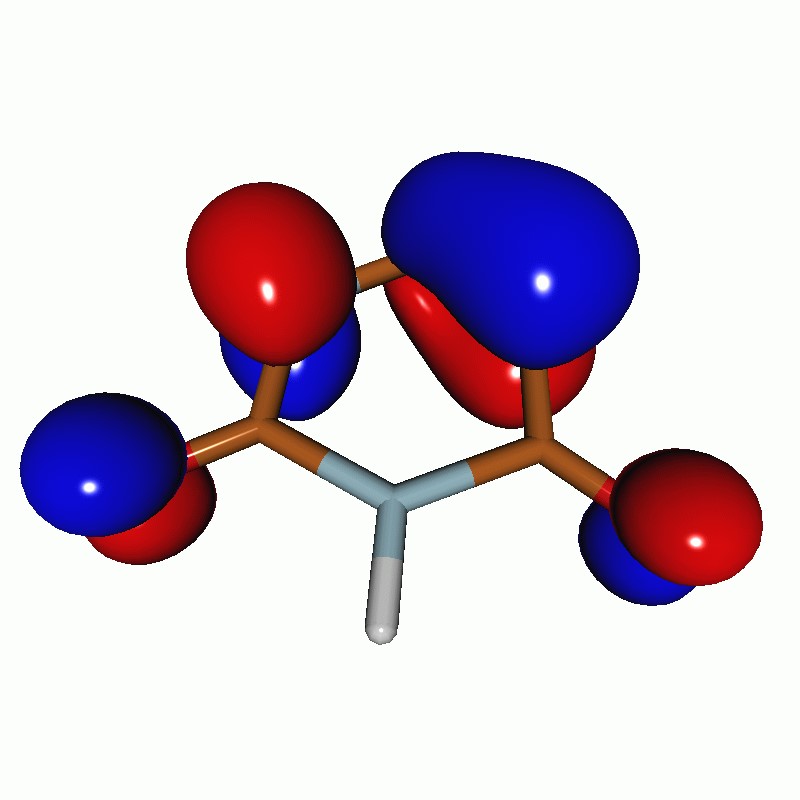}
     \end{minipage}
     & 0.85
     &  \begin{minipage}{0.06\textwidth}
         \centering
         \includegraphics[scale=0.05]{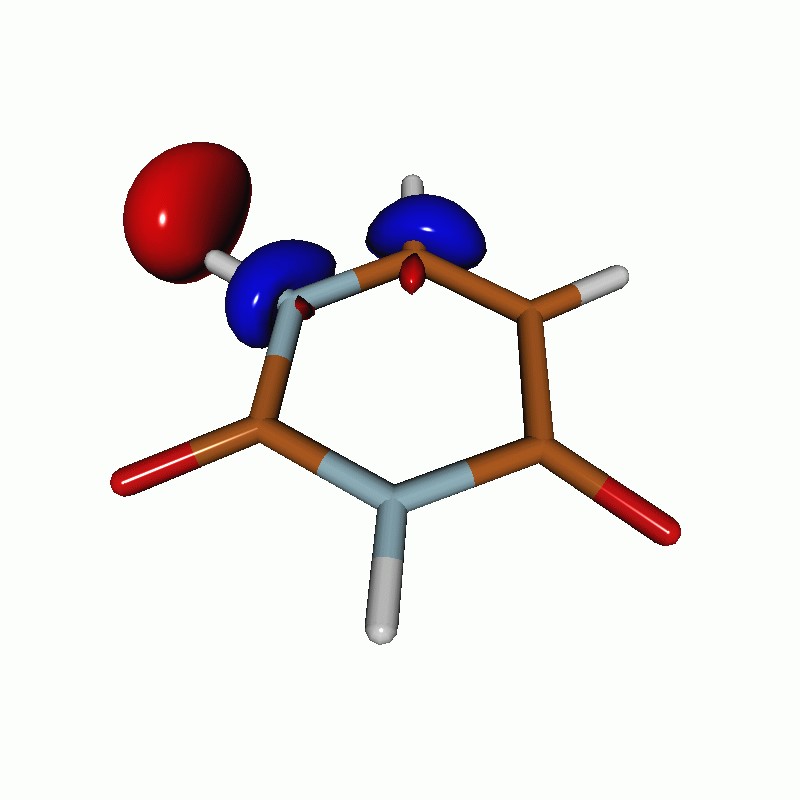}
     \end{minipage}
     \\
     \hline
     S$_5$(A',$\pi\pi^{\ast}$) & 6.74 & 0.0573 &
     \begin{minipage}{0.06\textwidth}
         \centering
         \includegraphics[scale=0.05]{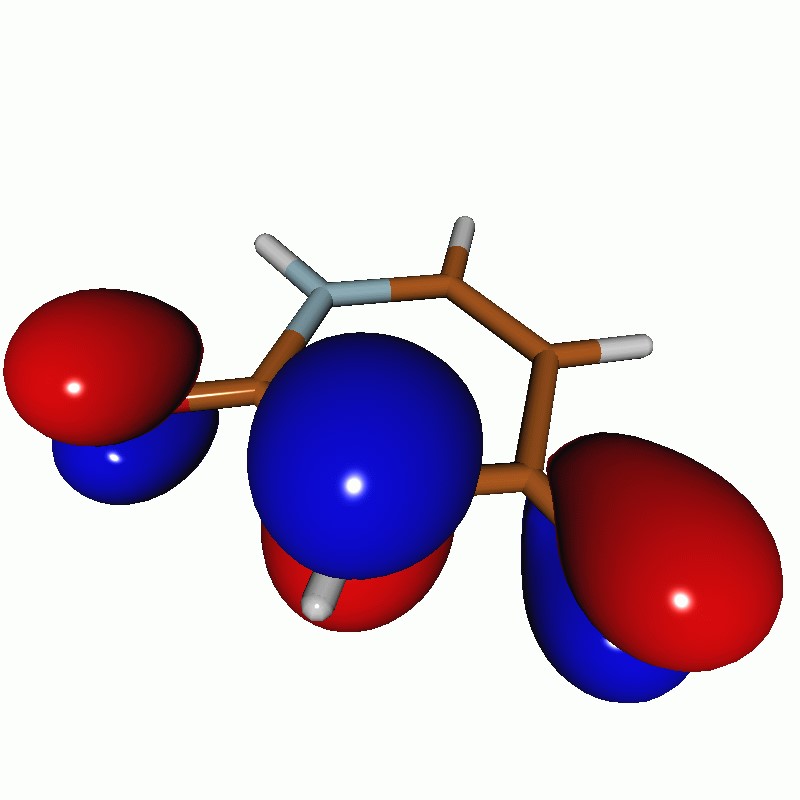}
     \end{minipage}
     & 0.73
     &  \begin{minipage}{0.06\textwidth}
         \centering
         \includegraphics[scale=0.05]{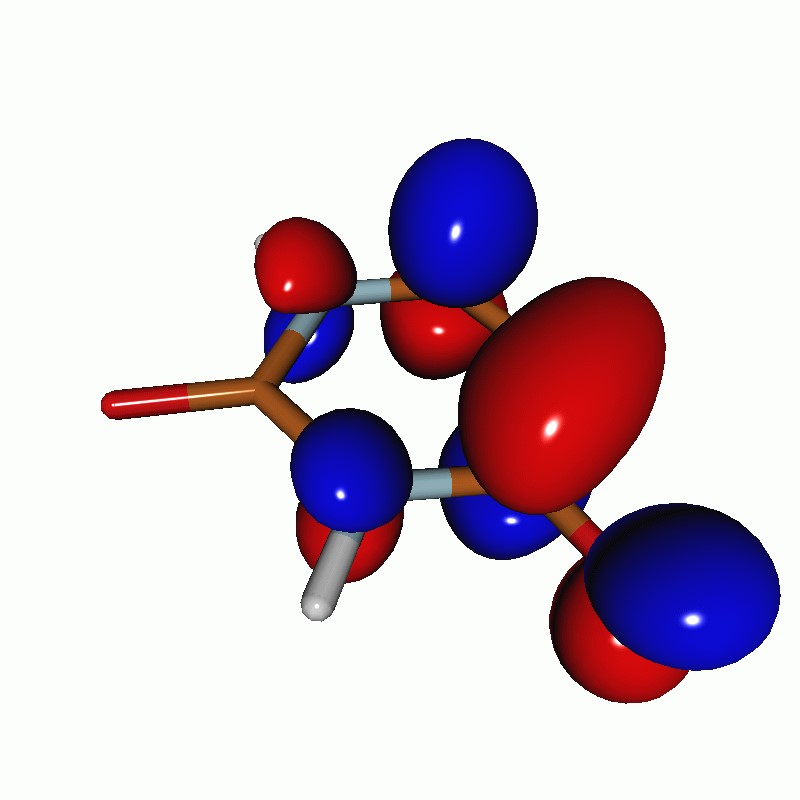}
     \end{minipage}
     \\
     \hline
     \end{tabular}
 \label{tab:uracil_NTO_gs2ex}
 \end{table}

Table~\ref{tab:thymine_valence_energy} shows the excitation energies of the five lowest triplet and singlet states of thymine, along with the experimental values obtained by EELS. \cite{Chernyshova2013} 
We did not find literature data for the UV absorption of thymine in the gas phase. 
The energetic order is based on EOM-EE-CCSD. 
Here, we reassign the peaks of the EELS spectra~\cite{Chernyshova2013} on the basis of the following considerations: $(i)$ optically bright transitions also exhibit strong peaks in the EELS spectra; $(ii)$ the excitation energy of a triplet state is lower than the excitation energy of the singlet state with the same electronic configuration; $(iii)$ the strengths of the transitions to triplet states are smaller than the strengths of the transitions to singlet states; $(iv)$ among the excitations enabled by  spin-orbit coupling, $\pi \to \pi^{\ast}$ transitions have relatively large transition moment.

Except for T$_1$($\pi\pi^{\ast}$), the ADC(2) excitation energies are red-shifted relative to EOM-CCSD. 
As a result, the ADC(2) excitation energies of the states considered here
are closest, in absolute values, to the experimental values from Table~\ref{tab:thymine_valence_energy}. 
However, the energy differences between the singlet states (S$_1$, S$_2$, S$_4$, and S$_5$) 
are much better reproduced by EOM-CCSD. 
TDDFT/B3LYP accurately reproduces the excitation energies of the T$_2$(n$\pi^{\ast}$), S$_1$(n$\pi^{\ast}$), and S$_2$($\pi\pi^{\ast}$) states.

\begin{table}[htb]
    \centering
    \caption{Thymine. Excitation energies (eV) at the FC geometry compared with the experimental values from EELS. \cite{Chernyshova2013} The oscillator strengths are from EOM-EE-CCSD, and used for the re-assignment.}
    \begin{tabular}{l|ccc|c|c}
        \hline
         & ADC(2) & EOM-CCSD & TDDFT & EELS & Osc. strength
         \\
         \hline
         T$_1$($\pi\pi^{\ast}$) & 3.70 & 3.63 & 3.19 & 3.66 & -
         \\
         \hline
         T$_2$(n$\pi^{\ast}$) & 4.39 & 4.81 & 4.25 & 4.20 & -
         \\
         \hline
         S$_1$(n$\pi^{\ast}$) & 4.60 & 5.08 & 4.64 & 4.61 & 0.0000
         \\
         \hline
         S$_2$($\pi\pi^{\ast}$) & 5.18 & 5.48 & 4.90 & 4.96 & 0.2289
         \\
         \hline
         T$_3$($\pi\pi^{\ast}$) & 5.27 & 5.32 & 4.61 & 5.41 & -
         \\
         \hline
         T$_4$($\pi\mathrm{Ryd}$) & 5.66 & 5.76 & 5.39 & - & -
         \\
         \hline
         S$_3$($\pi\mathrm{Ryd}$) & 5.71 & 5.82 & 5.46 & - & 0.0005
         \\
         \hline
         T$_5$($\pi\pi^{\ast}$) & 5.87 & 5.91 & 5.10 & 5.75 & -
         \\
         \hline
         S$_4$(n$\pi^{\ast}$) & 5.95 & 6.45 & 5.72 & 5.96 & 0.0000
         \\
         \hline
         S$_5$($\pi\pi^{\ast}$) & 6.15 & 6.63 & 5.87 & 6.17 & 0.0679
         \\
         \hline
    \end{tabular}
    \label{tab:thymine_valence_energy}
\end{table}

Table~\ref{tab:AcAc_valence_energy} shows  the excitation energies of the two lowest triplet and singlet states, and the lowest Rydberg states of acetylacetone, along with the 
experimental values obtained from EELS~\cite{Walzl} and UV absorption~\cite{UV_AcAc} (the exact state ordering of states in the singlet Rydberg manifold is unknown). 
Table~\ref{tab:AcAc_NTO_gs2ex} shows the NTOs obtained at the EOM-EE-CCSD/6-311++G** level.
Remarkably, for this molecule the excitation energies from EELS agree well with those from UV absorption. 
Note that the EELS spectra of acetylacetone were recorded with incident electron energies of 25 and 100 eV,~\cite{Walzl} whereas those for uracil~\cite{Chernyshova2012} were obtained with 0--8.0 eV. 
The higher incident electron energies reduce the effective acceptance angle of the electrons, which may hinder the detection of electrons that have undergone momentum transfer.
The transitions to the T$_1$($\pi\pi^{\ast}$) and T$_2$(n$\pi^{\ast}$) states  appeared only with the 25 eV incident electron energy and a scattering angle of 90\degree\  (see Fig. 3 of Ref.~\citenum{Walzl}). The peaks were broad and, furthermore, an order of magnitude less intense than the {S$_0 \to ~$S$_2$}($\pi\pi^{\ast}$) transition.  Consequently, it is difficult to resolve the excitation energies of T$_1$($\pi\pi^{\ast}$) and T$_2$(n$\pi^{\ast}$). ADC(2) yields the best match with the experimental results for acetylacetone.

\begin{table}[h]
    \centering
    \caption{Acetylacetone. Excitation energies (eV) at the FC geometry compared with the values obtained in EELS~\cite{Walzl} and UV absorption spectroscopy.~\cite{UV_AcAc}}
    \begin{tabular}{c|cccc|c|c}
        \hline
         & ADC(2) & ADC(2)-x & EOM-CCSD & TDDFT & EELS & UV
         \\
         \hline
         T$_1$($\pi\pi^{\ast}$) & 3.76 & 3.16 & 3.69 & 3.23 & 3.57? & -
         \\
         \hline
         T$_2$(n$\pi^{\ast}$) & 3.79 & 3.13 & 4.11 & 3.75 & ? & -
         \\
         \hline
         S$_1$(n$\pi^{\ast}$) & 4.03 & 3.29 & 4.39 & 4.18 & 4.04 & 4.2
         \\
         \hline
         S$_2$($\pi\pi^{\ast}$) & 4.96 & 4.28 & 5.24 & 5.08 & 4.70 & 4.72
         \\
         \hline
         T$_3(\pi\mathrm{Ryd})$ & 5.91 & 5.45 & 6.02 & 5.66 & 5.52 & -
         \\
         \hline
         S$_{3?}(\pi\mathrm{Ryd})$ & 5.98 & 5.53 & 6.13 & 5.72 & 5.84 & 5.85
         \\
         \hline
         S$_{5?}(\pi\mathrm{Ryd})$ & 6.87 & 6.30 & 7.06 & 6.64 & 6.52 & 6.61
         \\
         \hline
    \end{tabular}
    \label{tab:AcAc_valence_energy}
\end{table}

\begin{table}[h]
 \centering
 \caption{Acetylacetone. EOM-EE-CCSD/6-311++G** NTOs of the excitations from the ground state to the lowest-lying valence-excited states at the FC geometry (NTO isosurface is 0.03 for the Rydberg transitions and 0.05 for the rest).}
 \begin{tabular}{c|c|c|ccc}
     \hline
     Final state & $E^{\mathrm{ex}}$ (eV) & Osc. strength & Hole & $\sigma_K^2$ & Particle 
     \\
     \hline
     T$_1$(A',$\pi\pi^{\ast}$) & 3.69 & - &
     \begin{minipage}{0.06\textwidth}
         \centering
         \includegraphics[scale=0.04]{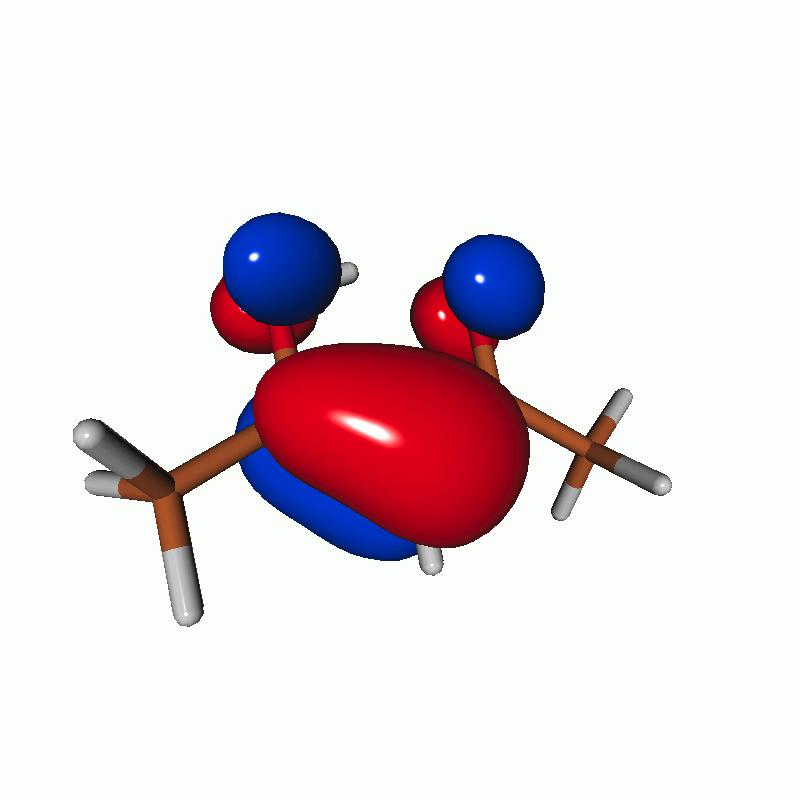}
     \end{minipage}
     & 0.82
     &  \begin{minipage}{0.06\textwidth}
         \centering
         \includegraphics[scale=0.04]{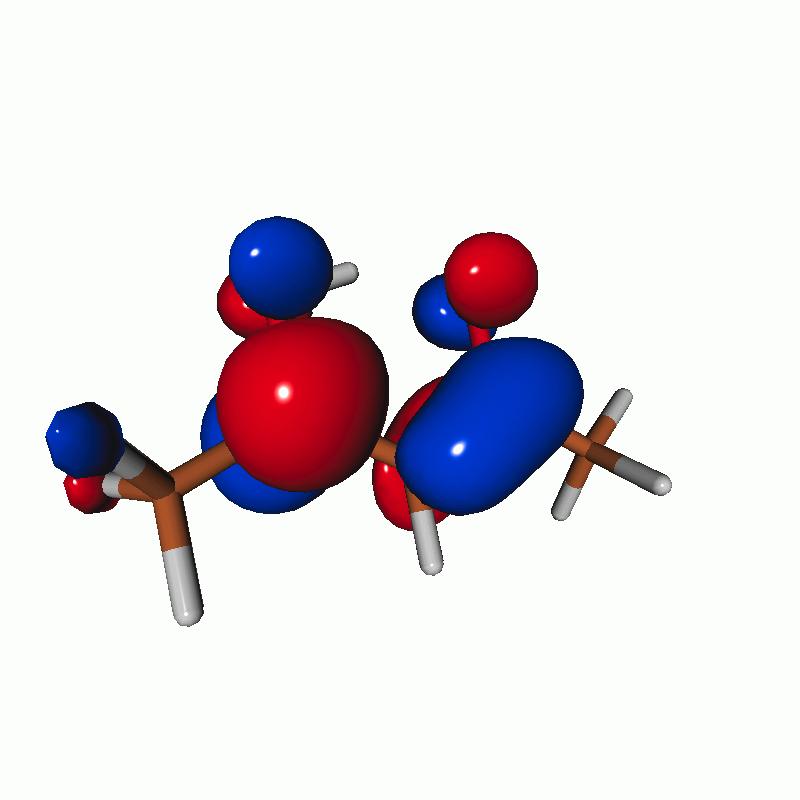}
     \end{minipage}
     \\
     \hline
     T$_2$(A",n$\pi^{\ast}$) & 4.11 & - &
     \begin{minipage}{0.06\textwidth}
         \centering
         \includegraphics[scale=0.04]{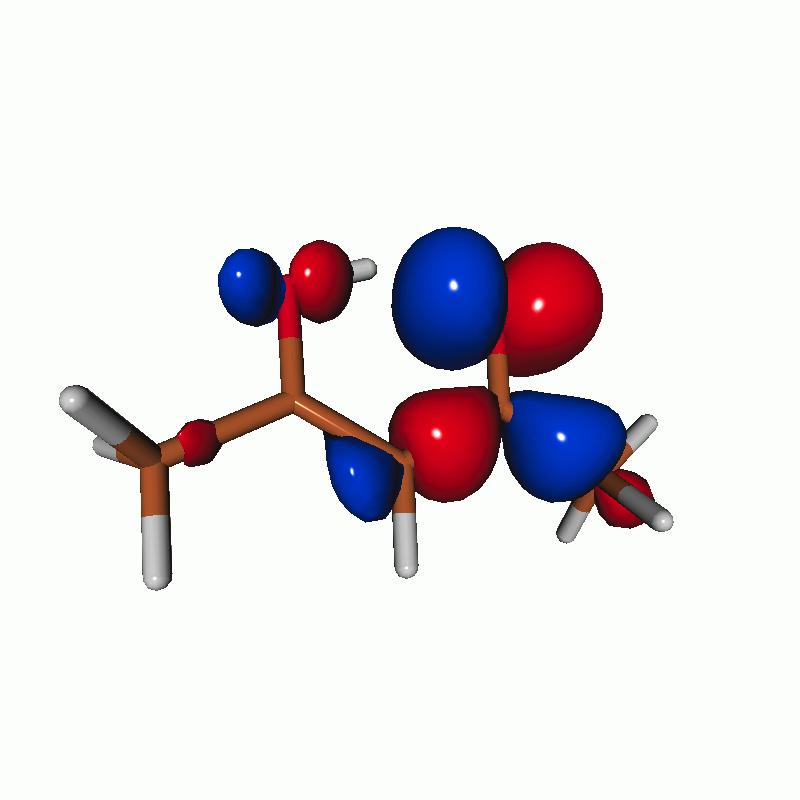}
     \end{minipage}
     & 0.82
     &  \begin{minipage}{0.06\textwidth}
         \centering
         \includegraphics[scale=0.04]{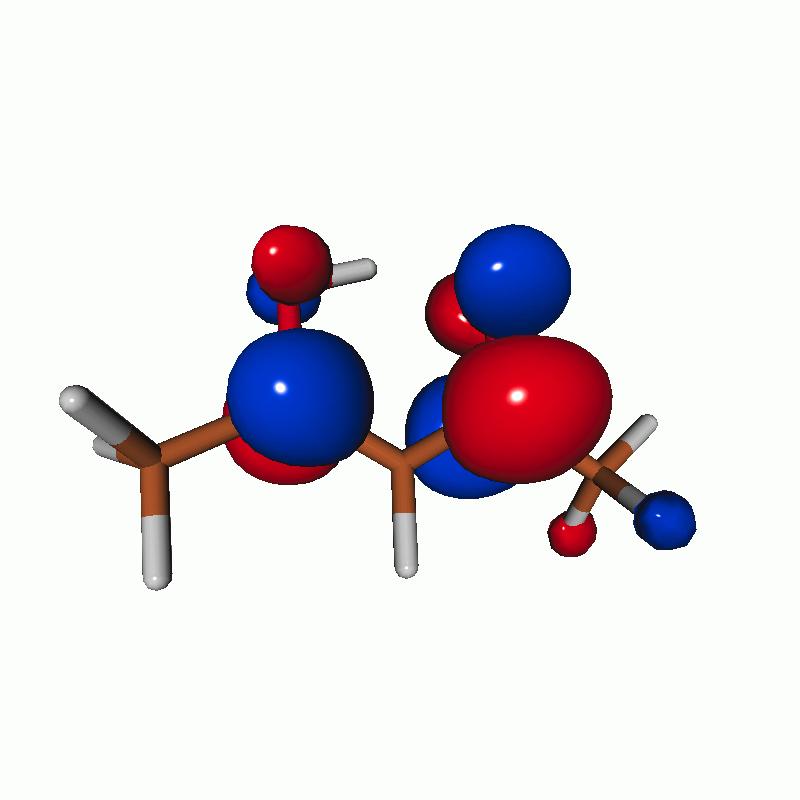}
     \end{minipage}
     \\
     \hline
     S$_1$(A",n$\pi^{\ast}$) & 4.39 & 0.0006 &
     \begin{minipage}{0.06\textwidth}
         \centering
         \includegraphics[scale=0.04]{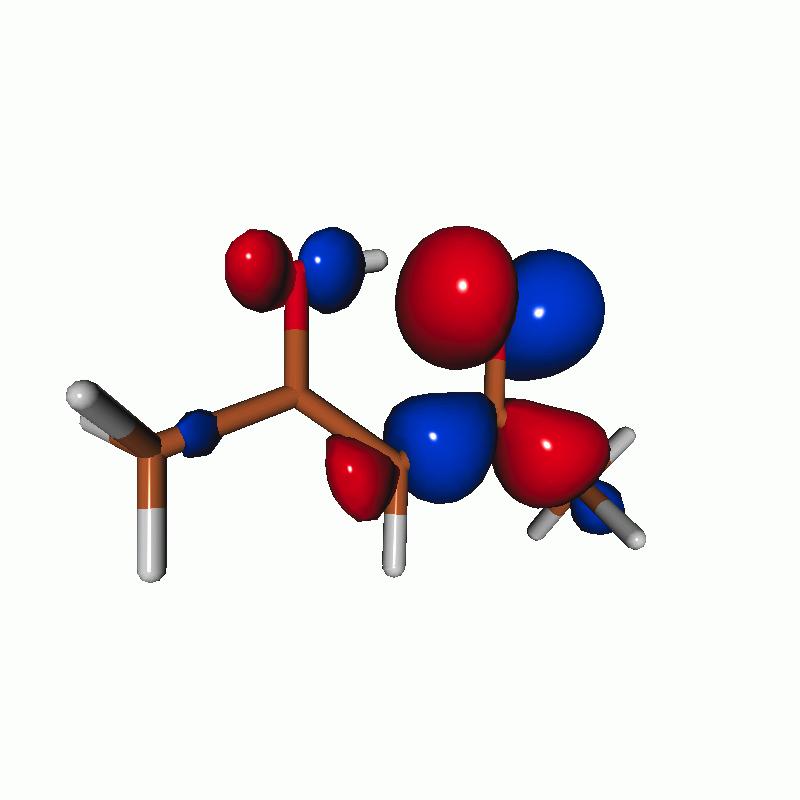}
     \end{minipage}
     & 0.81
     &  \begin{minipage}{0.06\textwidth}
         \centering
         \includegraphics[scale=0.04]{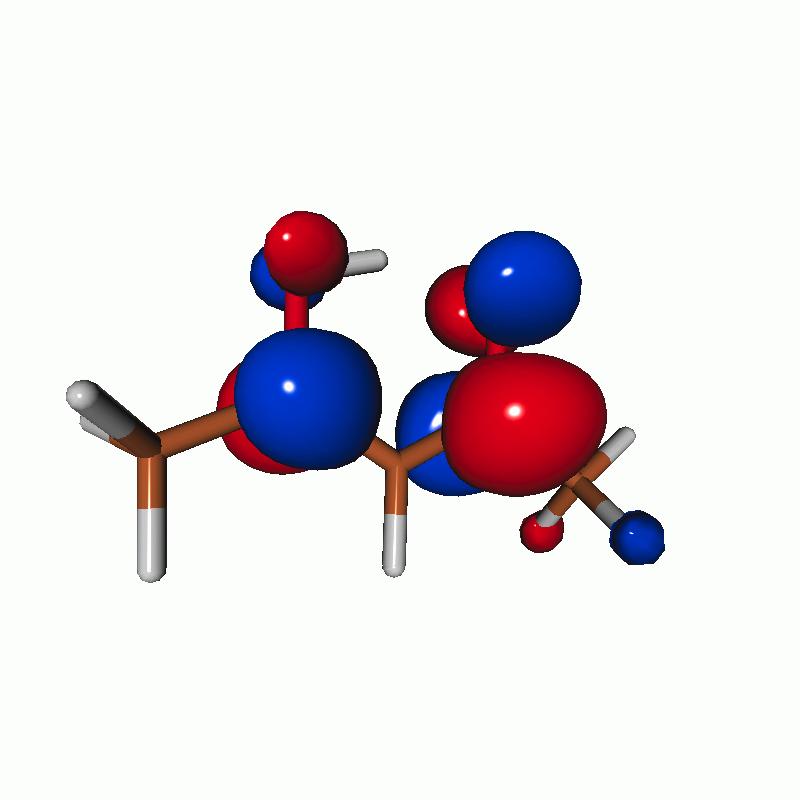}
     \end{minipage}
     \\
     \hline
     S$_2$(A',$\pi\pi^{\ast}$) & 5.24 & 0.3299 &
     \begin{minipage}{0.06\textwidth}
         \centering
         \includegraphics[scale=0.04]{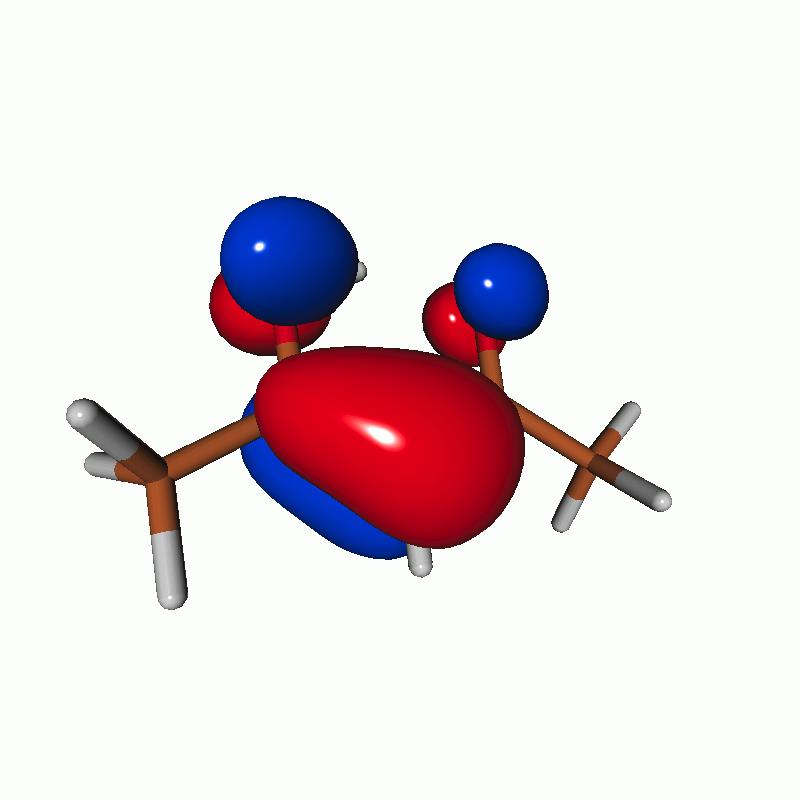}
     \end{minipage}
     & 0.77
     &  \begin{minipage}{0.06\textwidth}
         \centering
         \includegraphics[scale=0.04]{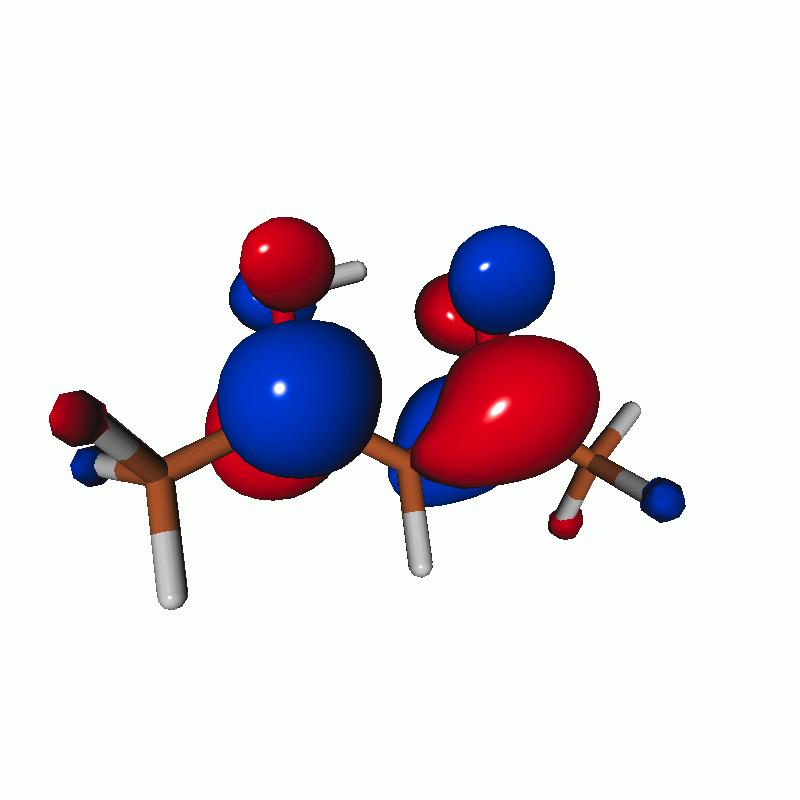}
     \end{minipage}
     \\
     \hline
     T$_3[\pi\mathrm{Ryd}(s)]$ & 6.02 & - &
     \begin{minipage}{0.06\textwidth}
         \centering
         \includegraphics[scale=0.04]{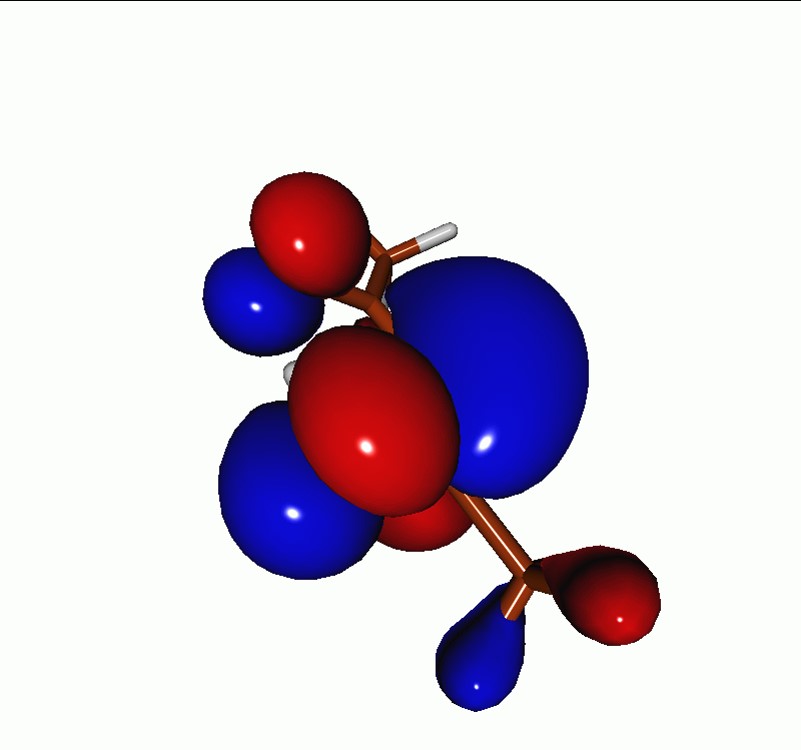}
     \end{minipage}
     & 0.86
     &
     \begin{minipage}{0.06\textwidth}
         \centering
         \includegraphics[scale=0.04]{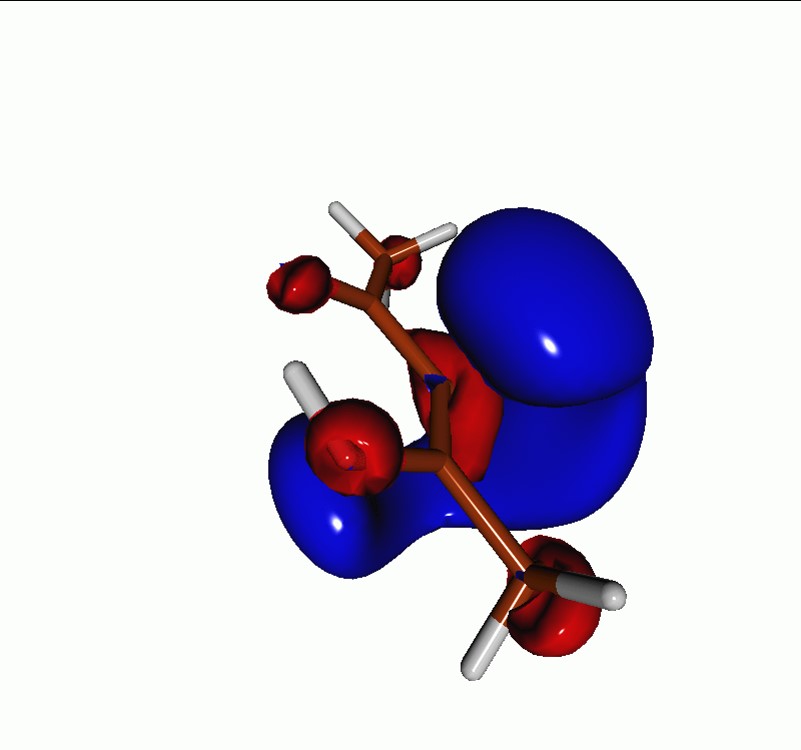}
     \end{minipage}
     \\
     \hline
     S$_{3?}[\pi\mathrm{Ryd}(s)]$ & 6.13 & 0.0072 &
     \begin{minipage}{0.06\textwidth}
         \centering
         \includegraphics[scale=0.04]{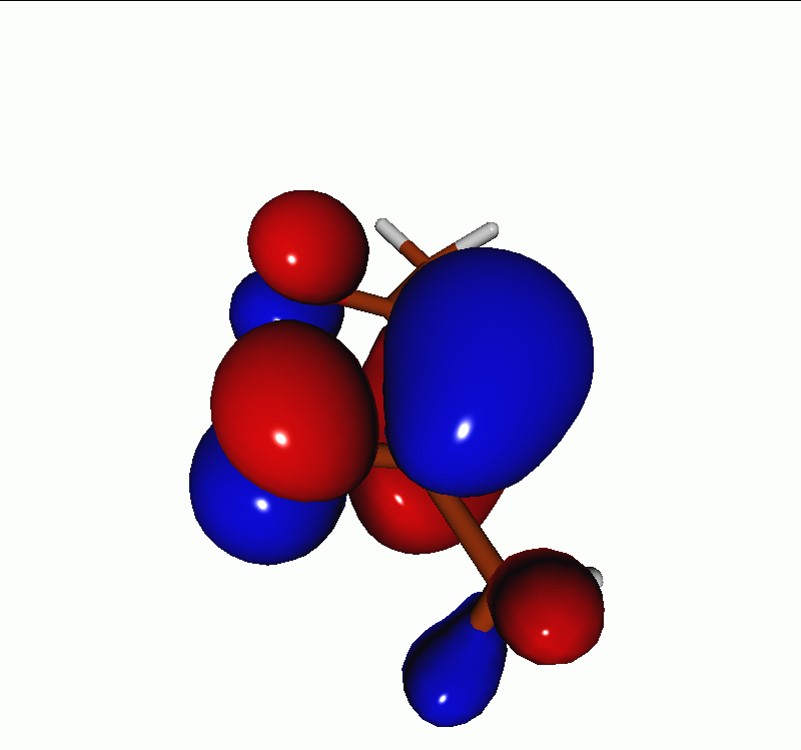}
     \end{minipage}
     & 0.86
     &
     \begin{minipage}{0.06\textwidth}
         \centering
         \includegraphics[scale=0.04]{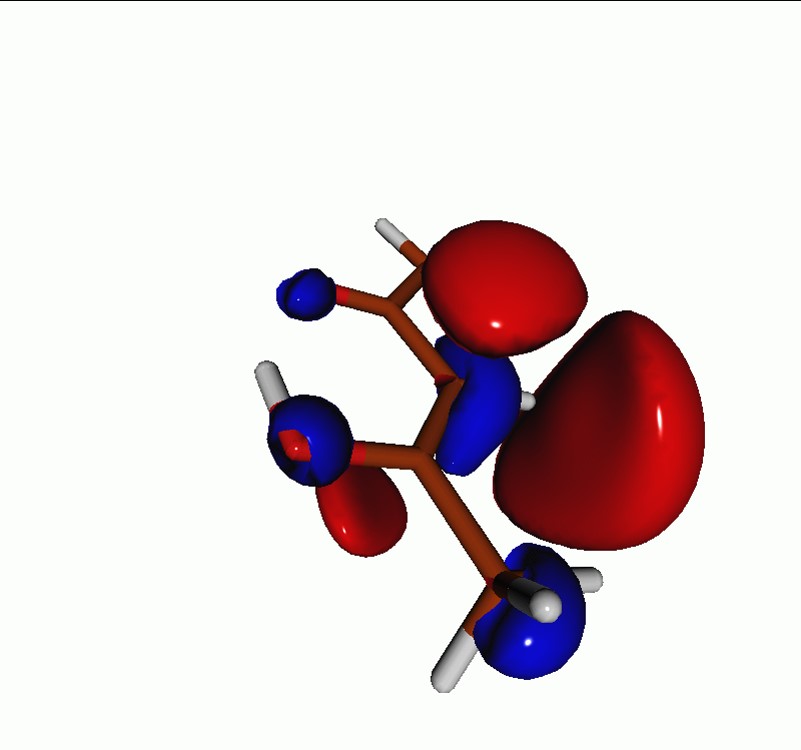}
     \end{minipage}
     \\
     \hline
     S$_{5?}[\pi\mathrm{Ryd}(p)]$ & 7.06 & 0.0571 &
     \begin{minipage}{0.06\textwidth}
         \centering
         \includegraphics[scale=0.04]{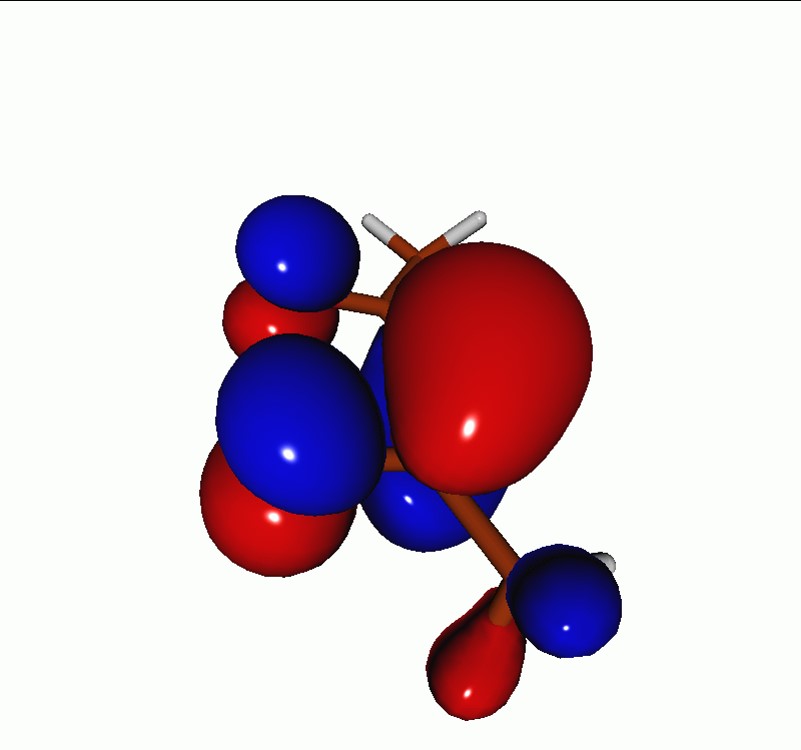}
     \end{minipage}
     & 0.85
     &
     \begin{minipage}{0.06\textwidth}
         \centering
         \includegraphics[scale=0.04]{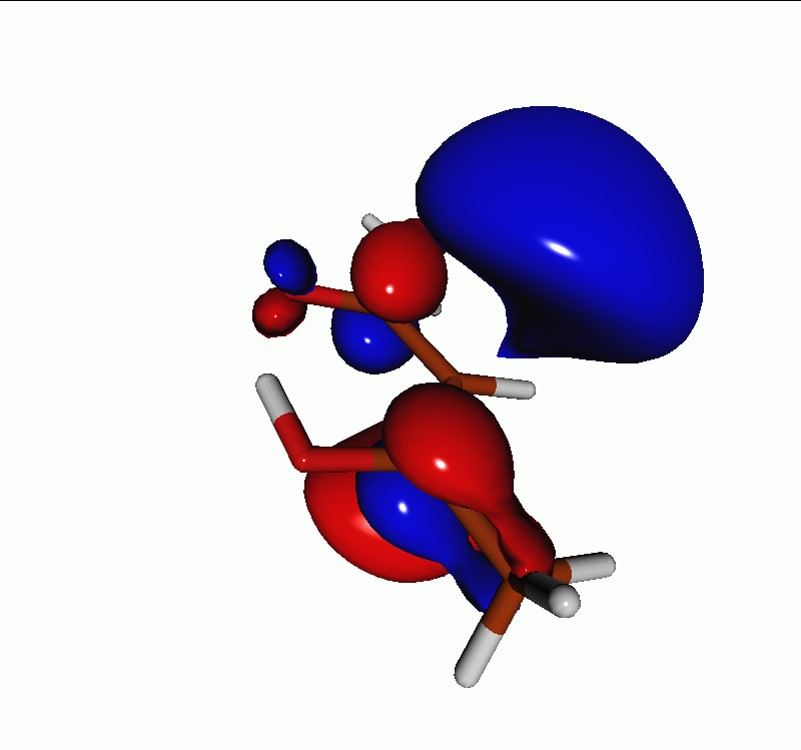}
     \end{minipage}
     \\
     \hline
     \end{tabular}
 \label{tab:AcAc_NTO_gs2ex}
 \end{table}

These results indicate that the excitation energies of the valence-excited states computed by EOM-EE-CCSD, ADC(2), and TDDFT/B3LYP are equally (in)accurate.
Which method yields the best match with experiment depends on the molecule.

\subsection{Core Excitations from the Valence-Excited States}
\label{subsec:valence_to_core}

In Secs.~\ref{subsec:gs_to_core}
and \ref{subsec:gs_to_valence}, 
we analyzed two of our three desiderata for a good electronic structure method for TR-XAS---that is, the ability to yield accurate results for ground-state XAS as well as for the valence-excited states involved in the dynamics. 
In this subsection, we focus on the remaining item, {\em{i.e.}}, the ability to yield accurate XAS of valence-excited states.

For uracil, we confirmed that EOM-CCSD and CVS-EOM-CCSD yield fairly accurate results for the valence-excited T$_1$($\pi\pi^{\ast}$), T$_2$($\mathrm{n}\pi^{\ast}$), S$_1$($\mathrm{n}\pi^{\ast}$), and S$_2$($\pi\pi^{\ast}$)  states 
and for the (final) singlet (O$_{1s}$) core-excited states at the FC geometry, respectively. 
It is thus reasonable to consider the oxygen K-edge XAS spectra of the S$_1$($\mathrm{n}\pi^{\ast}$) and S$_2$($\pi\pi^{\ast}$) states of uracil obtained from 
CVS-EOM-CCSD as our reference, 
even though 
CVS-EOM-CCSD only yields the peaks of the core-to-SOMO transitions. 

\begin{figure}[hbt]
    \centering
    \includegraphics[width=8cm,height=8cm,keepaspectratio]{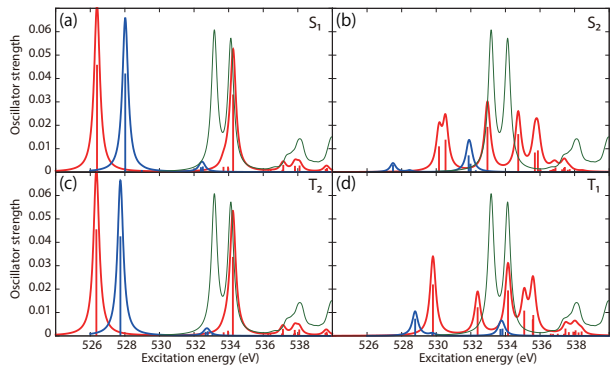}
    \caption{Uracil. Oxygen K-edge NEXAFS of the four lowest-lying valence states: (a) S$_1$($\mathrm{n}\pi^{\ast}$); (b) S$_2$($\pi\pi^{\ast}$); (c) T$_2$($\mathrm{n}\pi^{\ast}$); and (d) T$_1$($\pi\pi^{\ast}$)].
    The blue and red curves correspond to the CVS-EOM-CCSD and LSOR-CCSD results, respectively. Note that the HSOR spectra for S$_1$ and S$_2$ are identical to the LSOR-CCSD spectra for T$_2$ and T$_1$.
    Basis set: 6-311++G**. FC geometry. The ground state XAS (green  curve) is included for comparison.}
    \label{fig:uracil_FC}
\end{figure}

Fig.~\ref{fig:uracil_FC} shows the oxygen K-edge XAS of uracil in the 
(a)~S$_1$($\mathrm{n}\pi^{\ast}$), 
(b)~S$_2$($\pi\pi^{\ast}$), 
(c)~T$_2$($\mathrm{n}\pi^{\ast}$), and 
(d)~T$_1$($\pi\pi^{\ast}$) states, calculated using CVS-EOM-CCSD (blue curve) and LSOR-CCSD (red curve) at the FC geometry. 
Note that the HSOR-CCSD spectra of S$_1$($\mathrm{n}\pi^{\ast}$) and S$_2$($\pi\pi^{\ast}$) are identical to the LSOR-CCSD spectra for the T$_2$($\mathrm{n}\pi^{\ast}$) and T$_1$($\pi\pi^{\ast}$) states, respectively, because  their orbital electronic configuration are the same, see Table~\ref{tab:uracil_NTO_gs2ex}. 
The ground-state spectrum
(green curve) is included in all panels for comparison. 
The LSOR-CCSD NTOs of the transitions underlying the peaks in the S$_1$($\mathrm{n}\pi^{\ast}$), S$_2$($\pi\pi^{\ast}$) and T$_1$($\pi\pi^{\ast}$) spectra 
are given in Tabs.~\ref{tab:uracil_NTO_momS1_fc}, \ref{tab:uracil_NTO_momS2_fc}, and \ref{tab:uracil_NTO_momT1_fc}, respectively.

\begin{table}[hbt]
 \centering
 \caption{Uracil. LSOR-CCSD/6-311++G** NTOs of the O$_{1s}$ core excitations from the S$_1$ (n$\pi*$) 
 state at the FC geometry (NTO isosurface value is 0.05).}
 \begin{tabular}{c|c|c|ccc}
     \hline
     $E^{\mathrm{ex}}$ (eV) & Osc. strength & Spin & Hole & $\sigma_K^2$ & Particle 
     \\
     \hline
     526.39 & 0.0451 & $\alpha$ &
     \begin{minipage}{0.1\textwidth}
         \centering
         \includegraphics[scale=0.05]{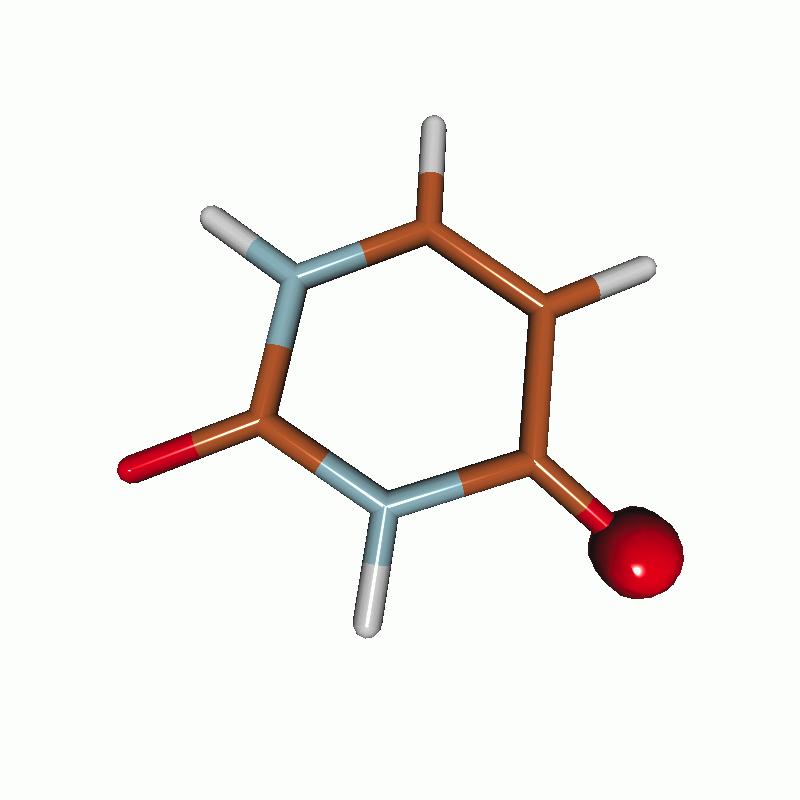}
     \end{minipage}
     & 0.86
     &  \begin{minipage}{0.1\textwidth}
         \centering
         \includegraphics[scale=0.05]{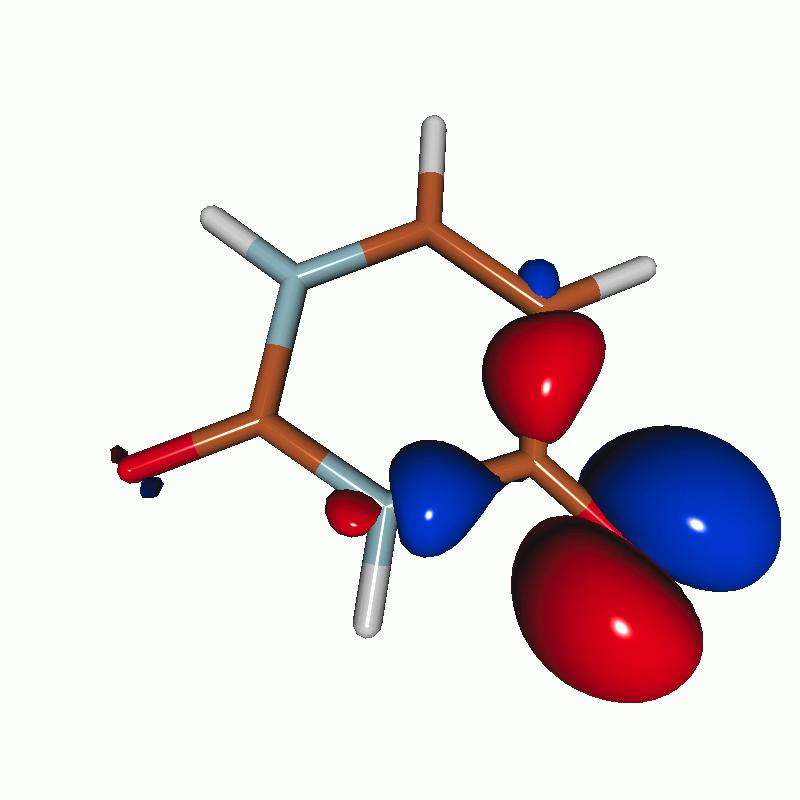}
     \end{minipage}
     \\
     \hline
     534.26 & 0.0323 & $\alpha$ &
     \begin{minipage}{0.1\textwidth}
         \centering
         \includegraphics[scale=0.05]{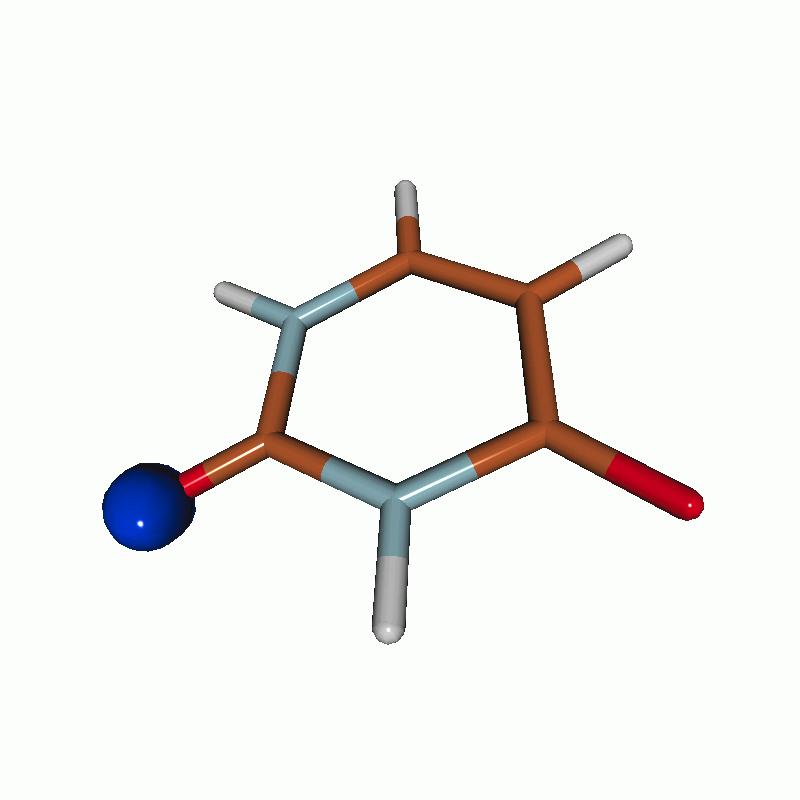}
     \end{minipage}
     & 0.56
     &  \begin{minipage}{0.1\textwidth}
         \centering
         \includegraphics[scale=0.05]{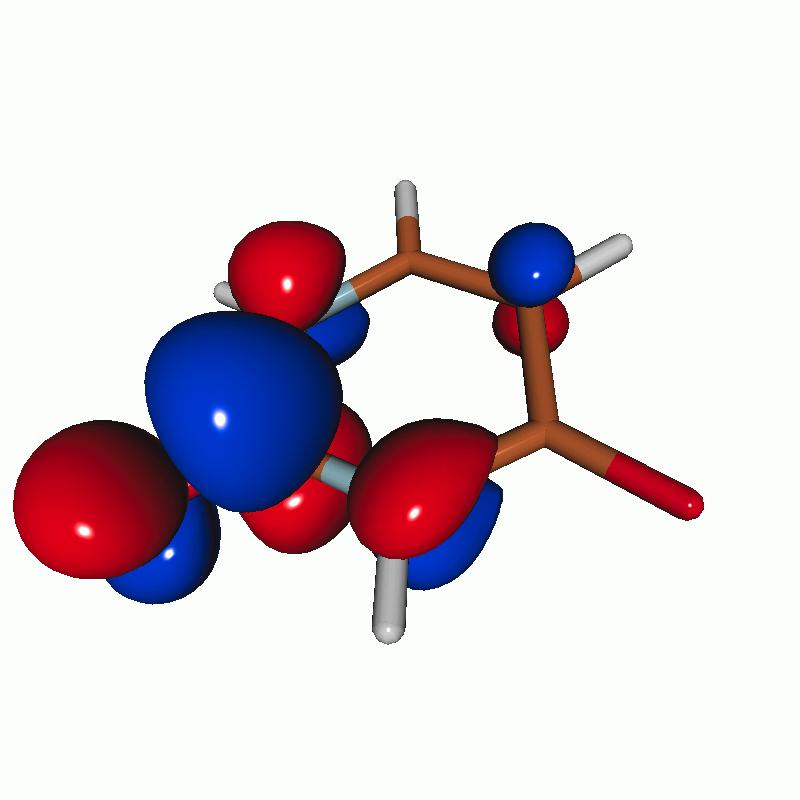}
     \end{minipage}
     \\
     & & $\beta$ &
     \begin{minipage}{0.1\textwidth}
         \centering
         \includegraphics[scale=0.05]{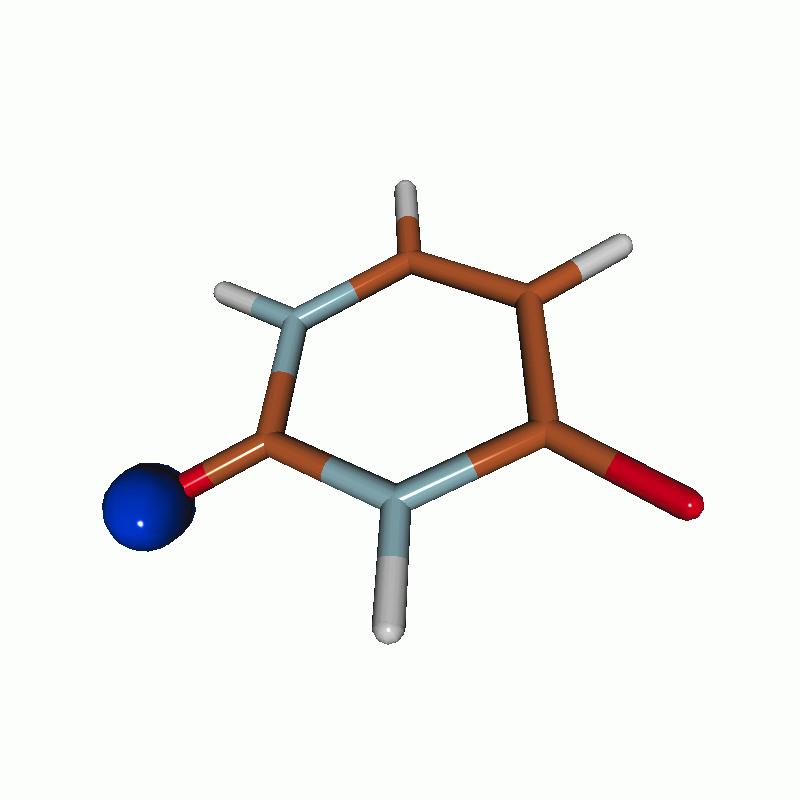}
     \end{minipage}
     & 0.23
     &  \begin{minipage}{0.1\textwidth}
         \centering
         \includegraphics[scale=0.05]{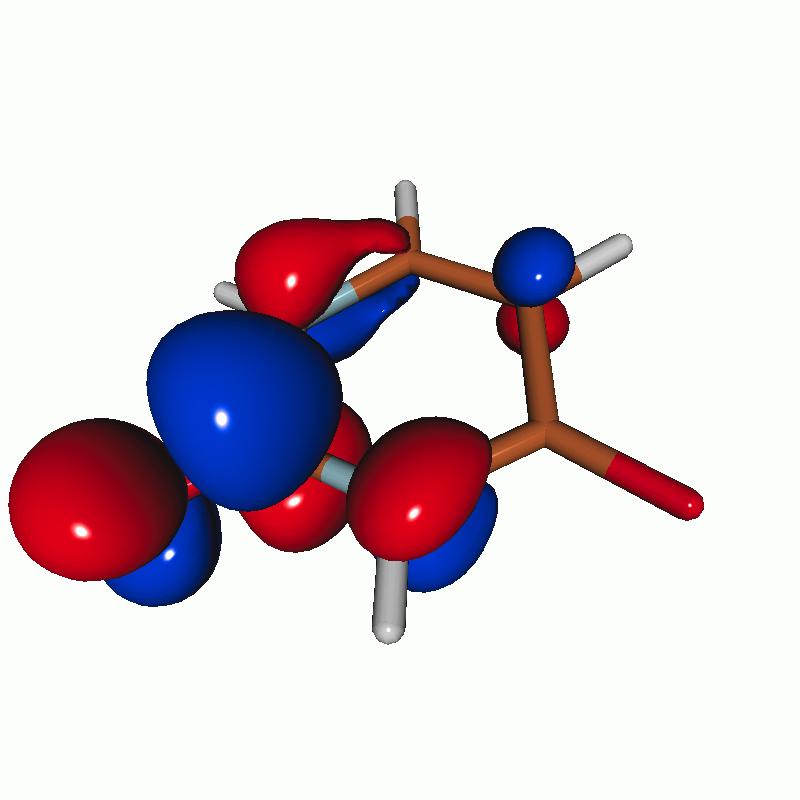}
     \end{minipage}
     \\
     \hline
     \end{tabular}
 \label{tab:uracil_NTO_momS1_fc}
 \end{table}

\begin{table}[hbt]
 \centering
 \caption{Uracil. LSOR-CCSD/6-311++G** NTOs of the O$_{1s}$ core excitations from the 
 S$_2$($\pi\pi^*$) state at the FC geometry 
 (NTO isosurface value is 0.05).}
 \begin{tabular}{c|c|c|ccc}
     \hline
     $E^{\mathrm{ex}}$ (eV) & Osc. strength & Spin & Hole & $\sigma_K^2$ & Particle  
     \\
     \hline
     530.16 & 0.0102 & $\alpha$ &
     \begin{minipage}{0.1\textwidth}
         \centering
         \includegraphics[scale=0.05]{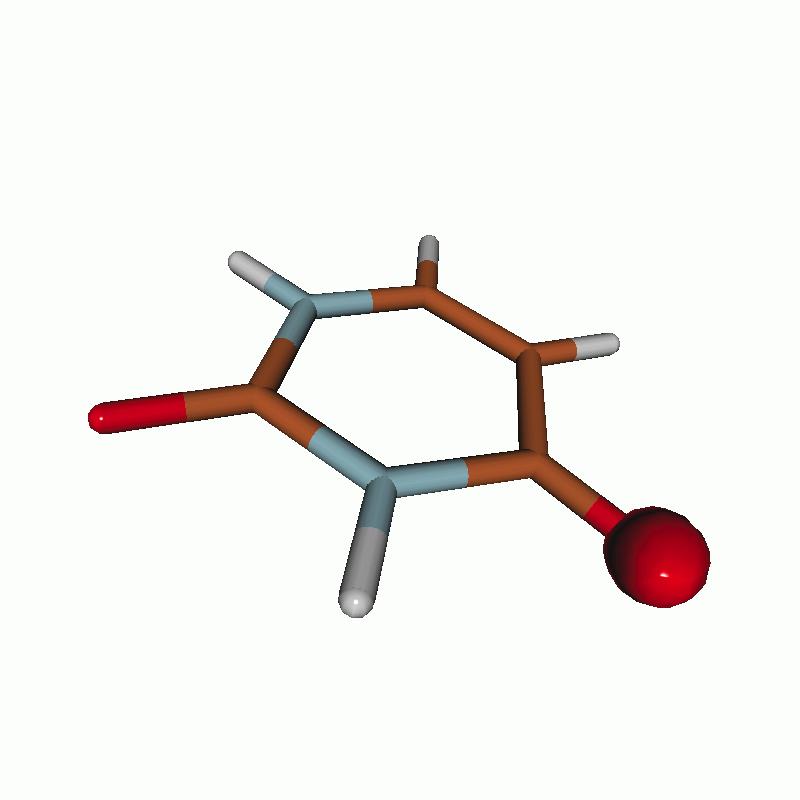}
     \end{minipage}
     & 0.68
     &  \begin{minipage}{0.1\textwidth}
         \centering
         \includegraphics[scale=0.05]{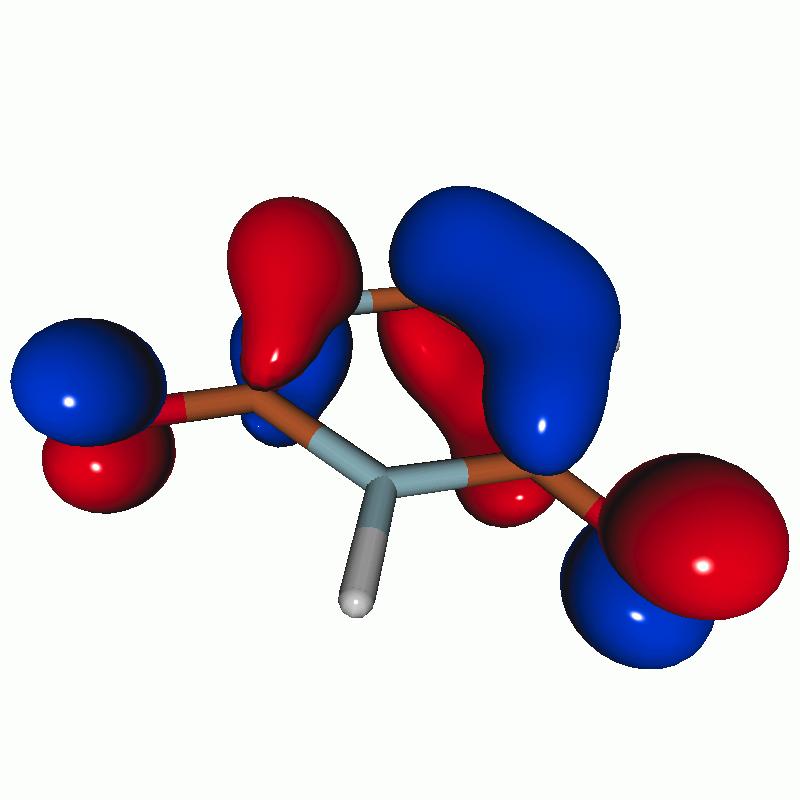}
     \end{minipage}
     \\
     \hline
     530.54 & 0.0131 & $\alpha$ &
     \begin{minipage}{0.1\textwidth}
         \centering
         \includegraphics[scale=0.05]{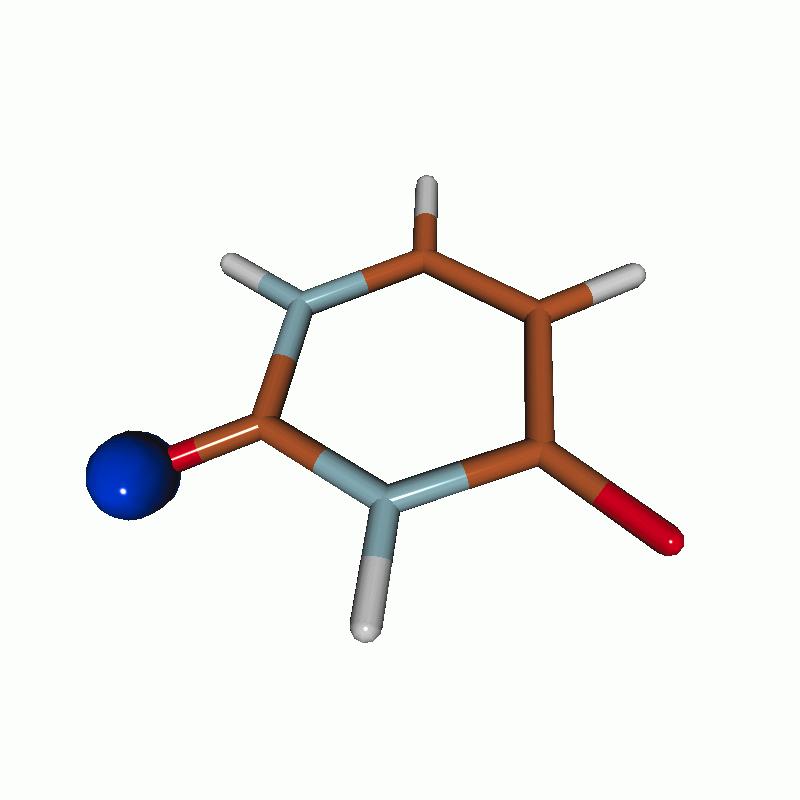}
     \end{minipage}
     & 0.67
     &  \begin{minipage}{0.1\textwidth}
         \centering
         \includegraphics[scale=0.05]{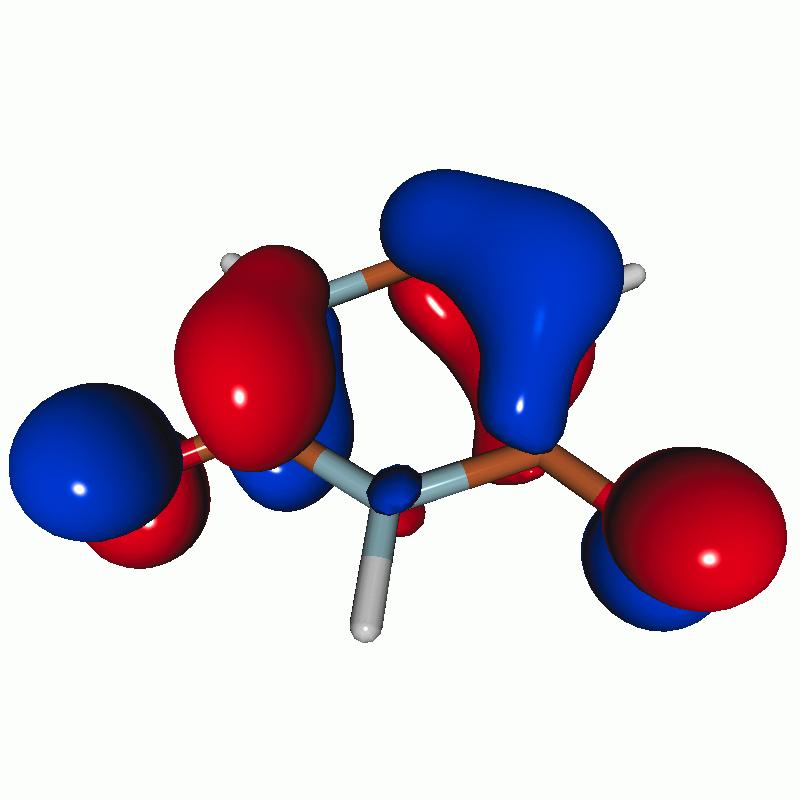}
     \end{minipage}
     \\
     \hline
     532.96 & 0.0186 & $\beta$ &
     \begin{minipage}{0.1\textwidth}
         \centering
         \includegraphics[scale=0.05]{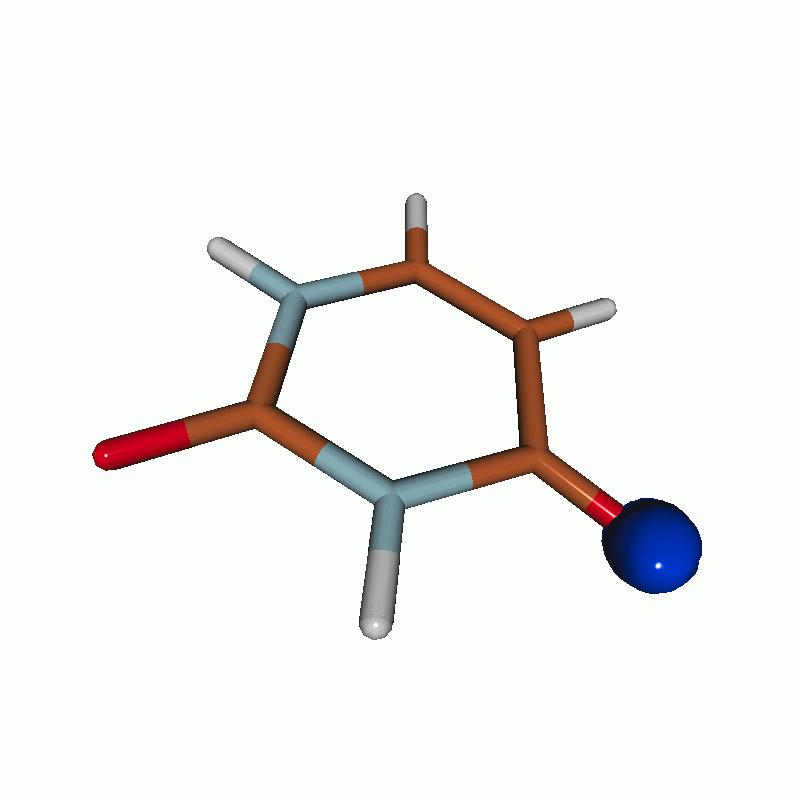}
     \end{minipage}
     & 0.74
     &  \begin{minipage}{0.1\textwidth}
         \centering
         \includegraphics[scale=0.05]{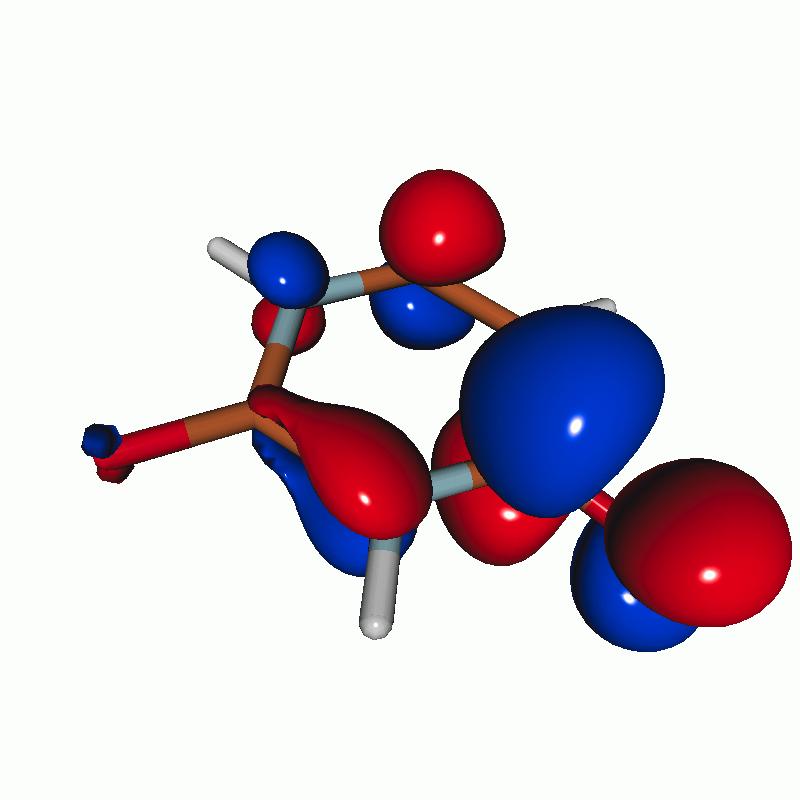}
     \end{minipage}
     \\
     \hline
     534.74 & 0.0155 & $\beta$ &
     \begin{minipage}{0.1\textwidth}
         \centering
         \includegraphics[scale=0.05]{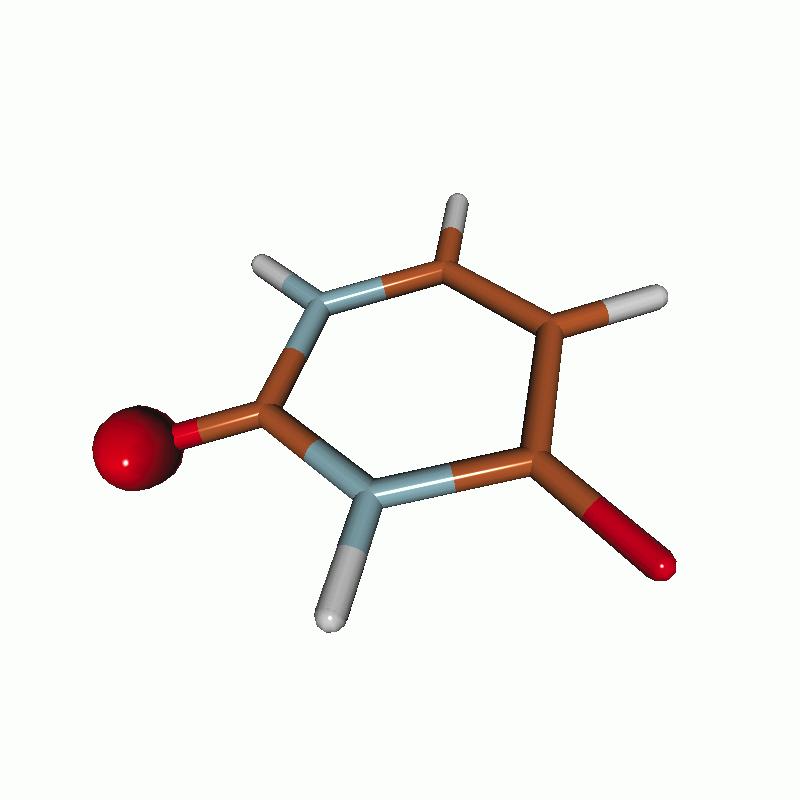}
     \end{minipage}
     & 0.80
     &  \begin{minipage}{0.1\textwidth}
         \centering
         \includegraphics[scale=0.05]{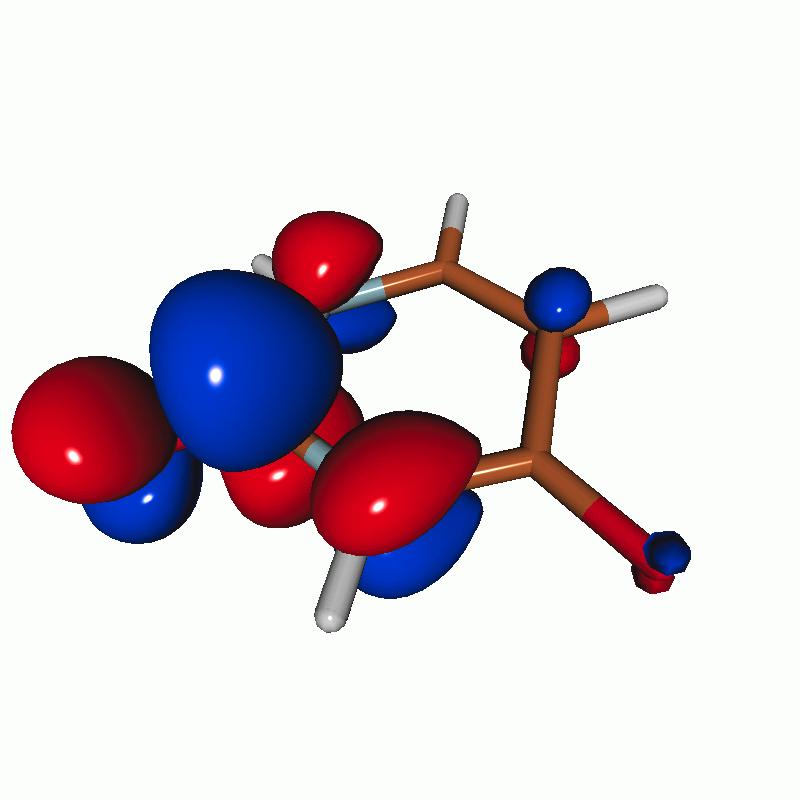}
     \end{minipage}
     \\
     \hline
     535.70 & 0.0076 & $\alpha$ &
     \begin{minipage}{0.1\textwidth}
         \centering
         \includegraphics[scale=0.05]{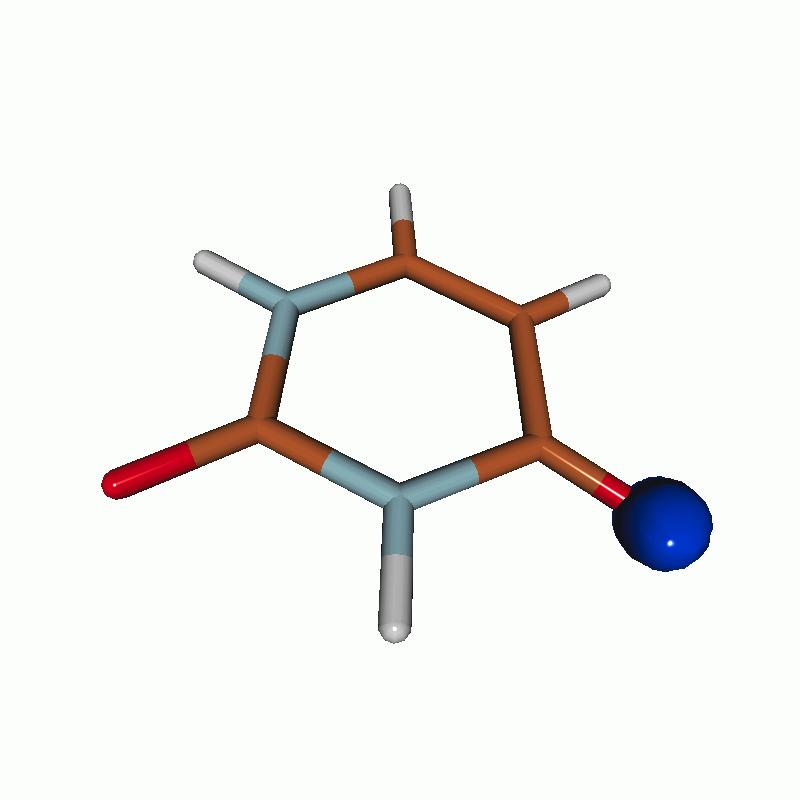}
     \end{minipage}
     & 0.77
     &  \begin{minipage}{0.1\textwidth}
         \centering
         \includegraphics[scale=0.05]{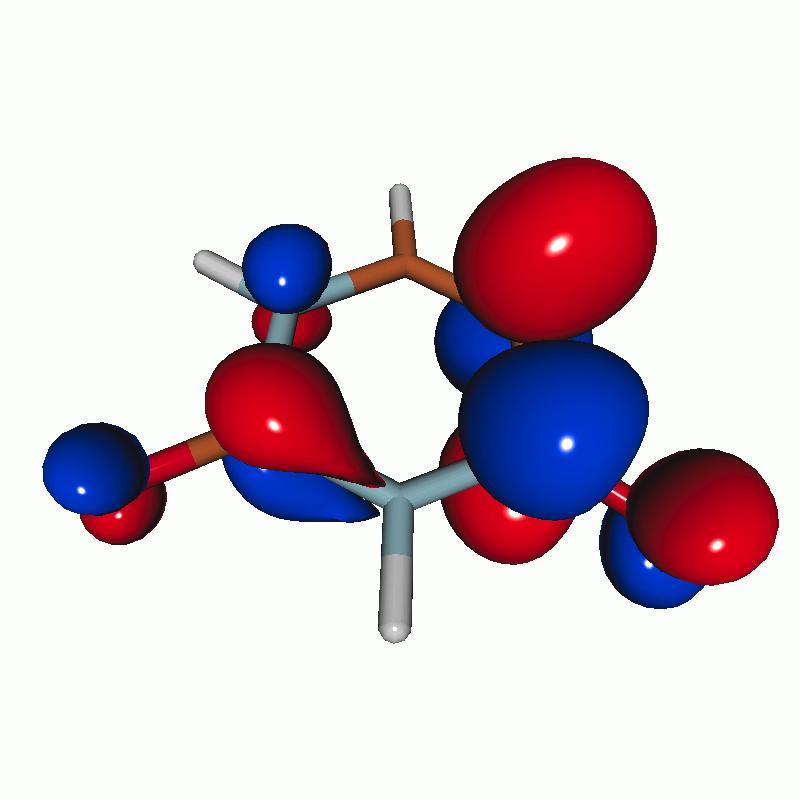}
     \end{minipage}
     \\
     \hline
     535.88 & 0.0085 & $\alpha$ &
     \begin{minipage}{0.1\textwidth}
         \centering
         \includegraphics[scale=0.05]{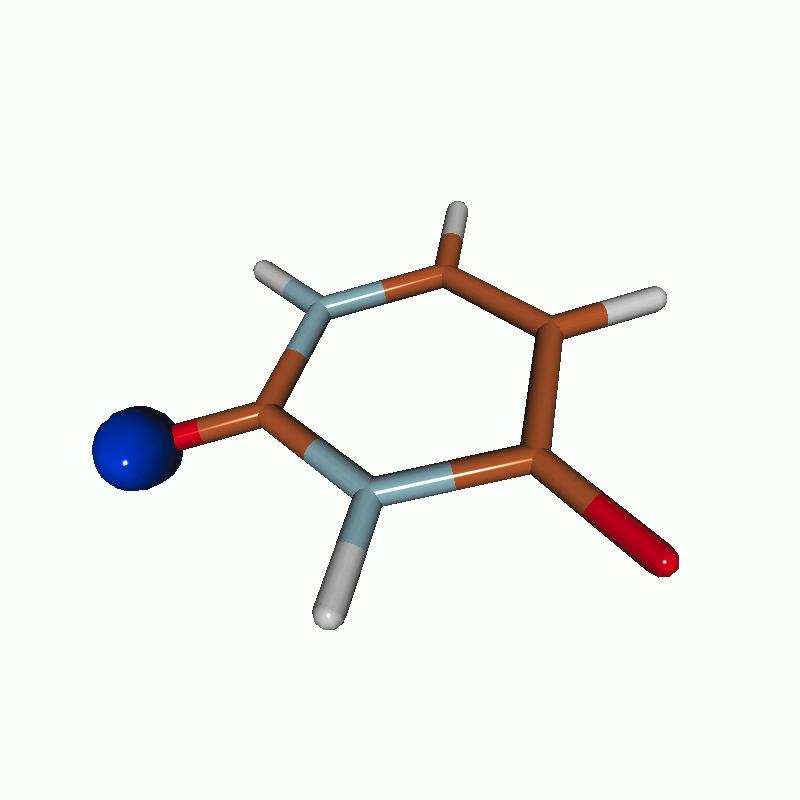}
     \end{minipage}
     & 0.76
     &  \begin{minipage}{0.1\textwidth}
         \centering
         \includegraphics[scale=0.05]{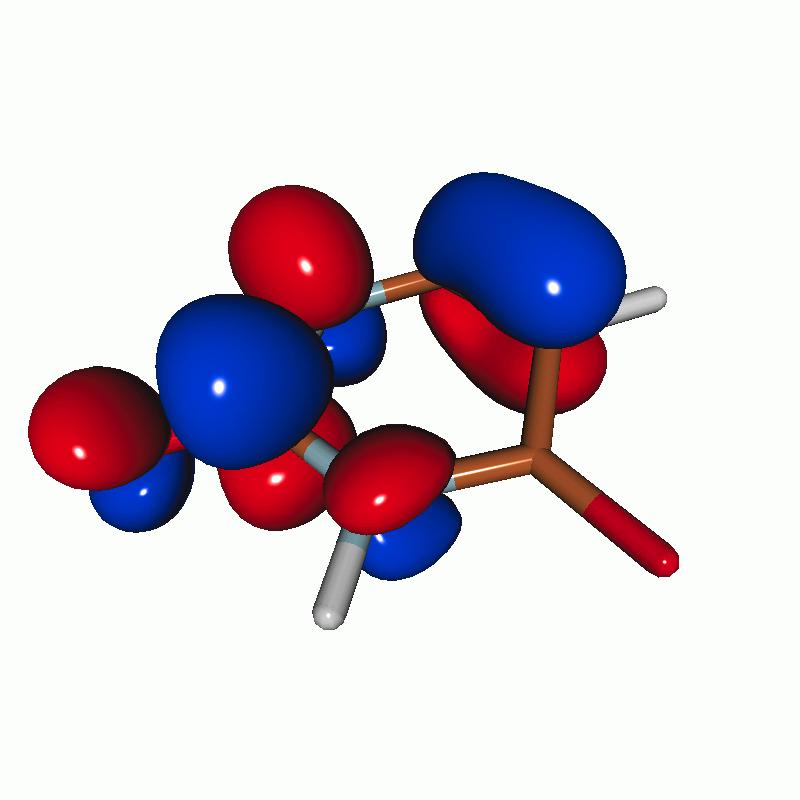}
     \end{minipage}
     \\
     \hline
     \end{tabular}
 \label{tab:uracil_NTO_momS2_fc}
 \end{table}

\begin{table}[hbt]
 \centering
 \caption{Uracil. LSOR-CCSD/6-311++G** 
 NTOs of the O$_{1s}$ core excitations from the 
 T$_1$($\pi\pi^*$) state at the FC geometry (NTO isosurface is 0.05).}
 \begin{tabular}{c|c|c|ccc}
    \hline
     $E^{\mathrm{ex}}$ (eV) & Osc. strength & Spin & Hole & $\sigma_K^2$ & Particle 
     \\
     \hline
     529.81 & 0.0212 & $\beta$ &
     \begin{minipage}{0.1\textwidth}
         \centering
         \includegraphics[scale=0.05]{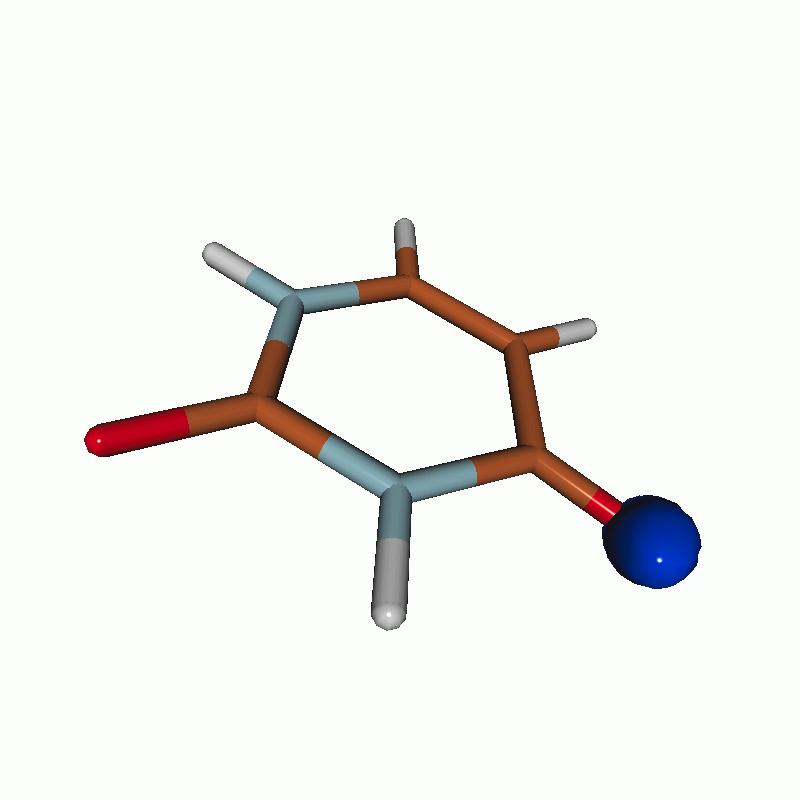}
     \end{minipage}
     & 0.79
     &  \begin{minipage}{0.1\textwidth}
         \centering
         \includegraphics[scale=0.05]{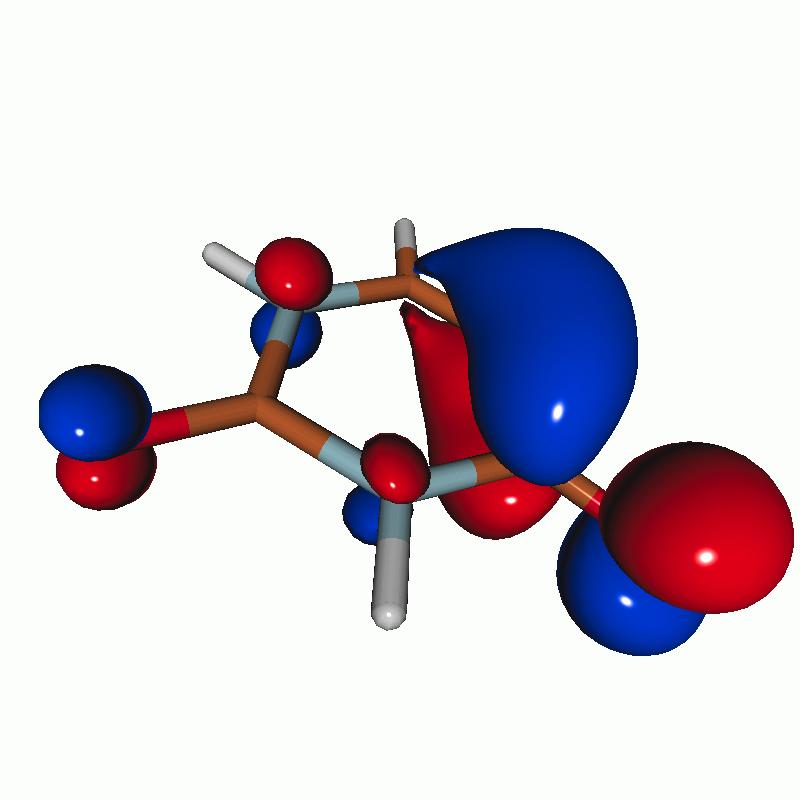}
     \end{minipage}
     \\
     \hline
     532.39 & 0.0115 & $\beta$ &
     \begin{minipage}{0.1\textwidth}
         \centering
         \includegraphics[scale=0.05]{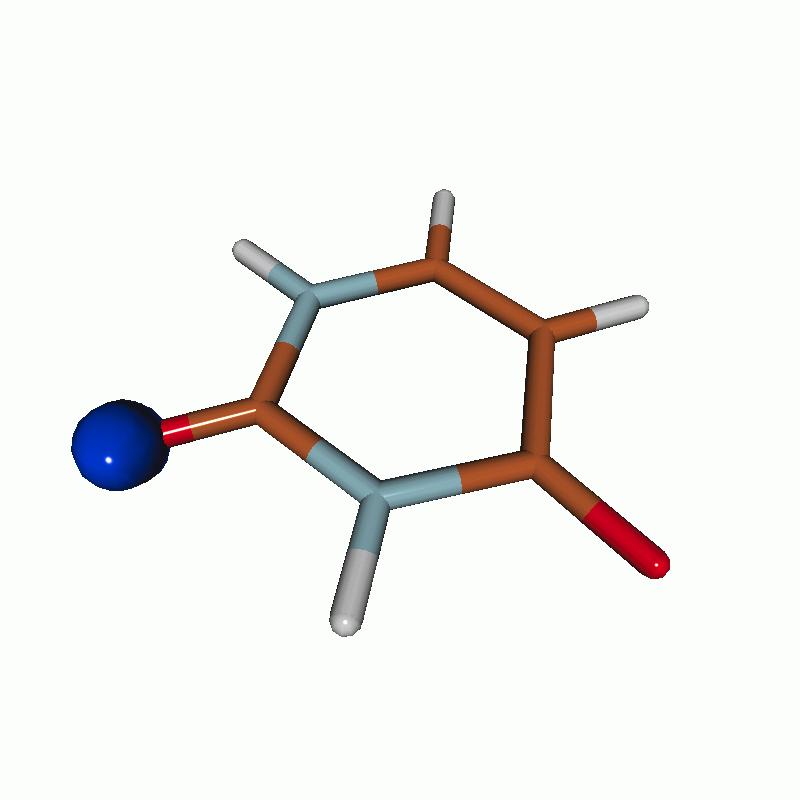}
     \end{minipage}
     & 0.78
     &  \begin{minipage}{0.1\textwidth}
         \centering
         \includegraphics[scale=0.05]{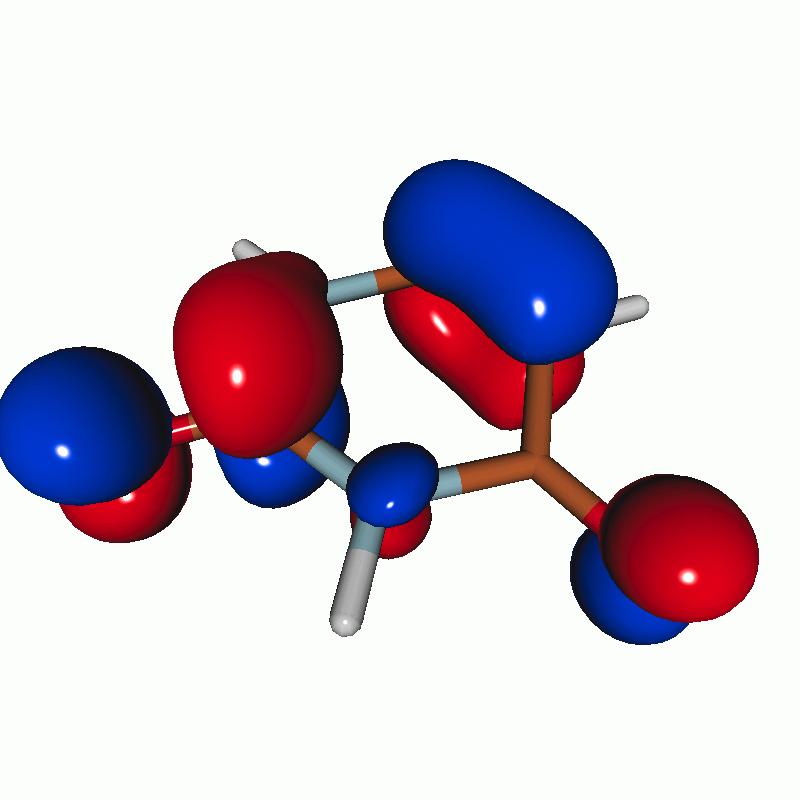}
     \end{minipage}
     \\
     \hline
     534.15 & 0.0187 & $\alpha$ &
     \begin{minipage}{0.1\textwidth}
         \centering
         \includegraphics[scale=0.05]{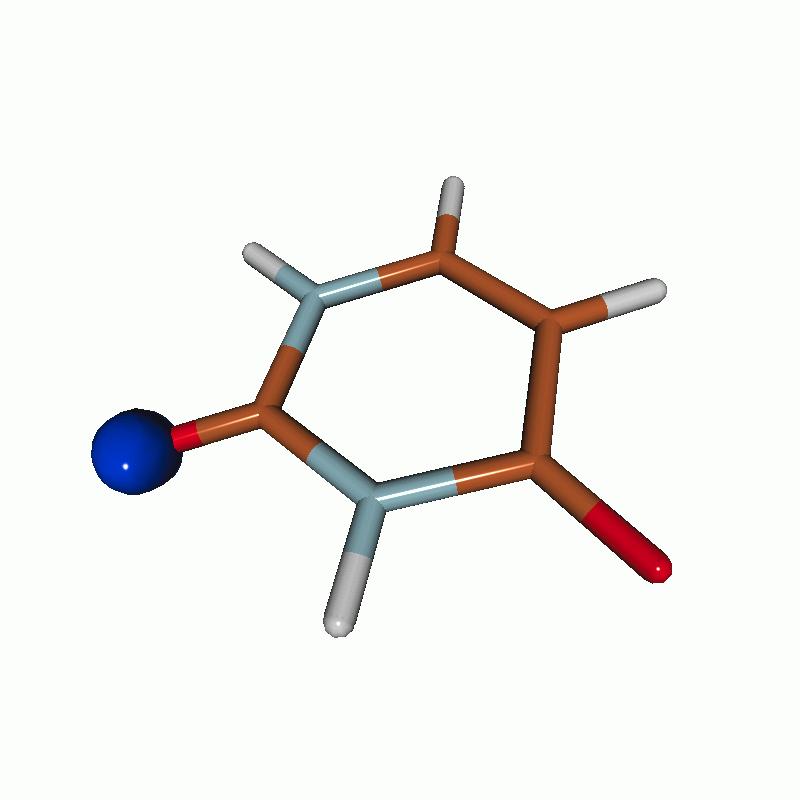}
     \end{minipage}
     & 0.76
     &  \begin{minipage}{0.1\textwidth}
         \centering
         \includegraphics[scale=0.05]{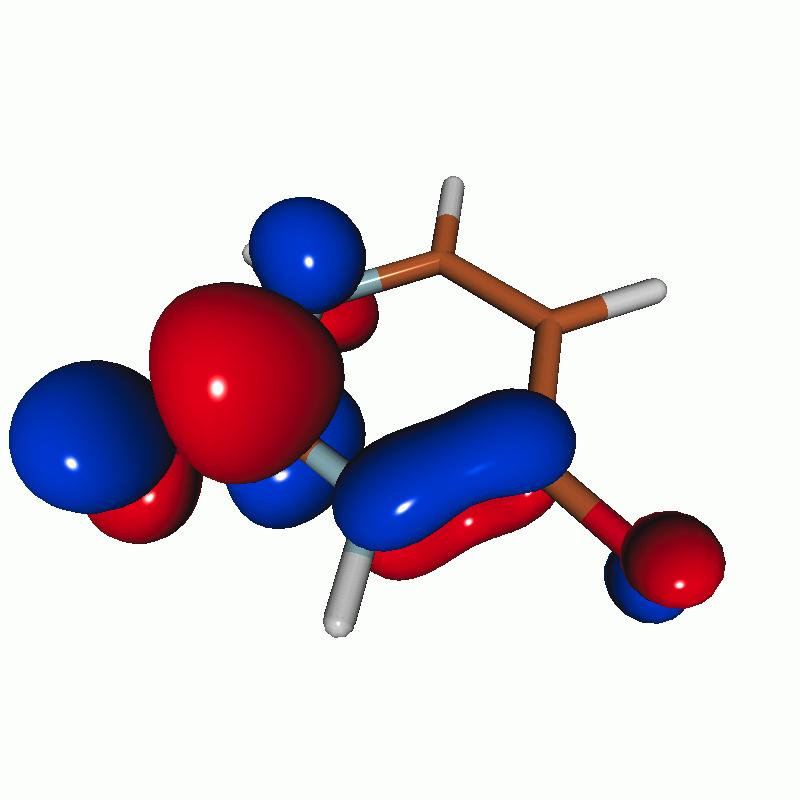}
     \end{minipage}
     \\
     \hline
     535.09 & 0.0100 & $\alpha$ &
     \begin{minipage}{0.1\textwidth}
         \centering
         \includegraphics[scale=0.05]{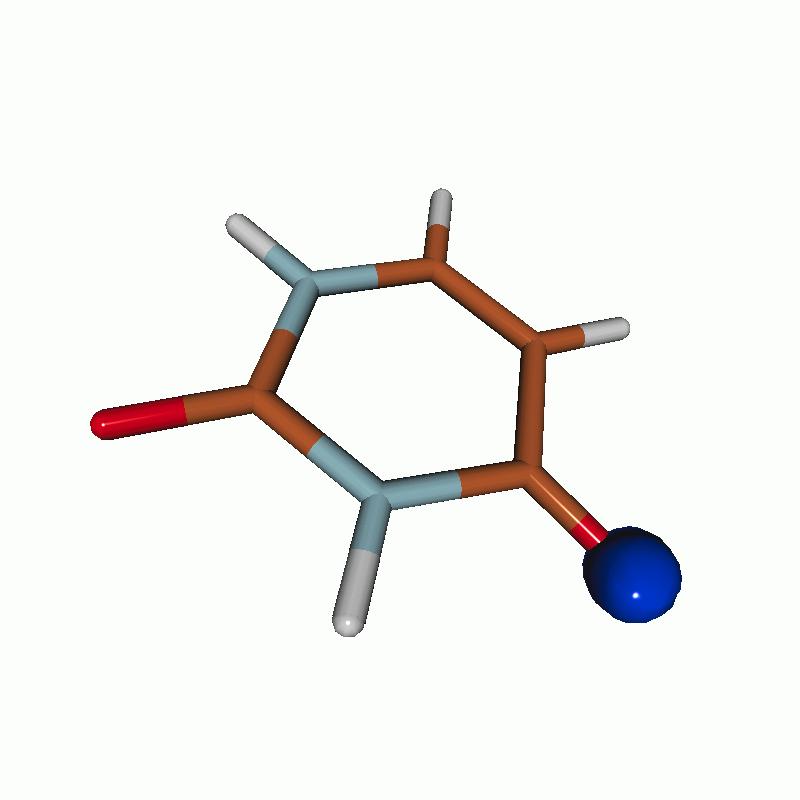}
     \end{minipage}
     & 0.73
     &  \begin{minipage}{0.1\textwidth}
         \centering
         \includegraphics[scale=0.05]{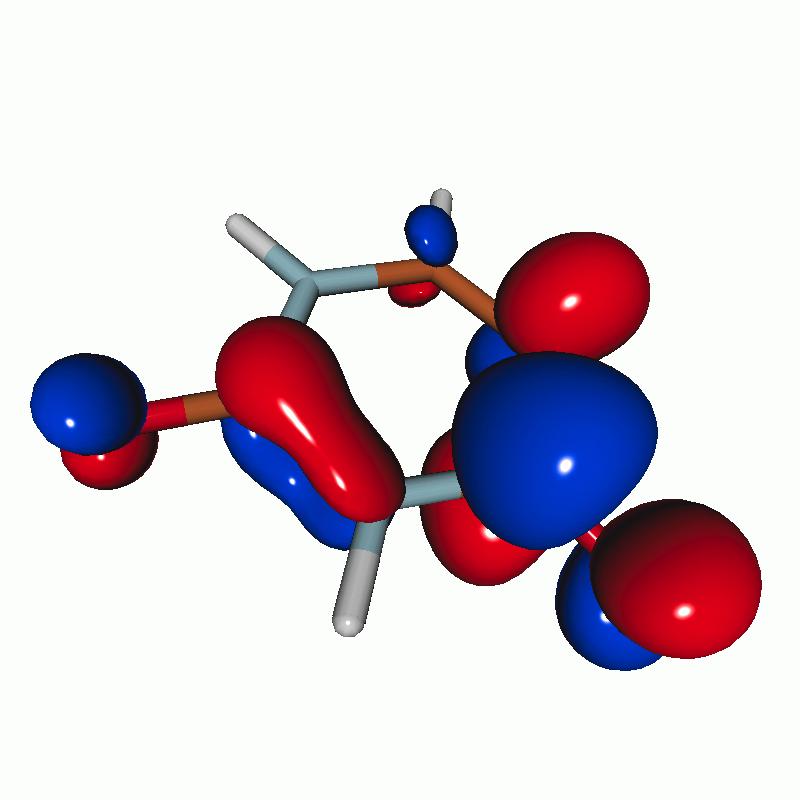}
     \end{minipage}
     \\
     \hline
     535.58 & 0.0062 & $\beta$ &
     \begin{minipage}{0.1\textwidth}
         \centering
         \includegraphics[scale=0.05]{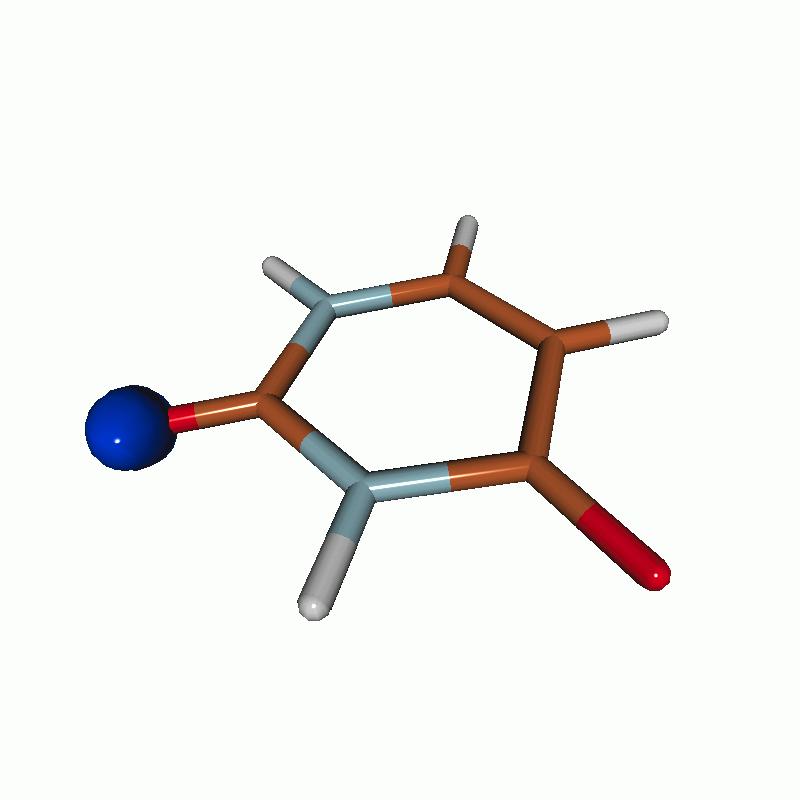}
     \end{minipage}
     & 0.77
     &  \begin{minipage}{0.1\textwidth}
         \centering
         \includegraphics[scale=0.05]{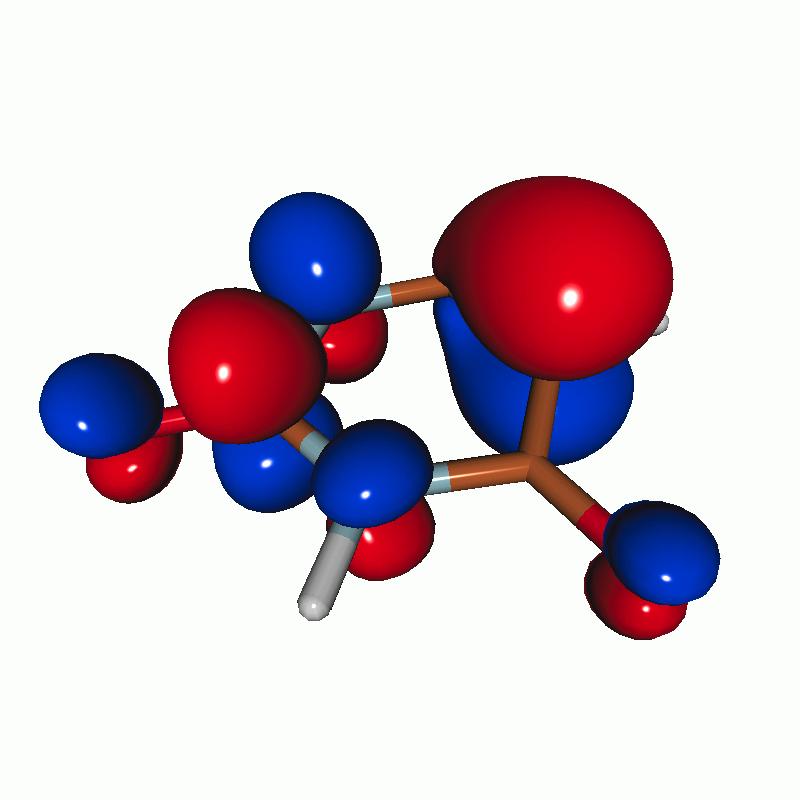}
     \end{minipage}
     \\
     \hline
     535.61 & 0.0081 & $\beta$ &
     \begin{minipage}{0.1\textwidth}
         \centering
         \includegraphics[scale=0.05]{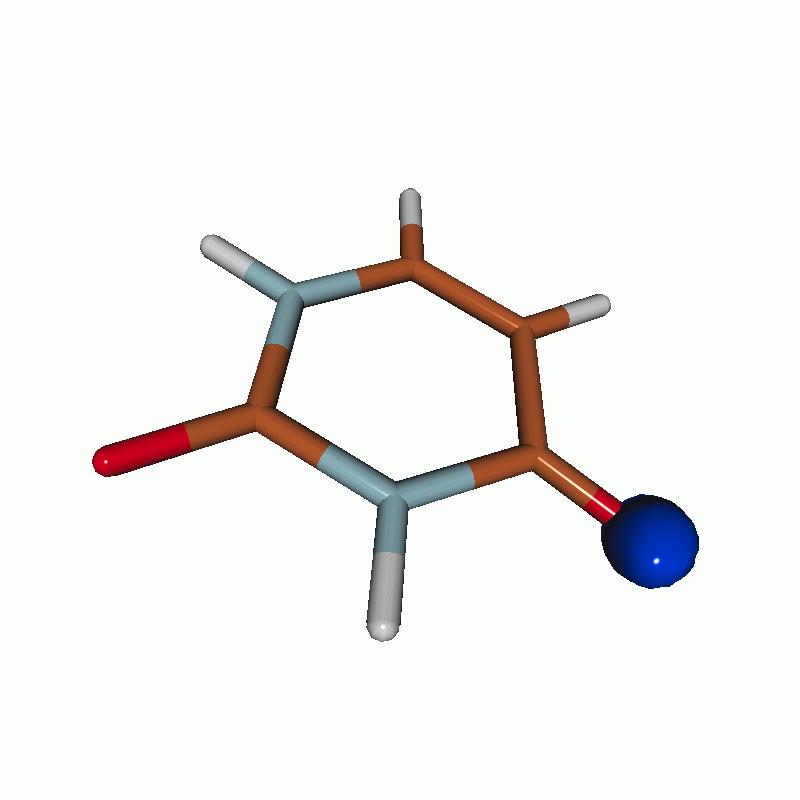}
     \end{minipage}
     & 0.72
     &  \begin{minipage}{0.1\textwidth}
         \centering
         \includegraphics[scale=0.05]{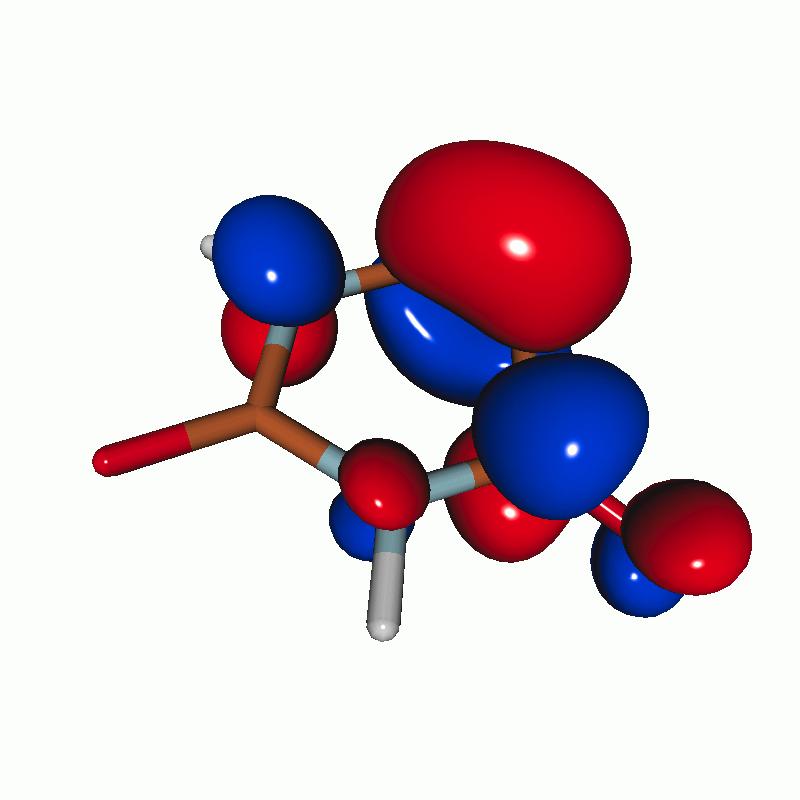}
     \end{minipage}
     \\
     \hline
     \end{tabular}
 \label{tab:uracil_NTO_momT1_fc}
 \end{table}
The CVS-EOM-CCSD spectrum of  S$_1$($\mathrm{n}\pi^{\ast}$) exhibits a relatively intense peak at 528.02 eV, and tiny peaks at 532.40 and 532.52 eV. The intense peak is due to transition from the 1$s$ orbital of O4 to SOMO, which is a lone-pair-type orbital localized on O4. The tiny peak at 532.40 eV is assigned to the transition to SOMO from the 1$s$ orbital of O2, whereas the peak at 532.52 eV is assigned to a transition with multiply excited chatacter. The LSOR-CCSD spectrum exhibits the strong core-to-SOMO transition peak at 526.39 eV, which is red-shifted from the corresponding CVS-EOM-CCSD one by 1.63 eV. As Table~\ref{tab:uracil_NTO_momS1_fc} shows, the peak at 534.26 eV is due to transition from the 1$s$ orbital of O2 to a $\pi^{\ast}$ orbital, and it corresponds to the second peak in the ground-state spectrum. In the S$_1$($\mathrm{n}\pi^{\ast}$) XAS spectrum there is no peak corresponding to the first band in the ground-state spectrum, there assigned to the
O4 1$s \to \pi^{\ast}$ transition. 
This suggests that this transition is suppressed  by the positive charge localized on O4 in the S$_1$($\mathrm{n}\pi^{\ast}$) state. 

The S$_1$($\mathrm{n}\pi^{\ast}$) state from LSOR-CCSD is spin-contaminated,
with $\langle S^2 \rangle = 1.033$.
The spectra of S$_1$($\mathrm{n}\pi^{\ast}$) yielded by LSOR-CCSD [panel (a)] and by HSOR-CCSD [panel (c)] are almost identical. This is not too surprising, as the spectra of S$_1$($\mathrm{n}\pi^{\ast}$) and T$_2$($\mathrm{n}\pi^{\ast}$) from CVS-EOM-CCSD are also almost identical. This is probably a consequence of small exchange interactions in the two states (the singlet and the triplet), due to negligible spatial overlap between the lone pair (n)  and $\pi^{\ast}$ orbitals.

In the CVS-EOM-CCSD spectrum of  S$_2$($\pi\pi^{\ast}$), see panel (b), the peaks due to the core-to-SOMO ($\pi$) transitions from O4 and O2 occur at 527.50 and 531.87 eV, respectively. The additional peak at 531.99 eV is assigned to a transition with multiple electronic excitation. In the LSOR-CCSD spectrum, the core-to-SOMO peaks appear at 530.16 and 530.54 eV, respectively. 

As shown in Table~\ref{tab:uracil_NTO_momS2_fc}, we assign the peaks at 532.96 and 534.74 eV in the LSOR-CCSD spectrum to transitions from the 1$s$ orbitals of the two oxygens to the $\pi^{\ast}$ orbital, which is half occupied in S$_2$($\pi\pi^{\ast}$). The NTO analysis reveals that they correspond to the first and second peak of the ground-state spectrum. Note that  
$\langle S^2 \rangle$ = 1.326 for the S$_2$($\pi\pi^{\ast}$) state 
obtained from LSOR-CCSD. 

In the HSOR-CCSD spectrum of the S$_2$($\pi\pi^{\ast}$) state [which is equal to the LSOR-CCSD spectrum of the T$_1$($\pi\pi^{\ast}$) state in panel (d)], the peaks of the core-to-SOMO ($\pi$) transitions 
from O4 and O2 appear at 529.81 and 532.39 eV, respectively (see Table~\ref{tab:uracil_NTO_momT1_fc}). They are followed by transitions to the half-occupied $\pi^{\ast}$ orbital at 534.15 and 535.09 eV, respectively. In contrast to what we observed in the S$_1$($\mathrm{n}\pi^{\ast}$) spectra, 
the LSOR-CCSD and HSOR-CCSD spectra of the S$_2$($\pi\pi^{\ast}$) state are qualitatively different. 
This can be explained, again, in terms of importance of the exchange interactions in the initial and final states. On one hand, there is a stabilization of the T$_1$($\pi\pi^{\ast}$) (initial) state over the S$_2$($\pi\pi^{\ast}$) state by exchange interaction, as the 
overlap between the $\pi$ and $\pi^{\ast}$ orbitals is not negligible. The exchange interaction between the strongly localized core-hole orbital and the half-occupied valence/virtual orbital in the final core-excited state, on the other hand, is expected to be small.

\begin{figure}[hbt]
    \centering
    \includegraphics[width=8.5cm,height=10cm,keepaspectratio]{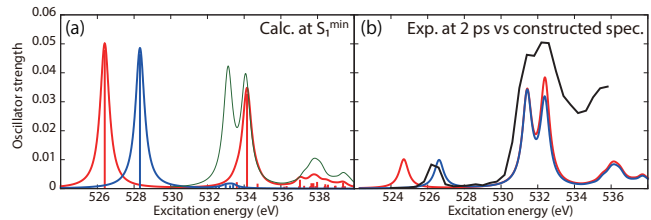}
    \caption{Thymine. (a) Oxygen K-edge NEXAFS in the S$_1$(n$\pi^{\ast}$) state at its potential energy minimum. Blue: CVS-EOM-CCSD. Red:  LSOR-CCSD. Thin green line:
    ground-state spectrum at the FC geometry. (b) Thick black: Experimental spectrum at the delay time of 2 ps.\cite{Wolf} Blue: computational spectrum made from the blue and green curves of (a), shifted by $-$1.7 eV. Red: computational spectrum made from the red and green curves of (a), shifted by $-$1.7 eV. The blue and red curves from (a) were scaled by 0.2 in (b). The ground-state spectrum from (a) was scaled by 0.8 in (b). FWHM of the Lorentzian convolution function is 0.6 eV.}
    \label{fig:thymine_S1min}
\end{figure}

To evaluate the accuracy of the excited-state XAS 
spectra from CVS-EOM-CCSD and LSOR-CCSD, 
we also calculated the XAS spectra of the S$_1$(n$\pi^{\ast}$) state of thymine at 
the potential energy minimum of 
S$_1$(n$\pi^{\ast}$), see panel (a) of
Fig.~\ref{fig:thymine_S1min}. For construction of the surface cut of the theoretical absorption spectra, we chose FWHM of 0.6 eV for the Lorentzian convolution function. Panel (b) shows 
the spectra of S$_1$(n$\pi^{\ast}$) multiplied by 0.2 and added to the ground-state spectrum multiplied by 0.8. These factors 0.2 and 0.8 were chosen for the best fit with the experimental spectrum. A surface cut of the experimental TR-NEXAFS spectrum at the delay time of 2 ps~\cite{Wolf} is also shown in panel (b) of Fig.~\ref{fig:thymine_S1min}. The reconstructed computational spectra are shifted by $-$1.7 eV. In the experimental spectrum, the core-to-SOMO transition peak occurs at 526.4 eV. In the reconstructed theoretical spectrum, the core-to-SOMO transition peaks appear at 526.62 and 524.70 eV, for CVS-EOM-CCSD and LSOR-CCSD, respectively. Thus, the CVS-EOM-CCSD superposed spectrum agrees slightly better with experiment than the LSOR-CCSD spectrum. Nonetheless, the accuracy of the LSOR-CCSD spectrum is quite reasonable, as compared with the experimental spectrum.

Due to the lack of experimental data, 
not much can be said about the accuracy of 
CVS-EOM-CCSD and LSOR-CCSD/HSOR-CCSD 
for core excitations from a triplet 
excited state in uracil and thymine.
Furthermore, we are unable to 
unambiguously clarify, 
using  uracil and thymine as model 
system, which
of the two methods, LSOR-CCSD or HSOR-CCSD, should be considered more reliable when they give qualitatively different spectra for the singlet excited states.

\begin{figure}
    \centering
    \includegraphics[width=8.5cm,height=8.5cm,keepaspectratio]{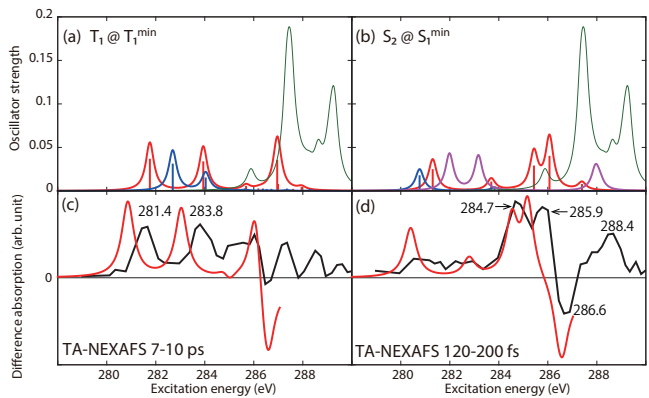}
    \caption{Acetylacetone. Carbon K-edge NEXAFS from the T$_1$($\pi\pi^{\ast}$) (a) and S$_2$($\pi\pi^{\ast}$) (b) states. 
    The spectra of T$_1$($\pi\pi^{\ast}$)  were computed at the potential energy minimum of  
    T$_1$($\pi\pi^{\ast}$). 
    The spectra of S$_2$($\pi\pi^{\ast}$) were computed at the potential energy minimum of  
    S$_1$(n$\pi^{\ast}$). Blue: CVS-EOM-CCSD. Red: LSOR-CCSD. Magenta: HSOR-CCSD. Green: ground-state spectrum at the FC geometry. 
    (c), (d) Black: Experimental transient absorption spectra at the delay times of 7-10 ps and 120-200 fs \cite{acac_ultrafast_ISC}, respectively. 
    Red: computational transient absorption spectra made from the red and the green curves of (a) and (b), respectively, shifted by $-$0.9 eV 
    as the spectrum of the ground state [see panel (b) of Fig.~\ref{fig:static_AcAc}].
    The red curves of panels (a) and (b) were scaled by 0.75 and from these, the green GS spectrum, scaled by 0.25, was subtracted. FWHM of the Lorentzian convolution function is 0.4 eV for panels (a) and (b), 0.6 eV for panels (c) and (d), respectively.
    Basis set: 6-311++G**. }
    \label{fig:AcAc_eachmin}
\end{figure}

Therefore, we turn our attention to the 
carbon K-edge spectra of acetylacetone and show, in
Fig.~\ref{fig:AcAc_eachmin}, the spectra obtained using CVS-EOM-CCSD (blue),
LSOR-CCSD (red), and HSOR-CCSD (magenta)
for the T$_1$($\pi\pi^{\ast}$) [panel (a)] and S$_2$($\pi\pi^{\ast}$) [panel (b)] states.
The T$_1$($\pi\pi^{\ast}$) spectra were obtained at the potential energy minimum of T$_1$($\pi\pi^{\ast}$). The spectra of S$_2$($\pi\pi^{\ast}$) were calculated at the potential energy minimum of the S$_1$(n$\pi^{\ast}$) state. In doing so, we assume that the nuclear wave packet propagates on the S$_2$($\pi\pi^{\ast}$) surface toward the potential energy minimum of the S$_1$(n$\pi^{\ast}$) surface. 
Note that CVS-EOM-CCSD does not describe all the core excitations from a valence-excited state (see Fig.~\ref{fig:schematic_ccsd}).
In panels (c) and (d), the LSOR-CCSD 
spectra were multiplied by 0.75 and subtracted from the ground-state spectrum, scaled by 0.25, and superposed to the surface cuts of the experimental transient-absorption NEXAFS at delay times 7-10 ps and 120-200 fs, respectively. 
The calculated transient-absorption spectra were shifted by $-$0.9 eV, i.e. by the same amount as the spectrum of the ground state [see panel (b) of Fig.~\ref{fig:static_AcAc}].
For construction of the surface cut of the theoretical transient-absorption spectra, we used FWHM of 0.6 eV for the Lorentzian convolution function. 
The scaling factors values 0.75 and 0.25 were chosen 
to yield the best fit with the experimental spectra. 
The NTOs of the core excitations from T$_1$($\pi\pi^{\ast}$) and S$_2$($\pi\pi^{\ast}$) are shown in Tabs.~\ref{tab:AcAc_NTO_momT1_T1min} and \ref{tab:AcAc_NTO_momS2_S1min}, respectively. In the experimental study,~\cite{acac_ultrafast_ISC} it was concluded  that
S$_2$($\pi\pi^{\ast}$) is populated at the shorter time scale, whereas at the longer time scale it is T$_1$($\pi\pi^{\ast}$) that becomes populated.

The surface cut of the experimental transient-absorption spectra at longer times (7-10 ps) features two peaks at 281.4 and 283.8 eV. 
In panel (a) of Fig.~\ref{fig:AcAc_eachmin}, the CVS-EOM-CCSD spectrum of T$_1$($\pi\pi^{\ast}$) shows the core-to-SOMO transition peaks at 282.69 and 284.04 eV, whereas the LSOR-CCSD ones appear at 281.76 and 283.94 eV. 
The LSOR-CCSD spectrum also shows a peak corresponding to a transition from C4 to the half-occupied $\pi^{\ast}$ orbital at 286.96 eV (see Table~\ref{tab:AcAc_NTO_momT1_T1min}). The separation of 2.4 eV between the two core-to-SOMO peaks in the experiment is well reproduced by LSOR-CCSD. Spin contamination is small,  $\langle S^2 \rangle$=2.004 for the T$_1$($\pi\pi^{\ast}$) state obtained using LSOR-CCSD. 
Therefore, it is safe to say that LSOR-CCSD accurately 
describes core excitations from the low-lying 
triplet states.

The surface cut of the transient-absorption spectra at shorter times, 120-240 fs, features relatively strong peaks at 284.7, 285.9 and a ground-state bleach at 286.6 eV.
The CVS-EOM-CCSD spectrum of the S$_2$($\pi\pi^{\ast}$) state shows the core-to-SOMO peak at 280.77. The LSOR-CCSD spectrum (red) has core-to-SOMO transition peaks at 281.30 and 283.69 eV, plus the peaks due to the transitions from the core of C2, C4 and C3 
to the half-occupied $\pi^{\ast}$ orbital at 285.43, 286.07 and 287.39 eV, respectively (see Table~\ref{tab:AcAc_NTO_momS2_S1min}). Note that the peaks at 285.43 and 286.07 eV 
correspond to the main degenerate peaks of the ground-state spectrum, 
as revealed by inspection of the NTOs.
The HSOR-CCSD spectrum (magenta) exhibits the core-to-SOMO transition peaks at 281.99 and 283.17 eV, followed by only one of the quasi-degenerate peaks corresponding to transitions to the 
half-occupied $\pi^{\ast}$ orbital, at 287.95 eV. Since the experimental surface-cut spectrum does not clearly show the core-to-SOMO transition peaks, it is difficult to assess the accuracy of these peaks as obtained in the calculations. 
When it comes to the experimental peaks at 284.7 and 285.9 eV, only LSOR-CCSD reproduces them with reasonable accuracy. 
The experimental peak at 288.4 eV is not reproduced. 
In the case of acetylacetone, the HSOR-CCSD approximation fails to correctly mimic the spectrum of S$_2$($\pi\pi^{\ast}$), 
since it does not give the peaks at 284.7 and 285.9 eV.
The differences between LSOR-CCSD and HSOR-CCSD spectra for S$_2$($\pi\pi^{\ast}$) can be rationalized as done for uracil.

We emphasize that the assignment  of the transient absorption  signal at shorter time to
S$_2$($\pi\pi^{\ast}$) is based on peaks assigned to transitions to the $\pi^\ast$ orbitals (almost degenerate in the ground state), which cannot be described by CVS-EOM-CCSD (see Fig.~\ref{fig:schematic_ccsd} in Sec.~\ref{subsec:protocols}). 

\begin{table}[h]
 \centering
 \caption{Acetylacetone. LSOR-CCSD/6-311++G** NTOs of the C$_{1s}$ core excitations from the T$_1$ state at the potential energy minimum (NTO isosurface is 0.05).}
 \begin{tabular}{c|c|c|ccc}
     \hline
     $E^{\mathrm{ex}}$ (eV) & Osc. strength & Spin & Hole & $\sigma_K^2$ & Particle 
     \\
     \hline
     281.76 & 0.0347 & $\beta$ &
     \begin{minipage}{0.1\textwidth}
         \centering
         \includegraphics[scale=0.05]{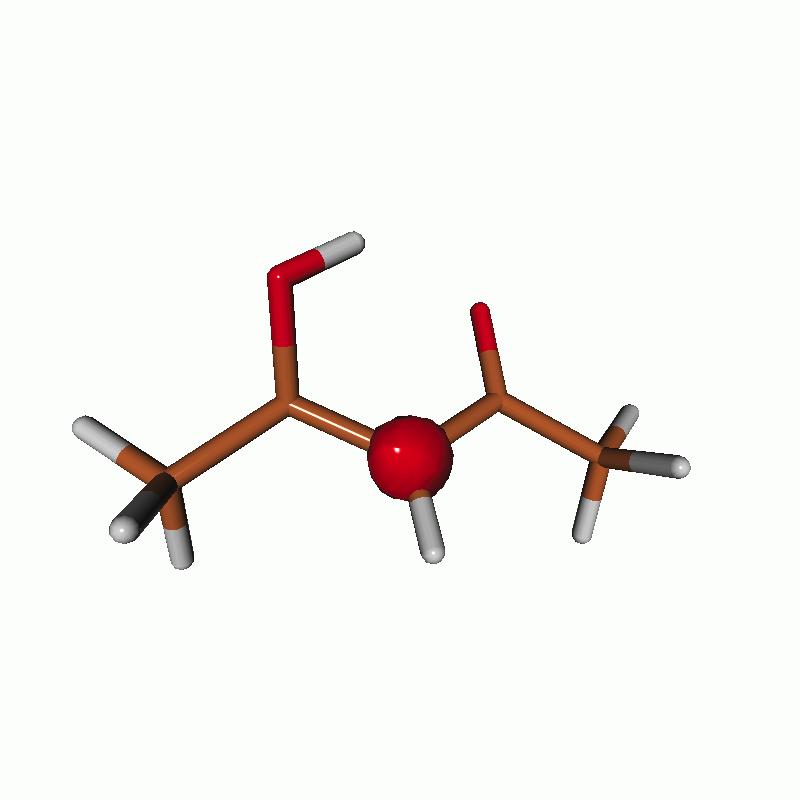}
     \end{minipage}
     & 0.86
     &  \begin{minipage}{0.1\textwidth}
         \centering
         \includegraphics[scale=0.05]{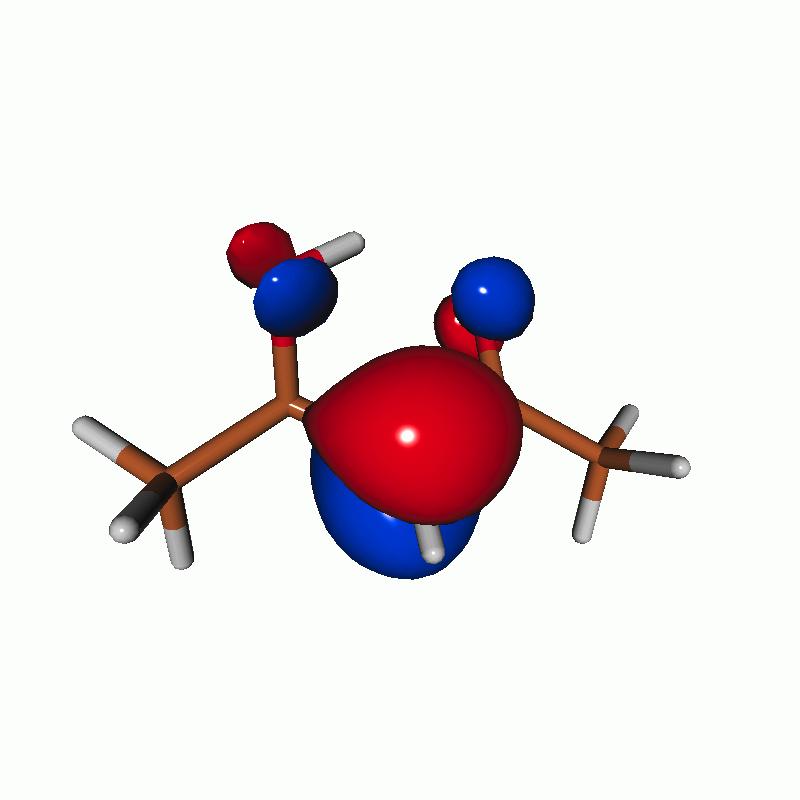}
     \end{minipage}
     \\
     \hline
     283.94 & 0.0318 & $\beta$ &
     \begin{minipage}{0.1\textwidth}
         \centering
         \includegraphics[scale=0.05]{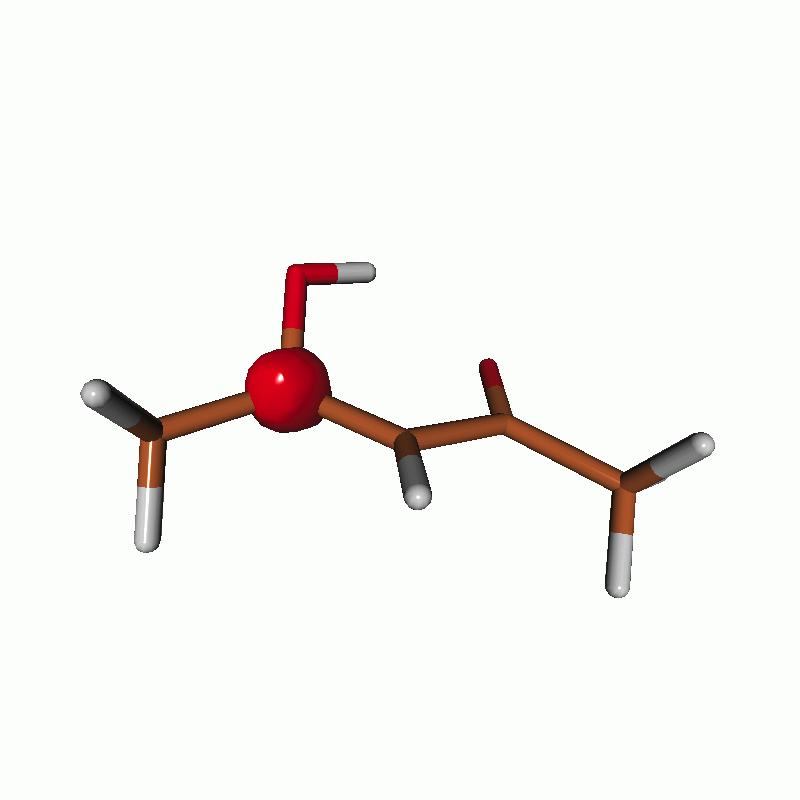}
     \end{minipage}
     & 0.84
     &  \begin{minipage}{0.1\textwidth}
         \centering
         \includegraphics[scale=0.05]{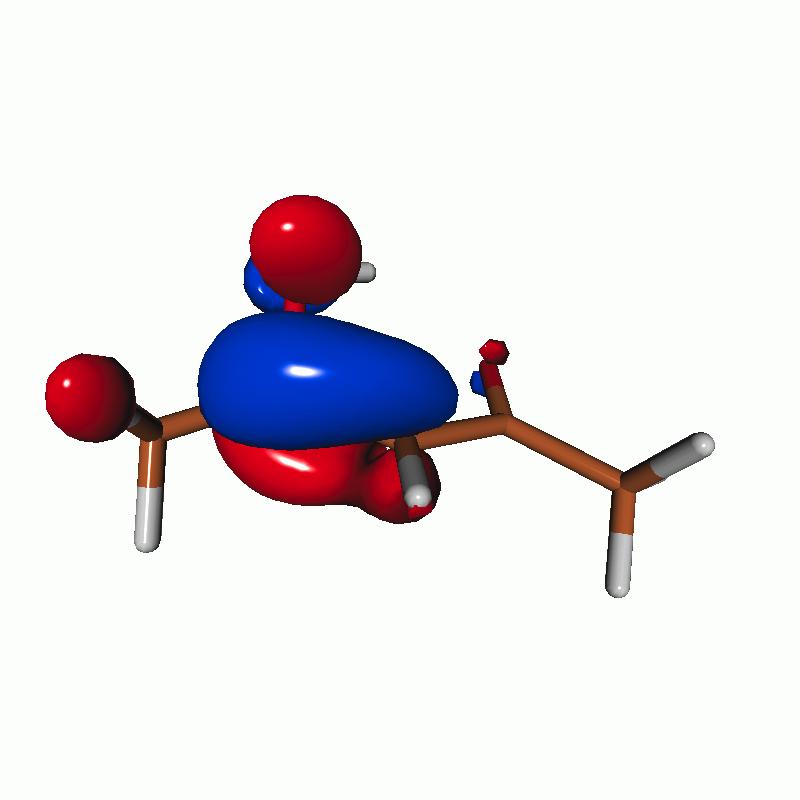}
     \end{minipage}
     \\
     \hline
     285.69 & 0.0036 & $\beta$ &
     \begin{minipage}{0.1\textwidth}
         \centering
         \includegraphics[scale=0.05]{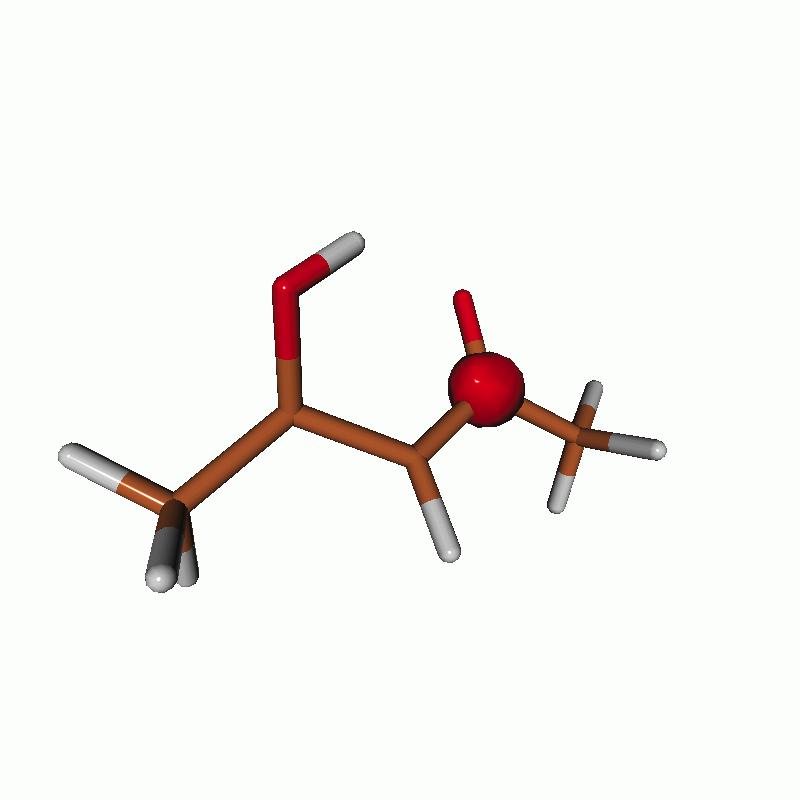}
     \end{minipage}
     & 0.72
     &  \begin{minipage}{0.1\textwidth}
         \centering
         \includegraphics[scale=0.05]{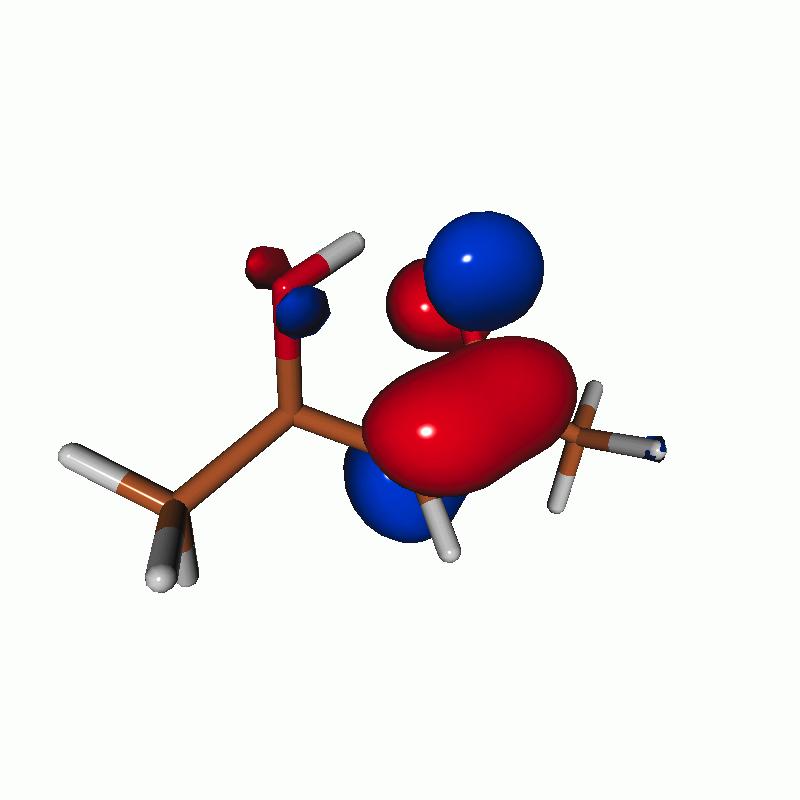}
     \end{minipage}
     \\
     \hline
     286.96 & 0.0334 & $\alpha$ &
     \begin{minipage}{0.1\textwidth}
         \centering
         \includegraphics[scale=0.05]{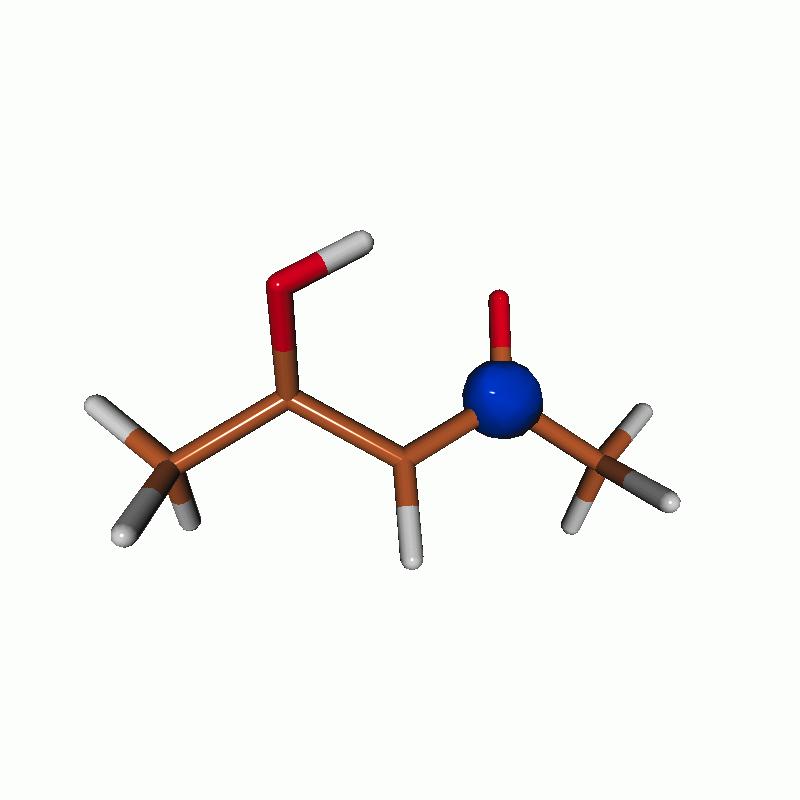}
     \end{minipage}
     & 0.65
     &  \begin{minipage}{0.1\textwidth}
         \centering
         \includegraphics[scale=0.05]{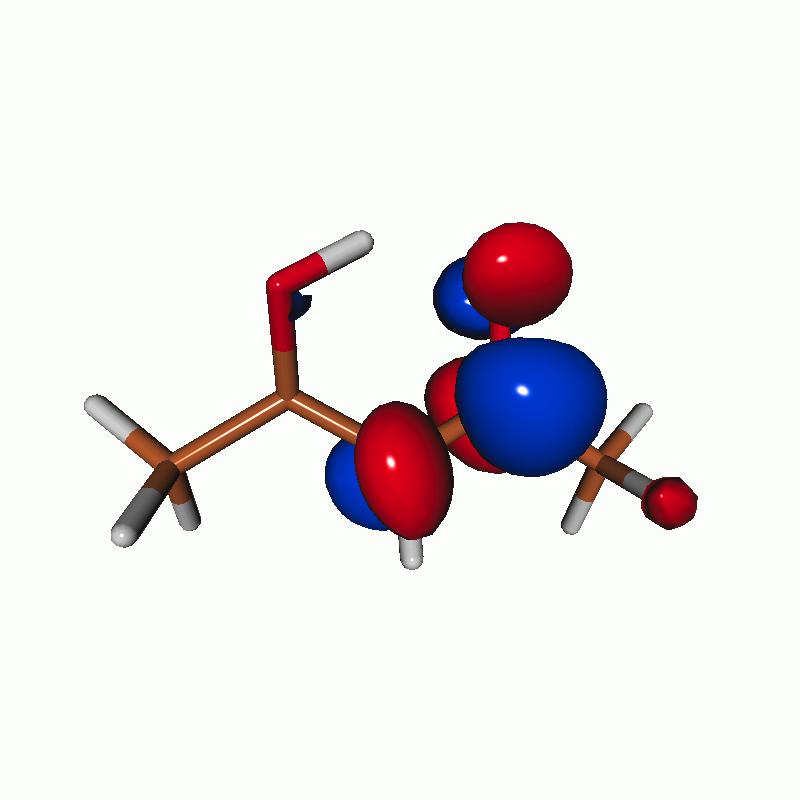}
     \end{minipage}
     \\
     & & $\beta$ &
     \begin{minipage}{0.1\textwidth}
         \centering
         \includegraphics[scale=0.05]{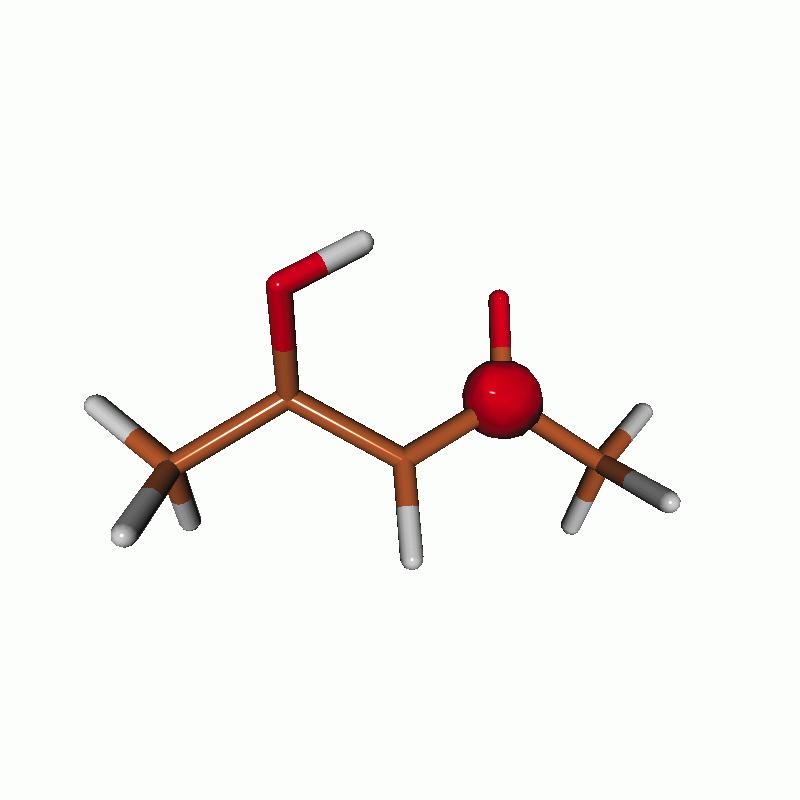}
     \end{minipage}
     & 0.14
     &  \begin{minipage}{0.1\textwidth}
         \centering
         \includegraphics[scale=0.05]{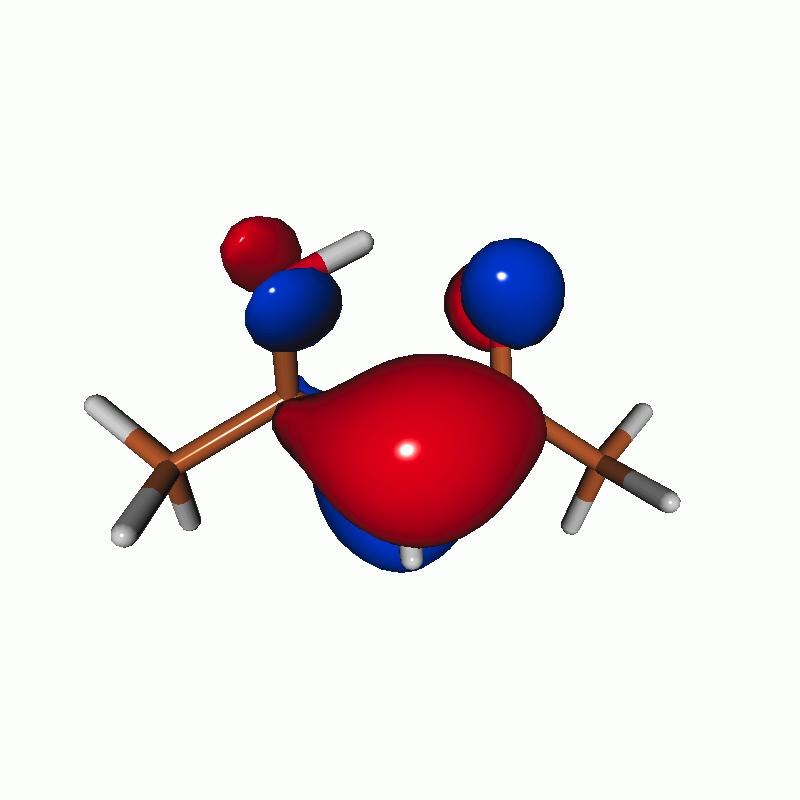}
     \end{minipage}
     \\
     \hline
     \end{tabular}
 \label{tab:AcAc_NTO_momT1_T1min}
 \end{table}

\begin{table}[h]
 \centering
 \caption{Acetylacetone. LSOR-CCSD/6-311++G** NTOs of the C$_{1s}$ core excitations from the S$_2$ state at the potential energy minimum of S$_1$ (NTO isosurface is 0.05).}
 \begin{tabular}{c|c|c|ccc}
     \hline
     $E^{\mathrm{ex}}$ (eV) & Osc. strength & Spin & Hole & $\sigma_K^2$ & Particle 
     \\
     \hline
     281.30 & 0.0228 & $\alpha$ &
     \begin{minipage}{0.1\textwidth}
         \centering
         \includegraphics[scale=0.05]{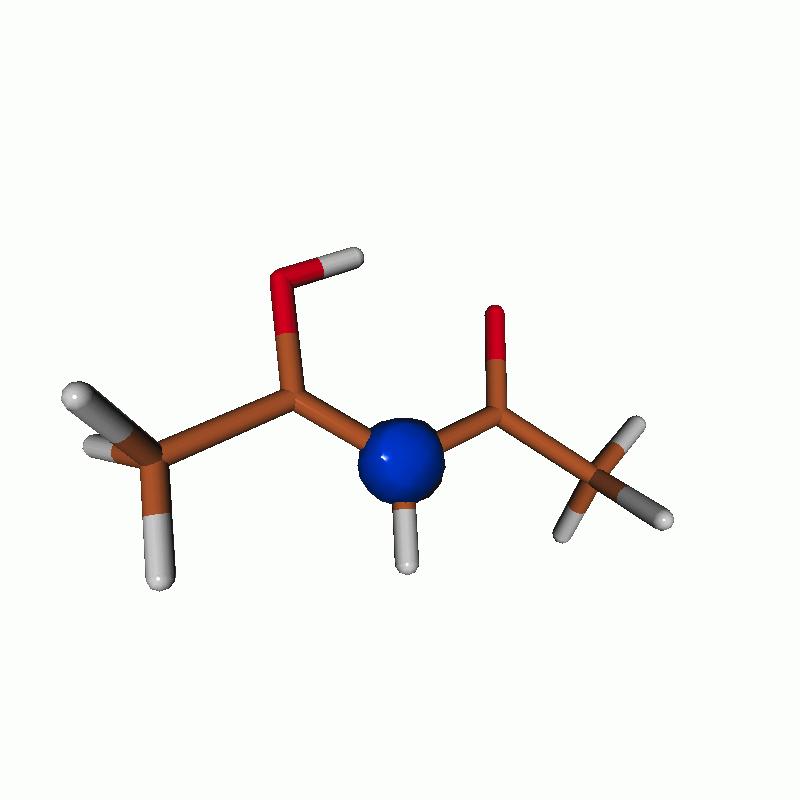}
     \end{minipage}
     & 0.77
     &  \begin{minipage}{0.1\textwidth}
         \centering
         \includegraphics[scale=0.05]{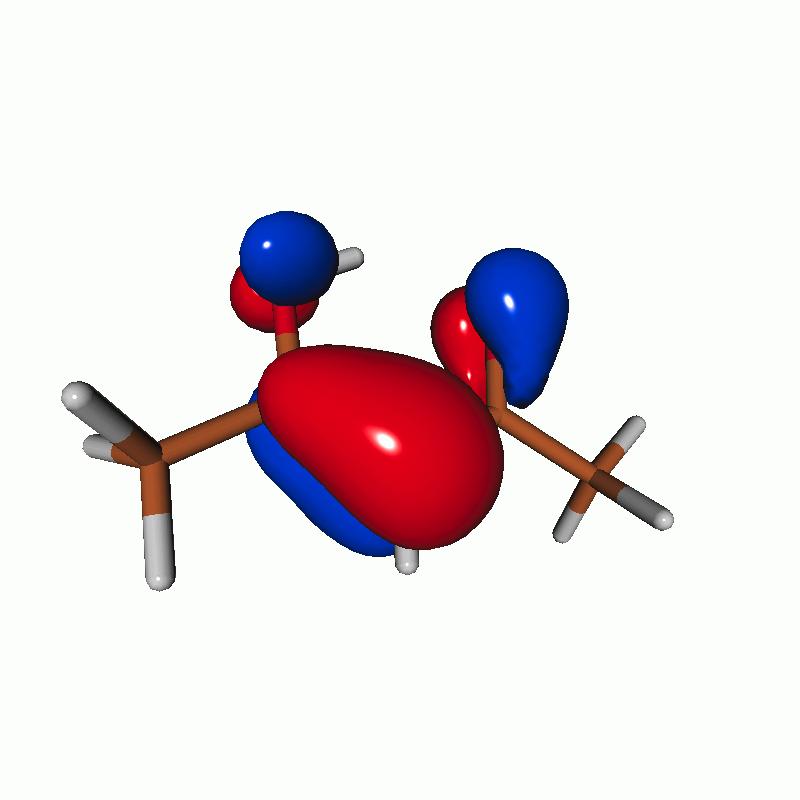}
     \end{minipage}
     \\
     \hline
     283.69 & 0.0085 & $\alpha$ &
     \begin{minipage}{0.1\textwidth}
         \centering
         \includegraphics[scale=0.05]{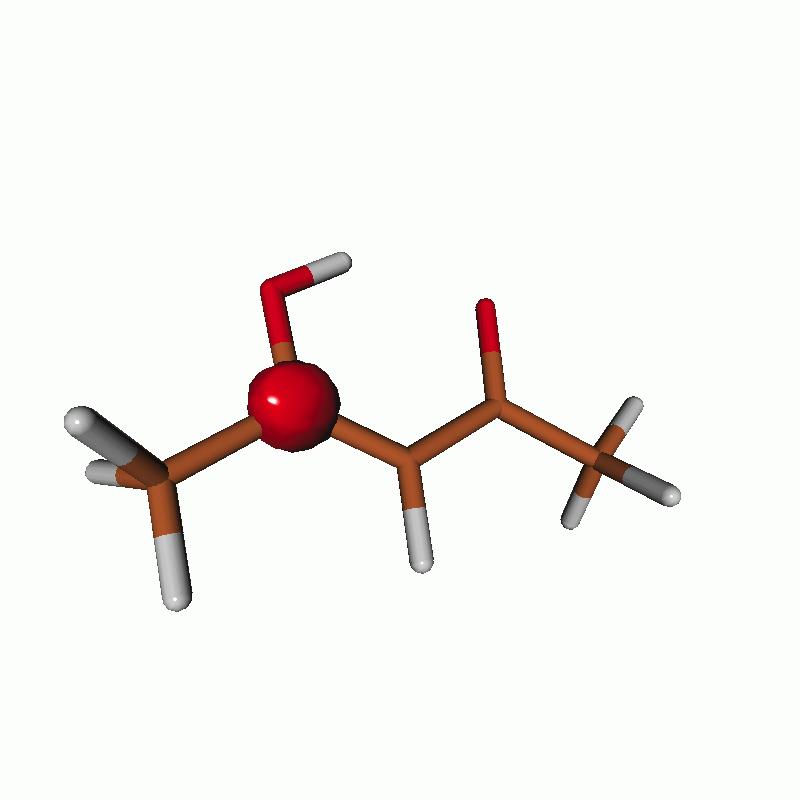}
     \end{minipage}
     & 0.71
     &  \begin{minipage}{0.1\textwidth}
         \centering
         \includegraphics[scale=0.05]{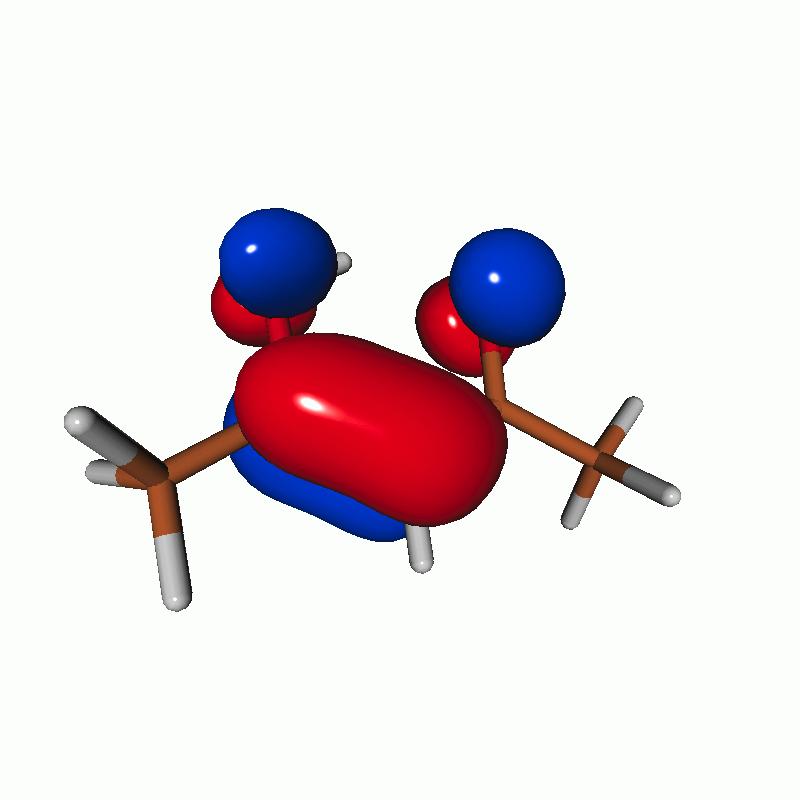}
     \end{minipage}
     \\
     \hline
     285.43 & 0.0269 & $\beta$ &
     \begin{minipage}{0.1\textwidth}
         \centering
         \includegraphics[scale=0.05]{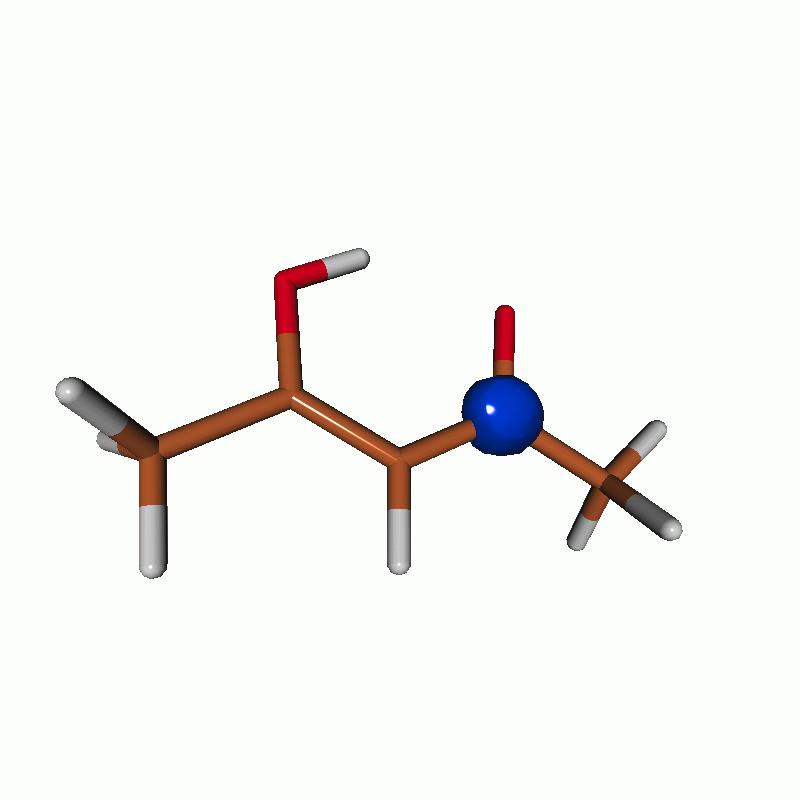}
     \end{minipage}
     & 0.76
     &  \begin{minipage}{0.1\textwidth}
         \centering
         \includegraphics[scale=0.05]{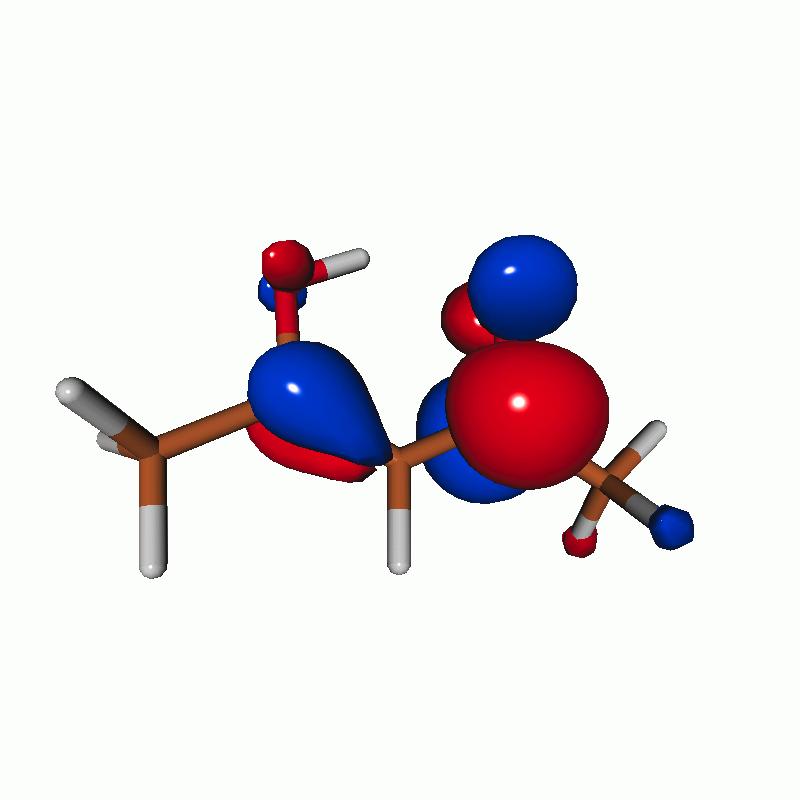}
     \end{minipage}
     \\
     \hline
     286.07 & 0.0381 & $\beta$ &
     \begin{minipage}{0.1\textwidth}
         \centering
         \includegraphics[scale=0.05]{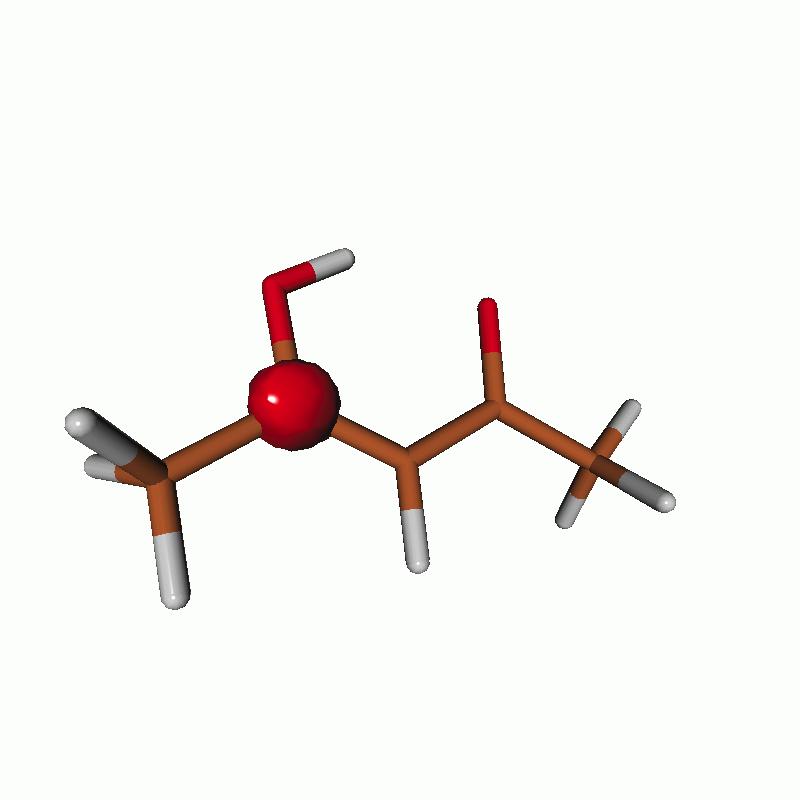}
     \end{minipage}
     & 0.76
     &  \begin{minipage}{0.1\textwidth}
         \centering
         \includegraphics[scale=0.05]{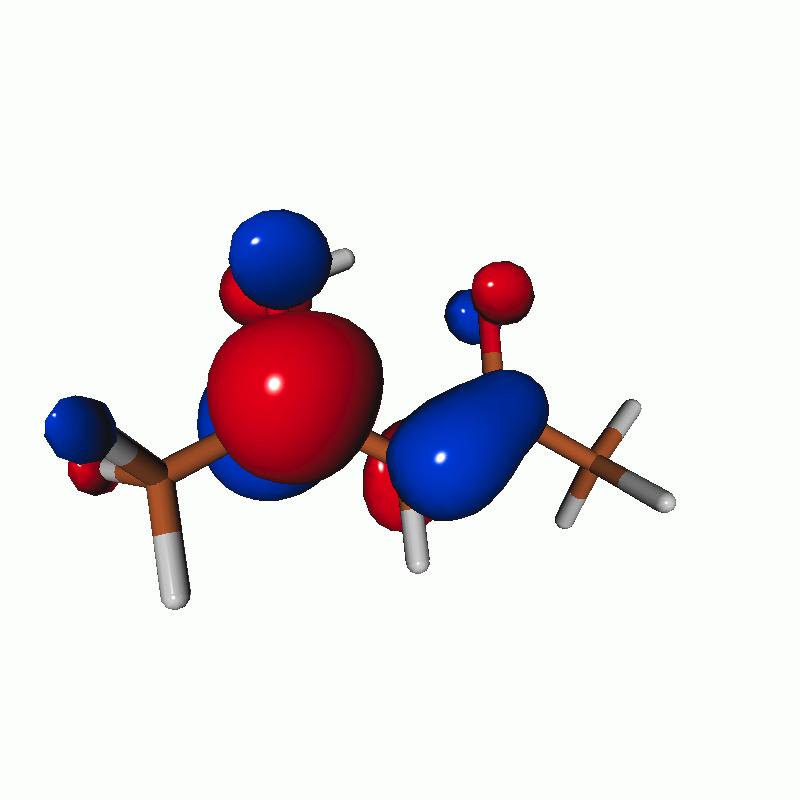}
     \end{minipage}
     \\
     \hline
     287.39 & 0.0057 & $\beta$ &
     \begin{minipage}{0.1\textwidth}
         \centering
         \includegraphics[scale=0.05]{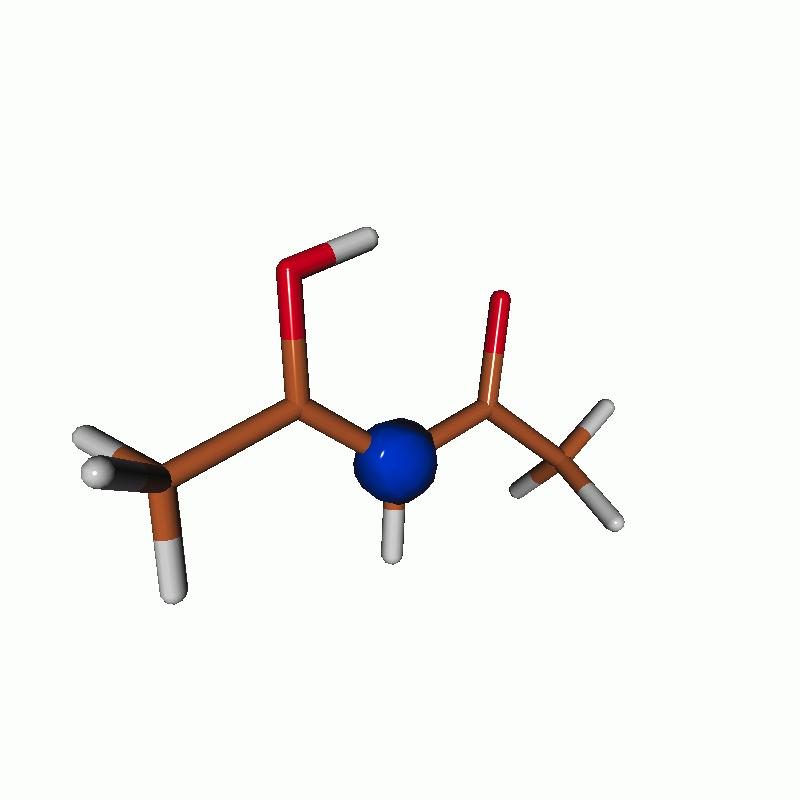}
     \end{minipage}
     & 0.64
     &  \begin{minipage}{0.1\textwidth}
         \centering
         \includegraphics[scale=0.05]{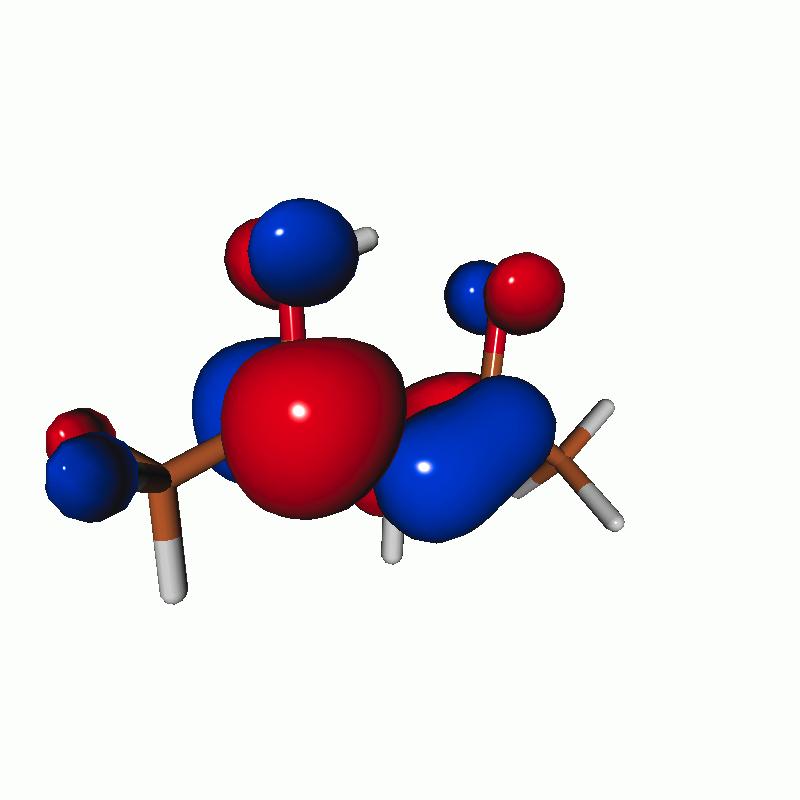}
     \end{minipage}
     \\
     \hline
     \end{tabular}
 \label{tab:AcAc_NTO_momS2_S1min}
 \end{table}

On the basis of the above analysis, we conclude that, despite spin contamination, LSOR-CCSD describes the XAS of singlet valence-excited states with reasonable accuracy. LSOR-CCSD could even be used as benchmark for other levels of  theory, especially when experimental TR-XAS spectra are not available. 

We conclude this section by analyzing the 
MOM-TDDFT results for the transient absorption.
As seen in Secs.~\ref{subsec:gs_to_core} and \ref{subsec:gs_to_valence}, 
ADC(2) and TDDFT/B3LYP yield reasonable results for the lowest-lying core-excited states and for the valence-excited states of interest in the nuclear dynamics. The next question is thus whether MOM-TDDFT/B3LYP
can reproduce the main peaks of the time-resolved spectra with reasonable accuracy. 
We attempt to answer this question by 
comparing the MOM-TDDFT/B3LYP spectra of thymine and acetylacetone with the surface cuts of the experimental spectra.

\begin{figure}[hbt]
    \centering
    \includegraphics[width=8.5cm,height=8.5cm,keepaspectratio]{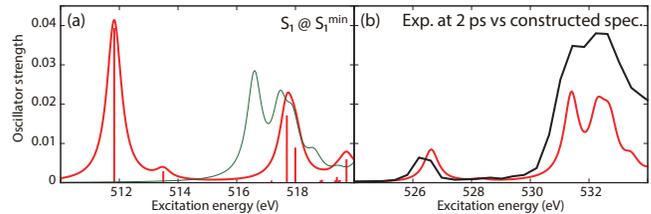}
    \caption{(a) Red: Oxygen K-edge NEXAFS for thymine in the S$_1$(n$\pi^{\ast}$) state calculated at the MOM-TDDFT/B3LYP/6-311++G** level at the potential energy minimum. Green: Ground-state spectrum. (b) Black: Experimental spectrum at the delay time of 2 ps \cite{Wolf}, Red: computational spectrum made from the red and the green curves of (a), shifted by $+$14.8 eV. The red curve of (a) was scaled by 0.2. The green curve of (a) was scaled by 0.8. FWHM of the Lorentzian convolution function is 0.6 eV.}
    \label{fig:thymine_S1min_B3LYP}
\end{figure}

The MOM-TDDFT/B3LYP O K-edge NEXAFS spectrum of thymine in the S$_1$(n$\pi^{\ast}$) state is shown in Fig.~\ref{fig:thymine_S1min_B3LYP}, panel (a). For construction of the surface cut of the theoretical absorption spectra, we used FWHM of 0.6 eV for the Lorentzian convolution function. A theoretical surface cut spectrum was constructed as sum of the MOM-TDDFT spectrum and the standard TDDFT spectrum of the ground state, scaled by 0.2 and 0.8, respectively. This is shown in panel (b), together with the experimental surface cut spectrum at 2 ps delay.\cite{Wolf} The MOM-TDDFT/B3LYP peaks due to the core transitions from O4 and O2 to SOMO (n) are found at 511.82 and 513.50 eV, respectively. The peak corresponding to the first main peak of the ground-state spectrum is missing, and the one corresponding to the second main peak in the GS appears at 517.71 eV. These features are equivalent to what we observed in the LSOR-CCSD case (see Fig.~\ref{fig:thymine_S1min}). Thus, the separation between the core-to-SOMO peak and the ground-state main peaks is accurately reproduced.

\begin{figure}[hbt]
    \centering
    \includegraphics[width=8.5cm,height=8.5cm,keepaspectratio]{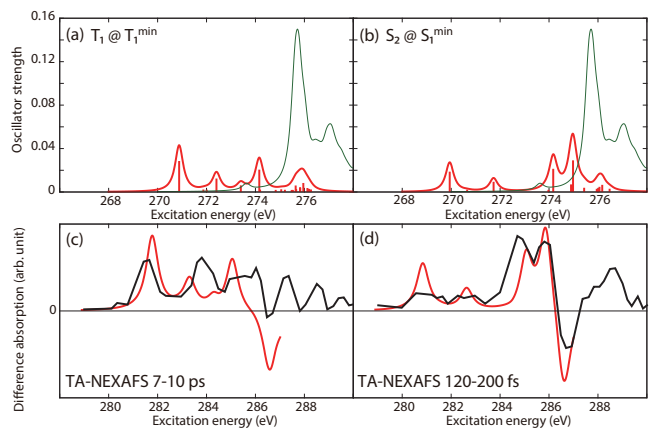}
    \caption{(a), (b) Carbon K-edge NEXAFS for acetylacetone in the T$_1$($\pi\pi^{\ast}$) and S$_2$($\pi\pi^{\ast}$) states calculated at the MOM-TDDFT/B3LYP/6-311++G** level at the potential energy minima of T$_1$($\pi\pi^{\ast}$)  and S$_1$(n$\pi^{\ast}$), respectively. The green curve is the ground-state spectrum. In panels (c) and (d) the experimental transient absorption spectra at delay times of 7-10 ps and 120-200 fs are reported with black lines.\cite{acac_ultrafast_ISC} 
    In red are the computational transient absorption spectra reconstructed from the red and green curves of panels (a) and (b), respectively, shifted by $+$10.9 eV. The red curves of (a) and (b) were scaled by 0.75, and subtracted from the green curves, which were scaled by 0.25. FWHM of the Lorentzian convolution function is 0.4 eV for panels (a) and (b), 0.6 eV for panels (c) and (d), respectively.}
    \label{fig:AcAc_T1_S2_B3LYP}
\end{figure}

Next, we consider the carbon K-edge spectra of acetylacetone in the T$_1$($\pi\pi^{\ast}$) [at the minimum of T$_1$($\pi\pi^{\ast}$)] and S$_2$($\pi\pi^{\ast}$) [at the minimum of S$_1$(n$\pi^{\ast}$)] states, as obtained from MOM-TDDFT. They are plotted in panels (a) and (b) of Fig.~\ref{fig:AcAc_T1_S2_B3LYP}, respectively. Surface cuts of the transient-absorption NEXAFS spectra were constructed by subtracting the TDDFT spectrum, scaled by 0.25, with the MOM-TDDFT spectra scaled by 0.75. For this construction, we convoluted the oscillator strengths with a Lorentzian function (FWHM = 0.6 eV), and chose the factors 0.75 and 0.25 for the best fit with the experimental spectra.
They are superposed with those from  experiment at delay times of 7-10 ps and 120-200 fs in Fig.~\ref{fig:AcAc_T1_S2_B3LYP}, panels (c) and (d). The MOM-TDDFT spectrum of T$_1$($\pi\pi^{\ast}$) exhibits the core-to-SOMO transition peaks at 270.88 and 272.41 eV. A  peak  due to the transition to the half-occupied $\pi^{\ast}$ orbital occurs at 274.16 eV. 
All peaks observed in the LSOR-CCSD spectrum were also obtained by MOM-TDDFT. The fine structure of the surface-cut transient absorption spectrum is qualitatively reproduced. 

The MOM-TDDFT spectrum of S$_2$($\pi\pi^{\ast}$) exhibits the core-to-SOMO($\pi^\ast$) transition peaks at 269.94 and 271.73 eV. The peaks due to the transitions to the half-occupied $\pi^{\ast}$ orbital appear at 274.17 and 274.98 eV. The reconstructed transient-absorption spectrum agrees well with the experimental surface-cut spectrum.

\section{Summary and Conclusions}

We have analyzed the performance of different single-reference electronic structure methods for excited-state XAS calculations. The analysis was carried out in three steps.
First, we compared the results for
the ground-state XAS spectra of uracil, thymine, and acetylacetone computed using CVS-ADC(2), CVS-EOM-CCSD, and TDDFT/B3LYP, and with the experimental spectra.
Second, we computed the
excitation energies of the valence-excited states
presumably involved in the dynamics
at ADC(2), EOM-EE-CCSD, and TDDFT/B3LYP levels, and compared them with the experimental data from EELS and UV absorption. 
Third, we analyzed different protocols for the XAS spectra of the lowest-lying valence-excited states based on the CCSD ansatz, 
namely, regular CVS-EOM-CCSD for transitions between excited states, and EOM-CCSD applied on the excited-state reference state optimized imposing the MOM constraint.
The results for thymine and acetylacetone 
were evaluated by comparison with the experimental time-resolved spectra. 
Finally, the performance of MOM-TDDFT/B3LYP for TR-XAS was evaluated, again on thymine and acetylacetone, by comparison with the LSOR-CCSD and the experimental spectra.

In the first step, we found that CVS-EOM-CCSD reproduces well the entire pre-edge region of the ground-state XAS spectra. On the other hand, CVS-ADC(2) and TDDFT/B3LYP only describe the lowest-lying core excitations with reasonable accuracy, while the Rydberg region is not captured.  In the second step, we observed that EOM-EE-CCSD, ADC(2), and TDDFT/B3LYP treat the valence-excited states with a comparable accuracy.

Among the methods analyzed in the third step, only LSOR-CCSD and MOM-TDDFT can reproduce  the entire pre-bleaching region of the excited-state XAS spectra for thymine and acetylacetone, despite spin contamination of the singlet excited states. 
LSOR-CCSD could be used as the reference when evaluating the performance of other electronic structure methods for excited-state XAS, especially if no experimental spectra are available. 
For the spectra of the spin-singlet states, CVS-EOM-CCSD yields slightly better core$\to$SOMO positions.

We note that the same procedure can be used to assess the performance of other xc-functional or post-HF methods for TR-XAS calculations. We also note that description of an initial state with the MOM algorithm is reasonably accurate only when the initial state has a single configurational wave-function character.
The low computational scaling and reasonable accuracy of MOM-TDDFT makes it rather attractive for the on-the-fly calculation of TR-XAS spectra in the excited-state nuclear dynamics simulations.

\section*{Supplementary material}

See supplementary material for the NTOs of all core and valence excitations.

\section*{Data availability statement}
The data that supports the findings of this study are available within the article and its supplementary material.

\section*{Acknowledgments}

This research was funded by the European
Union’s Horizon 2020 research and innovation program under the Marie Sk{\l}odowska-Curie Grant Agreements No. 713683 (COFUNDfellowsDTU) and 
No. 765739 (COSINE -- COmputational Spectroscopy
In Natural sciences and Engineering),
by DTU Chemistry, and by the Danish Council
for Independent Research 
(now Independent Research Fund
Denmark), Grant Nos. 7014-00258B, 4002-00272, 014-00258B, and 8021-00347B.
A.I.K. was supported by the U.S. National Science Foundation (No. CHE-1856342).

\section*{Conflicts of interest}
The authors declare the following competing financial interest(s): A. I. K. is president and part-owner of Q-Chem, Inc.

\section*{References}
\bibliography{abbr,shota}

\begin{thebibliography}{128}%
\makeatletter
\providecommand \@ifxundefined [1]{%
 \@ifx{#1\undefined}
}%
\providecommand \@ifnum [1]{%
 \ifnum #1\expandafter \@firstoftwo
 \else \expandafter \@secondoftwo
 \fi
}%
\providecommand \@ifx [1]{%
 \ifx #1\expandafter \@firstoftwo
 \else \expandafter \@secondoftwo
 \fi
}%
\providecommand \natexlab [1]{#1}%
\providecommand \enquote  [1]{``#1''}%
\providecommand \bibnamefont  [1]{#1}%
\providecommand \bibfnamefont [1]{#1}%
\providecommand \citenamefont [1]{#1}%
\providecommand \href@noop [0]{\@secondoftwo}%
\providecommand \href [0]{\begingroup \@sanitize@url \@href}%
\providecommand \@href[1]{\@@startlink{#1}\@@href}%
\providecommand \@@href[1]{\endgroup#1\@@endlink}%
\providecommand \@sanitize@url [0]{\catcode `\\12\catcode `\$12\catcode
  `\&12\catcode `\#12\catcode `\^12\catcode `\_12\catcode `\%12\relax}%
\providecommand \@@startlink[1]{}%
\providecommand \@@endlink[0]{}%
\providecommand \url  [0]{\begingroup\@sanitize@url \@url }%
\providecommand \@url [1]{\endgroup\@href {#1}{\urlprefix }}%
\providecommand \urlprefix  [0]{URL }%
\providecommand \Eprint [0]{\href }%
\providecommand \doibase [0]{http://dx.doi.org/}%
\providecommand \selectlanguage [0]{\@gobble}%
\providecommand \bibinfo  [0]{\@secondoftwo}%
\providecommand \bibfield  [0]{\@secondoftwo}%
\providecommand \translation [1]{[#1]}%
\providecommand \BibitemOpen [0]{}%
\providecommand \bibitemStop [0]{}%
\providecommand \bibitemNoStop [0]{.\EOS\space}%
\providecommand \EOS [0]{\spacefactor3000\relax}%
\providecommand \BibitemShut  [1]{\csname bibitem#1\endcsname}%
\let\auto@bib@innerbib\@empty
\bibitem [{\citenamefont {Scherer}\ \emph {et~al.}(1985)\citenamefont
  {Scherer}, \citenamefont {Knee}, \citenamefont {Smith},\ and\ \citenamefont
  {Zewail}}]{Zewail85}%
  \BibitemOpen
  \bibfield  {author} {\bibinfo {author} {\bibfnamefont {N.~F.}\ \bibnamefont
  {Scherer}}, \bibinfo {author} {\bibfnamefont {J.~L.}\ \bibnamefont {Knee}},
  \bibinfo {author} {\bibfnamefont {D.~D.}\ \bibnamefont {Smith}}, \ and\
  \bibinfo {author} {\bibfnamefont {A.~H.}\ \bibnamefont {Zewail}},\ }\bibfield
   {title} {\enquote {\bibinfo {title} {Femtosecond photofragment spectroscopy:
  The reaction {ICN} $\to$ {CN} + {I}},}\ }\href {\doibase 10.1021/j100270a001}
  {\bibfield  {journal} {\bibinfo  {journal} {J. Phys. Chem.}\ }\textbf
  {\bibinfo {volume} {89}},\ \bibinfo {pages} {5141--5143} (\bibinfo {year}
  {1985})}\BibitemShut {NoStop}%
\bibitem [{\citenamefont {Young}\ \emph {et~al.}(2018)\citenamefont {Young},
  \citenamefont {Ueda}, \citenamefont {G\"uhr}, \citenamefont {Bucksbaum},
  \citenamefont {Simon}, \citenamefont {Mukamel}, \citenamefont {Rohringer},
  \citenamefont {Prince}, \citenamefont {Masciovecchio}, \citenamefont {Meyer},
  \citenamefont {Rudenko}, \citenamefont {Rolles}, \citenamefont {Bostedt},
  \citenamefont {Fuchs}, \citenamefont {Reis}, \citenamefont {Santra},
  \citenamefont {Kapteyn}, \citenamefont {Murnane}, \citenamefont {Ibrahim},
  \citenamefont {L{\'{e}}gar{\'{e}}}, \citenamefont {Vrakking}, \citenamefont
  {Isinger}, \citenamefont {Kroon}, \citenamefont {Gisselbrecht}, \citenamefont
  {L'Huillier}, \citenamefont {W\"orner},\ and\ \citenamefont
  {Leone}}]{Roadmap_ultrafast_Xray_Young}%
  \BibitemOpen
  \bibfield  {author} {\bibinfo {author} {\bibfnamefont {L.}~\bibnamefont
  {Young}}, \bibinfo {author} {\bibfnamefont {K.}~\bibnamefont {Ueda}},
  \bibinfo {author} {\bibfnamefont {M.}~\bibnamefont {G\"uhr}}, \bibinfo
  {author} {\bibfnamefont {P.~H.}\ \bibnamefont {Bucksbaum}}, \bibinfo {author}
  {\bibfnamefont {M.}~\bibnamefont {Simon}}, \bibinfo {author} {\bibfnamefont
  {S.}~\bibnamefont {Mukamel}}, \bibinfo {author} {\bibfnamefont
  {N.}~\bibnamefont {Rohringer}}, \bibinfo {author} {\bibfnamefont {K.~C.}\
  \bibnamefont {Prince}}, \bibinfo {author} {\bibfnamefont {C.}~\bibnamefont
  {Masciovecchio}}, \bibinfo {author} {\bibfnamefont {M.}~\bibnamefont
  {Meyer}}, \bibinfo {author} {\bibfnamefont {A.}~\bibnamefont {Rudenko}},
  \bibinfo {author} {\bibfnamefont {D.}~\bibnamefont {Rolles}}, \bibinfo
  {author} {\bibfnamefont {C.}~\bibnamefont {Bostedt}}, \bibinfo {author}
  {\bibfnamefont {M.}~\bibnamefont {Fuchs}}, \bibinfo {author} {\bibfnamefont
  {D.~A.}\ \bibnamefont {Reis}}, \bibinfo {author} {\bibfnamefont
  {R.}~\bibnamefont {Santra}}, \bibinfo {author} {\bibfnamefont
  {H.}~\bibnamefont {Kapteyn}}, \bibinfo {author} {\bibfnamefont
  {M.}~\bibnamefont {Murnane}}, \bibinfo {author} {\bibfnamefont
  {H.}~\bibnamefont {Ibrahim}}, \bibinfo {author} {\bibfnamefont
  {F.}~\bibnamefont {L{\'{e}}gar{\'{e}}}}, \bibinfo {author} {\bibfnamefont
  {M.}~\bibnamefont {Vrakking}}, \bibinfo {author} {\bibfnamefont
  {M.}~\bibnamefont {Isinger}}, \bibinfo {author} {\bibfnamefont
  {D.}~\bibnamefont {Kroon}}, \bibinfo {author} {\bibfnamefont
  {M.}~\bibnamefont {Gisselbrecht}}, \bibinfo {author} {\bibfnamefont
  {A.}~\bibnamefont {L'Huillier}}, \bibinfo {author} {\bibfnamefont {H.~J.}\
  \bibnamefont {W\"orner}}, \ and\ \bibinfo {author} {\bibfnamefont {S.~R.}\
  \bibnamefont {Leone}},\ }\bibfield  {title} {\enquote {\bibinfo {title}
  {Roadmap of ultrafast x-ray atomic and molecular physics},}\ }\href {\doibase
  10.1088/1361-6455/aa9735} {\bibfield  {journal} {\bibinfo  {journal} {J.
  Phys. B}\ }\textbf {\bibinfo {volume} {51}},\ \bibinfo {pages} {032003}
  (\bibinfo {year} {2018})}\BibitemShut {NoStop}%
\bibitem [{\citenamefont {Chergui}\ and\ \citenamefont
  {Collet}(2017)}]{Chergui2017}%
  \BibitemOpen
  \bibfield  {author} {\bibinfo {author} {\bibfnamefont {M.}~\bibnamefont
  {Chergui}}\ and\ \bibinfo {author} {\bibfnamefont {E.}~\bibnamefont
  {Collet}},\ }\bibfield  {title} {\enquote {\bibinfo {title} {Photoinduced
  structural dynamics of molecular systems mapped by time-resolved x-ray
  methods},}\ }\href {\doibase 10.1021/acs.chemrev.6b00831} {\bibfield
  {journal} {\bibinfo  {journal} {Chem. Rev.}\ }\textbf {\bibinfo {volume}
  {117}},\ \bibinfo {pages} {11025--11065} (\bibinfo {year}
  {2017})}\BibitemShut {NoStop}%
\bibitem [{\citenamefont {Ueda}(2018)}]{Ueda}%
  \BibitemOpen
  \bibfield  {author} {\bibinfo {author} {\bibfnamefont {K.}~\bibnamefont
  {Ueda}},\ }\href {\doibase 10.3390/books978-3-03842-880-0} {\emph {\bibinfo
  {title} {X-ray Free Electron Lasers}}}\ (\bibinfo  {publisher} {MDPI},\
  \bibinfo {address} {Basel, Switzerland},\ \bibinfo {year} {2018})\BibitemShut
  {NoStop}%
\bibitem [{\citenamefont {Calegari}\ \emph {et~al.}(2016)\citenamefont
  {Calegari}, \citenamefont {Sansone}, \citenamefont {Stagira}, \citenamefont
  {Vozzi},\ and\ \citenamefont {Nisoli}}]{Calegari_2016}%
  \BibitemOpen
  \bibfield  {author} {\bibinfo {author} {\bibfnamefont {F.}~\bibnamefont
  {Calegari}}, \bibinfo {author} {\bibfnamefont {G.}~\bibnamefont {Sansone}},
  \bibinfo {author} {\bibfnamefont {S.}~\bibnamefont {Stagira}}, \bibinfo
  {author} {\bibfnamefont {C.}~\bibnamefont {Vozzi}}, \ and\ \bibinfo {author}
  {\bibfnamefont {M.}~\bibnamefont {Nisoli}},\ }\bibfield  {title} {\enquote
  {\bibinfo {title} {Advances in attosecond science},}\ }\href {\doibase
  10.1088/0953-4075/49/6/062001} {\bibfield  {journal} {\bibinfo  {journal} {J.
  Phys. B}\ }\textbf {\bibinfo {volume} {49}},\ \bibinfo {pages} {062001}
  (\bibinfo {year} {2016})}\BibitemShut {NoStop}%
\bibitem [{\citenamefont {Ramasesha}, \citenamefont {Leone},\ and\
  \citenamefont {Neumark}(2016)}]{Ramasesha}%
  \BibitemOpen
  \bibfield  {author} {\bibinfo {author} {\bibfnamefont {K.}~\bibnamefont
  {Ramasesha}}, \bibinfo {author} {\bibfnamefont {S.~R.}\ \bibnamefont
  {Leone}}, \ and\ \bibinfo {author} {\bibfnamefont {D.~M.}\ \bibnamefont
  {Neumark}},\ }\bibfield  {title} {\enquote {\bibinfo {title} {Real-time
  probing of electron dynamics using attosecond time-resolved spectroscopy},}\
  }\href {\doibase 10.1146/annurev-physchem-040215-112025} {\bibfield
  {journal} {\bibinfo  {journal} {Annu. Rev. Phys. Chem.}\ }\textbf {\bibinfo
  {volume} {67}},\ \bibinfo {pages} {41--63} (\bibinfo {year}
  {2016})}\BibitemShut {NoStop}%
\bibitem [{\citenamefont {Ischenko}, \citenamefont {Weber},\ and\ \citenamefont
  {Miller}(2017)}]{Ischenko}%
  \BibitemOpen
  \bibfield  {author} {\bibinfo {author} {\bibfnamefont {A.~A.}\ \bibnamefont
  {Ischenko}}, \bibinfo {author} {\bibfnamefont {P.~M.}\ \bibnamefont {Weber}},
  \ and\ \bibinfo {author} {\bibfnamefont {R.~J.~D.}\ \bibnamefont {Miller}},\
  }\bibfield  {title} {\enquote {\bibinfo {title} {Capturing chemistry in
  action with electrons: Realization of atomically resolved reaction
  dynamics},}\ }\href {\doibase 10.1021/acs.chemrev.6b00770} {\bibfield
  {journal} {\bibinfo  {journal} {Chem. Rev.}\ }\textbf {\bibinfo {volume}
  {117}},\ \bibinfo {pages} {11066--11124} (\bibinfo {year}
  {2017})}\BibitemShut {NoStop}%
\bibitem [{\citenamefont {Villeneuve}(2018)}]{Villeneuve}%
  \BibitemOpen
  \bibfield  {author} {\bibinfo {author} {\bibfnamefont {D.~M.}\ \bibnamefont
  {Villeneuve}},\ }\bibfield  {title} {\enquote {\bibinfo {title} {Attosecond
  science},}\ }\href {\doibase 10.1080/00107514.2017.1407093} {\bibfield
  {journal} {\bibinfo  {journal} {Contemp. Phys.}\ }\textbf {\bibinfo {volume}
  {59}},\ \bibinfo {pages} {47--61} (\bibinfo {year} {2018})}\BibitemShut
  {NoStop}%
\bibitem [{\citenamefont {Schuurman}\ and\ \citenamefont
  {Stolow}(2018)}]{Schuurman_Stolow}%
  \BibitemOpen
  \bibfield  {author} {\bibinfo {author} {\bibfnamefont {M.~S.}\ \bibnamefont
  {Schuurman}}\ and\ \bibinfo {author} {\bibfnamefont {A.}~\bibnamefont
  {Stolow}},\ }\bibfield  {title} {\enquote {\bibinfo {title} {Dynamics at
  conical intersections},}\ }\href {\doibase
  10.1146/annurev-physchem-052516-050721} {\bibfield  {journal} {\bibinfo
  {journal} {Ann. Rev. Phys. Chem.}\ }\textbf {\bibinfo {volume} {69}},\
  \bibinfo {pages} {427--450} (\bibinfo {year} {2018})}\BibitemShut {NoStop}%
\bibitem [{\citenamefont {Adachi}\ and\ \citenamefont
  {Suzuki}(2018)}]{Adachi_Suzuki}%
  \BibitemOpen
  \bibfield  {author} {\bibinfo {author} {\bibfnamefont {S.}~\bibnamefont
  {Adachi}}\ and\ \bibinfo {author} {\bibfnamefont {T.}~\bibnamefont
  {Suzuki}},\ }\bibfield  {title} {\enquote {\bibinfo {title} {{UV}-driven
  harmonic generation for time-resolved photoelectron spectroscopy of
  polyatomic molecules},}\ }\href {\doibase 10.3390/app8101784} {\bibfield
  {journal} {\bibinfo  {journal} {Appl. Sci.}\ }\textbf {\bibinfo {volume}
  {8}},\ \bibinfo {pages} {1784} (\bibinfo {year} {2018})}\BibitemShut
  {NoStop}%
\bibitem [{\citenamefont {Suzuki}(2019)}]{TSuzuki19}%
  \BibitemOpen
  \bibfield  {author} {\bibinfo {author} {\bibfnamefont {T.}~\bibnamefont
  {Suzuki}},\ }\bibfield  {title} {\enquote {\bibinfo {title} {Ultrafast
  photoelectron spectroscopy of aqueous solutions},}\ }\href {\doibase
  10.1063/1.5098402} {\bibfield  {journal} {\bibinfo  {journal} {J. Chem.
  Phys.}\ }\textbf {\bibinfo {volume} {151}},\ \bibinfo {pages} {090901}
  (\bibinfo {year} {2019})}\BibitemShut {NoStop}%
\bibitem [{\citenamefont {Liu}\ \emph {et~al.}(2020)\citenamefont {Liu},
  \citenamefont {Horton}, \citenamefont {Yang}, \citenamefont {Nunes},
  \citenamefont {Shen}, \citenamefont {Wolf}, \citenamefont {Forbes},
  \citenamefont {Cheng}, \citenamefont {Moore}, \citenamefont {Centurion},
  \citenamefont {Hegazy}, \citenamefont {Li}, \citenamefont {Lin},
  \citenamefont {Stolow}, \citenamefont {Hockett}, \citenamefont {Rozgonyi},
  \citenamefont {Marquetand}, \citenamefont {Wang},\ and\ \citenamefont
  {Weinacht}}]{TRPES_UED}%
  \BibitemOpen
  \bibfield  {author} {\bibinfo {author} {\bibfnamefont {Y.}~\bibnamefont
  {Liu}}, \bibinfo {author} {\bibfnamefont {S.~L.}\ \bibnamefont {Horton}},
  \bibinfo {author} {\bibfnamefont {J.}~\bibnamefont {Yang}}, \bibinfo {author}
  {\bibfnamefont {J.~P.~F.}\ \bibnamefont {Nunes}}, \bibinfo {author}
  {\bibfnamefont {X.}~\bibnamefont {Shen}}, \bibinfo {author} {\bibfnamefont
  {T.~J.~A.}\ \bibnamefont {Wolf}}, \bibinfo {author} {\bibfnamefont
  {R.}~\bibnamefont {Forbes}}, \bibinfo {author} {\bibfnamefont
  {C.}~\bibnamefont {Cheng}}, \bibinfo {author} {\bibfnamefont
  {B.}~\bibnamefont {Moore}}, \bibinfo {author} {\bibfnamefont
  {M.}~\bibnamefont {Centurion}}, \bibinfo {author} {\bibfnamefont
  {K.}~\bibnamefont {Hegazy}}, \bibinfo {author} {\bibfnamefont
  {R.}~\bibnamefont {Li}}, \bibinfo {author} {\bibfnamefont {M.-F.}\
  \bibnamefont {Lin}}, \bibinfo {author} {\bibfnamefont {A.}~\bibnamefont
  {Stolow}}, \bibinfo {author} {\bibfnamefont {P.}~\bibnamefont {Hockett}},
  \bibinfo {author} {\bibfnamefont {T.}~\bibnamefont {Rozgonyi}}, \bibinfo
  {author} {\bibfnamefont {P.}~\bibnamefont {Marquetand}}, \bibinfo {author}
  {\bibfnamefont {X.}~\bibnamefont {Wang}}, \ and\ \bibinfo {author}
  {\bibfnamefont {T.}~\bibnamefont {Weinacht}},\ }\bibfield  {title} {\enquote
  {\bibinfo {title} {Spectroscopic and structural probing of excited-state
  molecular dynamics with time-resolved photoelectron spectroscopy and
  ultrafast electron diffraction},}\ }\href {\doibase
  10.1103/PhysRevX.10.021016} {\bibfield  {journal} {\bibinfo  {journal} {Phys.
  Rev. X}\ }\textbf {\bibinfo {volume} {10}},\ \bibinfo {pages} {021016}
  (\bibinfo {year} {2020})}\BibitemShut {NoStop}%
\bibitem [{\citenamefont {Glownia}\ \emph {et~al.}(2016)\citenamefont
  {Glownia}, \citenamefont {Natan}, \citenamefont {Cryan}, \citenamefont
  {Hartsock}, \citenamefont {Kozina}, \citenamefont {Minitti}, \citenamefont
  {Nelson}, \citenamefont {Robinson}, \citenamefont {Sato}, \citenamefont {van
  Driel}, \citenamefont {Welch}, \citenamefont {Weninger}, \citenamefont
  {Zhu},\ and\ \citenamefont {Bucksbaum}}]{Bucksbaum2016}%
  \BibitemOpen
  \bibfield  {author} {\bibinfo {author} {\bibfnamefont {J.~M.}\ \bibnamefont
  {Glownia}}, \bibinfo {author} {\bibfnamefont {A.}~\bibnamefont {Natan}},
  \bibinfo {author} {\bibfnamefont {J.~P.}\ \bibnamefont {Cryan}}, \bibinfo
  {author} {\bibfnamefont {R.}~\bibnamefont {Hartsock}}, \bibinfo {author}
  {\bibfnamefont {M.}~\bibnamefont {Kozina}}, \bibinfo {author} {\bibfnamefont
  {M.~P.}\ \bibnamefont {Minitti}}, \bibinfo {author} {\bibfnamefont
  {S.}~\bibnamefont {Nelson}}, \bibinfo {author} {\bibfnamefont
  {J.}~\bibnamefont {Robinson}}, \bibinfo {author} {\bibfnamefont
  {T.}~\bibnamefont {Sato}}, \bibinfo {author} {\bibfnamefont {T.}~\bibnamefont
  {van Driel}}, \bibinfo {author} {\bibfnamefont {G.}~\bibnamefont {Welch}},
  \bibinfo {author} {\bibfnamefont {C.}~\bibnamefont {Weninger}}, \bibinfo
  {author} {\bibfnamefont {D.}~\bibnamefont {Zhu}}, \ and\ \bibinfo {author}
  {\bibfnamefont {P.~H.}\ \bibnamefont {Bucksbaum}},\ }\bibfield  {title}
  {\enquote {\bibinfo {title} {Self-referenced coherent diffraction x-ray movie
  of {{\r{A}}}ngstrom- and femtosecond-scale atomic motion},}\ }\href {\doibase
  10.1103/PhysRevLett.117.153003} {\bibfield  {journal} {\bibinfo  {journal}
  {Phys. Rev. Lett.}\ }\textbf {\bibinfo {volume} {117}},\ \bibinfo {pages}
  {153003} (\bibinfo {year} {2016})}\BibitemShut {NoStop}%
\bibitem [{\citenamefont {Stankus}\ \emph {et~al.}(2019)\citenamefont
  {Stankus}, \citenamefont {Yong}, \citenamefont {Zotev}, \citenamefont
  {Ruddock}, \citenamefont {Bellshaw}, \citenamefont {Lane}, \citenamefont
  {Liang}, \citenamefont {Boutet}, \citenamefont {Carbajo}, \citenamefont
  {Robinson}, \citenamefont {Du}, \citenamefont {Goff}, \citenamefont {Chang},
  \citenamefont {Koglin}, \citenamefont {Minitti}, \citenamefont {Kirrander},\
  and\ \citenamefont {Weber}}]{Weber2019}%
  \BibitemOpen
  \bibfield  {author} {\bibinfo {author} {\bibfnamefont {B.}~\bibnamefont
  {Stankus}}, \bibinfo {author} {\bibfnamefont {H.}~\bibnamefont {Yong}},
  \bibinfo {author} {\bibfnamefont {N.}~\bibnamefont {Zotev}}, \bibinfo
  {author} {\bibfnamefont {J.~M.}\ \bibnamefont {Ruddock}}, \bibinfo {author}
  {\bibfnamefont {D.}~\bibnamefont {Bellshaw}}, \bibinfo {author}
  {\bibfnamefont {T.~J.}\ \bibnamefont {Lane}}, \bibinfo {author}
  {\bibfnamefont {M.}~\bibnamefont {Liang}}, \bibinfo {author} {\bibfnamefont
  {S.}~\bibnamefont {Boutet}}, \bibinfo {author} {\bibfnamefont
  {S.}~\bibnamefont {Carbajo}}, \bibinfo {author} {\bibfnamefont {J.~S.}\
  \bibnamefont {Robinson}}, \bibinfo {author} {\bibfnamefont {W.}~\bibnamefont
  {Du}}, \bibinfo {author} {\bibfnamefont {N.}~\bibnamefont {Goff}}, \bibinfo
  {author} {\bibfnamefont {Y.}~\bibnamefont {Chang}}, \bibinfo {author}
  {\bibfnamefont {J.~E.}\ \bibnamefont {Koglin}}, \bibinfo {author}
  {\bibfnamefont {M.~P.}\ \bibnamefont {Minitti}}, \bibinfo {author}
  {\bibfnamefont {A.}~\bibnamefont {Kirrander}}, \ and\ \bibinfo {author}
  {\bibfnamefont {P.~M.}\ \bibnamefont {Weber}},\ }\bibfield  {title} {\enquote
  {\bibinfo {title} {Ultrafast {X}-ray scattering reveals vibrational coherence
  following {Rydberg} excitation},}\ }\href {\doibase
  10.1038/s41557-019-0291-0} {\bibfield  {journal} {\bibinfo  {journal} {Nat.
  Chem.}\ }\textbf {\bibinfo {volume} {11}},\ \bibinfo {pages} {716--721}
  (\bibinfo {year} {2019})}\BibitemShut {NoStop}%
\bibitem [{\citenamefont {Ruddock}\ \emph {et~al.}(2019)\citenamefont
  {Ruddock}, \citenamefont {Yong}, \citenamefont {Stankus}, \citenamefont {Du},
  \citenamefont {Goff}, \citenamefont {Chang}, \citenamefont {Odate},
  \citenamefont {Carrascosa}, \citenamefont {Bellshaw}, \citenamefont {Zotev},
  \citenamefont {Liang}, \citenamefont {Carbajo}, \citenamefont {Koglin},
  \citenamefont {Robinson}, \citenamefont {Boutet}, \citenamefont {Kirrander},
  \citenamefont {Minitti},\ and\ \citenamefont {Weber}}]{Ruddockeaax6625}%
  \BibitemOpen
  \bibfield  {author} {\bibinfo {author} {\bibfnamefont {J.~M.}\ \bibnamefont
  {Ruddock}}, \bibinfo {author} {\bibfnamefont {H.}~\bibnamefont {Yong}},
  \bibinfo {author} {\bibfnamefont {B.}~\bibnamefont {Stankus}}, \bibinfo
  {author} {\bibfnamefont {W.}~\bibnamefont {Du}}, \bibinfo {author}
  {\bibfnamefont {N.}~\bibnamefont {Goff}}, \bibinfo {author} {\bibfnamefont
  {Y.}~\bibnamefont {Chang}}, \bibinfo {author} {\bibfnamefont
  {A.}~\bibnamefont {Odate}}, \bibinfo {author} {\bibfnamefont {A.~M.}\
  \bibnamefont {Carrascosa}}, \bibinfo {author} {\bibfnamefont
  {D.}~\bibnamefont {Bellshaw}}, \bibinfo {author} {\bibfnamefont
  {N.}~\bibnamefont {Zotev}}, \bibinfo {author} {\bibfnamefont
  {M.}~\bibnamefont {Liang}}, \bibinfo {author} {\bibfnamefont
  {S.}~\bibnamefont {Carbajo}}, \bibinfo {author} {\bibfnamefont
  {J.}~\bibnamefont {Koglin}}, \bibinfo {author} {\bibfnamefont {J.~S.}\
  \bibnamefont {Robinson}}, \bibinfo {author} {\bibfnamefont {S.}~\bibnamefont
  {Boutet}}, \bibinfo {author} {\bibfnamefont {A.}~\bibnamefont {Kirrander}},
  \bibinfo {author} {\bibfnamefont {M.~P.}\ \bibnamefont {Minitti}}, \ and\
  \bibinfo {author} {\bibfnamefont {P.~M.}\ \bibnamefont {Weber}},\ }\bibfield
  {title} {\enquote {\bibinfo {title} {A deep {UV} trigger for ground-state
  ring-opening dynamics of 1,3-cyclohexadiene},}\ }\href {\doibase
  10.1126/sciadv.aax6625} {\bibfield  {journal} {\bibinfo  {journal} {Sci.
  Adv.}\ }\textbf {\bibinfo {volume} {5}},\ \bibinfo {pages} {eaax6625}
  (\bibinfo {year} {2019})}\BibitemShut {NoStop}%
\bibitem [{\citenamefont {Stankus}\ \emph {et~al.}(2020)\citenamefont
  {Stankus}, \citenamefont {Yong}, \citenamefont {Ruddock}, \citenamefont {Ma},
  \citenamefont {Carrascosa}, \citenamefont {Goff}, \citenamefont {Boutet},
  \citenamefont {Xu}, \citenamefont {Zotev}, \citenamefont {Kirrander},
  \citenamefont {Minitti},\ and\ \citenamefont {Weber}}]{Stankus_2020}%
  \BibitemOpen
  \bibfield  {author} {\bibinfo {author} {\bibfnamefont {B.}~\bibnamefont
  {Stankus}}, \bibinfo {author} {\bibfnamefont {H.}~\bibnamefont {Yong}},
  \bibinfo {author} {\bibfnamefont {J.}~\bibnamefont {Ruddock}}, \bibinfo
  {author} {\bibfnamefont {L.}~\bibnamefont {Ma}}, \bibinfo {author}
  {\bibfnamefont {A.~M.}\ \bibnamefont {Carrascosa}}, \bibinfo {author}
  {\bibfnamefont {N.}~\bibnamefont {Goff}}, \bibinfo {author} {\bibfnamefont
  {S.}~\bibnamefont {Boutet}}, \bibinfo {author} {\bibfnamefont
  {X.}~\bibnamefont {Xu}}, \bibinfo {author} {\bibfnamefont {N.}~\bibnamefont
  {Zotev}}, \bibinfo {author} {\bibfnamefont {A.}~\bibnamefont {Kirrander}},
  \bibinfo {author} {\bibfnamefont {M.~P.}\ \bibnamefont {Minitti}}, \ and\
  \bibinfo {author} {\bibfnamefont {P.~M.}\ \bibnamefont {Weber}},\ }\bibfield
  {title} {\enquote {\bibinfo {title} {Advances in ultrafast gas-phase x-ray
  scattering},}\ }\href {\doibase 10.1088/1361-6455/abbfea} {\bibfield
  {journal} {\bibinfo  {journal} {J. Phys. B}\ }\textbf {\bibinfo {volume}
  {53}},\ \bibinfo {pages} {234004} (\bibinfo {year} {2020})}\BibitemShut
  {NoStop}%
\bibitem [{\citenamefont {Yang}\ \emph {et~al.}(2016)\citenamefont {Yang},
  \citenamefont {Guehr}, \citenamefont {Shen}, \citenamefont {Li},
  \citenamefont {Vecchione}, \citenamefont {Coffee}, \citenamefont {Corbett},
  \citenamefont {Fry}, \citenamefont {Hartmann}, \citenamefont {Hast},
  \citenamefont {Hegazy}, \citenamefont {Jobe}, \citenamefont {Makasyuk},
  \citenamefont {Robinson}, \citenamefont {Robinson}, \citenamefont {Vetter},
  \citenamefont {Weathersby}, \citenamefont {Yoneda}, \citenamefont {Wang},\
  and\ \citenamefont {Centurion}}]{Centurion16}%
  \BibitemOpen
  \bibfield  {author} {\bibinfo {author} {\bibfnamefont {J.}~\bibnamefont
  {Yang}}, \bibinfo {author} {\bibfnamefont {M.}~\bibnamefont {Guehr}},
  \bibinfo {author} {\bibfnamefont {X.}~\bibnamefont {Shen}}, \bibinfo {author}
  {\bibfnamefont {R.}~\bibnamefont {Li}}, \bibinfo {author} {\bibfnamefont
  {T.}~\bibnamefont {Vecchione}}, \bibinfo {author} {\bibfnamefont
  {R.}~\bibnamefont {Coffee}}, \bibinfo {author} {\bibfnamefont
  {J.}~\bibnamefont {Corbett}}, \bibinfo {author} {\bibfnamefont
  {A.}~\bibnamefont {Fry}}, \bibinfo {author} {\bibfnamefont {N.}~\bibnamefont
  {Hartmann}}, \bibinfo {author} {\bibfnamefont {C.}~\bibnamefont {Hast}},
  \bibinfo {author} {\bibfnamefont {K.}~\bibnamefont {Hegazy}}, \bibinfo
  {author} {\bibfnamefont {K.}~\bibnamefont {Jobe}}, \bibinfo {author}
  {\bibfnamefont {I.}~\bibnamefont {Makasyuk}}, \bibinfo {author}
  {\bibfnamefont {J.}~\bibnamefont {Robinson}}, \bibinfo {author}
  {\bibfnamefont {M.~S.}\ \bibnamefont {Robinson}}, \bibinfo {author}
  {\bibfnamefont {S.}~\bibnamefont {Vetter}}, \bibinfo {author} {\bibfnamefont
  {S.}~\bibnamefont {Weathersby}}, \bibinfo {author} {\bibfnamefont
  {C.}~\bibnamefont {Yoneda}}, \bibinfo {author} {\bibfnamefont
  {X.}~\bibnamefont {Wang}}, \ and\ \bibinfo {author} {\bibfnamefont
  {M.}~\bibnamefont {Centurion}},\ }\bibfield  {title} {\enquote {\bibinfo
  {title} {Diffractive imaging of coherent nuclear motion in isolated
  molecules},}\ }\href {\doibase 10.1103/PhysRevLett.117.153002} {\bibfield
  {journal} {\bibinfo  {journal} {Phys. Rev. Lett.}\ }\textbf {\bibinfo
  {volume} {117}},\ \bibinfo {pages} {153002} (\bibinfo {year}
  {2016})}\BibitemShut {NoStop}%
\bibitem [{\citenamefont {St{\"o}hr}(1992)}]{Stohr}%
  \BibitemOpen
  \bibfield  {author} {\bibinfo {author} {\bibfnamefont {J.}~\bibnamefont
  {St{\"o}hr}},\ }\href@noop {} {\emph {\bibinfo {title} {NEXAFS
  Spectroscopy}}}\ (\bibinfo  {publisher} {Springer-Verlag},\ \bibinfo
  {address} {Berlin},\ \bibinfo {year} {1992})\BibitemShut {NoStop}%
\bibitem [{\citenamefont {Pertot}\ \emph {et~al.}(2017)\citenamefont {Pertot},
  \citenamefont {Schmidt}, \citenamefont {Matthews}, \citenamefont {Chauvet},
  \citenamefont {Huppert}, \citenamefont {Svoboda}, \citenamefont {von Conta},
  \citenamefont {Tehlar}, \citenamefont {Baykusheva}, \citenamefont {Wolf},\
  and\ \citenamefont {W{\"o}rner}}]{Pertot}%
  \BibitemOpen
  \bibfield  {author} {\bibinfo {author} {\bibfnamefont {Y.}~\bibnamefont
  {Pertot}}, \bibinfo {author} {\bibfnamefont {C.}~\bibnamefont {Schmidt}},
  \bibinfo {author} {\bibfnamefont {M.}~\bibnamefont {Matthews}}, \bibinfo
  {author} {\bibfnamefont {A.}~\bibnamefont {Chauvet}}, \bibinfo {author}
  {\bibfnamefont {M.}~\bibnamefont {Huppert}}, \bibinfo {author} {\bibfnamefont
  {V.}~\bibnamefont {Svoboda}}, \bibinfo {author} {\bibfnamefont
  {A.}~\bibnamefont {von Conta}}, \bibinfo {author} {\bibfnamefont
  {A.}~\bibnamefont {Tehlar}}, \bibinfo {author} {\bibfnamefont
  {D.}~\bibnamefont {Baykusheva}}, \bibinfo {author} {\bibfnamefont {J.-P.}\
  \bibnamefont {Wolf}}, \ and\ \bibinfo {author} {\bibfnamefont {H.~J.}\
  \bibnamefont {W{\"o}rner}},\ }\bibfield  {title} {\enquote {\bibinfo {title}
  {Time-resolved x-ray absorption spectroscopy with a water window
  high-harmonic source},}\ }\href {\doibase 10.1126/science.aah6114} {\bibfield
   {journal} {\bibinfo  {journal} {Science}\ }\textbf {\bibinfo {volume}
  {355}},\ \bibinfo {pages} {264--267} (\bibinfo {year} {2017})}\BibitemShut
  {NoStop}%
\bibitem [{\citenamefont {Attar}\ \emph {et~al.}(2017)\citenamefont {Attar},
  \citenamefont {Bhattacherjee}, \citenamefont {Pemmaraju}, \citenamefont
  {Schnorr}, \citenamefont {Closser}, \citenamefont {Prendergast},\ and\
  \citenamefont {Leone}}]{Attar}%
  \BibitemOpen
  \bibfield  {author} {\bibinfo {author} {\bibfnamefont {A.~R.}\ \bibnamefont
  {Attar}}, \bibinfo {author} {\bibfnamefont {A.}~\bibnamefont
  {Bhattacherjee}}, \bibinfo {author} {\bibfnamefont {C.~D.}\ \bibnamefont
  {Pemmaraju}}, \bibinfo {author} {\bibfnamefont {K.}~\bibnamefont {Schnorr}},
  \bibinfo {author} {\bibfnamefont {K.~D.}\ \bibnamefont {Closser}}, \bibinfo
  {author} {\bibfnamefont {D.}~\bibnamefont {Prendergast}}, \ and\ \bibinfo
  {author} {\bibfnamefont {S.~R.}\ \bibnamefont {Leone}},\ }\bibfield  {title}
  {\enquote {\bibinfo {title} {Femtosecond x-ray spectroscopy of an
  electrocyclic ring-opening reaction},}\ }\href {\doibase
  10.1126/science.aaj2198} {\bibfield  {journal} {\bibinfo  {journal}
  {Science}\ }\textbf {\bibinfo {volume} {356}},\ \bibinfo {pages} {54--59}
  (\bibinfo {year} {2017})}\BibitemShut {NoStop}%
\bibitem [{\citenamefont {Wolf}\ \emph {et~al.}(2017)\citenamefont {Wolf},
  \citenamefont {Myhre}, \citenamefont {Cryan}, \citenamefont {Coriani},
  \citenamefont {Squibb}, \citenamefont {Battistoni}, \citenamefont {Berrah},
  \citenamefont {Bostedt}, \citenamefont {Bucksbaum}, \citenamefont
  {Coslovich}, \citenamefont {Feifel}, \citenamefont {Gaffney}, \citenamefont
  {Grilj}, \citenamefont {Martinez}, \citenamefont {Miyabe}, \citenamefont
  {Moeller}, \citenamefont {Mucke}, \citenamefont {Natan}, \citenamefont
  {Obaid}, \citenamefont {Osipov}, \citenamefont {Plekan}, \citenamefont
  {Wang}, \citenamefont {Koch},\ and\ \citenamefont {G{\"u}hr}}]{Wolf}%
  \BibitemOpen
  \bibfield  {author} {\bibinfo {author} {\bibfnamefont {T.~J.~A.}\
  \bibnamefont {Wolf}}, \bibinfo {author} {\bibfnamefont {R.~H.}\ \bibnamefont
  {Myhre}}, \bibinfo {author} {\bibfnamefont {J.~P.}\ \bibnamefont {Cryan}},
  \bibinfo {author} {\bibfnamefont {S.}~\bibnamefont {Coriani}}, \bibinfo
  {author} {\bibfnamefont {R.~J.}\ \bibnamefont {Squibb}}, \bibinfo {author}
  {\bibfnamefont {A.}~\bibnamefont {Battistoni}}, \bibinfo {author}
  {\bibfnamefont {N.}~\bibnamefont {Berrah}}, \bibinfo {author} {\bibfnamefont
  {C.}~\bibnamefont {Bostedt}}, \bibinfo {author} {\bibfnamefont
  {P.}~\bibnamefont {Bucksbaum}}, \bibinfo {author} {\bibfnamefont
  {G.}~\bibnamefont {Coslovich}}, \bibinfo {author} {\bibfnamefont
  {R.}~\bibnamefont {Feifel}}, \bibinfo {author} {\bibfnamefont {K.~J.}\
  \bibnamefont {Gaffney}}, \bibinfo {author} {\bibfnamefont {J.}~\bibnamefont
  {Grilj}}, \bibinfo {author} {\bibfnamefont {T.~J.}\ \bibnamefont {Martinez}},
  \bibinfo {author} {\bibfnamefont {S.}~\bibnamefont {Miyabe}}, \bibinfo
  {author} {\bibfnamefont {S.~P.}\ \bibnamefont {Moeller}}, \bibinfo {author}
  {\bibfnamefont {M.}~\bibnamefont {Mucke}}, \bibinfo {author} {\bibfnamefont
  {A.}~\bibnamefont {Natan}}, \bibinfo {author} {\bibfnamefont
  {R.}~\bibnamefont {Obaid}}, \bibinfo {author} {\bibfnamefont
  {T.}~\bibnamefont {Osipov}}, \bibinfo {author} {\bibfnamefont
  {O.}~\bibnamefont {Plekan}}, \bibinfo {author} {\bibfnamefont
  {S.}~\bibnamefont {Wang}}, \bibinfo {author} {\bibfnamefont {H.}~\bibnamefont
  {Koch}}, \ and\ \bibinfo {author} {\bibfnamefont {M.}~\bibnamefont
  {G{\"u}hr}},\ }\bibfield  {title} {\enquote {\bibinfo {title} {Probing
  ultrafast $\pi\pi^\ast$/n$\pi^\ast$ internal conversion in organic
  chromophores via {K}-edge resonant absorption},}\ }\href {\doibase
  10.1038/s41467-017-00069-7} {\bibfield  {journal} {\bibinfo  {journal} {Nat.
  Commun.}\ }\textbf {\bibinfo {volume} {8}},\ \bibinfo {pages} {29} (\bibinfo
  {year} {2017})}\BibitemShut {NoStop}%
\bibitem [{\citenamefont {Bhattacherjee}\ \emph {et~al.}(2017)\citenamefont
  {Bhattacherjee}, \citenamefont {Pemmaraju}, \citenamefont {Schnorr},
  \citenamefont {Attar},\ and\ \citenamefont {Leone}}]{acac_ultrafast_ISC}%
  \BibitemOpen
  \bibfield  {author} {\bibinfo {author} {\bibfnamefont {A.}~\bibnamefont
  {Bhattacherjee}}, \bibinfo {author} {\bibfnamefont {C.~D.}\ \bibnamefont
  {Pemmaraju}}, \bibinfo {author} {\bibfnamefont {K.}~\bibnamefont {Schnorr}},
  \bibinfo {author} {\bibfnamefont {A.~R.}\ \bibnamefont {Attar}}, \ and\
  \bibinfo {author} {\bibfnamefont {S.~R.}\ \bibnamefont {Leone}},\ }\bibfield
  {title} {\enquote {\bibinfo {title} {Ultrafast intersystem crossing in
  acetylacetone via femtosecond {X-ray} transient absorption at the carbon
  {K}-edge},}\ }\href {\doibase 10.1021/jacs.7b07532} {\bibfield  {journal}
  {\bibinfo  {journal} {J. Am. Chem. Soc.}\ }\textbf {\bibinfo {volume}
  {139}},\ \bibinfo {pages} {16576--16583} (\bibinfo {year}
  {2017})}\BibitemShut {NoStop}%
\bibitem [{\citenamefont {Chen}, \citenamefont {Zhang},\ and\ \citenamefont
  {Shelby}(2014)}]{Chen}%
  \BibitemOpen
  \bibfield  {author} {\bibinfo {author} {\bibfnamefont {L.~X.}\ \bibnamefont
  {Chen}}, \bibinfo {author} {\bibfnamefont {X.}~\bibnamefont {Zhang}}, \ and\
  \bibinfo {author} {\bibfnamefont {M.~L.}\ \bibnamefont {Shelby}},\ }\bibfield
   {title} {\enquote {\bibinfo {title} {Recent advances on ultrafast {X}-ray
  spectroscopy in the chemical sciences},}\ }\href {\doibase
  10.1039/C4SC01333F} {\bibfield  {journal} {\bibinfo  {journal} {Chem. Sci.}\
  }\textbf {\bibinfo {volume} {5}},\ \bibinfo {pages} {4136--4152} (\bibinfo
  {year} {2014})}\BibitemShut {NoStop}%
\bibitem [{\citenamefont {Chergui}(2016)}]{Chergui2016}%
  \BibitemOpen
  \bibfield  {author} {\bibinfo {author} {\bibfnamefont {M.}~\bibnamefont
  {Chergui}},\ }\bibfield  {title} {\enquote {\bibinfo {title} {Time-resolved
  {X}-ray spectroscopies of chemical systems: New perspectives},}\ }\href
  {\doibase 10.1063/1.4953104} {\bibfield  {journal} {\bibinfo  {journal}
  {Struct. Dyn.}\ }\textbf {\bibinfo {volume} {3}},\ \bibinfo {pages} {031001}
  (\bibinfo {year} {2016})}\BibitemShut {NoStop}%
\bibitem [{\citenamefont {Wernet}(2019)}]{Wernet2019}%
  \BibitemOpen
  \bibfield  {author} {\bibinfo {author} {\bibfnamefont {P.}~\bibnamefont
  {Wernet}},\ }\bibfield  {title} {\enquote {\bibinfo {title} {Chemical
  interactions and dynamics with femtosecond {X}-ray spectroscopy and the role
  of {X}-ray free-electron lasers},}\ }\href {\doibase 10.1098/rsta.2017.0464}
  {\bibfield  {journal} {\bibinfo  {journal} {Phil. Trans. R. Soc. A}\ }\textbf
  {\bibinfo {volume} {377}},\ \bibinfo {pages} {20170464} (\bibinfo {year}
  {2019})}\BibitemShut {NoStop}%
\bibitem [{\citenamefont {Katayama}\ \emph {et~al.}(2019)\citenamefont
  {Katayama}, \citenamefont {Northey}, \citenamefont {Gawelda}, \citenamefont
  {Milne}, \citenamefont {Vank\'{o}}, \citenamefont {Lima}, \citenamefont
  {Bohinc}, \citenamefont {N\'{e}meth}, \citenamefont {Nozawa}, \citenamefont
  {Sato}, \citenamefont {Khakhulin}, \citenamefont {Szlachetko}, \citenamefont
  {Togashi}, \citenamefont {Owada}, \citenamefont {Adachi}, \citenamefont
  {Bressler}, \citenamefont {Yabashi},\ and\ \citenamefont
  {Penfold}}]{Katayama}%
  \BibitemOpen
  \bibfield  {author} {\bibinfo {author} {\bibfnamefont {T.}~\bibnamefont
  {Katayama}}, \bibinfo {author} {\bibfnamefont {T.}~\bibnamefont {Northey}},
  \bibinfo {author} {\bibfnamefont {W.}~\bibnamefont {Gawelda}}, \bibinfo
  {author} {\bibfnamefont {C.~J.}\ \bibnamefont {Milne}}, \bibinfo {author}
  {\bibfnamefont {G.}~\bibnamefont {Vank\'{o}}}, \bibinfo {author}
  {\bibfnamefont {F.~A.}\ \bibnamefont {Lima}}, \bibinfo {author}
  {\bibfnamefont {R.}~\bibnamefont {Bohinc}}, \bibinfo {author} {\bibfnamefont
  {Z.}~\bibnamefont {N\'{e}meth}}, \bibinfo {author} {\bibfnamefont
  {S.}~\bibnamefont {Nozawa}}, \bibinfo {author} {\bibfnamefont
  {T.}~\bibnamefont {Sato}}, \bibinfo {author} {\bibfnamefont {D.}~\bibnamefont
  {Khakhulin}}, \bibinfo {author} {\bibfnamefont {J.}~\bibnamefont
  {Szlachetko}}, \bibinfo {author} {\bibfnamefont {T.}~\bibnamefont {Togashi}},
  \bibinfo {author} {\bibfnamefont {S.}~\bibnamefont {Owada}}, \bibinfo
  {author} {\bibfnamefont {S.-i.}\ \bibnamefont {Adachi}}, \bibinfo {author}
  {\bibfnamefont {C.}~\bibnamefont {Bressler}}, \bibinfo {author}
  {\bibfnamefont {M.}~\bibnamefont {Yabashi}}, \ and\ \bibinfo {author}
  {\bibfnamefont {T.~J.}\ \bibnamefont {Penfold}},\ }\bibfield  {title}
  {\enquote {\bibinfo {title} {{Tracking multiple components of a nuclear
  wavepacket in photoexcited Cu(I)-phenanthroline complex using ultrafast X-ray
  spectroscopy}},}\ }\href {\doibase 10.1038/s41467-019-11499-w} {\bibfield
  {journal} {\bibinfo  {journal} {Nat. Commun.}\ }\textbf {\bibinfo {volume}
  {10}},\ \bibinfo {pages} {3606} (\bibinfo {year} {2019})}\BibitemShut
  {NoStop}%
\bibitem [{\citenamefont {Norman}\ and\ \citenamefont
  {Dreuw}(2018)}]{X-ray_calc_review}%
  \BibitemOpen
  \bibfield  {author} {\bibinfo {author} {\bibfnamefont {P.}~\bibnamefont
  {Norman}}\ and\ \bibinfo {author} {\bibfnamefont {A.}~\bibnamefont {Dreuw}},\
  }\bibfield  {title} {\enquote {\bibinfo {title} {Simulating x-ray
  spectroscopies and calculating core-excited states of molecules},}\ }\href
  {\doibase 10.1021/acs.chemrev.8b00156} {\bibfield  {journal} {\bibinfo
  {journal} {Chem. Rev.}\ }\textbf {\bibinfo {volume} {118}},\ \bibinfo {pages}
  {7208--7248} (\bibinfo {year} {2018})}\BibitemShut {NoStop}%
\bibitem [{\citenamefont {Bokarev}\ and\ \citenamefont
  {K\"{u}hn}(2020)}]{Bokarev_Kuhn}%
  \BibitemOpen
  \bibfield  {author} {\bibinfo {author} {\bibfnamefont {S.~I.}\ \bibnamefont
  {Bokarev}}\ and\ \bibinfo {author} {\bibfnamefont {O.}~\bibnamefont
  {K\"{u}hn}},\ }\bibfield  {title} {\enquote {\bibinfo {title} {Theoretical
  x-ray spectroscopy of transition metal compounds},}\ }\href {\doibase
  10.1002/wcms.1433} {\bibfield  {journal} {\bibinfo  {journal} {WIREs Comput.
  Mol. Sci.}\ }\textbf {\bibinfo {volume} {10}},\ \bibinfo {pages} {e1433}
  (\bibinfo {year} {2020})}\BibitemShut {NoStop}%
\bibitem [{\citenamefont {Triguero}, \citenamefont {Pettersson},\ and\
  \citenamefont {{\AA}gren}(1998)}]{Triguero}%
  \BibitemOpen
  \bibfield  {author} {\bibinfo {author} {\bibfnamefont {L.}~\bibnamefont
  {Triguero}}, \bibinfo {author} {\bibfnamefont {L.~G.~M.}\ \bibnamefont
  {Pettersson}}, \ and\ \bibinfo {author} {\bibfnamefont {H.}~\bibnamefont
  {{\AA}gren}},\ }\bibfield  {title} {\enquote {\bibinfo {title} {Calculations
  of near-edge x-ray-absorption spectra of gas-phase and chemisorbed molecules
  by means of density-functional and transition-potential theory},}\ }\href
  {\doibase 10.1103/PhysRevB.58.8097} {\bibfield  {journal} {\bibinfo
  {journal} {Phys. Rev. B}\ }\textbf {\bibinfo {volume} {58}},\ \bibinfo
  {pages} {8097--8110} (\bibinfo {year} {1998})}\BibitemShut {NoStop}%
\bibitem [{\citenamefont {Leetmaa}\ \emph {et~al.}(2010)\citenamefont
  {Leetmaa}, \citenamefont {Ljungberg}, \citenamefont {Lyubartsev},
  \citenamefont {Nilsson},\ and\ \citenamefont {Pettersson}}]{Leetmaa}%
  \BibitemOpen
  \bibfield  {author} {\bibinfo {author} {\bibfnamefont {M.}~\bibnamefont
  {Leetmaa}}, \bibinfo {author} {\bibfnamefont {M.}~\bibnamefont {Ljungberg}},
  \bibinfo {author} {\bibfnamefont {A.}~\bibnamefont {Lyubartsev}}, \bibinfo
  {author} {\bibfnamefont {A.}~\bibnamefont {Nilsson}}, \ and\ \bibinfo
  {author} {\bibfnamefont {L.~G.~M.}\ \bibnamefont {Pettersson}},\ }\bibfield
  {title} {\enquote {\bibinfo {title} {Theoretical approximations to {X}-ray
  absorption spectroscopy of liquid water and ice},}\ }\href {\doibase
  https://doi.org/10.1016/j.elspec.2010.02.004} {\bibfield  {journal} {\bibinfo
   {journal} {J. Electron Spectrosc. Relat. Phenom.}\ }\textbf {\bibinfo
  {volume} {177}},\ \bibinfo {pages} {135--157} (\bibinfo {year}
  {2010})}\BibitemShut {NoStop}%
\bibitem [{\citenamefont {Vall-llosera}\ \emph {et~al.}(2008)\citenamefont
  {Vall-llosera}, \citenamefont {Gao}, \citenamefont {Kivim\"{a}ki},
  \citenamefont {Coreno}, \citenamefont {\'{A}lvarez Ruiz}, \citenamefont
  {de~Simone}, \citenamefont {{\AA}gren},\ and\ \citenamefont
  {Rachlew}}]{Vall-llosera}%
  \BibitemOpen
  \bibfield  {author} {\bibinfo {author} {\bibfnamefont {G.}~\bibnamefont
  {Vall-llosera}}, \bibinfo {author} {\bibfnamefont {B.}~\bibnamefont {Gao}},
  \bibinfo {author} {\bibfnamefont {A.}~\bibnamefont {Kivim\"{a}ki}}, \bibinfo
  {author} {\bibfnamefont {M.}~\bibnamefont {Coreno}}, \bibinfo {author}
  {\bibfnamefont {J.}~\bibnamefont {\'{A}lvarez Ruiz}}, \bibinfo {author}
  {\bibfnamefont {M.}~\bibnamefont {de~Simone}}, \bibinfo {author}
  {\bibfnamefont {H.}~\bibnamefont {{\AA}gren}}, \ and\ \bibinfo {author}
  {\bibfnamefont {E.}~\bibnamefont {Rachlew}},\ }\bibfield  {title} {\enquote
  {\bibinfo {title} {The {C} 1$s$ and {N} 1$s$ near edge x-ray absorption fine
  structure spectra of five azabenzenes in the gas phase},}\ }\href {\doibase
  10.1063/1.2822985} {\bibfield  {journal} {\bibinfo  {journal} {J. Chem.
  Phys.}\ }\textbf {\bibinfo {volume} {128}},\ \bibinfo {pages} {044316}
  (\bibinfo {year} {2008})}\BibitemShut {NoStop}%
\bibitem [{\citenamefont {Perera}\ and\ \citenamefont
  {Urquhart}(2017)}]{Perera}%
  \BibitemOpen
  \bibfield  {author} {\bibinfo {author} {\bibfnamefont {S.~D.}\ \bibnamefont
  {Perera}}\ and\ \bibinfo {author} {\bibfnamefont {S.~G.}\ \bibnamefont
  {Urquhart}},\ }\bibfield  {title} {\enquote {\bibinfo {title} {Systematic
  investigation of $\pi$--$\pi$ interactions in near-edge x-ray fine structure
  ({NEXAFS}) spectroscopy of paracyclophanes},}\ }\href {\doibase
  10.1021/acs.jpca.7b03823} {\bibfield  {journal} {\bibinfo  {journal} {J.
  Phys. Chem. A}\ }\textbf {\bibinfo {volume} {121}},\ \bibinfo {pages}
  {4907--4913} (\bibinfo {year} {2017})}\BibitemShut {NoStop}%
\bibitem [{\citenamefont {Ehlert}, \citenamefont {G\"uhr},\ and\ \citenamefont
  {Saalfrank}(2018)}]{TPDFT_TRNEXAFS}%
  \BibitemOpen
  \bibfield  {author} {\bibinfo {author} {\bibfnamefont {C.}~\bibnamefont
  {Ehlert}}, \bibinfo {author} {\bibfnamefont {M.}~\bibnamefont {G\"uhr}}, \
  and\ \bibinfo {author} {\bibfnamefont {P.}~\bibnamefont {Saalfrank}},\
  }\bibfield  {title} {\enquote {\bibinfo {title} {An efficient first
  principles method for molecular pump-probe {NEXAFS} spectra: Application to
  thymine and azobenzene},}\ }\href@noop {} {\bibfield  {journal} {\bibinfo
  {journal} {J. Chem. Phys.}\ }\textbf {\bibinfo {volume} {149}},\ \bibinfo
  {pages} {144112} (\bibinfo {year} {2018})}\BibitemShut {NoStop}%
\bibitem [{\citenamefont {Ehlert}\ and\ \citenamefont
  {Klamroth}(2020)}]{PSIXAS}%
  \BibitemOpen
  \bibfield  {author} {\bibinfo {author} {\bibfnamefont {C.}~\bibnamefont
  {Ehlert}}\ and\ \bibinfo {author} {\bibfnamefont {T.}~\bibnamefont
  {Klamroth}},\ }\bibfield  {title} {\enquote {\bibinfo {title} {{PSIXAS}: A
  {P}si4 plugin for efficient simulations of {X}-ray absorption spectra based
  on the transition-potential and {$\Delta$}-{K}ohn–{S}ham method},}\ }\href
  {\doibase 10.1002/jcc.26219} {\bibfield  {journal} {\bibinfo  {journal} {J.
  Comput. Chem.}\ }\textbf {\bibinfo {volume} {41}},\ \bibinfo {pages}
  {1781--1789} (\bibinfo {year} {2020})}\BibitemShut {NoStop}%
\bibitem [{\citenamefont {Michelitsch}\ and\ \citenamefont
  {Reuter}(2019)}]{Michelitsch}%
  \BibitemOpen
  \bibfield  {author} {\bibinfo {author} {\bibfnamefont {G.~S.}\ \bibnamefont
  {Michelitsch}}\ and\ \bibinfo {author} {\bibfnamefont {K.}~\bibnamefont
  {Reuter}},\ }\bibfield  {title} {\enquote {\bibinfo {title} {Efficient
  simulation of near-edge x-ray absorption fine structure ({NEXAFS}) in
  density-functional theory: Comparison of core-level constraining
  approaches},}\ }\href {\doibase 10.1063/1.5083618} {\bibfield  {journal}
  {\bibinfo  {journal} {J. Chem. Phys.}\ }\textbf {\bibinfo {volume} {150}},\
  \bibinfo {pages} {074104} (\bibinfo {year} {2019})}\BibitemShut {NoStop}%
\bibitem [{\citenamefont {Dreuw}\ and\ \citenamefont
  {Head-Gordon}(2005)}]{review_TDDFT_Dreuw_Head-Gordon}%
  \BibitemOpen
  \bibfield  {author} {\bibinfo {author} {\bibfnamefont {A.}~\bibnamefont
  {Dreuw}}\ and\ \bibinfo {author} {\bibfnamefont {M.}~\bibnamefont
  {Head-Gordon}},\ }\bibfield  {title} {\enquote {\bibinfo {title}
  {Single-reference ab initio methods for the calculation of excited states of
  large molecules},}\ }\href {\doibase 10.1021/cr0505627} {\bibfield  {journal}
  {\bibinfo  {journal} {Chem. Rev.}\ }\textbf {\bibinfo {volume} {105}},\
  \bibinfo {pages} {4009--4037} (\bibinfo {year} {2005})}\BibitemShut {NoStop}%
\bibitem [{\citenamefont {Luzanov}\ and\ \citenamefont
  {Zhikol}(2012)}]{Luzanov2012}%
  \BibitemOpen
  \bibfield  {author} {\bibinfo {author} {\bibfnamefont {A.~V.}\ \bibnamefont
  {Luzanov}}\ and\ \bibinfo {author} {\bibfnamefont {O.~A.}\ \bibnamefont
  {Zhikol}},\ }\enquote {\bibinfo {title} {Excited state structural analysis:
  {TDDFT} and related models},}\ in\ \href
  {https://doi-org.proxy.findit.dtu.dk/10.1007/978-94-007-0919-5} {\emph
  {\bibinfo {booktitle} {Practical Aspects of Computational Chemistry}}},\
  Vol.~\bibinfo {volume} {I},\ \bibinfo {editor} {edited by\ \bibinfo {editor}
  {\bibfnamefont {J.}~\bibnamefont {Leszczynski}}\ and\ \bibinfo {editor}
  {\bibfnamefont {M.~K.}\ \bibnamefont {Shukla}}}\ (\bibinfo  {publisher}
  {Springer},\ \bibinfo {address} {Heidelberg, Germany},\ \bibinfo {year}
  {2012})\ pp.\ \bibinfo {pages} {415 -- 449}\BibitemShut {NoStop}%
\bibitem [{\citenamefont {Laurent}\ and\ \citenamefont
  {Jacquemin}(2013)}]{Laurent_Jacquemine}%
  \BibitemOpen
  \bibfield  {author} {\bibinfo {author} {\bibfnamefont {A.~D.}\ \bibnamefont
  {Laurent}}\ and\ \bibinfo {author} {\bibfnamefont {D.}~\bibnamefont
  {Jacquemin}},\ }\bibfield  {title} {\enquote {\bibinfo {title} {{TD-DFT}
  benchmarks: A review},}\ }\href {\doibase 10.1002/qua.24438} {\bibfield
  {journal} {\bibinfo  {journal} {Int. J. Quantum Chem.}\ }\textbf {\bibinfo
  {volume} {113}},\ \bibinfo {pages} {2019--2039} (\bibinfo {year}
  {2013})}\BibitemShut {NoStop}%
\bibitem [{\citenamefont {Ferr\'{e}}, \citenamefont {Filatov},\ and\
  \citenamefont {Huix-Rotllant}(2016)}]{Ferre16}%
  \BibitemOpen
  \bibinfo {editor} {\bibfnamefont {N.}~\bibnamefont {Ferr\'{e}}}, \bibinfo
  {editor} {\bibfnamefont {M.}~\bibnamefont {Filatov}}, \ and\ \bibinfo
  {editor} {\bibfnamefont {M.}~\bibnamefont {Huix-Rotllant}},\ eds.,\ \href
  {\doibase 10.1007/978-3-319-22081-9} {\emph {\bibinfo {title}
  {Density-Functional Methods for Excited States}}}\ (\bibinfo  {publisher}
  {Springer},\ \bibinfo {address} {Cham, Switzerland},\ \bibinfo {year}
  {2016})\BibitemShut {NoStop}%
\bibitem [{\citenamefont {Stener}, \citenamefont {Fronzoni},\ and\
  \citenamefont {de~Simone}(2003)}]{Stener2003}%
  \BibitemOpen
  \bibfield  {author} {\bibinfo {author} {\bibfnamefont {M.}~\bibnamefont
  {Stener}}, \bibinfo {author} {\bibfnamefont {G.}~\bibnamefont {Fronzoni}}, \
  and\ \bibinfo {author} {\bibfnamefont {M.}~\bibnamefont {de~Simone}},\
  }\bibfield  {title} {\enquote {\bibinfo {title} {Time dependent density
  functional theory of core electrons excitations},}\ }\href@noop {} {\bibfield
   {journal} {\bibinfo  {journal} {Chem. Phys. Lett.}\ }\textbf {\bibinfo
  {volume} {373}},\ \bibinfo {pages} {115--123} (\bibinfo {year}
  {2003})}\BibitemShut {NoStop}%
\bibitem [{\citenamefont {Besley}\ and\ \citenamefont
  {Asmuruf}(2010)}]{core_TDDFT}%
  \BibitemOpen
  \bibfield  {author} {\bibinfo {author} {\bibfnamefont {N.~A.}\ \bibnamefont
  {Besley}}\ and\ \bibinfo {author} {\bibfnamefont {F.~A.}\ \bibnamefont
  {Asmuruf}},\ }\bibfield  {title} {\enquote {\bibinfo {title} {Time-dependent
  density functional theory calculations of the spectroscopy of core
  electrons},}\ }\href {\doibase 10.1039/C002207A} {\bibfield  {journal}
  {\bibinfo  {journal} {Phys. Chem. Chem. Phys.}\ }\textbf {\bibinfo {volume}
  {12}},\ \bibinfo {pages} {12024--12039} (\bibinfo {year} {2010})}\BibitemShut
  {NoStop}%
\bibitem [{\citenamefont {Cederbaum}, \citenamefont {Domcke},\ and\
  \citenamefont {Schirmer}(1980)}]{Cederbaum1980}%
  \BibitemOpen
  \bibfield  {author} {\bibinfo {author} {\bibfnamefont {L.~S.}\ \bibnamefont
  {Cederbaum}}, \bibinfo {author} {\bibfnamefont {W.}~\bibnamefont {Domcke}}, \
  and\ \bibinfo {author} {\bibfnamefont {J.}~\bibnamefont {Schirmer}},\
  }\bibfield  {title} {\enquote {\bibinfo {title} {Many-body theory of core
  holes},}\ }\href@noop {} {\bibfield  {journal} {\bibinfo  {journal} {Phys.
  Rev. A}\ }\textbf {\bibinfo {volume} {22}},\ \bibinfo {pages} {206--222}
  (\bibinfo {year} {1980})}\BibitemShut {NoStop}%
\bibitem [{\citenamefont {Besley}(2004)}]{BESLEY2004124}%
  \BibitemOpen
  \bibfield  {author} {\bibinfo {author} {\bibfnamefont {N.~A.}\ \bibnamefont
  {Besley}},\ }\bibfield  {title} {\enquote {\bibinfo {title} {Calculation of
  the electronic spectra of molecules in solution and on surfaces},}\ }\href
  {\doibase https://doi.org/10.1016/j.cplett.2004.04.004} {\bibfield  {journal}
  {\bibinfo  {journal} {Chem. Phys. Lett.}\ }\textbf {\bibinfo {volume}
  {390}},\ \bibinfo {pages} {124 -- 129} (\bibinfo {year} {2004})}\BibitemShut
  {NoStop}%
\bibitem [{\citenamefont {Becke}(1993)}]{Becke_B3LYP}%
  \BibitemOpen
  \bibfield  {author} {\bibinfo {author} {\bibfnamefont {A.~D.}\ \bibnamefont
  {Becke}},\ }\bibfield  {title} {\enquote {\bibinfo {title}
  {Density-functional thermochemistry. {III}. {T}he role of exact exchange},}\
  }\href {\doibase 10.1063/1.464913} {\bibfield  {journal} {\bibinfo  {journal}
  {J. Chem. Phys.}\ }\textbf {\bibinfo {volume} {98}},\ \bibinfo {pages}
  {5648--5652} (\bibinfo {year} {1993})}\BibitemShut {NoStop}%
\bibitem [{\citenamefont {Hait}\ and\ \citenamefont
  {Head-Gordon}(2020{\natexlab{a}})}]{Hait_Head-Gordon}%
  \BibitemOpen
  \bibfield  {author} {\bibinfo {author} {\bibfnamefont {D.}~\bibnamefont
  {Hait}}\ and\ \bibinfo {author} {\bibfnamefont {M.}~\bibnamefont
  {Head-Gordon}},\ }\bibfield  {title} {\enquote {\bibinfo {title} {Highly
  accurate prediction of core spectra of molecules at density functional theory
  cost: Attaining sub-electronvolt error from a restricted open-shell
  {K}ohn–{S}ham approach},}\ }\href {\doibase 10.1021/acs.jpclett.9b03661}
  {\bibfield  {journal} {\bibinfo  {journal} {J. Phys. Chem. Lett.}\ }\textbf
  {\bibinfo {volume} {11}},\ \bibinfo {pages} {775--786} (\bibinfo {year}
  {2020}{\natexlab{a}})}\BibitemShut {NoStop}%
\bibitem [{\citenamefont {Gilbert}, \citenamefont {Besley},\ and\ \citenamefont
  {Gill}(2008)}]{Gilbert2008}%
  \BibitemOpen
  \bibfield  {author} {\bibinfo {author} {\bibfnamefont {A.~T.~B.}\
  \bibnamefont {Gilbert}}, \bibinfo {author} {\bibfnamefont {N.~A.}\
  \bibnamefont {Besley}}, \ and\ \bibinfo {author} {\bibfnamefont {P.~M.~W.}\
  \bibnamefont {Gill}},\ }\bibfield  {title} {\enquote {\bibinfo {title}
  {Self-consistent field calculations of excited states using the maximum
  overlap method ({MOM})},}\ }\href {\doibase 10.1021/jp801738f} {\bibfield
  {journal} {\bibinfo  {journal} {J. Phys. Chem. A}\ }\textbf {\bibinfo
  {volume} {112}},\ \bibinfo {pages} {13164--13171} (\bibinfo {year}
  {2008})}\BibitemShut {NoStop}%
\bibitem [{\citenamefont {Northey}\ \emph {et~al.}(2020)\citenamefont
  {Northey}, \citenamefont {Norell}, \citenamefont {Fouda}, \citenamefont
  {Besley}, \citenamefont {Odelius},\ and\ \citenamefont {Penfold}}]{Northey2}%
  \BibitemOpen
  \bibfield  {author} {\bibinfo {author} {\bibfnamefont {T.}~\bibnamefont
  {Northey}}, \bibinfo {author} {\bibfnamefont {J.}~\bibnamefont {Norell}},
  \bibinfo {author} {\bibfnamefont {A.~E.~A.}\ \bibnamefont {Fouda}}, \bibinfo
  {author} {\bibfnamefont {N.~A.}\ \bibnamefont {Besley}}, \bibinfo {author}
  {\bibfnamefont {M.}~\bibnamefont {Odelius}}, \ and\ \bibinfo {author}
  {\bibfnamefont {T.~J.}\ \bibnamefont {Penfold}},\ }\bibfield  {title}
  {\enquote {\bibinfo {title} {Ultrafast nonadiabatic dynamics probed by
  nitrogen {K}-edge absorption spectroscopy},}\ }\href {\doibase
  10.1039/C9CP03019K} {\bibfield  {journal} {\bibinfo  {journal} {Phys. Chem.
  Chem. Phys.}\ }\textbf {\bibinfo {volume} {22}},\ \bibinfo {pages}
  {2667--2676} (\bibinfo {year} {2020})}\BibitemShut {NoStop}%
\bibitem [{\citenamefont {Hait}\ and\ \citenamefont
  {Head-Gordon}(2020{\natexlab{b}})}]{SGM}%
  \BibitemOpen
  \bibfield  {author} {\bibinfo {author} {\bibfnamefont {D.}~\bibnamefont
  {Hait}}\ and\ \bibinfo {author} {\bibfnamefont {M.}~\bibnamefont
  {Head-Gordon}},\ }\bibfield  {title} {\enquote {\bibinfo {title} {Excited
  state orbital optimization via minimizing the square of the gradient: General
  approach and application to singly and doubly excited states via density
  functional theory},}\ }\href {\doibase 10.1021/acs.jctc.9b01127} {\bibfield
  {journal} {\bibinfo  {journal} {J. Chem. Theory Comput.}\ }\textbf {\bibinfo
  {volume} {16}},\ \bibinfo {pages} {1699--1710} (\bibinfo {year}
  {2020}{\natexlab{b}})}\BibitemShut {NoStop}%
\bibitem [{\citenamefont {Hait}\ \emph {et~al.}(2020)\citenamefont {Hait},
  \citenamefont {Haugen}, \citenamefont {Yang}, \citenamefont {Oosterbaan},
  \citenamefont {Leone},\ and\ \citenamefont {Head-Gordon}}]{Hait_radical}%
  \BibitemOpen
  \bibfield  {author} {\bibinfo {author} {\bibfnamefont {D.}~\bibnamefont
  {Hait}}, \bibinfo {author} {\bibfnamefont {E.~A.}\ \bibnamefont {Haugen}},
  \bibinfo {author} {\bibfnamefont {Z.}~\bibnamefont {Yang}}, \bibinfo {author}
  {\bibfnamefont {K.~J.}\ \bibnamefont {Oosterbaan}}, \bibinfo {author}
  {\bibfnamefont {S.~R.}\ \bibnamefont {Leone}}, \ and\ \bibinfo {author}
  {\bibfnamefont {M.}~\bibnamefont {Head-Gordon}},\ }\bibfield  {title}
  {\enquote {\bibinfo {title} {Accurate prediction of core-level spectra of
  radicals at density functional theory cost via square gradient minimization
  and recoupling of mixed configurations},}\ }\href {\doibase
  10.1063/5.0018833} {\bibfield  {journal} {\bibinfo  {journal} {J. Chem.
  Phys.}\ }\textbf {\bibinfo {volume} {153}},\ \bibinfo {pages} {134108}
  (\bibinfo {year} {2020})}\BibitemShut {NoStop}%
\bibitem [{\citenamefont {Bartlett}\ and\ \citenamefont
  {Musia\l{}}(2007)}]{Bartlett2007}%
  \BibitemOpen
  \bibfield  {author} {\bibinfo {author} {\bibfnamefont {R.~J.}\ \bibnamefont
  {Bartlett}}\ and\ \bibinfo {author} {\bibfnamefont {M.}~\bibnamefont
  {Musia\l{}}},\ }\bibfield  {title} {\enquote {\bibinfo {title}
  {Coupled-cluster theory in quantum chemistry},}\ }\href {\doibase
  10.1103/RevModPhys.79.291} {\bibfield  {journal} {\bibinfo  {journal} {Rev.
  Mod. Phys.}\ }\textbf {\bibinfo {volume} {79}},\ \bibinfo {pages} {291--352}
  (\bibinfo {year} {2007})}\BibitemShut {NoStop}%
\bibitem [{\citenamefont {Koch}\ and\ \citenamefont
  {J{\o}rgensen}(1990)}]{koch1990}%
  \BibitemOpen
  \bibfield  {author} {\bibinfo {author} {\bibfnamefont {H.}~\bibnamefont
  {Koch}}\ and\ \bibinfo {author} {\bibfnamefont {P.}~\bibnamefont
  {J{\o}rgensen}},\ }\bibfield  {title} {\enquote {\bibinfo {title} {Coupled
  cluster response functions},}\ }\href@noop {} {\bibfield  {journal} {\bibinfo
   {journal} {J. Chem. Phys.}\ }\textbf {\bibinfo {volume} {93}},\ \bibinfo
  {pages} {3333--3344} (\bibinfo {year} {1990})}\BibitemShut {NoStop}%
\bibitem [{\citenamefont {Christiansen}, \citenamefont {J{\o}rgensen},\ and\
  \citenamefont {H{\"a}ttig}(1998)}]{Christiansen_IJQC}%
  \BibitemOpen
  \bibfield  {author} {\bibinfo {author} {\bibfnamefont {O.}~\bibnamefont
  {Christiansen}}, \bibinfo {author} {\bibfnamefont {P.}~\bibnamefont
  {J{\o}rgensen}}, \ and\ \bibinfo {author} {\bibfnamefont {C.}~\bibnamefont
  {H{\"a}ttig}},\ }\bibfield  {title} {\enquote {\bibinfo {title} {Response
  functions from {F}ourier component variational perturbation theory applied to
  a time-averaged quasienergy},}\ }\href@noop {} {\bibfield  {journal}
  {\bibinfo  {journal} {Int. J. Quantum Chem.}\ }\textbf {\bibinfo {volume}
  {68}},\ \bibinfo {pages} {1--52} (\bibinfo {year} {1998})}\BibitemShut
  {NoStop}%
\bibitem [{\citenamefont {Sneskov}\ and\ \citenamefont
  {Christiansen}(2012)}]{Sneskov_Christiansen}%
  \BibitemOpen
  \bibfield  {author} {\bibinfo {author} {\bibfnamefont {K.}~\bibnamefont
  {Sneskov}}\ and\ \bibinfo {author} {\bibfnamefont {O.}~\bibnamefont
  {Christiansen}},\ }\bibfield  {title} {\enquote {\bibinfo {title} {Excited
  state coupled cluster methods},}\ }\href {\doibase 10.1002/wcms.99}
  {\bibfield  {journal} {\bibinfo  {journal} {WIREs Comput. Mol. Sci.}\
  }\textbf {\bibinfo {volume} {2}},\ \bibinfo {pages} {566--584} (\bibinfo
  {year} {2012})}\BibitemShut {NoStop}%
\bibitem [{\citenamefont {Stanton}\ and\ \citenamefont
  {Bartlett}(1993)}]{Stanton1993}%
  \BibitemOpen
  \bibfield  {author} {\bibinfo {author} {\bibfnamefont {J.~F.}\ \bibnamefont
  {Stanton}}\ and\ \bibinfo {author} {\bibfnamefont {R.~J.}\ \bibnamefont
  {Bartlett}},\ }\bibfield  {title} {\enquote {\bibinfo {title} {The equation
  of motion coupled-cluster method. {A} systematic biorthogonal approach to
  molecular excitation energies, transition probabilities, and excited state
  properties},}\ }\href {\doibase 10.1063/1.464746} {\bibfield  {journal}
  {\bibinfo  {journal} {J. Chem. Phys.}\ }\textbf {\bibinfo {volume} {98}},\
  \bibinfo {pages} {7029--7039} (\bibinfo {year} {1993})}\BibitemShut {NoStop}%
\bibitem [{\citenamefont {Krylov}(2008)}]{krylov_eom_2008}%
  \BibitemOpen
  \bibfield  {author} {\bibinfo {author} {\bibfnamefont {A.~I.}\ \bibnamefont
  {Krylov}},\ }\bibfield  {title} {\enquote {\bibinfo {title}
  {{Equation-of-Motion Coupled-Cluster Methods for Open-Shell and
  Electronically Excited Species: The Hitchhiker's Guide to Fock Space}},}\
  }\href {\doibase 10.1146/annurev.physchem.59.032607.093602} {\bibfield
  {journal} {\bibinfo  {journal} {Ann.~Rev.~Phys.~Chem.}\ }\textbf {\bibinfo
  {volume} {59}},\ \bibinfo {pages} {433--462} (\bibinfo {year}
  {2008})}\BibitemShut {NoStop}%
\bibitem [{\citenamefont {Bartlett}(2012)}]{CC_EOMCC_Bartlett}%
  \BibitemOpen
  \bibfield  {author} {\bibinfo {author} {\bibfnamefont {R.~J.}\ \bibnamefont
  {Bartlett}},\ }\bibfield  {title} {\enquote {\bibinfo {title}
  {Coupled-cluster theory and its equation-of-motion extensions},}\ }\href
  {\doibase 10.1002/wcms.76} {\bibfield  {journal} {\bibinfo  {journal} {WIREs
  Comput. Mol. Sci.}\ }\textbf {\bibinfo {volume} {2}},\ \bibinfo {pages}
  {126--138} (\bibinfo {year} {2012})}\BibitemShut {NoStop}%
\bibitem [{\citenamefont {Coriani}\ \emph {et~al.}(2016)\citenamefont
  {Coriani}, \citenamefont {Paw{\l}owski}, \citenamefont {Olsen},\ and\
  \citenamefont {J{\o}rgensen}}]{corianiEOMRSP}%
  \BibitemOpen
  \bibfield  {author} {\bibinfo {author} {\bibfnamefont {S.}~\bibnamefont
  {Coriani}}, \bibinfo {author} {\bibfnamefont {F.}~\bibnamefont
  {Paw{\l}owski}}, \bibinfo {author} {\bibfnamefont {J.}~\bibnamefont {Olsen}},
  \ and\ \bibinfo {author} {\bibfnamefont {P.}~\bibnamefont {J{\o}rgensen}},\
  }\bibfield  {title} {\enquote {\bibinfo {title} {Molecular response
  properties in equation of motion coupled cluster theory: A time-dependent
  perspective},}\ }\href {\doibase 10.1063/1.4939183} {\bibfield  {journal}
  {\bibinfo  {journal} {J. Chem. Phys.}\ }\textbf {\bibinfo {volume} {144}},\
  \bibinfo {pages} {024102} (\bibinfo {year} {2016})}\BibitemShut {NoStop}%
\bibitem [{\citenamefont {Coriani}\ and\ \citenamefont
  {Koch}(2015)}]{coriani2015jcp}%
  \BibitemOpen
  \bibfield  {author} {\bibinfo {author} {\bibfnamefont {S.}~\bibnamefont
  {Coriani}}\ and\ \bibinfo {author} {\bibfnamefont {H.}~\bibnamefont {Koch}},\
  }\bibfield  {title} {\enquote {\bibinfo {title} {Communication: X-ray
  absorption spectra and core-ionization potentials within a core-valence
  separated coupled cluster framework},}\ }\href@noop {} {\bibfield  {journal}
  {\bibinfo  {journal} {J. Chem. Phys.}\ }\textbf {\bibinfo {volume} {143}},\
  \bibinfo {pages} {181103} (\bibinfo {year} {2015})}\BibitemShut {NoStop}%
\bibitem [{\citenamefont {Vidal}\ \emph {et~al.}(2019)\citenamefont {Vidal},
  \citenamefont {Feng}, \citenamefont {Epifanovsky}, \citenamefont {Krylov},\
  and\ \citenamefont {Coria\-ni}}]{Vidal1}%
  \BibitemOpen
  \bibfield  {author} {\bibinfo {author} {\bibfnamefont {M.~L.}\ \bibnamefont
  {Vidal}}, \bibinfo {author} {\bibfnamefont {X.}~\bibnamefont {Feng}},
  \bibinfo {author} {\bibfnamefont {E.}~\bibnamefont {Epifanovsky}}, \bibinfo
  {author} {\bibfnamefont {A.~I.}\ \bibnamefont {Krylov}}, \ and\ \bibinfo
  {author} {\bibfnamefont {S.}~\bibnamefont {Coria\-ni}},\ }\bibfield  {title}
  {\enquote {\bibinfo {title} {New and efficient equation-of-motion
  coupled-cluster framework for core-excited and core-ionized states},}\ }\href
  {\doibase 10.1021/acs.jctc.9b00039} {\bibfield  {journal} {\bibinfo
  {journal} {J. Chem. Theory Comput.}\ }\textbf {\bibinfo {volume} {15}},\
  \bibinfo {pages} {3117--3133} (\bibinfo {year} {2019})}\BibitemShut {NoStop}%
\bibitem [{\citenamefont {Tsuru}\ \emph {et~al.}(2019)\citenamefont {Tsuru},
  \citenamefont {Vidal}, \citenamefont {P{\'a}pai}, \citenamefont {Krylov},
  \citenamefont {M{\o}ller},\ and\ \citenamefont {Coria\-ni}}]{tsuru}%
  \BibitemOpen
  \bibfield  {author} {\bibinfo {author} {\bibfnamefont {S.}~\bibnamefont
  {Tsuru}}, \bibinfo {author} {\bibfnamefont {M.~L.}\ \bibnamefont {Vidal}},
  \bibinfo {author} {\bibfnamefont {M.}~\bibnamefont {P{\'a}pai}}, \bibinfo
  {author} {\bibfnamefont {A.~I.}\ \bibnamefont {Krylov}}, \bibinfo {author}
  {\bibfnamefont {K.~B.}\ \bibnamefont {M{\o}ller}}, \ and\ \bibinfo {author}
  {\bibfnamefont {S.}~\bibnamefont {Coria\-ni}},\ }\bibfield  {title} {\enquote
  {\bibinfo {title} {{Time-resolved near-edge X-ray absorption fine structure
  of pyrazine from electronic structure and nuclear wave packet dynamics
  simulations}},}\ }\href {\doibase 10.1063/1.5115154} {\bibfield  {journal}
  {\bibinfo  {journal} {J. Chem. Phys.}\ }\textbf {\bibinfo {volume} {151}},\
  \bibinfo {pages} {124114} (\bibinfo {year} {2019})}\BibitemShut {NoStop}%
\bibitem [{\citenamefont {Faber}\ \emph {et~al.}(2019)\citenamefont {Faber},
  \citenamefont {Kj{\o}nstad}, \citenamefont {Koch},\ and\ \citenamefont
  {Coriani}}]{Faber19}%
  \BibitemOpen
  \bibfield  {author} {\bibinfo {author} {\bibfnamefont {R.}~\bibnamefont
  {Faber}}, \bibinfo {author} {\bibfnamefont {E.~F.}\ \bibnamefont
  {Kj{\o}nstad}}, \bibinfo {author} {\bibfnamefont {H.}~\bibnamefont {Koch}}, \
  and\ \bibinfo {author} {\bibfnamefont {S.}~\bibnamefont {Coriani}},\
  }\bibfield  {title} {\enquote {\bibinfo {title} {Spin adapted implementation
  of {EOM-CCSD} for triplet excited states: Probing intersystem crossings of
  acetylacetone at the carbon and oxygen {K}-edges},}\ }\href {\doibase
  10.1063/1.5112164} {\bibfield  {journal} {\bibinfo  {journal} {J. Chem.
  Phys.}\ }\textbf {\bibinfo {volume} {151}},\ \bibinfo {pages} {144107}
  (\bibinfo {year} {2019})}\BibitemShut {NoStop}%
\bibitem [{\citenamefont {Nanda}\ \emph {et~al.}(2020)\citenamefont {Nanda},
  \citenamefont {Vidal}, \citenamefont {Faber}, \citenamefont {Coriani},\ and\
  \citenamefont {Krylov}}]{Nanda:RIXS:2020}%
  \BibitemOpen
  \bibfield  {author} {\bibinfo {author} {\bibfnamefont {K.~D.}\ \bibnamefont
  {Nanda}}, \bibinfo {author} {\bibfnamefont {M.~L.}\ \bibnamefont {Vidal}},
  \bibinfo {author} {\bibfnamefont {R.}~\bibnamefont {Faber}}, \bibinfo
  {author} {\bibfnamefont {S.}~\bibnamefont {Coriani}}, \ and\ \bibinfo
  {author} {\bibfnamefont {A.~I.}\ \bibnamefont {Krylov}},\ }\bibfield  {title}
  {\enquote {\bibinfo {title} {How to stay out of trouble in {RIXS}
  calculations within the equation-of-motion coupled-cluster damped response
  theory? {S}afe hitchhiking in the excitation manifold by means of
  core-valence separation},}\ }\href@noop {} {\bibfield  {journal} {\bibinfo
  {journal} {Phys. Chem. Chem. Phys.}\ }\textbf {\bibinfo {volume} {22}},\
  \bibinfo {pages} {2629--2641} (\bibinfo {year} {2020})}\BibitemShut {NoStop}%
\bibitem [{\citenamefont {Faber}\ and\ \citenamefont
  {Coriani}(2020)}]{Coriani:RIXS-CVS:2020}%
  \BibitemOpen
  \bibfield  {author} {\bibinfo {author} {\bibfnamefont {R.}~\bibnamefont
  {Faber}}\ and\ \bibinfo {author} {\bibfnamefont {S.}~\bibnamefont
  {Coriani}},\ }\bibfield  {title} {\enquote {\bibinfo {title}
  {Core–valence-separated coupled-cluster-singles-and-doubles
  complex-polarization-propagator approach to {X}-ray spectroscopies},}\
  }\href@noop {} {\bibfield  {journal} {\bibinfo  {journal} {Phys. Chem. Chem.
  Phys.}\ }\textbf {\bibinfo {volume} {22}},\ \bibinfo {pages} {2642--2647}
  (\bibinfo {year} {2020})}\BibitemShut {NoStop}%
\bibitem [{\citenamefont {Vidal}, \citenamefont {Krylov},\ and\ \citenamefont
  {Coriani}(2020{\natexlab{a}})}]{vidal2}%
  \BibitemOpen
  \bibfield  {author} {\bibinfo {author} {\bibfnamefont {M.~L.}\ \bibnamefont
  {Vidal}}, \bibinfo {author} {\bibfnamefont {A.~I.}\ \bibnamefont {Krylov}}, \
  and\ \bibinfo {author} {\bibfnamefont {S.}~\bibnamefont {Coriani}},\
  }\bibfield  {title} {\enquote {\bibinfo {title} {Dyson orbitals within the
  fc-{CVS-EOM-CCSD} framework: theory and application to {X}-ray photoelectron
  spectroscopy of ground and excited states},}\ }\href {\doibase
  10.1039/C9CP03695D} {\bibfield  {journal} {\bibinfo  {journal} {Phys. Chem.
  Chem. Phys.}\ }\textbf {\bibinfo {volume} {22}},\ \bibinfo {pages}
  {2693--2703} (\bibinfo {year} {2020}{\natexlab{a}})}\BibitemShut {NoStop}%
\bibitem [{\citenamefont {Vidal}\ \emph
  {et~al.}(2020{\natexlab{a}})\citenamefont {Vidal}, \citenamefont {Pokhilko},
  \citenamefont {Krylov},\ and\ \citenamefont {Coriani}}]{Vidal:L-edge}%
  \BibitemOpen
  \bibfield  {author} {\bibinfo {author} {\bibfnamefont {M.~L.}\ \bibnamefont
  {Vidal}}, \bibinfo {author} {\bibfnamefont {P.}~\bibnamefont {Pokhilko}},
  \bibinfo {author} {\bibfnamefont {A.~I.}\ \bibnamefont {Krylov}}, \ and\
  \bibinfo {author} {\bibfnamefont {S.}~\bibnamefont {Coriani}},\ }\bibfield
  {title} {\enquote {\bibinfo {title} {{Equation-of-Motion Coupled-Cluster
  Theory to Model L-Edge X-ray Absorption and Photoelectron Spectra}},}\ }\href
  {\doibase 10.1021/acs.jpclett.0c02027} {\bibfield  {journal} {\bibinfo
  {journal} {J. Phys. Chem. Lett.}\ }\textbf {\bibinfo {volume} {11}},\
  \bibinfo {pages} {8314--8321} (\bibinfo {year}
  {2020}{\natexlab{a}})}\BibitemShut {NoStop}%
\bibitem [{\citenamefont {Coriani}\ \emph {et~al.}(2012)\citenamefont
  {Coriani}, \citenamefont {Christiansen}, \citenamefont {Fransson},\ and\
  \citenamefont {Norman}}]{coriani2012pra}%
  \BibitemOpen
  \bibfield  {author} {\bibinfo {author} {\bibfnamefont {S.}~\bibnamefont
  {Coriani}}, \bibinfo {author} {\bibfnamefont {O.}~\bibnamefont
  {Christiansen}}, \bibinfo {author} {\bibfnamefont {T.}~\bibnamefont
  {Fransson}}, \ and\ \bibinfo {author} {\bibfnamefont {P.}~\bibnamefont
  {Norman}},\ }\bibfield  {title} {\enquote {\bibinfo {title} {{Coupled-Cluster
  Response Theory for Near-Edge X-Ray-Absorption Fine Structure of Atoms and
  Molecules}},}\ }\href {\doibase 10.1103/PhysRevA.85.022507} {\bibfield
  {journal} {\bibinfo  {journal} {Phys. Rev. A}\ }\textbf {\bibinfo {volume}
  {85}},\ \bibinfo {pages} {022507} (\bibinfo {year} {2012})}\BibitemShut
  {NoStop}%
\bibitem [{\citenamefont {Frati}\ \emph {et~al.}(2019)\citenamefont {Frati},
  \citenamefont {de~Groot}, \citenamefont {Cerezo}, \citenamefont {Santoro},
  \citenamefont {Cheng}, \citenamefont {Faber},\ and\ \citenamefont
  {Coriani}}]{Frati:2019}%
  \BibitemOpen
  \bibfield  {author} {\bibinfo {author} {\bibfnamefont {F.}~\bibnamefont
  {Frati}}, \bibinfo {author} {\bibfnamefont {F.}~\bibnamefont {de~Groot}},
  \bibinfo {author} {\bibfnamefont {J.}~\bibnamefont {Cerezo}}, \bibinfo
  {author} {\bibfnamefont {F.}~\bibnamefont {Santoro}}, \bibinfo {author}
  {\bibfnamefont {L.}~\bibnamefont {Cheng}}, \bibinfo {author} {\bibfnamefont
  {R.}~\bibnamefont {Faber}}, \ and\ \bibinfo {author} {\bibfnamefont
  {S.}~\bibnamefont {Coriani}},\ }\bibfield  {title} {\enquote {\bibinfo
  {title} {Coupled cluster study of the x-ray absorption spectra of
  formaldehyde derivatives at the oxygen, carbon, and fluorine {K}-edges},}\
  }\href@noop {} {\bibfield  {journal} {\bibinfo  {journal} {J. Chem. Phys.}\
  }\textbf {\bibinfo {volume} {151}},\ \bibinfo {pages} {064107} (\bibinfo
  {year} {2019})}\BibitemShut {NoStop}%
\bibitem [{\citenamefont {Carbone}\ \emph {et~al.}(2019)\citenamefont
  {Carbone}, \citenamefont {Cheng}, \citenamefont {Myhre}, \citenamefont
  {Matthews}, \citenamefont {Koch},\ and\ \citenamefont
  {Coriani}}]{Carbone:2019}%
  \BibitemOpen
  \bibfield  {author} {\bibinfo {author} {\bibfnamefont {J.~P.}\ \bibnamefont
  {Carbone}}, \bibinfo {author} {\bibfnamefont {L.}~\bibnamefont {Cheng}},
  \bibinfo {author} {\bibfnamefont {R.~H.}\ \bibnamefont {Myhre}}, \bibinfo
  {author} {\bibfnamefont {D.}~\bibnamefont {Matthews}}, \bibinfo {author}
  {\bibfnamefont {H.}~\bibnamefont {Koch}}, \ and\ \bibinfo {author}
  {\bibfnamefont {S.}~\bibnamefont {Coriani}},\ }\bibfield  {title} {\enquote
  {\bibinfo {title} {An analysis of the performance of coupled cluster methods
  for {K}-edge core excitations and ionizations using standard basis sets},}\
  }\href@noop {} {\bibfield  {journal} {\bibinfo  {journal} {Adv. Quant.
  Chem.}\ }\textbf {\bibinfo {volume} {79}},\ \bibinfo {pages} {241--261}
  (\bibinfo {year} {2019})}\BibitemShut {NoStop}%
\bibitem [{\citenamefont {Peng}\ \emph {et~al.}(2015)\citenamefont {Peng},
  \citenamefont {Lestrange}, \citenamefont {Goings}, \citenamefont {Caricato},\
  and\ \citenamefont {Li}}]{Peng2015}%
  \BibitemOpen
  \bibfield  {author} {\bibinfo {author} {\bibfnamefont {B.}~\bibnamefont
  {Peng}}, \bibinfo {author} {\bibfnamefont {P.~J.}\ \bibnamefont {Lestrange}},
  \bibinfo {author} {\bibfnamefont {J.~J.}\ \bibnamefont {Goings}}, \bibinfo
  {author} {\bibfnamefont {M.}~\bibnamefont {Caricato}}, \ and\ \bibinfo
  {author} {\bibfnamefont {X.}~\bibnamefont {Li}},\ }\bibfield  {title}
  {\enquote {\bibinfo {title} {{Energy-Specific Equation-of-Motion
  Coupled-Cluster Methods for High-Energy Excited States: Application to
  $K$-Edge X-Ray Absorption Spectroscopy}},}\ }\href@noop {} {\bibfield
  {journal} {\bibinfo  {journal} {J. Chem. Theory Comput.}\ }\textbf {\bibinfo
  {volume} {11}},\ \bibinfo {pages} {4146--4153} (\bibinfo {year}
  {2015})}\BibitemShut {NoStop}%
\bibitem [{\citenamefont {Fransson}\ \emph {et~al.}(2013)\citenamefont
  {Fransson}, \citenamefont {Coriani}, \citenamefont {Christiansen},\ and\
  \citenamefont {Norman}}]{fransson2013jcp}%
  \BibitemOpen
  \bibfield  {author} {\bibinfo {author} {\bibfnamefont {T.}~\bibnamefont
  {Fransson}}, \bibinfo {author} {\bibfnamefont {S.}~\bibnamefont {Coriani}},
  \bibinfo {author} {\bibfnamefont {O.}~\bibnamefont {Christiansen}}, \ and\
  \bibinfo {author} {\bibfnamefont {P.}~\bibnamefont {Norman}},\ }\bibfield
  {title} {\enquote {\bibinfo {title} {Carbon x-ray absorption spectra of
  fluoroethenes and acetone: A study at the coupled cluster, density
  functional, and static-exchange levels of theory},}\ }\href@noop {}
  {\bibfield  {journal} {\bibinfo  {journal} {J. Chem. Phys.}\ }\textbf
  {\bibinfo {volume} {138}},\ \bibinfo {pages} {124311} (\bibinfo {year}
  {2013})}\BibitemShut {NoStop}%
\bibitem [{\citenamefont {Sarangi}\ \emph {et~al.}(2020)\citenamefont
  {Sarangi}, \citenamefont {Vidal}, \citenamefont {Coriani},\ and\
  \citenamefont {Krylov}}]{xpsbasis2020}%
  \BibitemOpen
  \bibfield  {author} {\bibinfo {author} {\bibfnamefont {R.}~\bibnamefont
  {Sarangi}}, \bibinfo {author} {\bibfnamefont {M.~L.}\ \bibnamefont {Vidal}},
  \bibinfo {author} {\bibfnamefont {S.}~\bibnamefont {Coriani}}, \ and\
  \bibinfo {author} {\bibfnamefont {A.~I.}\ \bibnamefont {Krylov}},\ }\bibfield
   {title} {\enquote {\bibinfo {title} {On the basis set selection for
  calculations of core-level states: different strategies to balance cost and
  accuracy},}\ }\href {\doibase 10.1080/00268976.2020.1769872} {\bibfield
  {journal} {\bibinfo  {journal} {Mol. Phys.}\ }\textbf {\bibinfo {volume}
  {118}},\ \bibinfo {pages} {e1769872} (\bibinfo {year} {2020})}\BibitemShut
  {NoStop}%
\bibitem [{\citenamefont {Tenorio}\ \emph {et~al.}(2019)\citenamefont
  {Tenorio}, \citenamefont {Moitra}, \citenamefont {Nascimento}, \citenamefont
  {Rocha},\ and\ \citenamefont {Coriani}}]{Bruno-cvs}%
  \BibitemOpen
  \bibfield  {author} {\bibinfo {author} {\bibfnamefont {B.~N.~C.}\
  \bibnamefont {Tenorio}}, \bibinfo {author} {\bibfnamefont {T.}~\bibnamefont
  {Moitra}}, \bibinfo {author} {\bibfnamefont {M.~A.~C.}\ \bibnamefont
  {Nascimento}}, \bibinfo {author} {\bibfnamefont {A.~B.}\ \bibnamefont
  {Rocha}}, \ and\ \bibinfo {author} {\bibfnamefont {S.}~\bibnamefont
  {Coriani}},\ }\bibfield  {title} {\enquote {\bibinfo {title} {Molecular
  inner-shell photoabsorption/photoionization cross sections at
  core-valence-separated coupled cluster level: Theory and examples},}\
  }\href@noop {} {\bibfield  {journal} {\bibinfo  {journal} {J. Chem. Phys.}\
  }\textbf {\bibinfo {volume} {150}},\ \bibinfo {pages} {224104} (\bibinfo
  {year} {2019})}\BibitemShut {NoStop}%
\bibitem [{\citenamefont {Moitra}\ \emph {et~al.}(2020)\citenamefont {Moitra},
  \citenamefont {Madsen}, \citenamefont {Christiansen},\ and\ \citenamefont
  {Coriani}}]{Moitra:Vib:core}%
  \BibitemOpen
  \bibfield  {author} {\bibinfo {author} {\bibfnamefont {T.}~\bibnamefont
  {Moitra}}, \bibinfo {author} {\bibfnamefont {D.}~\bibnamefont {Madsen}},
  \bibinfo {author} {\bibfnamefont {O.}~\bibnamefont {Christiansen}}, \ and\
  \bibinfo {author} {\bibfnamefont {S.}~\bibnamefont {Coriani}},\ }\bibfield
  {title} {\enquote {\bibinfo {title} {{Vibrationally resolved coupled-cluster
  x-ray absorption spectra from vibrational configuration interaction
  anharmonic calculations}},}\ }\href@noop {} {\bibfield  {journal} {\bibinfo
  {journal} {J. Chem. Phys.}\ }\textbf {\bibinfo {volume} {153}},\ \bibinfo
  {pages} {234111} (\bibinfo {year} {2020})}\BibitemShut {NoStop}%
\bibitem [{\citenamefont {Vidal}\ \emph
  {et~al.}(2020{\natexlab{b}})\citenamefont {Vidal}, \citenamefont {Epshtein},
  \citenamefont {Scutelnic}, \citenamefont {Yang}, \citenamefont {Xue},
  \citenamefont {Leone}, \citenamefont {Krylov},\ and\ \citenamefont
  {Coriani}}]{Vidal:Benzene+}%
  \BibitemOpen
  \bibfield  {author} {\bibinfo {author} {\bibfnamefont {M.~L.}\ \bibnamefont
  {Vidal}}, \bibinfo {author} {\bibfnamefont {M.}~\bibnamefont {Epshtein}},
  \bibinfo {author} {\bibfnamefont {V.}~\bibnamefont {Scutelnic}}, \bibinfo
  {author} {\bibfnamefont {Z.}~\bibnamefont {Yang}}, \bibinfo {author}
  {\bibfnamefont {T.}~\bibnamefont {Xue}}, \bibinfo {author} {\bibfnamefont
  {S.~R.}\ \bibnamefont {Leone}}, \bibinfo {author} {\bibfnamefont {A.~I.}\
  \bibnamefont {Krylov}}, \ and\ \bibinfo {author} {\bibfnamefont
  {S.}~\bibnamefont {Coriani}},\ }\bibfield  {title} {\enquote {\bibinfo
  {title} {{Interplay of Open-Shell Spin-Coupling and Jahn-Teller Distortion in
  Benzene Radical Cation Probed by X-ray Spectroscopy}},}\ }\href {\doibase
  10.1021/acs.jpca.0c08732} {\bibfield  {journal} {\bibinfo  {journal} {J.
  Phys. Chem. A}\ }\textbf {\bibinfo {volume} {124}},\ \bibinfo {pages}
  {9532--9541} (\bibinfo {year} {2020}{\natexlab{b}})}\BibitemShut {NoStop}%
\bibitem [{\citenamefont {Costantini}\ \emph {et~al.}(2019)\citenamefont
  {Costantini}, \citenamefont {Faber}, \citenamefont {Cossaro}, \citenamefont
  {Floreano}, \citenamefont {Verdini}, \citenamefont {H{\"a}ttig},
  \citenamefont {Morgante}, \citenamefont {Coriani},\ and\ \citenamefont
  {{Dell'Angela}}}]{martina2019}%
  \BibitemOpen
  \bibfield  {author} {\bibinfo {author} {\bibfnamefont {R.}~\bibnamefont
  {Costantini}}, \bibinfo {author} {\bibfnamefont {R.}~\bibnamefont {Faber}},
  \bibinfo {author} {\bibfnamefont {A.}~\bibnamefont {Cossaro}}, \bibinfo
  {author} {\bibfnamefont {L.}~\bibnamefont {Floreano}}, \bibinfo {author}
  {\bibfnamefont {A.}~\bibnamefont {Verdini}}, \bibinfo {author} {\bibfnamefont
  {C.}~\bibnamefont {H{\"a}ttig}}, \bibinfo {author} {\bibfnamefont
  {A.}~\bibnamefont {Morgante}}, \bibinfo {author} {\bibfnamefont
  {S.}~\bibnamefont {Coriani}}, \ and\ \bibinfo {author} {\bibfnamefont
  {M.}~\bibnamefont {{Dell'Angela}}},\ }\bibfield  {title} {\enquote {\bibinfo
  {title} {Picosecond timescale tracking of pentacene triplet excitons with
  chemical sensitivity},}\ }\href@noop {} {\bibfield  {journal} {\bibinfo
  {journal} {Commun. Phys.}\ }\textbf {\bibinfo {volume} {2}},\ \bibinfo
  {pages} {56} (\bibinfo {year} {2019})}\BibitemShut {NoStop}%
\bibitem [{\citenamefont {Myhre}\ \emph {et~al.}(2018)\citenamefont {Myhre},
  \citenamefont {Wolf}, \citenamefont {Cheng}, \citenamefont {Nandi},
  \citenamefont {Coriani}, \citenamefont {G{\"u}hr},\ and\ \citenamefont
  {Koch}}]{N2_Myhre_2018}%
  \BibitemOpen
  \bibfield  {author} {\bibinfo {author} {\bibfnamefont {R.~H.}\ \bibnamefont
  {Myhre}}, \bibinfo {author} {\bibfnamefont {T.~J.~A.}\ \bibnamefont {Wolf}},
  \bibinfo {author} {\bibfnamefont {L.}~\bibnamefont {Cheng}}, \bibinfo
  {author} {\bibfnamefont {S.}~\bibnamefont {Nandi}}, \bibinfo {author}
  {\bibfnamefont {S.}~\bibnamefont {Coriani}}, \bibinfo {author} {\bibfnamefont
  {M.}~\bibnamefont {G{\"u}hr}}, \ and\ \bibinfo {author} {\bibfnamefont
  {H.}~\bibnamefont {Koch}},\ }\bibfield  {title} {\enquote {\bibinfo {title}
  {A theoretical and experimental benchmark study of core-excited states in
  nitrogen},}\ }\href@noop {} {\bibfield  {journal} {\bibinfo  {journal} {J.
  Chem. Phys.}\ }\textbf {\bibinfo {volume} {148}},\ \bibinfo {pages} {064106}
  (\bibinfo {year} {2018})}\BibitemShut {NoStop}%
\bibitem [{\citenamefont {Folkestad}\ \emph {et~al.}(2020)\citenamefont
  {Folkestad}, \citenamefont {Kj{\o}nstad}, \citenamefont {Myhre},
  \citenamefont {Andersen}, \citenamefont {Balbi}, \citenamefont {Coriani},
  \citenamefont {Giovannini}, \citenamefont {Goletto}, \citenamefont
  {Haugland}, \citenamefont {Hutcheson}, \citenamefont {H{\o}yvik},
  \citenamefont {Moitra}, \citenamefont {Paul}, \citenamefont {Scavino},
  \citenamefont {Skeidsvoll}, \citenamefont {Tveten},\ and\ \citenamefont
  {Koch}}]{et2020}%
  \BibitemOpen
  \bibfield  {author} {\bibinfo {author} {\bibfnamefont {S.~D.}\ \bibnamefont
  {Folkestad}}, \bibinfo {author} {\bibfnamefont {E.~F.}\ \bibnamefont
  {Kj{\o}nstad}}, \bibinfo {author} {\bibfnamefont {R.~H.}\ \bibnamefont
  {Myhre}}, \bibinfo {author} {\bibfnamefont {J.~H.}\ \bibnamefont {Andersen}},
  \bibinfo {author} {\bibfnamefont {A.}~\bibnamefont {Balbi}}, \bibinfo
  {author} {\bibfnamefont {S.}~\bibnamefont {Coriani}}, \bibinfo {author}
  {\bibfnamefont {T.}~\bibnamefont {Giovannini}}, \bibinfo {author}
  {\bibfnamefont {L.}~\bibnamefont {Goletto}}, \bibinfo {author} {\bibfnamefont
  {T.~S.}\ \bibnamefont {Haugland}}, \bibinfo {author} {\bibfnamefont
  {A.}~\bibnamefont {Hutcheson}}, \bibinfo {author} {\bibfnamefont {I.-M.}\
  \bibnamefont {H{\o}yvik}}, \bibinfo {author} {\bibfnamefont {T.}~\bibnamefont
  {Moitra}}, \bibinfo {author} {\bibfnamefont {A.~C.}\ \bibnamefont {Paul}},
  \bibinfo {author} {\bibfnamefont {M.}~\bibnamefont {Scavino}}, \bibinfo
  {author} {\bibfnamefont {A.~S.}\ \bibnamefont {Skeidsvoll}}, \bibinfo
  {author} {\bibfnamefont {{\r{A}}.~H.}\ \bibnamefont {Tveten}}, \ and\
  \bibinfo {author} {\bibfnamefont {H.}~\bibnamefont {Koch}},\ }\bibfield
  {title} {\enquote {\bibinfo {title} {{$e^{T}$ 1.0: An open source electronic
  structure program with emphasis on coupled cluster and multilevel
  methods}},}\ }\href@noop {} {\bibfield  {journal} {\bibinfo  {journal} {J.
  Chem. Phys.}\ }\textbf {\bibinfo {volume} {152}},\ \bibinfo {pages} {184103}
  (\bibinfo {year} {2020})}\BibitemShut {NoStop}%
\bibitem [{\citenamefont {Paul}, \citenamefont {Myhre},\ and\ \citenamefont
  {Koch}(2021)}]{Alex:newcc3}%
  \BibitemOpen
  \bibfield  {author} {\bibinfo {author} {\bibfnamefont {A.~C.}\ \bibnamefont
  {Paul}}, \bibinfo {author} {\bibfnamefont {R.~H.}\ \bibnamefont {Myhre}}, \
  and\ \bibinfo {author} {\bibfnamefont {H.}~\bibnamefont {Koch}},\ }\bibfield
  {title} {\enquote {\bibinfo {title} {{New and Efficient Implementation of
  CC3}},}\ }\href {\doibase 10.1021/acs.jctc.0c00686} {\bibfield  {journal}
  {\bibinfo  {journal} {J. Chem. Theory Comput.}\ }\textbf {\bibinfo {volume}
  {17}},\ \bibinfo {pages} {117--126} (\bibinfo {year} {2021})}\BibitemShut
  {NoStop}%
\bibitem [{\citenamefont {Matthews}(2020)}]{daMatthews}%
  \BibitemOpen
  \bibfield  {author} {\bibinfo {author} {\bibfnamefont {D.~A.}\ \bibnamefont
  {Matthews}},\ }\bibfield  {title} {\enquote {\bibinfo {title} {{EOM-CC
  methods with approximate triple excitations applied to core excitation and
  ionisation energies}},}\ }\href {\doibase 10.1080/00268976.2020.1771448}
  {\bibfield  {journal} {\bibinfo  {journal} {Mol. Phys.}\ }\textbf {\bibinfo
  {volume} {118}},\ \bibinfo {pages} {e1771448} (\bibinfo {year}
  {2020})}\BibitemShut {NoStop}%
\bibitem [{\citenamefont {Peng}, \citenamefont {Copan},\ and\ \citenamefont
  {Sokolov}(2019)}]{Peng_LR-DCT}%
  \BibitemOpen
  \bibfield  {author} {\bibinfo {author} {\bibfnamefont {R.}~\bibnamefont
  {Peng}}, \bibinfo {author} {\bibfnamefont {A.~V.}\ \bibnamefont {Copan}}, \
  and\ \bibinfo {author} {\bibfnamefont {A.~Y.}\ \bibnamefont {Sokolov}},\
  }\bibfield  {title} {\enquote {\bibinfo {title} {Simulating x-ray absorption
  spectra with linear-response density cumulant theory},}\ }\href {\doibase
  10.1021/acs.jpca.8b12259} {\bibfield  {journal} {\bibinfo  {journal} {J.
  Phys. Chem. A}\ }\textbf {\bibinfo {volume} {123}},\ \bibinfo {pages}
  {1840--1850} (\bibinfo {year} {2019})}\BibitemShut {NoStop}%
\bibitem [{\citenamefont {Schirmer}(1982)}]{ADC82}%
  \BibitemOpen
  \bibfield  {author} {\bibinfo {author} {\bibfnamefont {J.}~\bibnamefont
  {Schirmer}},\ }\bibfield  {title} {\enquote {\bibinfo {title} {Beyond the
  random-phase approximation: A new approximation scheme for the polarization
  propagator},}\ }\href {\doibase 10.1103/PhysRevA.26.2395} {\bibfield
  {journal} {\bibinfo  {journal} {Phys. Rev. A}\ }\textbf {\bibinfo {volume}
  {26}},\ \bibinfo {pages} {2395--2416} (\bibinfo {year} {1982})}\BibitemShut
  {NoStop}%
\bibitem [{\citenamefont {Dreuw}\ and\ \citenamefont
  {Wormit}(2015)}]{ADC_review}%
  \BibitemOpen
  \bibfield  {author} {\bibinfo {author} {\bibfnamefont {A.}~\bibnamefont
  {Dreuw}}\ and\ \bibinfo {author} {\bibfnamefont {M.}~\bibnamefont {Wormit}},\
  }\bibfield  {title} {\enquote {\bibinfo {title} {The algebraic diagrammatic
  construction scheme for the polarization propagator for the calculation of
  excited states},}\ }\href@noop {} {\bibfield  {journal} {\bibinfo  {journal}
  {WIREs Comput. Mol. Sci.}\ }\textbf {\bibinfo {volume} {5}},\ \bibinfo
  {pages} {82--95} (\bibinfo {year} {2015})}\BibitemShut {NoStop}%
\bibitem [{\citenamefont {Barth}\ and\ \citenamefont
  {Schirmer}(1985)}]{Barth1985}%
  \BibitemOpen
  \bibfield  {author} {\bibinfo {author} {\bibfnamefont {A.}~\bibnamefont
  {Barth}}\ and\ \bibinfo {author} {\bibfnamefont {J.}~\bibnamefont
  {Schirmer}},\ }\bibfield  {title} {\enquote {\bibinfo {title} {Theoretical
  core-level excitation spectra of {N$_2$} and {CO} by a new polarisation
  propagator method},}\ }\href@noop {} {\bibfield  {journal} {\bibinfo
  {journal} {J. Phys. B}\ }\textbf {\bibinfo {volume} {18}},\ \bibinfo {pages}
  {867--885} (\bibinfo {year} {1985})}\BibitemShut {NoStop}%
\bibitem [{\citenamefont {Christiansen}, \citenamefont {Koch},\ and\
  \citenamefont {J{\o}rgensen}(1995)}]{CHRISTIANSEN_CC2}%
  \BibitemOpen
  \bibfield  {author} {\bibinfo {author} {\bibfnamefont {O.}~\bibnamefont
  {Christiansen}}, \bibinfo {author} {\bibfnamefont {H.}~\bibnamefont {Koch}},
  \ and\ \bibinfo {author} {\bibfnamefont {P.}~\bibnamefont {J{\o}rgensen}},\
  }\bibfield  {title} {\enquote {\bibinfo {title} {The second-order approximate
  coupled cluster singles and doubles model {CC2}},}\ }\href {\doibase
  https://doi.org/10.1016/0009-2614(95)00841-Q} {\bibfield  {journal} {\bibinfo
   {journal} {Chem. Phys. Lett.}\ }\textbf {\bibinfo {volume} {243}},\ \bibinfo
  {pages} {409 -- 418} (\bibinfo {year} {1995})}\BibitemShut {NoStop}%
\bibitem [{\citenamefont {Schirmer}\ \emph {et~al.}(1993)\citenamefont
  {Schirmer}, \citenamefont {Trofimov}, \citenamefont {Randall}, \citenamefont
  {Feldhaus}, \citenamefont {Bradshaw}, \citenamefont {Ma}, \citenamefont
  {Chen},\ and\ \citenamefont {Sette}}]{NEXAFS_H2O_NH3_CH4_Exp}%
  \BibitemOpen
  \bibfield  {author} {\bibinfo {author} {\bibfnamefont {J.}~\bibnamefont
  {Schirmer}}, \bibinfo {author} {\bibfnamefont {A.~B.}\ \bibnamefont
  {Trofimov}}, \bibinfo {author} {\bibfnamefont {K.~J.}\ \bibnamefont
  {Randall}}, \bibinfo {author} {\bibfnamefont {J.}~\bibnamefont {Feldhaus}},
  \bibinfo {author} {\bibfnamefont {A.~M.}\ \bibnamefont {Bradshaw}}, \bibinfo
  {author} {\bibfnamefont {Y.}~\bibnamefont {Ma}}, \bibinfo {author}
  {\bibfnamefont {C.~T.}\ \bibnamefont {Chen}}, \ and\ \bibinfo {author}
  {\bibfnamefont {F.}~\bibnamefont {Sette}},\ }\bibfield  {title} {\enquote
  {\bibinfo {title} {{$K$-shell excitation of the water, ammonia, and methane
  molecules using high-resolution photoabsorption spectroscopy}},}\ }\href@noop
  {} {\bibfield  {journal} {\bibinfo  {journal} {Phys. Rev. A}\ }\textbf
  {\bibinfo {volume} {47}},\ \bibinfo {pages} {1136} (\bibinfo {year}
  {1993})}\BibitemShut {NoStop}%
\bibitem [{\citenamefont {Wenzel}, \citenamefont {Wormit},\ and\ \citenamefont
  {Dreuw}(2014)}]{wenzel2014}%
  \BibitemOpen
  \bibfield  {author} {\bibinfo {author} {\bibfnamefont {J.}~\bibnamefont
  {Wenzel}}, \bibinfo {author} {\bibfnamefont {M.}~\bibnamefont {Wormit}}, \
  and\ \bibinfo {author} {\bibfnamefont {A.}~\bibnamefont {Dreuw}},\ }\bibfield
   {title} {\enquote {\bibinfo {title} {Calculating core-level excitations and
  x-ray absorption spectra of medium-sized closed-shell molecules with the
  algebraic-diagrammatic construction scheme for the polarization
  propagator},}\ }\href@noop {} {\bibfield  {journal} {\bibinfo  {journal} {J.
  Comput. Chem.}\ }\textbf {\bibinfo {volume} {35}},\ \bibinfo {pages}
  {1900--1915} (\bibinfo {year} {2014})}\BibitemShut {NoStop}%
\bibitem [{\citenamefont {Plasser}\ \emph
  {et~al.}(2014{\natexlab{a}})\citenamefont {Plasser}, \citenamefont
  {Crespo-Otero}, \citenamefont {Pederzoli}, \citenamefont {Pittner},
  \citenamefont {Lischka},\ and\ \citenamefont {Barbatti}}]{Plasser_Barbatti}%
  \BibitemOpen
  \bibfield  {author} {\bibinfo {author} {\bibfnamefont {F.}~\bibnamefont
  {Plasser}}, \bibinfo {author} {\bibfnamefont {R.}~\bibnamefont
  {Crespo-Otero}}, \bibinfo {author} {\bibfnamefont {M.}~\bibnamefont
  {Pederzoli}}, \bibinfo {author} {\bibfnamefont {J.}~\bibnamefont {Pittner}},
  \bibinfo {author} {\bibfnamefont {H.}~\bibnamefont {Lischka}}, \ and\
  \bibinfo {author} {\bibfnamefont {M.}~\bibnamefont {Barbatti}},\ }\bibfield
  {title} {\enquote {\bibinfo {title} {Surface hopping dynamics with correlated
  single-reference methods: 9{H}-adenine as a case study},}\ }\href {\doibase
  10.1021/ct4011079} {\bibfield  {journal} {\bibinfo  {journal} {J. Chem.
  Theory Comput.}\ }\textbf {\bibinfo {volume} {10}},\ \bibinfo {pages}
  {1395--1405} (\bibinfo {year} {2014}{\natexlab{a}})}\BibitemShut {NoStop}%
\bibitem [{\citenamefont {Neville}\ \emph
  {et~al.}(2016{\natexlab{a}})\citenamefont {Neville}, \citenamefont
  {Averbukh}, \citenamefont {Patchkovskii}, \citenamefont {Ruberti},
  \citenamefont {Yun}, \citenamefont {Chergui}, \citenamefont {Stolow},\ and\
  \citenamefont {Schuurman}}]{Neville_ADC_TRNEXAFS}%
  \BibitemOpen
  \bibfield  {author} {\bibinfo {author} {\bibfnamefont {S.~P.}\ \bibnamefont
  {Neville}}, \bibinfo {author} {\bibfnamefont {V.}~\bibnamefont {Averbukh}},
  \bibinfo {author} {\bibfnamefont {S.}~\bibnamefont {Patchkovskii}}, \bibinfo
  {author} {\bibfnamefont {M.}~\bibnamefont {Ruberti}}, \bibinfo {author}
  {\bibfnamefont {R.}~\bibnamefont {Yun}}, \bibinfo {author} {\bibfnamefont
  {M.}~\bibnamefont {Chergui}}, \bibinfo {author} {\bibfnamefont
  {A.}~\bibnamefont {Stolow}}, \ and\ \bibinfo {author} {\bibfnamefont {M.~S.}\
  \bibnamefont {Schuurman}},\ }\bibfield  {title} {\enquote {\bibinfo {title}
  {Beyond structure: ultrafast {X}-ray absorption spectroscopy as a probe of
  non-adiabatic wavepacket dynamics},}\ }\href {\doibase 10.1039/C6FD00117C}
  {\bibfield  {journal} {\bibinfo  {journal} {Faraday Discuss.}\ }\textbf
  {\bibinfo {volume} {194}},\ \bibinfo {pages} {117--145} (\bibinfo {year}
  {2016}{\natexlab{a}})}\BibitemShut {NoStop}%
\bibitem [{\citenamefont {Neville}\ \emph
  {et~al.}(2016{\natexlab{b}})\citenamefont {Neville}, \citenamefont
  {Averbukh}, \citenamefont {Ruberti}, \citenamefont {Yun}, \citenamefont
  {Patchkovskii}, \citenamefont {Chergui}, \citenamefont {Stolow},\ and\
  \citenamefont {Schuurman}}]{Neville_TRXAS}%
  \BibitemOpen
  \bibfield  {author} {\bibinfo {author} {\bibfnamefont {S.~P.}\ \bibnamefont
  {Neville}}, \bibinfo {author} {\bibfnamefont {V.}~\bibnamefont {Averbukh}},
  \bibinfo {author} {\bibfnamefont {M.}~\bibnamefont {Ruberti}}, \bibinfo
  {author} {\bibfnamefont {R.}~\bibnamefont {Yun}}, \bibinfo {author}
  {\bibfnamefont {S.}~\bibnamefont {Patchkovskii}}, \bibinfo {author}
  {\bibfnamefont {M.}~\bibnamefont {Chergui}}, \bibinfo {author} {\bibfnamefont
  {A.}~\bibnamefont {Stolow}}, \ and\ \bibinfo {author} {\bibfnamefont {M.~S.}\
  \bibnamefont {Schuurman}},\ }\bibfield  {title} {\enquote {\bibinfo {title}
  {Excited state {X}-ray absorption spectroscopy: Probing both electronic and
  structural dynamics},}\ }\href {\doibase 10.1063/1.4964369} {\bibfield
  {journal} {\bibinfo  {journal} {J. Chem. Phys.}\ }\textbf {\bibinfo {volume}
  {145}},\ \bibinfo {pages} {144307} (\bibinfo {year}
  {2016}{\natexlab{b}})}\BibitemShut {NoStop}%
\bibitem [{\citenamefont {Neville}\ \emph {et~al.}(2018)\citenamefont
  {Neville}, \citenamefont {Chergui}, \citenamefont {Stolow},\ and\
  \citenamefont {Schuurman}}]{Neville_TRXAS_at_CI}%
  \BibitemOpen
  \bibfield  {author} {\bibinfo {author} {\bibfnamefont {S.~P.}\ \bibnamefont
  {Neville}}, \bibinfo {author} {\bibfnamefont {M.}~\bibnamefont {Chergui}},
  \bibinfo {author} {\bibfnamefont {A.}~\bibnamefont {Stolow}}, \ and\ \bibinfo
  {author} {\bibfnamefont {M.~S.}\ \bibnamefont {Schuurman}},\ }\bibfield
  {title} {\enquote {\bibinfo {title} {Ultrafast x-ray spectroscopy of conical
  intersections},}\ }\href {\doibase 10.1103/PhysRevLett.120.243001} {\bibfield
   {journal} {\bibinfo  {journal} {Phys. Rev. Lett.}\ }\textbf {\bibinfo
  {volume} {120}},\ \bibinfo {pages} {243001} (\bibinfo {year}
  {2018})}\BibitemShut {NoStop}%
\bibitem [{\citenamefont {Neville}\ and\ \citenamefont
  {Schuurman}(2018)}]{Neville_Schuurman_autocorrelation}%
  \BibitemOpen
  \bibfield  {author} {\bibinfo {author} {\bibfnamefont {S.~P.}\ \bibnamefont
  {Neville}}\ and\ \bibinfo {author} {\bibfnamefont {M.~S.}\ \bibnamefont
  {Schuurman}},\ }\bibfield  {title} {\enquote {\bibinfo {title} {A general
  approach for the calculation and characterization of x-ray absorption
  spectra},}\ }\href {\doibase 10.1063/1.5048520} {\bibfield  {journal}
  {\bibinfo  {journal} {J. Chem. Phys.}\ }\textbf {\bibinfo {volume} {149}},\
  \bibinfo {pages} {154111} (\bibinfo {year} {2018})}\BibitemShut {NoStop}%
\bibitem [{\citenamefont {Trofimov}\ and\ \citenamefont
  {Schirmer}(1995)}]{Trofimov_1995}%
  \BibitemOpen
  \bibfield  {author} {\bibinfo {author} {\bibfnamefont {A.~B.}\ \bibnamefont
  {Trofimov}}\ and\ \bibinfo {author} {\bibfnamefont {J.}~\bibnamefont
  {Schirmer}},\ }\bibfield  {title} {\enquote {\bibinfo {title} {An efficient
  polarization propagator approach to valence electron excitation spectra},}\
  }\href {\doibase 10.1088/0953-4075/28/12/003} {\bibfield  {journal} {\bibinfo
   {journal} {J. Phys. B}\ }\textbf {\bibinfo {volume} {28}},\ \bibinfo {pages}
  {2299--2324} (\bibinfo {year} {1995})}\BibitemShut {NoStop}%
\bibitem [{\citenamefont {Plekan}\ \emph {et~al.}(2008)\citenamefont {Plekan},
  \citenamefont {Feyer}, \citenamefont {Richter}, \citenamefont {Coreno},
  \citenamefont {de~Simone}, \citenamefont {Prince}, \citenamefont {Trofimov},
  \citenamefont {Gromov}, \citenamefont {Zaytseva},\ and\ \citenamefont
  {Schirmer}}]{nexafs_thymine_adenine}%
  \BibitemOpen
  \bibfield  {author} {\bibinfo {author} {\bibfnamefont {O.}~\bibnamefont
  {Plekan}}, \bibinfo {author} {\bibfnamefont {V.}~\bibnamefont {Feyer}},
  \bibinfo {author} {\bibfnamefont {R.}~\bibnamefont {Richter}}, \bibinfo
  {author} {\bibfnamefont {M.}~\bibnamefont {Coreno}}, \bibinfo {author}
  {\bibfnamefont {M.}~\bibnamefont {de~Simone}}, \bibinfo {author}
  {\bibfnamefont {K.~C.}\ \bibnamefont {Prince}}, \bibinfo {author}
  {\bibfnamefont {A.~B.}\ \bibnamefont {Trofimov}}, \bibinfo {author}
  {\bibfnamefont {E.~V.}\ \bibnamefont {Gromov}}, \bibinfo {author}
  {\bibfnamefont {I.~L.}\ \bibnamefont {Zaytseva}}, \ and\ \bibinfo {author}
  {\bibfnamefont {J.}~\bibnamefont {Schirmer}},\ }\bibfield  {title} {\enquote
  {\bibinfo {title} {{A theoretical and experimental study of the near edge
  X-ray absorption fine structure (NEXAFS) and X-ray photoelectron spectra
  (XPS) of nucleobases: Thymine and adenine}},}\ }\href@noop {} {\bibfield
  {journal} {\bibinfo  {journal} {Chem. Phys.}\ }\textbf {\bibinfo {volume}
  {347}},\ \bibinfo {pages} {360--375} (\bibinfo {year} {2008})}\BibitemShut
  {NoStop}%
\bibitem [{\citenamefont {List}\ \emph {et~al.}(2020)\citenamefont {List},
  \citenamefont {Dempwolff}, \citenamefont {Dreuw}, \citenamefont {Norman},\
  and\ \citenamefont {Mart{\'\i}nez}}]{NHList:TRXAS}%
  \BibitemOpen
  \bibfield  {author} {\bibinfo {author} {\bibfnamefont {N.~H.}\ \bibnamefont
  {List}}, \bibinfo {author} {\bibfnamefont {A.~L.}\ \bibnamefont {Dempwolff}},
  \bibinfo {author} {\bibfnamefont {A.}~\bibnamefont {Dreuw}}, \bibinfo
  {author} {\bibfnamefont {P.}~\bibnamefont {Norman}}, \ and\ \bibinfo {author}
  {\bibfnamefont {T.~J.}\ \bibnamefont {Mart{\'\i}nez}},\ }\bibfield  {title}
  {\enquote {\bibinfo {title} {{Probing competing relaxation pathways in
  malonaldehyde with transient X-ray absorption spectroscopy}},}\ }\href
  {\doibase 10.1039/D0SC00840K} {\bibfield  {journal} {\bibinfo  {journal}
  {Chem. Sci.}\ }\textbf {\bibinfo {volume} {11}},\ \bibinfo {pages}
  {4180--4193} (\bibinfo {year} {2020})}\BibitemShut {NoStop}%
\bibitem [{\citenamefont {Seidu}\ \emph {et~al.}(2019)\citenamefont {Seidu},
  \citenamefont {Neville}, \citenamefont {Kleinschmidt}, \citenamefont {Heil},
  \citenamefont {Marian},\ and\ \citenamefont {Schuurman}}]{Seidu19}%
  \BibitemOpen
  \bibfield  {author} {\bibinfo {author} {\bibfnamefont {I.}~\bibnamefont
  {Seidu}}, \bibinfo {author} {\bibfnamefont {S.~P.}\ \bibnamefont {Neville}},
  \bibinfo {author} {\bibfnamefont {M.}~\bibnamefont {Kleinschmidt}}, \bibinfo
  {author} {\bibfnamefont {A.}~\bibnamefont {Heil}}, \bibinfo {author}
  {\bibfnamefont {C.~M.}\ \bibnamefont {Marian}}, \ and\ \bibinfo {author}
  {\bibfnamefont {M.~S.}\ \bibnamefont {Schuurman}},\ }\bibfield  {title}
  {\enquote {\bibinfo {title} {The simulation of {X}-ray absorption spectra
  from ground and excited electronic states using core-valence separated
  {DFT/MRCI}},}\ }\href {\doibase 10.1063/1.5110418} {\bibfield  {journal}
  {\bibinfo  {journal} {J. Chem. Phys.}\ }\textbf {\bibinfo {volume} {151}},\
  \bibinfo {pages} {144104} (\bibinfo {year} {2019})}\BibitemShut {NoStop}%
\bibitem [{\citenamefont {Lyskov}, \citenamefont {Kleinschmidt},\ and\
  \citenamefont {Marian}(2016)}]{Lyskov_DFT/MRCI}%
  \BibitemOpen
  \bibfield  {author} {\bibinfo {author} {\bibfnamefont {I.}~\bibnamefont
  {Lyskov}}, \bibinfo {author} {\bibfnamefont {M.}~\bibnamefont
  {Kleinschmidt}}, \ and\ \bibinfo {author} {\bibfnamefont {C.~M.}\
  \bibnamefont {Marian}},\ }\bibfield  {title} {\enquote {\bibinfo {title}
  {{Redesign of the DFT/MRCI Hamiltonian}},}\ }\href {\doibase
  10.1063/1.4940036} {\bibfield  {journal} {\bibinfo  {journal} {J. Chem.
  Phys.}\ }\textbf {\bibinfo {volume} {144}},\ \bibinfo {pages} {034104}
  (\bibinfo {year} {2016})}\BibitemShut {NoStop}%
\bibitem [{\citenamefont {Olsen}\ \emph {et~al.}(1988)\citenamefont {Olsen},
  \citenamefont {Roos}, \citenamefont {J{\o}rgensen},\ and\ \citenamefont
  {Jensen}}]{RASSCF88}%
  \BibitemOpen
  \bibfield  {author} {\bibinfo {author} {\bibfnamefont {J.}~\bibnamefont
  {Olsen}}, \bibinfo {author} {\bibfnamefont {B.~O.}\ \bibnamefont {Roos}},
  \bibinfo {author} {\bibfnamefont {P.}~\bibnamefont {J{\o}rgensen}}, \ and\
  \bibinfo {author} {\bibfnamefont {H.~J.~A.}\ \bibnamefont {Jensen}},\
  }\bibfield  {title} {\enquote {\bibinfo {title} {Determinant based
  configuration interaction algorithms for complete and restricted
  configuration interaction spaces},}\ }\href {\doibase 10.1063/1.455063}
  {\bibfield  {journal} {\bibinfo  {journal} {J. Chem. Phys.}\ }\textbf
  {\bibinfo {volume} {89}},\ \bibinfo {pages} {2185--2192} (\bibinfo {year}
  {1988})}\BibitemShut {NoStop}%
\bibitem [{\citenamefont {Malmqvist}, \citenamefont {Rendell},\ and\
  \citenamefont {Roos}(1990)}]{RASSCF90}%
  \BibitemOpen
  \bibfield  {author} {\bibinfo {author} {\bibfnamefont {P.~{\AA}.}\
  \bibnamefont {Malmqvist}}, \bibinfo {author} {\bibfnamefont {A.}~\bibnamefont
  {Rendell}}, \ and\ \bibinfo {author} {\bibfnamefont {B.~O.}\ \bibnamefont
  {Roos}},\ }\bibfield  {title} {\enquote {\bibinfo {title} {The restricted
  active space self-consistent-field method, implemented with a split graph
  unitary group approach},}\ }\href {\doibase 10.1021/j100377a011} {\bibfield
  {journal} {\bibinfo  {journal} {J. Phys. Chem.}\ }\textbf {\bibinfo {volume}
  {94}},\ \bibinfo {pages} {5477--5482} (\bibinfo {year} {1990})}\BibitemShut
  {NoStop}%
\bibitem [{\citenamefont {Malmqvist}\ \emph {et~al.}(2008)\citenamefont
  {Malmqvist}, \citenamefont {Pierloot}, \citenamefont {Shahi}, \citenamefont
  {Cramer},\ and\ \citenamefont {Gagliardi}}]{RASPT2_Malmqvist}%
  \BibitemOpen
  \bibfield  {author} {\bibinfo {author} {\bibfnamefont {P.~{\AA}.}\
  \bibnamefont {Malmqvist}}, \bibinfo {author} {\bibfnamefont {K.}~\bibnamefont
  {Pierloot}}, \bibinfo {author} {\bibfnamefont {A.~R.~M.}\ \bibnamefont
  {Shahi}}, \bibinfo {author} {\bibfnamefont {C.~J.}\ \bibnamefont {Cramer}}, \
  and\ \bibinfo {author} {\bibfnamefont {L.}~\bibnamefont {Gagliardi}},\
  }\bibfield  {title} {\enquote {\bibinfo {title} {The restricted active space
  followed by second-order perturbation theory method: Theory and application
  to the study of {CuO$_2$} and {Cu$_2$O$_2$} systems},}\ }\href {\doibase
  10.1063/1.2920188} {\bibfield  {journal} {\bibinfo  {journal} {J. Chem.
  Phys.}\ }\textbf {\bibinfo {volume} {128}},\ \bibinfo {pages} {204109}
  (\bibinfo {year} {2008})}\BibitemShut {NoStop}%
\bibitem [{\citenamefont {Delcey}\ \emph {et~al.}(2019)\citenamefont {Delcey},
  \citenamefont {S{\o}rensen}, \citenamefont {Vacher}, \citenamefont {Couto},\
  and\ \citenamefont {Lundberg}}]{Delcey2019}%
  \BibitemOpen
  \bibfield  {author} {\bibinfo {author} {\bibfnamefont {M.~G.}\ \bibnamefont
  {Delcey}}, \bibinfo {author} {\bibfnamefont {L.~K.}\ \bibnamefont
  {S{\o}rensen}}, \bibinfo {author} {\bibfnamefont {M.}~\bibnamefont {Vacher}},
  \bibinfo {author} {\bibfnamefont {R.~C.}\ \bibnamefont {Couto}}, \ and\
  \bibinfo {author} {\bibfnamefont {M.}~\bibnamefont {Lundberg}},\ }\bibfield
  {title} {\enquote {\bibinfo {title} {Efficient calculations of a large number
  of highly excited states for multiconfigurational wavefunctions},}\ }\href
  {\doibase 10.1002/jcc.25832} {\bibfield  {journal} {\bibinfo  {journal} {J.
  Comput. Chem.}\ }\textbf {\bibinfo {volume} {40}},\ \bibinfo {pages}
  {1789--1799} (\bibinfo {year} {2019})}\BibitemShut {NoStop}%
\bibitem [{\citenamefont {Hua}, \citenamefont {Mukamel},\ and\ \citenamefont
  {Luo}(2019)}]{Mukamel_uracil}%
  \BibitemOpen
  \bibfield  {author} {\bibinfo {author} {\bibfnamefont {W.}~\bibnamefont
  {Hua}}, \bibinfo {author} {\bibfnamefont {S.}~\bibnamefont {Mukamel}}, \ and\
  \bibinfo {author} {\bibfnamefont {Y.}~\bibnamefont {Luo}},\ }\bibfield
  {title} {\enquote {\bibinfo {title} {Transient x-ray absorption spectral
  fingerprints of the {S$_1$} dark state in uracil},}\ }\href {\doibase
  10.1021/acs.jpclett.9b02692} {\bibfield  {journal} {\bibinfo  {journal} {J.
  Phys. Chem. Lett.}\ }\textbf {\bibinfo {volume} {10}},\ \bibinfo {pages}
  {7172--7178} (\bibinfo {year} {2019})}\BibitemShut {NoStop}%
\bibitem [{\citenamefont {Segatta}\ \emph {et~al.}(2020)\citenamefont
  {Segatta}, \citenamefont {Nenov}, \citenamefont {Orlandi}, \citenamefont
  {Arcioni}, \citenamefont {Mukamel},\ and\ \citenamefont
  {Garavelli}}]{Segatta}%
  \BibitemOpen
  \bibfield  {author} {\bibinfo {author} {\bibfnamefont {F.}~\bibnamefont
  {Segatta}}, \bibinfo {author} {\bibfnamefont {A.}~\bibnamefont {Nenov}},
  \bibinfo {author} {\bibfnamefont {S.}~\bibnamefont {Orlandi}}, \bibinfo
  {author} {\bibfnamefont {A.}~\bibnamefont {Arcioni}}, \bibinfo {author}
  {\bibfnamefont {S.}~\bibnamefont {Mukamel}}, \ and\ \bibinfo {author}
  {\bibfnamefont {M.}~\bibnamefont {Garavelli}},\ }\bibfield  {title} {\enquote
  {\bibinfo {title} {Exploring the capabilities of optical pump {X}-ray probe
  {NEXAFS} spectroscopy to track photo-induced dynamics mediated by conical
  intersections},}\ }\href {\doibase 10.1039/C9FD00073A} {\bibfield  {journal}
  {\bibinfo  {journal} {Faraday Discuss.}\ }\textbf {\bibinfo {volume} {221}},\
  \bibinfo {pages} {245--264} (\bibinfo {year} {2020})}\BibitemShut {NoStop}%
\bibitem [{\citenamefont {Schweigert}\ and\ \citenamefont
  {Mukamel}(2007)}]{Schweigert}%
  \BibitemOpen
  \bibfield  {author} {\bibinfo {author} {\bibfnamefont {I.~V.}\ \bibnamefont
  {Schweigert}}\ and\ \bibinfo {author} {\bibfnamefont {S.}~\bibnamefont
  {Mukamel}},\ }\bibfield  {title} {\enquote {\bibinfo {title} {Coherent
  ultrafast core-hole correlation spectroscopy: X-ray analogues of
  multidimensional {NMR}},}\ }\href {\doibase 10.1103/PhysRevLett.99.163001}
  {\bibfield  {journal} {\bibinfo  {journal} {Phys. Rev. Lett.}\ }\textbf
  {\bibinfo {volume} {99}},\ \bibinfo {pages} {163001} (\bibinfo {year}
  {2007})}\BibitemShut {NoStop}%
\bibitem [{\citenamefont {Lee}, \citenamefont {Small},\ and\ \citenamefont
  {Head-Gordon}(2019)}]{Lee_Head-Gordon}%
  \BibitemOpen
  \bibfield  {author} {\bibinfo {author} {\bibfnamefont {J.}~\bibnamefont
  {Lee}}, \bibinfo {author} {\bibfnamefont {D.~W.}\ \bibnamefont {Small}}, \
  and\ \bibinfo {author} {\bibfnamefont {M.}~\bibnamefont {Head-Gordon}},\
  }\bibfield  {title} {\enquote {\bibinfo {title} {Excited states via coupled
  cluster theory without equation-of-motion methods: Seeking higher roots with
  application to doubly excited states and double core hole states},}\ }\href
  {\doibase 10.1063/1.5128795} {\bibfield  {journal} {\bibinfo  {journal} {J.
  Chem. Phys.}\ }\textbf {\bibinfo {volume} {151}},\ \bibinfo {pages} {214103}
  (\bibinfo {year} {2019})}\BibitemShut {NoStop}%
\bibitem [{\citenamefont {Feyer}\ \emph {et~al.}(2010)\citenamefont {Feyer},
  \citenamefont {Plekan}, \citenamefont {Richter}, \citenamefont {Coreno},
  \citenamefont {de~Simone}, \citenamefont {Prince}, \citenamefont {Trofimov},
  \citenamefont {Zaytseva},\ and\ \citenamefont
  {Schirmer}}]{nexafs_cytosine_uracil}%
  \BibitemOpen
  \bibfield  {author} {\bibinfo {author} {\bibfnamefont {V.}~\bibnamefont
  {Feyer}}, \bibinfo {author} {\bibfnamefont {O.}~\bibnamefont {Plekan}},
  \bibinfo {author} {\bibfnamefont {R.}~\bibnamefont {Richter}}, \bibinfo
  {author} {\bibfnamefont {M.}~\bibnamefont {Coreno}}, \bibinfo {author}
  {\bibfnamefont {M.}~\bibnamefont {de~Simone}}, \bibinfo {author}
  {\bibfnamefont {K.~C.}\ \bibnamefont {Prince}}, \bibinfo {author}
  {\bibfnamefont {A.~B.}\ \bibnamefont {Trofimov}}, \bibinfo {author}
  {\bibfnamefont {I.~L.}\ \bibnamefont {Zaytseva}}, \ and\ \bibinfo {author}
  {\bibfnamefont {J.}~\bibnamefont {Schirmer}},\ }\bibfield  {title} {\enquote
  {\bibinfo {title} {{Tautomerism in Cytosine and Uracil: A Theoretical and
  Experimental X-ray Absorption and Resonant Auger Study}},}\ }\href@noop {}
  {\bibfield  {journal} {\bibinfo  {journal} {J. Phys. Chem. A}\ }\textbf
  {\bibinfo {volume} {114}},\ \bibinfo {pages} {10270--10276} (\bibinfo {year}
  {2010})}\BibitemShut {NoStop}%
\bibitem [{\citenamefont {Trajmar}(1980)}]{Trajmar}%
  \BibitemOpen
  \bibfield  {author} {\bibinfo {author} {\bibfnamefont {S.}~\bibnamefont
  {Trajmar}},\ }\bibfield  {title} {\enquote {\bibinfo {title} {Electron impact
  spectroscopy},}\ }\href {\doibase 10.1021/ar50145a003} {\bibfield  {journal}
  {\bibinfo  {journal} {Acc. Chem. Res}\ }\textbf {\bibinfo {volume} {13}},\
  \bibinfo {pages} {14--20} (\bibinfo {year} {1980})}\BibitemShut {NoStop}%
\bibitem [{\citenamefont {Christiansen}\ \emph {et~al.}(1996)\citenamefont
  {Christiansen}, \citenamefont {Koch}, \citenamefont {Halkier}, \citenamefont
  {J{\o}rgensen}, \citenamefont {Helgaker},\ and\ \citenamefont {S\'{a}nchez~de
  Mer\'{a}s}}]{Christiansen_benzene}%
  \BibitemOpen
  \bibfield  {author} {\bibinfo {author} {\bibfnamefont {O.}~\bibnamefont
  {Christiansen}}, \bibinfo {author} {\bibfnamefont {H.}~\bibnamefont {Koch}},
  \bibinfo {author} {\bibfnamefont {A.}~\bibnamefont {Halkier}}, \bibinfo
  {author} {\bibfnamefont {P.}~\bibnamefont {J{\o}rgensen}}, \bibinfo {author}
  {\bibfnamefont {T.}~\bibnamefont {Helgaker}}, \ and\ \bibinfo {author}
  {\bibfnamefont {A.}~\bibnamefont {S\'{a}nchez~de Mer\'{a}s}},\ }\bibfield
  {title} {\enquote {\bibinfo {title} {Large-scale calculations of excitation
  energies in coupled cluster theory: The singlet excited states of benzene},}\
  }\href {\doibase 10.1063/1.471985} {\bibfield  {journal} {\bibinfo  {journal}
  {J. Chem. Phys.}\ }\textbf {\bibinfo {volume} {105}},\ \bibinfo {pages}
  {6921--6939} (\bibinfo {year} {1996})}\BibitemShut {NoStop}%
\bibitem [{\citenamefont {Hald}, \citenamefont {H\"{a}ttig},\ and\
  \citenamefont {J{\o}rgensen}(2000)}]{Hald}%
  \BibitemOpen
  \bibfield  {author} {\bibinfo {author} {\bibfnamefont {K.}~\bibnamefont
  {Hald}}, \bibinfo {author} {\bibfnamefont {C.}~\bibnamefont {H\"{a}ttig}}, \
  and\ \bibinfo {author} {\bibfnamefont {P.}~\bibnamefont {J{\o}rgensen}},\
  }\bibfield  {title} {\enquote {\bibinfo {title} {Triplet excitation energies
  in the coupled cluster singles and doubles model using an explicit triplet
  spin coupled excitation space},}\ }\href {\doibase 10.1063/1.1316033}
  {\bibfield  {journal} {\bibinfo  {journal} {J. Chem. Phys.}\ }\textbf
  {\bibinfo {volume} {113}},\ \bibinfo {pages} {7765--7772} (\bibinfo {year}
  {2000})}\BibitemShut {NoStop}%
\bibitem [{\citenamefont {Vidal}, \citenamefont {Krylov},\ and\ \citenamefont
  {Coriani}(2020{\natexlab{b}})}]{vidal2:correction}%
  \BibitemOpen
  \bibfield  {author} {\bibinfo {author} {\bibfnamefont {M.~L.}\ \bibnamefont
  {Vidal}}, \bibinfo {author} {\bibfnamefont {A.~I.}\ \bibnamefont {Krylov}}, \
  and\ \bibinfo {author} {\bibfnamefont {S.}~\bibnamefont {Coriani}},\
  }\bibfield  {title} {\enquote {\bibinfo {title} {Correction: Dyson orbitals
  within the fc-{CVS-EOM-CCSD} framework: theory and application to {X}-ray
  photoelectron spectroscopy of ground and excited states},}\ }\href {\doibase
  10.1039/D0CP90012E} {\bibfield  {journal} {\bibinfo  {journal} {Phys. Chem.
  Chem. Phys.}\ }\textbf {\bibinfo {volume} {22}},\ \bibinfo {pages}
  {3744--3747} (\bibinfo {year} {2020}{\natexlab{b}})}\BibitemShut {NoStop}%
\bibitem [{\citenamefont {Krylov}(2017)}]{krylov_open-shell_2017}%
  \BibitemOpen
  \bibfield  {author} {\bibinfo {author} {\bibfnamefont {A.~I.}\ \bibnamefont
  {Krylov}},\ }\enquote {\bibinfo {title} {The quantum chemistry of open-shell
  species},}\ in\ \href {\doibase 10.1002/9781119356059.ch4} {\emph {\bibinfo
  {booktitle} {Reviews in Computational Chemistry}}}\ (\bibinfo  {publisher}
  {John Wiley \& Sons, Ltd},\ \bibinfo {year} {2017})\ Chap.~\bibinfo {chapter}
  {4}, pp.\ \bibinfo {pages} {151--224}\BibitemShut {NoStop}%
\bibitem [{\citenamefont {Casanova}\ and\ \citenamefont
  {Krylov}(2020)}]{casanova_krylov_2020}%
  \BibitemOpen
  \bibfield  {author} {\bibinfo {author} {\bibfnamefont {D.}~\bibnamefont
  {Casanova}}\ and\ \bibinfo {author} {\bibfnamefont {A.~I.}\ \bibnamefont
  {Krylov}},\ }\bibfield  {title} {\enquote {\bibinfo {title} {Spin-flip
  methods in quantum chemistry},}\ }\href {\doibase 10.1039/C9CP06507E}
  {\bibfield  {journal} {\bibinfo  {journal} {Phys. Chem. Chem. Phys.}\
  }\textbf {\bibinfo {volume} {22}},\ \bibinfo {pages} {4326--4342} (\bibinfo
  {year} {2020})}\BibitemShut {NoStop}%
\bibitem [{\citenamefont {Luzanov}, \citenamefont {Sukhorukov},\ and\
  \citenamefont {Umanskii}(1976)}]{Luzanov1976}%
  \BibitemOpen
  \bibfield  {author} {\bibinfo {author} {\bibfnamefont {A.~V.}\ \bibnamefont
  {Luzanov}}, \bibinfo {author} {\bibfnamefont {A.~A.}\ \bibnamefont
  {Sukhorukov}}, \ and\ \bibinfo {author} {\bibfnamefont {V.~{\'E}.}\
  \bibnamefont {Umanskii}},\ }\bibfield  {title} {\enquote {\bibinfo {title}
  {Application of transition density matrix for analysis of excited states},}\
  }\href {\doibase 10.1007/BF00526670} {\bibfield  {journal} {\bibinfo
  {journal} {Theor. Exp. Chem.}\ }\textbf {\bibinfo {volume} {10}},\ \bibinfo
  {pages} {354--361} (\bibinfo {year} {1976})}\BibitemShut {NoStop}%
\bibitem [{\citenamefont {Martin}(2003)}]{Martin2003}%
  \BibitemOpen
  \bibfield  {author} {\bibinfo {author} {\bibfnamefont {R.~L.}\ \bibnamefont
  {Martin}},\ }\bibfield  {title} {\enquote {\bibinfo {title} {{Natural
  transition orbitals}},}\ }\href@noop {} {\bibfield  {journal} {\bibinfo
  {journal} {J. Chem. Phys.}\ }\textbf {\bibinfo {volume} {118}},\ \bibinfo
  {pages} {4775--4777} (\bibinfo {year} {2003})}\BibitemShut {NoStop}%
\bibitem [{\citenamefont {Plasser}, \citenamefont {Wormit},\ and\ \citenamefont
  {Dreuw}(2014)}]{Dreuw:ESSAImpl:14}%
  \BibitemOpen
  \bibfield  {author} {\bibinfo {author} {\bibfnamefont {F.}~\bibnamefont
  {Plasser}}, \bibinfo {author} {\bibfnamefont {M.}~\bibnamefont {Wormit}}, \
  and\ \bibinfo {author} {\bibfnamefont {A.}~\bibnamefont {Dreuw}},\ }\bibfield
   {title} {\enquote {\bibinfo {title} {New tools for the systematic analysis
  and visualization of electronic excitations. {I. Formalism}},}\ }\href@noop
  {} {\bibfield  {journal} {\bibinfo  {journal} {J. Chem. Phys.}\ }\textbf
  {\bibinfo {volume} {141}},\ \bibinfo {pages} {024106} (\bibinfo {year}
  {2014})}\BibitemShut {NoStop}%
\bibitem [{\citenamefont {Plasser}\ \emph
  {et~al.}(2014{\natexlab{b}})\citenamefont {Plasser}, \citenamefont
  {B{\"a}ppler}, \citenamefont {Wormit},\ and\ \citenamefont
  {Dreuw}}]{Dreuw:ESSAImpl-2:14}%
  \BibitemOpen
  \bibfield  {author} {\bibinfo {author} {\bibfnamefont {F.}~\bibnamefont
  {Plasser}}, \bibinfo {author} {\bibfnamefont {S.~A.}\ \bibnamefont
  {B{\"a}ppler}}, \bibinfo {author} {\bibfnamefont {M.}~\bibnamefont {Wormit}},
  \ and\ \bibinfo {author} {\bibfnamefont {A.}~\bibnamefont {Dreuw}},\
  }\bibfield  {title} {\enquote {\bibinfo {title} {New tools for the systematic
  analysis and visualization of electronic excitations. {II. Applications}},}\
  }\href@noop {} {\bibfield  {journal} {\bibinfo  {journal} {J. Chem. Phys.}\
  }\textbf {\bibinfo {volume} {141}},\ \bibinfo {pages} {024107} (\bibinfo
  {year} {2014}{\natexlab{b}})}\BibitemShut {NoStop}%
\bibitem [{\citenamefont {B{\"a}ppler}\ \emph {et~al.}(2014)\citenamefont
  {B{\"a}ppler}, \citenamefont {Plasser}, \citenamefont {Wormit},\ and\
  \citenamefont {Dreuw}}]{Dreuw:ESSA:14}%
  \BibitemOpen
  \bibfield  {author} {\bibinfo {author} {\bibfnamefont {S.~A.}\ \bibnamefont
  {B{\"a}ppler}}, \bibinfo {author} {\bibfnamefont {F.}~\bibnamefont
  {Plasser}}, \bibinfo {author} {\bibfnamefont {M.}~\bibnamefont {Wormit}}, \
  and\ \bibinfo {author} {\bibfnamefont {A.}~\bibnamefont {Dreuw}},\ }\bibfield
   {title} {\enquote {\bibinfo {title} {Exciton analysis of many-body wave
  functions: Bridging the gap between the quasiparticle and molecular orbital
  pictures},}\ }\href@noop {} {\bibfield  {journal} {\bibinfo  {journal} {Phys.
  Rev. A}\ }\textbf {\bibinfo {volume} {90}},\ \bibinfo {pages} {052521}
  (\bibinfo {year} {2014})}\BibitemShut {NoStop}%
\bibitem [{\citenamefont {Mewes}\ \emph {et~al.}(2018)\citenamefont {Mewes},
  \citenamefont {Plasser}, \citenamefont {Krylov},\ and\ \citenamefont
  {Dreuw}}]{Krylov:Libwfa:18}%
  \BibitemOpen
  \bibfield  {author} {\bibinfo {author} {\bibfnamefont {S.}~\bibnamefont
  {Mewes}}, \bibinfo {author} {\bibfnamefont {F.}~\bibnamefont {Plasser}},
  \bibinfo {author} {\bibfnamefont {A.~I.}\ \bibnamefont {Krylov}}, \ and\
  \bibinfo {author} {\bibfnamefont {A.}~\bibnamefont {Dreuw}},\ }\bibfield
  {title} {\enquote {\bibinfo {title} {Benchmarking excited-state calculations
  using exciton properties},}\ }\href@noop {} {\bibfield  {journal} {\bibinfo
  {journal} {J. Chem. Theory Comput.}\ }\textbf {\bibinfo {volume} {14}},\
  \bibinfo {pages} {710--725} (\bibinfo {year} {2018})}\BibitemShut {NoStop}%
\bibitem [{\citenamefont {Kimber}\ and\ \citenamefont
  {Plasser}(2020)}]{Plasser:NTOfeature:2020}%
  \BibitemOpen
  \bibfield  {author} {\bibinfo {author} {\bibfnamefont {P.}~\bibnamefont
  {Kimber}}\ and\ \bibinfo {author} {\bibfnamefont {F.}~\bibnamefont
  {Plasser}},\ }\bibfield  {title} {\enquote {\bibinfo {title} {Toward an
  understanding of electronic excitation energies beyond the molecular orbital
  picture},}\ }\href@noop {} {\bibfield  {journal} {\bibinfo  {journal} {Phys.
  Chem. Chem. Phys.}\ }\textbf {\bibinfo {volume} {22}},\ \bibinfo {pages}
  {6058--6080} (\bibinfo {year} {2020})}\BibitemShut {NoStop}%
\bibitem [{\citenamefont {Krylov}(2020)}]{Krylov:Orbitals}%
  \BibitemOpen
  \bibfield  {author} {\bibinfo {author} {\bibfnamefont {A.~I.}\ \bibnamefont
  {Krylov}},\ }\bibfield  {title} {\enquote {\bibinfo {title} {From orbitals to
  observables and back},}\ }\href@noop {} {\bibfield  {journal} {\bibinfo
  {journal} {J. Chem. Phys.}\ }\textbf {\bibinfo {volume} {153}},\ \bibinfo
  {pages} {080901} (\bibinfo {year} {2020})}\BibitemShut {NoStop}%
\bibitem [{\citenamefont {Shao}\ \emph {et~al.}(2015)\citenamefont {Shao},
  \citenamefont {Gan}, \citenamefont {Epifanovsky}, \citenamefont {Gilbert},
  \citenamefont {Wormit}, \citenamefont {Kussmann}, \citenamefont {Lange},
  \citenamefont {Behn}, \citenamefont {Deng}, \citenamefont {Feng},
  \citenamefont {Ghosh}, \citenamefont {Goldey}, \citenamefont {Horn},
  \citenamefont {Jacobson}, \citenamefont {Kaliman}, \citenamefont
  {Khaliullin}, \citenamefont {Ku{\'s}}, \citenamefont {Landau}, \citenamefont
  {Liu}, \citenamefont {Proynov}, \citenamefont {Rhee}, \citenamefont
  {Richard}, \citenamefont {Rohrdanz}, \citenamefont {Steele}, \citenamefont
  {Sundstrom}, \citenamefont {Woodcock~III}, \citenamefont {Zimmerman},
  \citenamefont {Zuev}, \citenamefont {Albrecht}, \citenamefont {Alguire},
  \citenamefont {Austin}, \citenamefont {Beran}, \citenamefont {Bernard},
  \citenamefont {Berquist}, \citenamefont {Brandhorst}, \citenamefont
  {Bravaya}, \citenamefont {Brown}, \citenamefont {Casanova}, \citenamefont
  {Chang}, \citenamefont {Chen}, \citenamefont {Chien}, \citenamefont
  {Closser}, \citenamefont {Crittenden}, \citenamefont {Diedenhofen},
  \citenamefont {DiStasio~Jr.}, \citenamefont {Do}, \citenamefont {Dutoi},
  \citenamefont {Edgar}, \citenamefont {Fatehi}, \citenamefont {Fusti-Molnar},
  \citenamefont {Ghysels}, \citenamefont {Golubeva-Zadorozhnaya}, \citenamefont
  {Gomes}, \citenamefont {Hanson-Heine}, \citenamefont {Harbach}, \citenamefont
  {Hauser}, \citenamefont {Hohenstein}, \citenamefont {Holden}, \citenamefont
  {Jagau}, \citenamefont {Ji}, \citenamefont {Kaduk}, \citenamefont
  {Khistyaev}, \citenamefont {Kim}, \citenamefont {Kim}, \citenamefont {King},
  \citenamefont {Klunzinger}, \citenamefont {Kosenkov}, \citenamefont
  {Kowalczyk}, \citenamefont {Krauter}, \citenamefont {Lao}, \citenamefont
  {Laurent}, \citenamefont {Lawler}, \citenamefont {Levchenko}, \citenamefont
  {Lin}, \citenamefont {Liu}, \citenamefont {Livshits}, \citenamefont {Lochan},
  \citenamefont {Luenser}, \citenamefont {Manohar}, \citenamefont {Manzer},
  \citenamefont {Mao}, \citenamefont {Mardirossian}, \citenamefont {Marenich},
  \citenamefont {Maurer}, \citenamefont {Mayhall}, \citenamefont {Neuscamman},
  \citenamefont {Oana}, \citenamefont {Olivares-Amaya}, \citenamefont
  {O'Neill}, \citenamefont {Parkhill}, \citenamefont {Perrine}, \citenamefont
  {Peverati}, \citenamefont {Prociuk}, \citenamefont {Rehn}, \citenamefont
  {Rosta}, \citenamefont {Russ}, \citenamefont {Sharada}, \citenamefont
  {Sharma}, \citenamefont {Small}, \citenamefont {Sodt}, \citenamefont {Stein},
  \citenamefont {St\"uck}, \citenamefont {Su}, \citenamefont {Thom},
  \citenamefont {Tsuchimochi}, \citenamefont {Vanovschi}, \citenamefont {Vogt},
  \citenamefont {Vydrov}, \citenamefont {Wang}, \citenamefont {Watson},
  \citenamefont {Wenzel}, \citenamefont {White}, \citenamefont {Williams},
  \citenamefont {Yang}, \citenamefont {Yeganeh}, \citenamefont {Yost},
  \citenamefont {You}, \citenamefont {Zhang}, \citenamefont {Zhang},
  \citenamefont {Zhao}, \citenamefont {Brooks}, \citenamefont {Chan},
  \citenamefont {Chipman}, \citenamefont {Cramer}, \citenamefont {Goddard~III},
  \citenamefont {Gordon}, \citenamefont {Hehre}, \citenamefont {Klamt},
  \citenamefont {Schaefer~III}, \citenamefont {Schmidt}, \citenamefont
  {Sherrill}, \citenamefont {Truhlar}, \citenamefont {Warshel}, \citenamefont
  {Xu}, \citenamefont {Aspuru-Guzik}, \citenamefont {Baer}, \citenamefont
  {Bell}, \citenamefont {Besley}, \citenamefont {Chai}, \citenamefont {Dreuw},
  \citenamefont {Dunietz}, \citenamefont {Furlani}, \citenamefont {Gwaltney},
  \citenamefont {Hsu}, \citenamefont {Jung}, \citenamefont {Kong},
  \citenamefont {Lambrecht}, \citenamefont {Liang}, \citenamefont {Ochsenfeld},
  \citenamefont {Rassolov}, \citenamefont {Slipchenko}, \citenamefont
  {Subotnik}, \citenamefont {Van~Voorhis}, \citenamefont {Herbert},
  \citenamefont {Krylov}, \citenamefont {Gill},\ and\ \citenamefont
  {Head-Gordon}}]{Qchem_MP_paper}%
  \BibitemOpen
  \bibfield  {author} {\bibinfo {author} {\bibfnamefont {Y.}~\bibnamefont
  {Shao}}, \bibinfo {author} {\bibfnamefont {Z.}~\bibnamefont {Gan}}, \bibinfo
  {author} {\bibfnamefont {E.}~\bibnamefont {Epifanovsky}}, \bibinfo {author}
  {\bibfnamefont {A.~T.~B.}\ \bibnamefont {Gilbert}}, \bibinfo {author}
  {\bibfnamefont {M.}~\bibnamefont {Wormit}}, \bibinfo {author} {\bibfnamefont
  {J.}~\bibnamefont {Kussmann}}, \bibinfo {author} {\bibfnamefont {A.~W.}\
  \bibnamefont {Lange}}, \bibinfo {author} {\bibfnamefont {A.}~\bibnamefont
  {Behn}}, \bibinfo {author} {\bibfnamefont {J.}~\bibnamefont {Deng}}, \bibinfo
  {author} {\bibfnamefont {X.}~\bibnamefont {Feng}}, \bibinfo {author}
  {\bibfnamefont {D.}~\bibnamefont {Ghosh}}, \bibinfo {author} {\bibfnamefont
  {M.}~\bibnamefont {Goldey}}, \bibinfo {author} {\bibfnamefont {P.~R.}\
  \bibnamefont {Horn}}, \bibinfo {author} {\bibfnamefont {L.~D.}\ \bibnamefont
  {Jacobson}}, \bibinfo {author} {\bibfnamefont {I.}~\bibnamefont {Kaliman}},
  \bibinfo {author} {\bibfnamefont {R.~Z.}\ \bibnamefont {Khaliullin}},
  \bibinfo {author} {\bibfnamefont {T.}~\bibnamefont {Ku{\'s}}}, \bibinfo
  {author} {\bibfnamefont {A.}~\bibnamefont {Landau}}, \bibinfo {author}
  {\bibfnamefont {J.}~\bibnamefont {Liu}}, \bibinfo {author} {\bibfnamefont
  {E.~I.}\ \bibnamefont {Proynov}}, \bibinfo {author} {\bibfnamefont {Y.~M.}\
  \bibnamefont {Rhee}}, \bibinfo {author} {\bibfnamefont {R.~M.}\ \bibnamefont
  {Richard}}, \bibinfo {author} {\bibfnamefont {M.~A.}\ \bibnamefont
  {Rohrdanz}}, \bibinfo {author} {\bibfnamefont {R.~P.}\ \bibnamefont
  {Steele}}, \bibinfo {author} {\bibfnamefont {E.~J.}\ \bibnamefont
  {Sundstrom}}, \bibinfo {author} {\bibfnamefont {H.~L.}\ \bibnamefont
  {Woodcock~III}}, \bibinfo {author} {\bibfnamefont {P.~M.}\ \bibnamefont
  {Zimmerman}}, \bibinfo {author} {\bibfnamefont {D.}~\bibnamefont {Zuev}},
  \bibinfo {author} {\bibfnamefont {B.}~\bibnamefont {Albrecht}}, \bibinfo
  {author} {\bibfnamefont {E.}~\bibnamefont {Alguire}}, \bibinfo {author}
  {\bibfnamefont {B.}~\bibnamefont {Austin}}, \bibinfo {author} {\bibfnamefont
  {G.~J.~O.}\ \bibnamefont {Beran}}, \bibinfo {author} {\bibfnamefont {Y.~A.}\
  \bibnamefont {Bernard}}, \bibinfo {author} {\bibfnamefont {E.}~\bibnamefont
  {Berquist}}, \bibinfo {author} {\bibfnamefont {K.}~\bibnamefont
  {Brandhorst}}, \bibinfo {author} {\bibfnamefont {K.~B.}\ \bibnamefont
  {Bravaya}}, \bibinfo {author} {\bibfnamefont {S.~T.}\ \bibnamefont {Brown}},
  \bibinfo {author} {\bibfnamefont {D.}~\bibnamefont {Casanova}}, \bibinfo
  {author} {\bibfnamefont {C.-M.}\ \bibnamefont {Chang}}, \bibinfo {author}
  {\bibfnamefont {Y.}~\bibnamefont {Chen}}, \bibinfo {author} {\bibfnamefont
  {S.~H.}\ \bibnamefont {Chien}}, \bibinfo {author} {\bibfnamefont {K.~D.}\
  \bibnamefont {Closser}}, \bibinfo {author} {\bibfnamefont {D.~L.}\
  \bibnamefont {Crittenden}}, \bibinfo {author} {\bibfnamefont
  {M.}~\bibnamefont {Diedenhofen}}, \bibinfo {author} {\bibfnamefont {R.~A.}\
  \bibnamefont {DiStasio~Jr.}}, \bibinfo {author} {\bibfnamefont
  {H.}~\bibnamefont {Do}}, \bibinfo {author} {\bibfnamefont {A.~D.}\
  \bibnamefont {Dutoi}}, \bibinfo {author} {\bibfnamefont {R.~G.}\ \bibnamefont
  {Edgar}}, \bibinfo {author} {\bibfnamefont {S.}~\bibnamefont {Fatehi}},
  \bibinfo {author} {\bibfnamefont {L.}~\bibnamefont {Fusti-Molnar}}, \bibinfo
  {author} {\bibfnamefont {A.}~\bibnamefont {Ghysels}}, \bibinfo {author}
  {\bibfnamefont {A.}~\bibnamefont {Golubeva-Zadorozhnaya}}, \bibinfo {author}
  {\bibfnamefont {J.}~\bibnamefont {Gomes}}, \bibinfo {author} {\bibfnamefont
  {M.~W.~D.}\ \bibnamefont {Hanson-Heine}}, \bibinfo {author} {\bibfnamefont
  {P.~H.~P.}\ \bibnamefont {Harbach}}, \bibinfo {author} {\bibfnamefont
  {A.~W.}\ \bibnamefont {Hauser}}, \bibinfo {author} {\bibfnamefont {E.~G.}\
  \bibnamefont {Hohenstein}}, \bibinfo {author} {\bibfnamefont {Z.~C.}\
  \bibnamefont {Holden}}, \bibinfo {author} {\bibfnamefont {T.-C.}\
  \bibnamefont {Jagau}}, \bibinfo {author} {\bibfnamefont {H.}~\bibnamefont
  {Ji}}, \bibinfo {author} {\bibfnamefont {B.}~\bibnamefont {Kaduk}}, \bibinfo
  {author} {\bibfnamefont {K.}~\bibnamefont {Khistyaev}}, \bibinfo {author}
  {\bibfnamefont {J.}~\bibnamefont {Kim}}, \bibinfo {author} {\bibfnamefont
  {J.}~\bibnamefont {Kim}}, \bibinfo {author} {\bibfnamefont {R.~A.}\
  \bibnamefont {King}}, \bibinfo {author} {\bibfnamefont {P.}~\bibnamefont
  {Klunzinger}}, \bibinfo {author} {\bibfnamefont {D.}~\bibnamefont
  {Kosenkov}}, \bibinfo {author} {\bibfnamefont {T.}~\bibnamefont {Kowalczyk}},
  \bibinfo {author} {\bibfnamefont {C.~M.}\ \bibnamefont {Krauter}}, \bibinfo
  {author} {\bibfnamefont {K.~U.}\ \bibnamefont {Lao}}, \bibinfo {author}
  {\bibfnamefont {A.~D.}\ \bibnamefont {Laurent}}, \bibinfo {author}
  {\bibfnamefont {K.~V.}\ \bibnamefont {Lawler}}, \bibinfo {author}
  {\bibfnamefont {S.~V.}\ \bibnamefont {Levchenko}}, \bibinfo {author}
  {\bibfnamefont {C.~Y.}\ \bibnamefont {Lin}}, \bibinfo {author} {\bibfnamefont
  {F.}~\bibnamefont {Liu}}, \bibinfo {author} {\bibfnamefont {E.}~\bibnamefont
  {Livshits}}, \bibinfo {author} {\bibfnamefont {R.~C.}\ \bibnamefont
  {Lochan}}, \bibinfo {author} {\bibfnamefont {A.}~\bibnamefont {Luenser}},
  \bibinfo {author} {\bibfnamefont {P.}~\bibnamefont {Manohar}}, \bibinfo
  {author} {\bibfnamefont {S.~F.}\ \bibnamefont {Manzer}}, \bibinfo {author}
  {\bibfnamefont {S.-P.}\ \bibnamefont {Mao}}, \bibinfo {author} {\bibfnamefont
  {N.}~\bibnamefont {Mardirossian}}, \bibinfo {author} {\bibfnamefont {A.~V.}\
  \bibnamefont {Marenich}}, \bibinfo {author} {\bibfnamefont {S.~A.}\
  \bibnamefont {Maurer}}, \bibinfo {author} {\bibfnamefont {N.~J.}\
  \bibnamefont {Mayhall}}, \bibinfo {author} {\bibfnamefont {E.}~\bibnamefont
  {Neuscamman}}, \bibinfo {author} {\bibfnamefont {C.~M.}\ \bibnamefont
  {Oana}}, \bibinfo {author} {\bibfnamefont {R.}~\bibnamefont
  {Olivares-Amaya}}, \bibinfo {author} {\bibfnamefont {D.~P.}\ \bibnamefont
  {O'Neill}}, \bibinfo {author} {\bibfnamefont {J.~A.}\ \bibnamefont
  {Parkhill}}, \bibinfo {author} {\bibfnamefont {T.~M.}\ \bibnamefont
  {Perrine}}, \bibinfo {author} {\bibfnamefont {R.}~\bibnamefont {Peverati}},
  \bibinfo {author} {\bibfnamefont {A.}~\bibnamefont {Prociuk}}, \bibinfo
  {author} {\bibfnamefont {D.~R.}\ \bibnamefont {Rehn}}, \bibinfo {author}
  {\bibfnamefont {E.}~\bibnamefont {Rosta}}, \bibinfo {author} {\bibfnamefont
  {N.~J.}\ \bibnamefont {Russ}}, \bibinfo {author} {\bibfnamefont {S.~M.}\
  \bibnamefont {Sharada}}, \bibinfo {author} {\bibfnamefont {S.}~\bibnamefont
  {Sharma}}, \bibinfo {author} {\bibfnamefont {D.~W.}\ \bibnamefont {Small}},
  \bibinfo {author} {\bibfnamefont {A.}~\bibnamefont {Sodt}}, \bibinfo {author}
  {\bibfnamefont {T.}~\bibnamefont {Stein}}, \bibinfo {author} {\bibfnamefont
  {D.}~\bibnamefont {St\"uck}}, \bibinfo {author} {\bibfnamefont {Y.-C.}\
  \bibnamefont {Su}}, \bibinfo {author} {\bibfnamefont {A.~J.~W.}\ \bibnamefont
  {Thom}}, \bibinfo {author} {\bibfnamefont {T.}~\bibnamefont {Tsuchimochi}},
  \bibinfo {author} {\bibfnamefont {V.}~\bibnamefont {Vanovschi}}, \bibinfo
  {author} {\bibfnamefont {L.}~\bibnamefont {Vogt}}, \bibinfo {author}
  {\bibfnamefont {O.}~\bibnamefont {Vydrov}}, \bibinfo {author} {\bibfnamefont
  {T.}~\bibnamefont {Wang}}, \bibinfo {author} {\bibfnamefont {M.~A.}\
  \bibnamefont {Watson}}, \bibinfo {author} {\bibfnamefont {J.}~\bibnamefont
  {Wenzel}}, \bibinfo {author} {\bibfnamefont {A.}~\bibnamefont {White}},
  \bibinfo {author} {\bibfnamefont {C.~F.}\ \bibnamefont {Williams}}, \bibinfo
  {author} {\bibfnamefont {J.}~\bibnamefont {Yang}}, \bibinfo {author}
  {\bibfnamefont {S.}~\bibnamefont {Yeganeh}}, \bibinfo {author} {\bibfnamefont
  {S.~R.}\ \bibnamefont {Yost}}, \bibinfo {author} {\bibfnamefont {Z.-Q.}\
  \bibnamefont {You}}, \bibinfo {author} {\bibfnamefont {I.~Y.}\ \bibnamefont
  {Zhang}}, \bibinfo {author} {\bibfnamefont {X.}~\bibnamefont {Zhang}},
  \bibinfo {author} {\bibfnamefont {Y.}~\bibnamefont {Zhao}}, \bibinfo {author}
  {\bibfnamefont {B.~R.}\ \bibnamefont {Brooks}}, \bibinfo {author}
  {\bibfnamefont {G.~K.~L.}\ \bibnamefont {Chan}}, \bibinfo {author}
  {\bibfnamefont {D.~M.}\ \bibnamefont {Chipman}}, \bibinfo {author}
  {\bibfnamefont {C.~J.}\ \bibnamefont {Cramer}}, \bibinfo {author}
  {\bibfnamefont {W.~A.}\ \bibnamefont {Goddard~III}}, \bibinfo {author}
  {\bibfnamefont {M.~S.}\ \bibnamefont {Gordon}}, \bibinfo {author}
  {\bibfnamefont {W.~J.}\ \bibnamefont {Hehre}}, \bibinfo {author}
  {\bibfnamefont {A.}~\bibnamefont {Klamt}}, \bibinfo {author} {\bibfnamefont
  {H.~F.}\ \bibnamefont {Schaefer~III}}, \bibinfo {author} {\bibfnamefont
  {M.~W.}\ \bibnamefont {Schmidt}}, \bibinfo {author} {\bibfnamefont {C.~D.}\
  \bibnamefont {Sherrill}}, \bibinfo {author} {\bibfnamefont {D.~G.}\
  \bibnamefont {Truhlar}}, \bibinfo {author} {\bibfnamefont {A.}~\bibnamefont
  {Warshel}}, \bibinfo {author} {\bibfnamefont {X.}~\bibnamefont {Xu}},
  \bibinfo {author} {\bibfnamefont {A.}~\bibnamefont {Aspuru-Guzik}}, \bibinfo
  {author} {\bibfnamefont {R.}~\bibnamefont {Baer}}, \bibinfo {author}
  {\bibfnamefont {A.~T.}\ \bibnamefont {Bell}}, \bibinfo {author}
  {\bibfnamefont {N.~A.}\ \bibnamefont {Besley}}, \bibinfo {author}
  {\bibfnamefont {J.-D.}\ \bibnamefont {Chai}}, \bibinfo {author}
  {\bibfnamefont {A.}~\bibnamefont {Dreuw}}, \bibinfo {author} {\bibfnamefont
  {B.~D.}\ \bibnamefont {Dunietz}}, \bibinfo {author} {\bibfnamefont {T.~R.}\
  \bibnamefont {Furlani}}, \bibinfo {author} {\bibfnamefont {S.~R.}\
  \bibnamefont {Gwaltney}}, \bibinfo {author} {\bibfnamefont {C.-P.}\
  \bibnamefont {Hsu}}, \bibinfo {author} {\bibfnamefont {Y.}~\bibnamefont
  {Jung}}, \bibinfo {author} {\bibfnamefont {J.}~\bibnamefont {Kong}}, \bibinfo
  {author} {\bibfnamefont {D.~S.}\ \bibnamefont {Lambrecht}}, \bibinfo {author}
  {\bibfnamefont {W.}~\bibnamefont {Liang}}, \bibinfo {author} {\bibfnamefont
  {C.}~\bibnamefont {Ochsenfeld}}, \bibinfo {author} {\bibfnamefont {V.~A.}\
  \bibnamefont {Rassolov}}, \bibinfo {author} {\bibfnamefont {L.~V.}\
  \bibnamefont {Slipchenko}}, \bibinfo {author} {\bibfnamefont {J.~E.}\
  \bibnamefont {Subotnik}}, \bibinfo {author} {\bibfnamefont {T.}~\bibnamefont
  {Van~Voorhis}}, \bibinfo {author} {\bibfnamefont {J.~M.}\ \bibnamefont
  {Herbert}}, \bibinfo {author} {\bibfnamefont {A.~I.}\ \bibnamefont {Krylov}},
  \bibinfo {author} {\bibfnamefont {P.~M.~W.}\ \bibnamefont {Gill}}, \ and\
  \bibinfo {author} {\bibfnamefont {M.}~\bibnamefont {Head-Gordon}},\
  }\bibfield  {title} {\enquote {\bibinfo {title} {Advances in molecular
  quantum chemistry contained in the q-chem 4 program package},}\ }\href
  {\doibase 10.1080/00268976.2014.952696} {\bibfield  {journal} {\bibinfo
  {journal} {Mol. Phys.}\ }\textbf {\bibinfo {volume} {113}},\ \bibinfo {pages}
  {184--215} (\bibinfo {year} {2015})}\BibitemShut {NoStop}%
\bibitem [{\citenamefont {Barca}, \citenamefont {Gilbert},\ and\ \citenamefont
  {Gill}(2018)}]{IMOM}%
  \BibitemOpen
  \bibfield  {author} {\bibinfo {author} {\bibfnamefont {G.~M.~J.}\
  \bibnamefont {Barca}}, \bibinfo {author} {\bibfnamefont {A.~T.~B.}\
  \bibnamefont {Gilbert}}, \ and\ \bibinfo {author} {\bibfnamefont {P.~M.~W.}\
  \bibnamefont {Gill}},\ }\bibfield  {title} {\enquote {\bibinfo {title}
  {Simple models for difficult electronic excitations},}\ }\href {\doibase
  10.1021/acs.jctc.7b00994} {\bibfield  {journal} {\bibinfo  {journal} {J.
  Chem. Theory Comput.}\ }\textbf {\bibinfo {volume} {14}},\ \bibinfo {pages}
  {1501--1509} (\bibinfo {year} {2018})}\BibitemShut {NoStop}%
\bibitem [{\citenamefont {Rehr}\ \emph {et~al.}(1978)\citenamefont {Rehr},
  \citenamefont {Stern}, \citenamefont {Martin},\ and\ \citenamefont
  {Davidson}}]{Rehr1978}%
  \BibitemOpen
  \bibfield  {author} {\bibinfo {author} {\bibfnamefont {J.~J.}\ \bibnamefont
  {Rehr}}, \bibinfo {author} {\bibfnamefont {E.~A.}\ \bibnamefont {Stern}},
  \bibinfo {author} {\bibfnamefont {R.~L.}\ \bibnamefont {Martin}}, \ and\
  \bibinfo {author} {\bibfnamefont {E.~R.}\ \bibnamefont {Davidson}},\
  }\bibfield  {title} {\enquote {\bibinfo {title} {Extended x-ray-absorption
  fine-structure amplitudes---{W}ave-function relaxation and chemical
  effects},}\ }\href {\doibase 10.1103/PhysRevB.17.560} {\bibfield  {journal}
  {\bibinfo  {journal} {Phys. Rev. B}\ }\textbf {\bibinfo {volume} {17}},\
  \bibinfo {pages} {560--565} (\bibinfo {year} {1978})}\BibitemShut {NoStop}%
\bibitem [{\citenamefont {Chernyshova}\ \emph {et~al.}(2012)\citenamefont
  {Chernyshova}, \citenamefont {Kontros}, \citenamefont {Markush},\ and\
  \citenamefont {Shpenik}}]{Chernyshova2012}%
  \BibitemOpen
  \bibfield  {author} {\bibinfo {author} {\bibfnamefont {I.~V.}\ \bibnamefont
  {Chernyshova}}, \bibinfo {author} {\bibfnamefont {J.~E.}\ \bibnamefont
  {Kontros}}, \bibinfo {author} {\bibfnamefont {P.~P.}\ \bibnamefont
  {Markush}}, \ and\ \bibinfo {author} {\bibfnamefont {O.~B.}\ \bibnamefont
  {Shpenik}},\ }\bibfield  {title} {\enquote {\bibinfo {title} {Excitation of
  lowest electronic states of the uracil molecule by slow electrons},}\ }\href
  {\doibase 10.1134/S0030400X12070077} {\bibfield  {journal} {\bibinfo
  {journal} {Opt. Spectrosc.}\ }\textbf {\bibinfo {volume} {113}},\ \bibinfo
  {pages} {5--8} (\bibinfo {year} {2012})}\BibitemShut {NoStop}%
\bibitem [{\citenamefont {Clark}, \citenamefont {Peschel},\ and\ \citenamefont
  {Tinoco}(1965)}]{UV_uracil}%
  \BibitemOpen
  \bibfield  {author} {\bibinfo {author} {\bibfnamefont {L.~B.}\ \bibnamefont
  {Clark}}, \bibinfo {author} {\bibfnamefont {G.~G.}\ \bibnamefont {Peschel}},
  \ and\ \bibinfo {author} {\bibfnamefont {I.}~\bibnamefont {Tinoco}},\
  }\bibfield  {title} {\enquote {\bibinfo {title} {Vapor spectra and heats of
  vaporization of some purine and pyrimidine bases},}\ }\href@noop {}
  {\bibfield  {journal} {\bibinfo  {journal} {J. Phys. Chem.}\ }\textbf
  {\bibinfo {volume} {69}},\ \bibinfo {pages} {3615--3618} (\bibinfo {year}
  {1965})}\BibitemShut {NoStop}%
\bibitem [{\citenamefont {Fedotov}\ \emph {et~al.}(2020)\citenamefont
  {Fedotov}, \citenamefont {Paul}, \citenamefont {Posocco}, \citenamefont
  {Santoro}, \citenamefont {Garavelli}, \citenamefont {Koch}, \citenamefont
  {Coriani},\ and\ \citenamefont {Improta}}]{Fedotov:uracil}%
  \BibitemOpen
  \bibfield  {author} {\bibinfo {author} {\bibfnamefont {D.~A.}\ \bibnamefont
  {Fedotov}}, \bibinfo {author} {\bibfnamefont {A.~C.}\ \bibnamefont {Paul}},
  \bibinfo {author} {\bibfnamefont {P.}~\bibnamefont {Posocco}}, \bibinfo
  {author} {\bibfnamefont {F.}~\bibnamefont {Santoro}}, \bibinfo {author}
  {\bibfnamefont {M.}~\bibnamefont {Garavelli}}, \bibinfo {author}
  {\bibfnamefont {H.}~\bibnamefont {Koch}}, \bibinfo {author} {\bibfnamefont
  {S.}~\bibnamefont {Coriani}}, \ and\ \bibinfo {author} {\bibfnamefont
  {R.}~\bibnamefont {Improta}},\ }\bibfield  {title} {\enquote {\bibinfo
  {title} {{Excited State Absorption of Uracil in the Gas Phase: Mapping the
  Main Decay Paths by Different Electronic Structure Methods}},}\ }\href
  {\doibase 10.26434/chemrxiv.13176554.v1} {\  (\bibinfo {year} {2020}),\
  10.26434/chemrxiv.13176554.v1},\ \bibinfo {note} {accepted in {\em J. Chem.
  Theory Comput.}}\BibitemShut {Stop}%
\bibitem [{\citenamefont {Chernyshova}\ \emph {et~al.}(2013)\citenamefont
  {Chernyshova}, \citenamefont {Kontros}, \citenamefont {Markush},\ and\
  \citenamefont {Shpenik}}]{Chernyshova2013}%
  \BibitemOpen
  \bibfield  {author} {\bibinfo {author} {\bibfnamefont {I.~V.}\ \bibnamefont
  {Chernyshova}}, \bibinfo {author} {\bibfnamefont {J.~E.}\ \bibnamefont
  {Kontros}}, \bibinfo {author} {\bibfnamefont {P.~P.}\ \bibnamefont
  {Markush}}, \ and\ \bibinfo {author} {\bibfnamefont {O.~B.}\ \bibnamefont
  {Shpenik}},\ }\bibfield  {title} {\enquote {\bibinfo {title} {Excitations of
  lowest electronic states of thymine by slow electrons},}\ }\href {\doibase
  10.1134/S0030400X13110040} {\bibfield  {journal} {\bibinfo  {journal} {Opt.
  Spectrosc.}\ }\textbf {\bibinfo {volume} {115}},\ \bibinfo {pages} {645--650}
  (\bibinfo {year} {2013})}\BibitemShut {NoStop}%
\bibitem [{\citenamefont {Walzl}, \citenamefont {Xavier},\ and\ \citenamefont
  {Kuppermann}(1987)}]{Walzl}%
  \BibitemOpen
  \bibfield  {author} {\bibinfo {author} {\bibfnamefont {K.~N.}\ \bibnamefont
  {Walzl}}, \bibinfo {author} {\bibfnamefont {I.~M.}\ \bibnamefont {Xavier}}, \
  and\ \bibinfo {author} {\bibfnamefont {A.}~\bibnamefont {Kuppermann}},\
  }\bibfield  {title} {\enquote {\bibinfo {title} {Electron-impact spectroscopy
  of various diketone compounds},}\ }\href {\doibase 10.1063/1.452418}
  {\bibfield  {journal} {\bibinfo  {journal} {J. Chem. Phys.}\ }\textbf
  {\bibinfo {volume} {86}},\ \bibinfo {pages} {6701--6706} (\bibinfo {year}
  {1987})}\BibitemShut {NoStop}%
\bibitem [{\citenamefont {Nakanishi}, \citenamefont {Morita},\ and\
  \citenamefont {Nagakura}(1977)}]{UV_AcAc}%
  \BibitemOpen
  \bibfield  {author} {\bibinfo {author} {\bibfnamefont {H.}~\bibnamefont
  {Nakanishi}}, \bibinfo {author} {\bibfnamefont {H.}~\bibnamefont {Morita}}, \
  and\ \bibinfo {author} {\bibfnamefont {S.}~\bibnamefont {Nagakura}},\
  }\bibfield  {title} {\enquote {\bibinfo {title} {Electronic structures and
  spectra of the keto and enol forms of acetylacetone},}\ }\href {\doibase
  10.1246/bcsj.50.2255} {\bibfield  {journal} {\bibinfo  {journal} {Bull. Chem.
  Soc. Jpn.}\ }\textbf {\bibinfo {volume} {50}},\ \bibinfo {pages} {2255--2261}
  (\bibinfo {year} {1977})}\BibitemShut {NoStop}%
\end{thebibliography}%

\end{document}